 \documentclass[review,authoryear,3p,times,10pt]{elsarticle}

\usepackage{multirow,setspace,times,amssymb,amsmath,graphicx,color,rotating,subfigure,url}
\usepackage{lineno,color}
\usepackage{natbib}
\usepackage{booktabs}%
\usepackage{longtable}%
\usepackage{rotating}
\usepackage{lscape}
\usepackage{pdflscape}
\usepackage{ulem}
\usepackage{threeparttable}
\usepackage{bm}

\graphicspath{{Figures/}{Figures_Appendix/}}
\usepackage[table]{xcolor}
\usepackage{tabularx}
\usepackage{graphicx} %use graph format
\usepackage{epstopdf}
\usepackage{mathrsfs}
\usepackage{makecell}
\usepackage[bookmarks=true,colorlinks,linkcolor=blue,anchorcolor=blue,citecolor=blue,unicode]{hyperref}
\usepackage{bookmark}

\usepackage{dcolumn} % 添加到导言区
\usepackage{todonotes}
\usepackage{ragged2e}
\usepackage[font=small]{caption}
\hypersetup{CJKbookmarks=true}%
\begin{document}
% \begin{CJK*}{GBK}{Song} % Use default fonts from CJK (see below)
% \linenumbers

\begin{frontmatter}
%\newpage
\title{Import dependence and per capita production are main determinants of economies' food supply robustness under production shocks}

\author[SB,RCE]{Han-Yu Zhu}
\author[PM]{Maria Cristina Rulli\corref{cor}}
\ead{mariacristina.rulli@polimi.it}
\author[SB,RCE,Math]{Wei-Xing Zhou\corref{cor}}
\ead{wxzhou@ecust.edu.cn} 
\cortext[cor]{Corresponding authors}
%  Address: 130 Meilong Road, P.O. Box 114, School of Business, East China University of Science and Technology, Shanghai 200237, China.}%Phone: +86-15216879656.}
\address[SB]{School of Business, East China University of Science and Technology, Shanghai 200237, China}
\address[RCE]{Research Center for Econophysics, East China University of Science and Technology, Shanghai 200237, China}
\address[PM]{Department of Civil and Environmental Engineering, Politecnico di Milano, Milan, Italy}
\address[Math]{School of Mathematics, East China University of Science and Technology, Shanghai 200237, China}

\begin{abstract}
Food supply shocks in major producing economies can propagate through trade networks and generate uneven impacts across the global food system. This study examines the robustness of economies’ food supply under production shocks to major producers in the global staple food system. Using 2023 production, reserve, and bilateral trade data for wheat, rice, maize, and soybean, we construct a calorie-based global food supply network across economies. We extend a dynamic shock propagation framework and then simulate production shocks to major producing economies, tracing how supply losses propagate.
The results show substantial heterogeneity in robustness across crops and economies. Wheat exhibits the highest overall robustness, whereas soybean shows the lowest. Economies with high robustness tend to be either relatively isolated from the trade network or actively engaged in trade while maintaining strong and stable domestic production, whereas low-robustness economies are predominantly those with high import dependence.
Import dependence and per capita production emerge as the most important determinants of robustness. 
Based on these findings, we design two counterfactual policies targeting highly import-dependent economies: increasing reserve availability and adjusting trade linkages. 
Counterfactual experiments show that the two policies yield only modest overall improvements, with effects varying substantially across crops. Both policies improve robustness in the aggregated system and wheat, trade adjustment is more effective for rice, and it brings limited or even negative effects for maize and soybean.
\end{abstract}


\end{frontmatter}

% \newpage
% \section*{Acknowledgment}

% This work was partly supported by the National Natural Science Foundation of China (72171083) and the Fundamental Research Funds for the Central Universities.

% \end{document}

\section{Introduction}

Food security is of fundamental importance to every economy. Yet the global situation remains far from resolved. According to \textit{The State of Food Security and Nutrition in the World} (SOFI), despite signs of improvement at the global level, around 2.3 billion people still experienced moderate or severe food insecurity in 2024, and the burden remains unevenly distributed across regions.
Sudden shocks from extreme weather, geopolitical conflicts, and other disruptions increasingly challenge the resilience of food supply systems \citep{Huang-Forero-WagnerMedina-Diaz-Tremma-Fargetton-LowenbergDeBoer-2025-GlobFoodSecur-AgricPolicy}.
In this context, we investigate staple food supply across economies for four key crops: wheat, rice, soybeans, and maize, which are among the most important sources of calories for human consumption.

\begin{figure}[h!]
    % \centering
    % \includegraphics[width=0.30\linewidth]{Figures/Aggregated_Production_Proportion_2023.pdf}
    \includegraphics[width=0.24\linewidth]{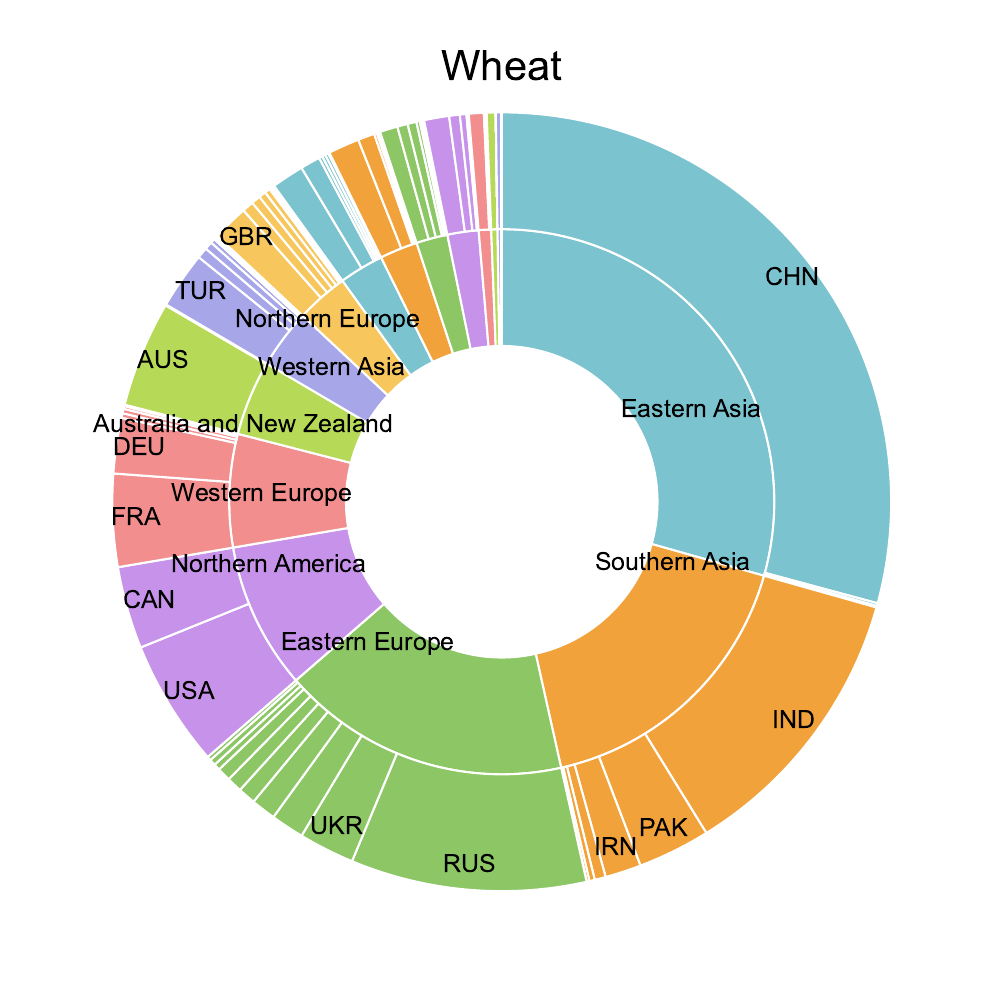}
    \includegraphics[width=0.24\linewidth]{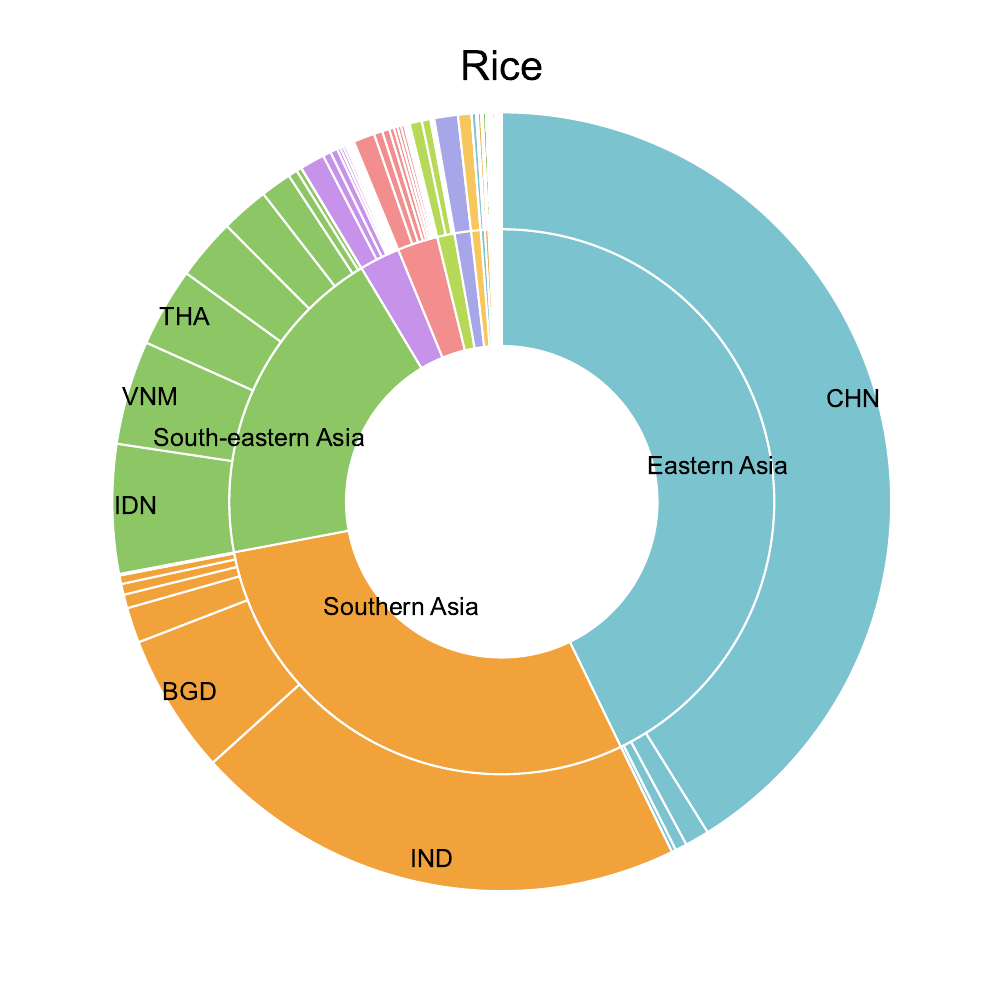}
    \includegraphics[width=0.24\linewidth]{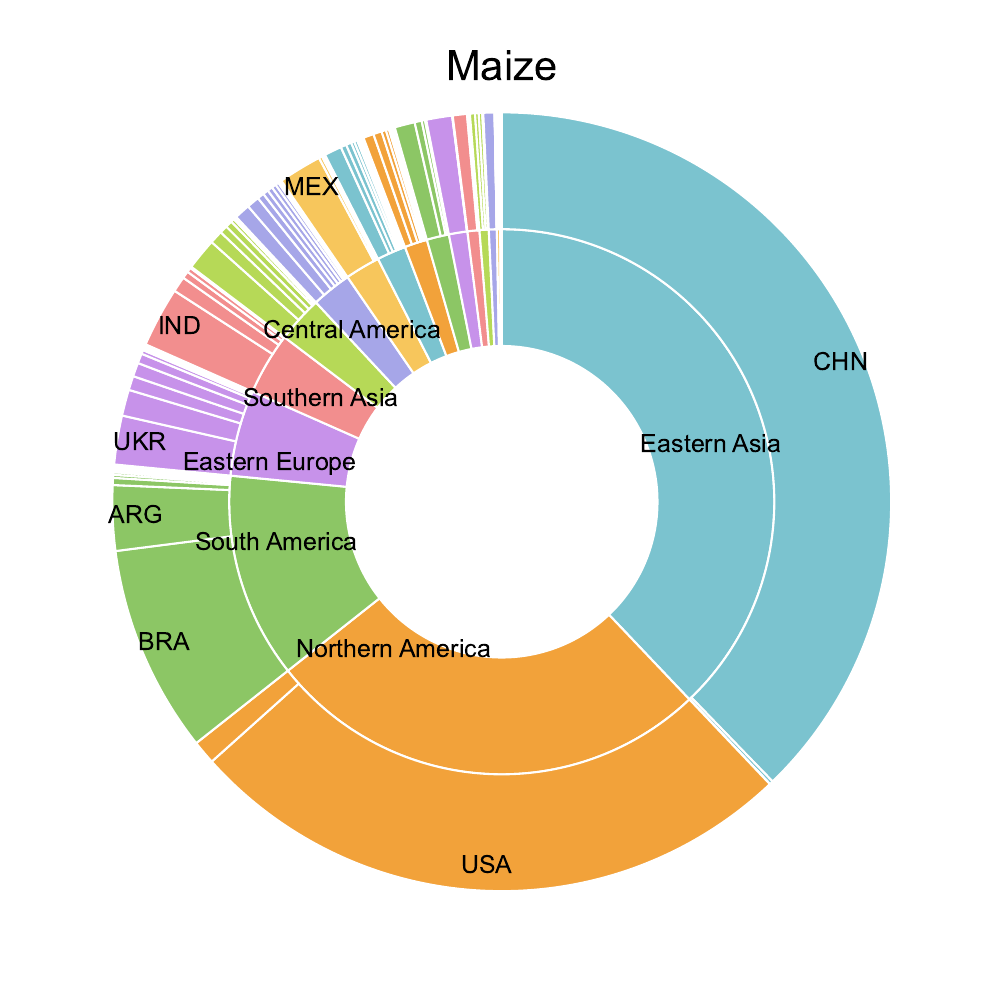}
     \includegraphics[width=0.24\linewidth]{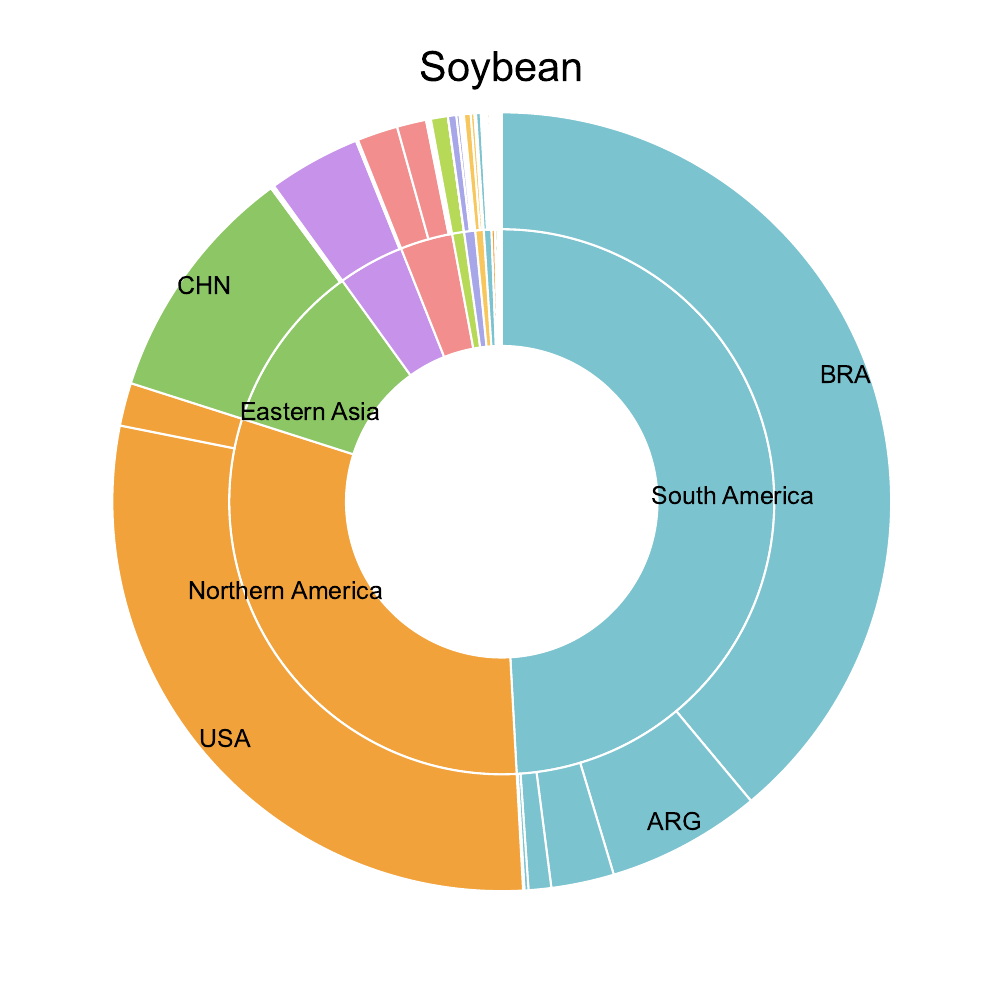}
    \caption{Geographical distribution of staple crop production (2023).}
    \label{Fig:Production_distribution}
\end{figure}

\begin{figure}[h!]
    \centering
    \includegraphics[width=0.45\linewidth]{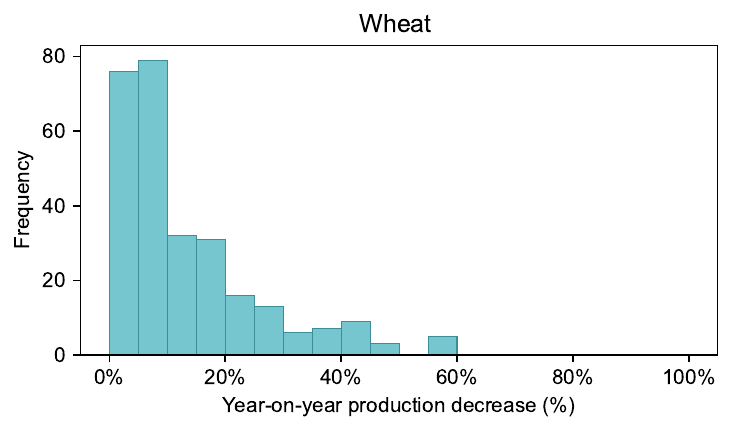}
    \includegraphics[width=0.45\linewidth]{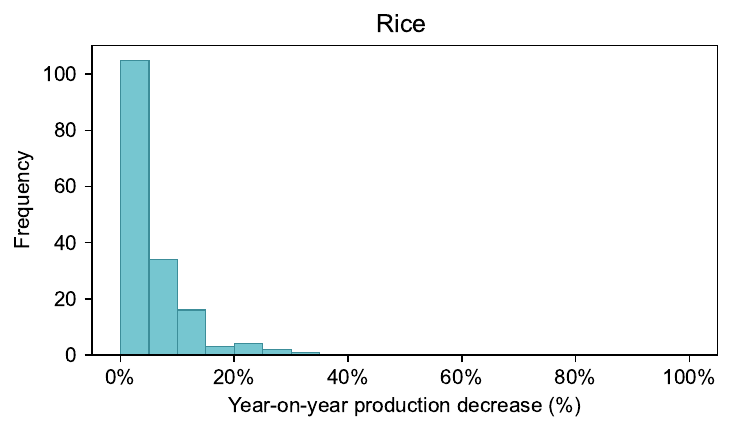}\\
    \includegraphics[width=0.45\linewidth]{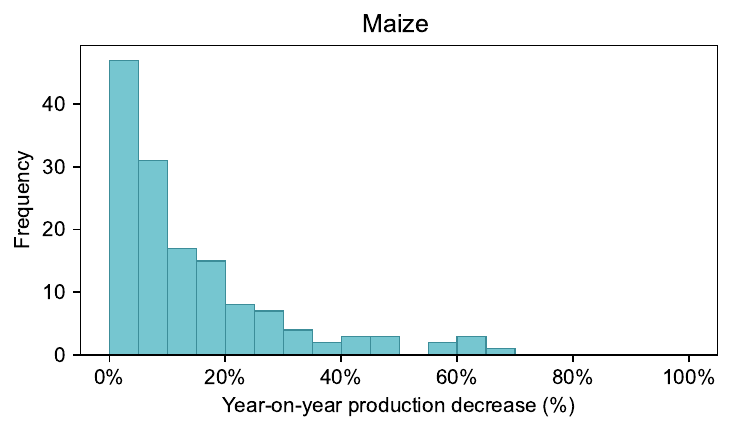}
    \includegraphics[width=0.45\linewidth]{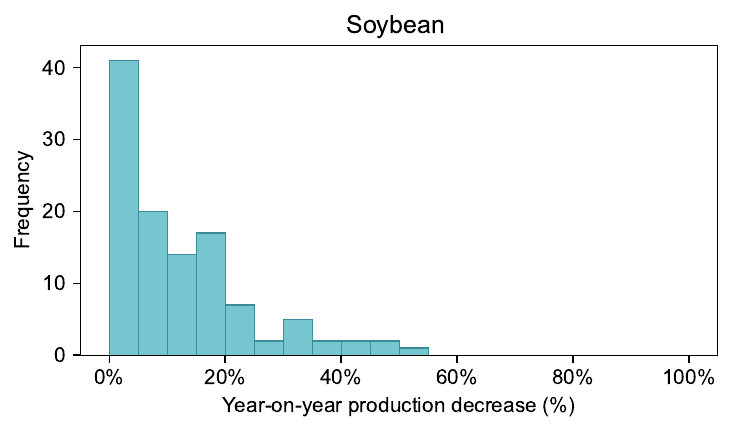}
    \caption{Distribution of year-on-year production declines (1986--2023) among major producers ($\geq 1\%$ of global production in 2023).}
    \label{Fig:Production_drop_summary}
\end{figure}

In the FAO food balance sheet framework, domestic food availability is determined by domestic production, international trade and stock variation. 
Production is the foundational source but also a primary origin of supply shocks. As shown in Fig.~\ref{Fig:Production_distribution}, global staple output is highly concentrated: the annotated economies account for 80\% of total production, so disruptions in a few major producers can have outsized effects on global availability and importers’ exposure. Production is sensitive to climate hazards, which have been shown to induce crop losses and persistent output variability \citep{Lobell-Schlenker-CostaRoberts-2011-Science}. 
Large shortfalls have occurred repeatedly, including the 2010 decline in Russia’s wheat production by  34\% \citep{Hunt-Femia-Werrell-Christian-Otkin-Basara-Anderson-White-Hain-Randall-McGaughey-2021-WeatherClimExtremes} and the 2012 U.S. drought that reduced corn production by 13\% \citep{Church-Haigh-Widhalm-deJalon-Babin-Carlton-Dunn-Fagan-Knutson-Prokopy-2017-CLIMRISKMANAG}. 
Beyond climate, geopolitical tensions can also depress output in major producers \citep{Chen-Tu-An-Wu-Lin-Gong-2024-CommunEarthEnviron}. Consistent with these mechanisms, defining major producers as economies with more than 1\% of global output in 2023, we compile production declines for 1986-2023 and find that year-on-year drops among major producers are frequent (Fig.~\ref{Fig:Production_drop_summary}).
This evidence indicates that major producing economies are not only highly concentrated in global supply, but also subject to substantial production instability.
% \begin{figure}[h!]
%     \centering
%     \includegraphics[width=0.48\linewidth]{Figures/Wheat_global_production_trade_stock.pdf}
%     \includegraphics[width=0.48\linewidth]{Figures/Rice_global_production_trade_stock.pdf}\\
    
%     \includegraphics[width=0.48\linewidth]{Figures/Corn_global_production_trade_stock.pdf}
%     \includegraphics[width=0.48\linewidth]{Figures/Soybean_global_production_trade_stock.pdf}
%     \caption{Production, stock and trade volumes of four staples.}
%     \label{Fig:P_S_T_4staple}
% \end{figure}

Reserves and trade both serve as key buffers when production shocks occur \citep{Marchand-Carr-Dell'Angelo-Fader-Gephart-Kummu-Magliocca-Porkka-Puma-Ratajczak-Rulli-Seekell-Suweis-Tavoni-D'Odorico-2016-EnvironResLett}. 
Stocks provide an additional cushion: releasing food stocks can temporarily supplement domestic availability, reduce immediate shortfalls when production or imports decline, and help stabilize prices, thereby enhancing food security \citep{Wright-2011-ApplEconPerspectPolicy}. However, maintaining stocks is costly, so stockholding levels vary widely across economies, implying that the same shock can translate into very different availability outcomes across importers. 
Trade offers another buffering channel by reallocating supplies across partners, yet it can also transmit shocks \citep{Tu-Suweis-D'Odorico-2019-NatSustain}. When disruptions arise in major producing economies, import-dependent countries may face shortfalls not only through direct price and quantity effects, but also through reduced export availability, trade diversion, and export-restricting policy responses under stress \citep{Bouet-Debucquet-2012-RevWorldEcon}. Stock use and trade reallocation therefore interact in shaping exposure during out-of-equilibrium adjustments, rather than acting as independent stabilizers.

Taken together, these considerations motivate the core focus of this study. When staple production is concentrated and  production shocks occur, economies’ supply outcomes depend on how the shock is absorbed domestically and how shortfalls are transmitted and reallocated through trade linkages. We therefore simulate large shocks to major producing economies and examine the resulting economy-level food supply responses. We aim to (1) quantify food supply robustness across economies, (2) explain cross-economy differences in robustness and identify key determinants, and (3) assess whether changes in stocks and trade reallocation can improve robustness.

The rest of this paper is organized as follows. Section~\ref{S2:LitRev} provides the literature review. Section~\ref{S3:Methodology} describes the data and methods. Section~\ref{S4:EmpAnal1} presents economy-level robustness and its classification under production shocks to major producing economies, and analyzes key determinants of robustness. Section~\ref{S5:EmpAnal2} conducts counterfactual analyses. Section~\ref{S6:Conclude} concludes.

\section{Literature review}
\label{S2:LitRev}

Global food security is commonly understood as ensuring that all people, at all times, have access to sufficient, safe, and nutritious food \citep{Simon-1996-FoodPolicy}. From an availability standpoint, supply shocks are disturbances that reduce food availability. Prior studies have examined multiple sources of such shocks, including climate extremes, geopolitical conflicts, and policy interventions \citep{Lesk-Rowhani-Ramankutty-2016-Nature,Kafando-Sakurai-2025-JAgricEcon,Giordani-Rocha-Ruta-2016-JIntEcon}.
In subsequent research, production shortfalls are often treated as the initiating shock, while trade and policy responses are modeled as transmission mechanisms that shape the impacts of the shock.

Broadly, existing studies typically follow two complementary approaches: one uses observed data to examine how shocks affect food supply at the global level and across economies, while the other relies on computational experiments to simulate shock transmission and quantify impacts under specified assumptions.

For studies based on observed shocks, the focus is typically on quantifying their impacts, explaining cross-economy differences in outcomes and underlying drivers, and examining the role of trade linkages and policy responses in shock transmission.
Studies of the Russia–Ukraine conflict show that the conflict disrupted food production and trade and generated uneven food security impacts across regions, with import-dependent and lower-income economies facing particularly high risks \citep{Jia-Xia-Li-Yu-Wu-Li-Su-Wang-Chen-Liu-2024-CommunEarthEnviron,Abay-Breisinger-Glauber-Kurdi-Laborde-Siddig-2023-GlobFoodSecur-AgricPolicy}. 
By examining how domestic and global trade shocks have propagated across countries historically, \cite{Distefano-Laio-Ridolfi-Schiavo-2018-PLoSOne} show that per capita income plays a key role in the transmission of trade shocks.
\cite{Headey-2011-FoodPolicy} use trade and international price data to conduct historical event analysis, finding that trade shocks were key drivers of the 2007-2008 food price spike.
In the context of COVID-19-related disruptions, \cite{Falkendal-Otto-Schewe-Jagermeyr-Konar-Kummu-Watkins-Puma-2021-NatFood} show that export restrictions can substantially amplify supply shocks.
Other studies likewise show that export restrictions can raise grain prices or intensify price volatility, thereby worsening food insecurity risks \citep{An-Qiu-Zheng-2016-FoodPolicy,Rude-An-2015-FoodPolicy}. 
Beyond external shocks and policy responses, economies’ exposure is also shaped by their structural characteristics.
\cite{Kummu-Kinnunen-Lehikoinen-Porkka-Queiroz-Roos-Troell-Well-2020-GlobFoodSecur-AgricPolicy} find that high food import dependence combined with limited trade partners may increase vulnerability to food supply shocks. 
\cite{Wassenius-Porkka-Nystrom-Jorgensen-2023-GlobFoodSecur-AgricPolicy} further highlight the role of domestic production capacity and potential self-sufficiency in shaping food supply risks.
Overall, these studies show that food supply shocks have uneven effects across economies and that trade and policy responses are important channels of international transmission. However, less attention has been paid to systematically comparing economies’ food supply robustness under major production shocks and explaining the structural factors behind these differences.

Unlike studies based on historical events and observational data, another strand uses computational experiments to simulate how localized production or supply shocks propagate through the global food system and to quantify cross-economy differences in supply losses, exposure, and vulnerability.
Early studies in this strand typically relied on fixed international trade networks and a set of predefined rules governing how shocks were absorbed and how trade adjusted. 
\cite{Marchand-Carr-Dell'Angelo-Fader-Gephart-Kummu-Magliocca-Porkka-Puma-Ratajczak-Rulli-Seekell-Suweis-Tavoni-D'Odorico-2016-EnvironResLett} simulated the short-term response to food supply shocks and examined how reserves, consumption, and trade adjustment jointly shape cross-country differences in exposure. \cite{Heslin-Puma-Marchand-Carr-Dell'Angelo-D'Odorico-Gephart-Kummu-Porkka-Rulli-Seekell-Suweis-Tavoni-2020-FrontSustainFoodSyst} later applied a similar Food Shock Cascade model to a contemporary US Dust Bowl scenario, focusing on the propagation of a severe US wheat production shock through the global trade network.
Subsequent studies further incorporated policy and behavioural responses, such as export restrictions, into the propagation process, showing that shocks do not simply spread passively through a static network but can generate broader cascade effects through countries’ responses. 
\cite{Burkholz-Schweitzer-2019-EnvironResLett}, for example, show that when countries impose export restrictions, shocks can continue to propagate through higher-order dependency structures and amplify systemic risk.
Building on this framework, \cite{Grassia-Mangioni-Schiavo-Traverso-2022-SciRep} provide a more systematic analysis of how country characteristics shape shock transmission and vulnerability.
Later work adopted agent-based models to represent richer behavioural rules and adaptive responses, further extending this line of research toward dynamic frameworks that link production, trade, stocks, and consumption under different shock scenarios \citep{Fonteijn-vanOort-Hengeveld-2024-JASSS,Kuhla-Kubiczek-Otto-2025-EcolEcon}.
While this literature has considerably advanced understanding of shock propagation, it has devoted less effort to counterfactual policy evaluation aimed at identifying which interventions could improve food supply robustness under major production shocks. Building on this gap, this study examines economies’ food supply robustness under major production shocks, investigates the factors shaping cross-economy differences, and conducts counterfactual policy analysis to assess potential interventions.

\section{Data and methodology}
\label{S3:Methodology}

Our research utilizes production and bilateral trade data from the Food and Agriculture Organization (FAO, {\url{http://www.fao.org}}), complemented by ending stocks data provided by the United States Department of Agriculture’s Foreign Agricultural Service (USDA-PSD, {\url{https://apps.fas.usda.gov/psdonline/}}). 
We convert production, bilateral trade, and stock quantities from mass units into caloric equivalents using crop specific caloric conversion factors, thereby expressing all food quantities on a common basis.
To ensure the reliability of our results, we restrict our sample to economies with a population exceeding 500,000 and documented trade records for the specific crop under study.
Summary statistics for these variables are presented in Table~\ref{Table: Statsitic_Description}.

\begin{table}[htbp]
\centering
\caption{Statistic description of production, stock and trade network of different staple foods (2023).}
\begin{tabular*}{\textwidth}{@{\extracolsep{\fill}}lccccccc@{}}
\toprule
Staple food & $\sum{P}$ (kcal) & $\sum{S}$ (kcal) & $\sum{F}$ (kcal) & $N$ & $L$ & $\rho$ & $\langle d \rangle$ \\
\midrule
Aggregated & $1.27\times 10^{12}$ & $2.76\times 10^{11}$ & $1.96\times 10^{11}$ & 169 & 5433 & 0.191  & 1.72 \\
Wheat      & $3.12\times 10^{11}$ & $8.63\times 10^{10}$ & $6.05\times 10^{10}$ & 163 & 1885 & 0.0714 & 2.11 \\
Rice       & $2.82\times 10^{11}$ & $5.03\times 10^{10}$ & $1.35\times 10^{10}$ & 169 & 3751 & 0.132  & 1.80 \\
Maize      & $5.45\times 10^{11}$ & $1.06\times 10^{11}$ & $6.65\times 10^{10}$ & 168 & 2718 & 0.0969 & 1.97 \\
Soybean    & $1.31\times 10^{11}$ & $3.37\times 10^{10}$ & $5.57\times 10^{10}$ & 162 & 1629 & 0.0625 & 2.17 \\
\bottomrule
\end{tabular*}
 \begin{tablenotes}   
        \footnotesize        
        \item Note: $P=$ production, $S=$ stocks, $F=$ tradevolume, $N=$ numbers of economics in network, $\rho=$ density of trade network, $\langle d \rangle=$ average shortest path length.
      \end{tablenotes}  
\label{Table: Statsitic_Description}
\end{table}

Following the model design by \cite{Marchand-Carr-Dell'Angelo-Fader-Gephart-Kummu-Magliocca-Porkka-Puma-Ratajczak-Rulli-Seekell-Suweis-Tavoni-D'Odorico-2016-EnvironResLett}, we incorporate a new behavioral rule to account for domestic supply stabilization. In our model, economies prioritize their own food security under stress. Consequently, any economy that draws upon its internal stocks is labeled as shocked and is prohibited from expanding its export volume to the global market. 

\begin{figure}[h!]
    \centering
    \includegraphics[width=0.9\linewidth]{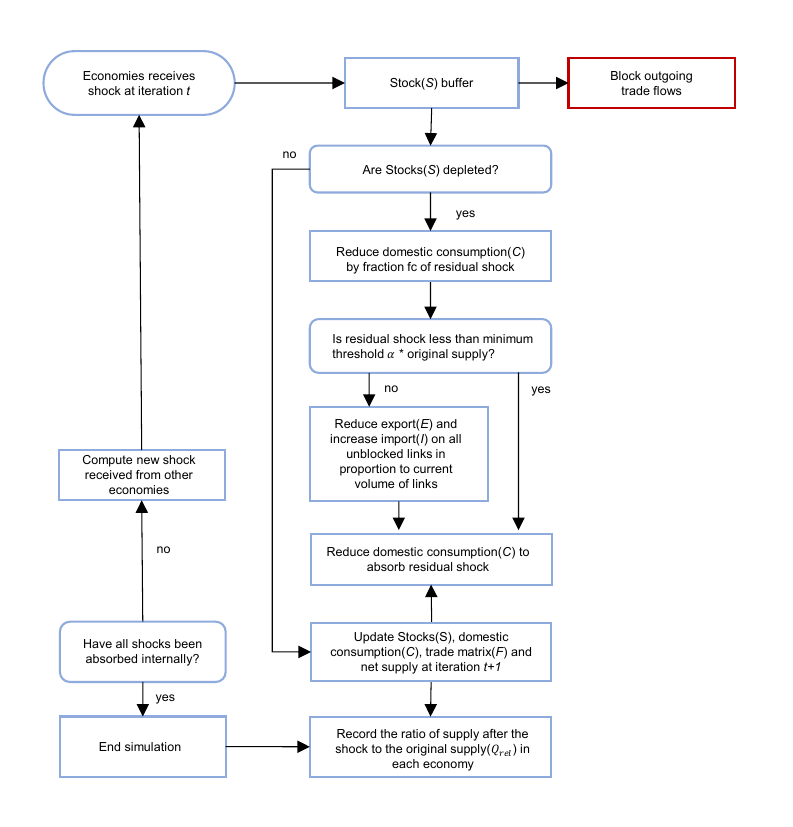}
    \caption{The flow chart of the simulation model. The simulation model is the same as \cite{Marchand-Carr-Dell'Angelo-Fader-Gephart-Kummu-Magliocca-Porkka-Puma-Ratajczak-Rulli-Seekell-Suweis-Tavoni-D'Odorico-2016-EnvironResLett}, except that we incorporate domestic supply stabilization (the red part). Economies utilizing internal stocks are classified as shocked and cannot increase exports.}
    \label{Fig:Simuilation}
\end{figure}

The model structure is illustrated in Fig.~\ref{Fig:Simuilation}.

\begin{table}[t]
\centering
\caption{List of variables and parameters}
\label{tab:variables}
\begin{tabular}{ll}
\toprule
Symbol & Description \\
\midrule
$P$ & Production by economy \\
$S$ & Stock by economy \\
$F$ & Trade matrix  \\
$E$ & Export by economy\\
$I$ & Import by economy\\
$V$ & Bilateral trade flow matrix, where $V_{ij}$ denotes exports from economy $i$ to $j$ \\
$C$ & Domestic consumption by economy\\
$Q$ & Net supply, $Q = P + I - E$ \\
$R$ & Robustness by economy \\
$\alpha$ & Minimum threshold (as a fraction of $Q$)\\
$f_c$ & Fraction of residual shock absorbed by $C$ if $R$ is depleted \\
$f_s$ & Fraction of actual stocks that are available to absorb shocks \\
$f_p$ & Magnitude of the initial shock as a fraction of global production\\
\bottomrule
\end{tabular}
\label{Table: Variables}
\end{table}
Table~\ref{Table: Variables} lists the variables and parameters used in the production shock model. 
% In our baseline specification, we set $f_c = 0.1$ and $f_s = 0.2$. These values are chosen to reflect the characteristics of a short-term shock model: in the short run, consumption exhibit relative rigidity, and not all reserves are immediately accessible for deployment. Furthermore, the threshold $\alpha$ is established at 0.001\%.
For economy $i$, baseline net supply is defined as
\begin{equation}
Q_i^0 = P_i^0 + I_i^0 - E_i^0,
\label{Eq:Q}
\end{equation}
This study focuses on production shocks to major producing economies. Therefore, only economies accounting for more than 1\% of global total production are selected as candidates for the initial shock. For a shocked major producing economy $i$, the production shock is specified as
\begin{equation}
\Delta{P_i} = -f_p P_{Total},
\end{equation}
where $P_{Total}$ denotes total global production. The initial net supply shock is given by
\begin{equation}
\Delta{Q_i^{0}} = \Delta{P_i}.
\end{equation}
After being hit by a shock, an economy first uses its stocks as a buffer. The available stock balance for economy $i$ is defined as
\begin{equation}
A_i^{0} = f_s S_i^0,
\end{equation}
where  $f_s$ is the share of initial stocks that can be used to absorb the shock, in our study,we set $f_s = 0.2$.
In iteration $k$, the change in stocks for economy $i$ is given by
\begin{equation}
\Delta{S_i^{k}} = \max \left( \Delta{Q_i^{k}},\,-A_i^{k} \right).
\end{equation}
After stock adjustment, the residual shock for economy $i$ in iteration $k$ is defined as
\begin{equation}
res_i^{k} = \Delta{Q_i^{k}} - \Delta{S_i^{k}}.
\end{equation}

This residual shock is first absorbed by domestic consumption at a fixed proportion $cfrac$:
\begin{equation}
\Delta{C_i^{(k,1)}} = f_c  res_i^{(k)}.
\end{equation}
We set $f_c = 0.1$. The remaining shock not absorbed by consumption is then given by
\begin{equation}
res_i^{\prime k} = (1-f_c)\, res_i^{k}.
\end{equation}
If $\left| res_i^{\prime k} \right| < \alpha_{\min} \cdot Q_i^{k}$, no further trade adjustment is triggered, and the remaining shock is fully absorbed through domestic consumption. The threshold $\alpha$ is established at 0.001\%.
Otherwise, trade adjustment is triggered. If an economy satisfies
\begin{equation}
\Delta{Q_i^{k}} < 0
\end{equation}
in iteration $k$, it is marked as \textit{shocked} and is no longer allowed to help absorb shocks in other economies by expanding its exports. 

The adjustable trade volume for economy $i$ is defined as
\begin{equation}
Tvol_i^{k} = E_i^{k} + \sum_{m \notin shocked} V_{mi}^{k},
\end{equation}
where the first term is the economy's current total exports and the second term is the volume of imports from exporters that are not marked as shocked. The corresponding trade adjustment is
\begin{equation}
\Delta{Tvol_i^{k}} = \max \left( res_i^{\prime k},\,-Tvol_i^{k} \right).
\end{equation}
If $Tvol_i^{k} > 0$, the proportional change in total trade volume for each economy is computed as
\begin{equation}
prop\Delta{T_i^{k}} = \frac{\Delta{Tvol}_i^{k}}{Tvol_i^{k}}.
\end{equation}
Trade flow adjustments are then allocated across bilateral links as
\begin{equation}
\Delta{V_{ij}}^{k} =
\left[
prop\Delta{T_i^{k}} - \mathbf{1}(i \notin shocked)\, prop\Delta{T_j^{k}}
\right] V_{ij}^{k}, \qquad i \neq j.
\end{equation}

The net supply shock in the next iteration is given by
\begin{equation}
\Delta{Q_i^{k+1}}=\sum_m \Delta{V_{mi}^{k}}-\sum_j \Delta{V_{ij}}^{k}+\Delta{Tvol_i^{k}}.
\end{equation}
If $\Delta{Q^{k+1}} \approx 0$,
the iteration terminates.

In our model, initial shocks are applied exclusively to major producing economies to observe the resulting feedback from the others. These shocks are set at a uniform absolute level and scaled up gradually from 0\% to 1\% of global production. During this process, we derive relative supply $Q_{rel}$, the ratio of post-shock supply to original supply for each economy:
\begin{equation}
Q_{rel,i}^{(j)}(f_p)
=
\frac{Q_i^{(j)}(f_p)}{Q_i^{(j)}(0)}, 
\label{Eq:Qrel}
\end{equation}
where $j$ represents the shock scenario and $i$ represents the affected economy. The average impact across all shock scenarios is calculated for each economy:
\begin{equation}
\bar{Q}_i(f_p)
=
\frac{1}{\lvert \mathcal{J}_i \rvert}
\sum_{j \in \mathcal{J}_i}
Q_{rel,i}^{(j)}(f_p), 
\label{Eq:Qbar}
\end{equation}
Here, $\mathcal{J}_i$ denotes the set of scenarios for economy $i$. If $i$ is a major producing economy, its scenario set excludes the case of its own production shock; otherwise, the set comprises scenarios involving shocks to all designated major producing economies.
After calculating the average relative supply, we introduce the economic robustness index, $R_i$. While \cite{Schneider-Moreira-Andrade-Havlin-Herrmann-2011-ProcNatlAcadSciUSA} define network robustness as the average size of the largest connected component during a progressive node failure process, we adapt this fundamental concept to our context. Specifically, we substitute network connectivity with the relative domestic supply level to evaluate economic robustness, formulating the index $R_i$ as follows:
\begin{equation}
R_i
=
\frac{1}{f_p^{\max}}
\int_{0}^{f_p^{\max}}
\bar{Q}_i\!\left(f_p\right)\,\mathrm{d}f_p. 
\end{equation}

\section{Results}
\label{S4:EmpAnal1}

\subsection{Economic robustness to production shocks in major producer economies}

Table~\ref{tab:robustness_rank} presents the top ten and bottom ten economies in terms of robustness under these scenarios for different crops. Overall, economies that consistently rank among the most robust tend to be those relatively isolated from international trade, such as North Korea and certain African nations. By contrast, the least robust economies are mainly small open and highly import-dependent economies, including many island economies, whose food supply is more vulnerable to disruptions originating abroad.
Robustness also varies markedly across crop types. The aggregated ranking includes not only relatively isolated economies, but also some more developed economies, such as France, Sweden, Denmark. However, this pattern is not uniform across individual crops. In particular, several higher-income European economies appear among the bottom ten in the rice ranking, and some economies that perform strongly in the aggregated results rank much lower for specific crops. This indicates that aggregate rankings may conceal important crop-specific differences in robustness, highlighting the need for crop-specific analysis.
\begin{table}[h!]
\centering
\caption{Robustness ranking of economies under production shocks to major producer economies.}
\label{tab:robustness_rank}
\begin{tabular*}{\textwidth}{@{\extracolsep{\fill}}>{\centering\arraybackslash}p{1.7cm}>{\centering\arraybackslash}p{1.4cm}cccc@{}}
\hline
Top 10 & Aggregated & Wheat & Rice & Maize & Soybean \\
\hline
1  & PRK & PRK & PRK & LAO & PRK \\
2  & TCD & TCD & MWI & TJK & MLI \\
3  & ERI & PSE & KAZ & PRK & COD \\
4  & LSO & NER & TZA & CAF & ECU \\
5  & MNE & MNE & MKD & TCD & CMR \\
6  & FRA & BLR & SUR & NGA & NIC \\
7  & SWE & MNG & MLI & MLI & KHM \\
8  & HUN & IND & NGA & MNE & MWI \\
9  & DNK & FRA & BOL & GMB & SYR \\
10 & BIH & DNK & TCD & MUS & HND \\
\hline
Bottom 10 & Aggregated & Wheat & Rice & Maize & Soybean \\
\hline
1  & BHR & SUR & LUX & SLB & GNB \\
2  & FJI & GIN & HUN & SGP & GNQ \\
3  & MDV & BWA & SVK & GNQ & LBY \\
4  & COM & LBY & EST & PSE & BHR \\
5  & SLB & LSO & FIN & MDV & MLT \\
6  & KWT & MDG & AUT & ARM & SWE \\
7  & SOM & CPV & HRV & PRY & DNK \\
8  & CPV & BEN & BWA & BWA & BLR \\
9  & GAB & COD & LSO & MAC & TLS \\
10 & PNG & TUN & SWE & SWZ & QAT \\
\hline
\end{tabular*}
\end{table}

\begin{figure}[h!]
\centering
    \includegraphics[width=0.48\linewidth]{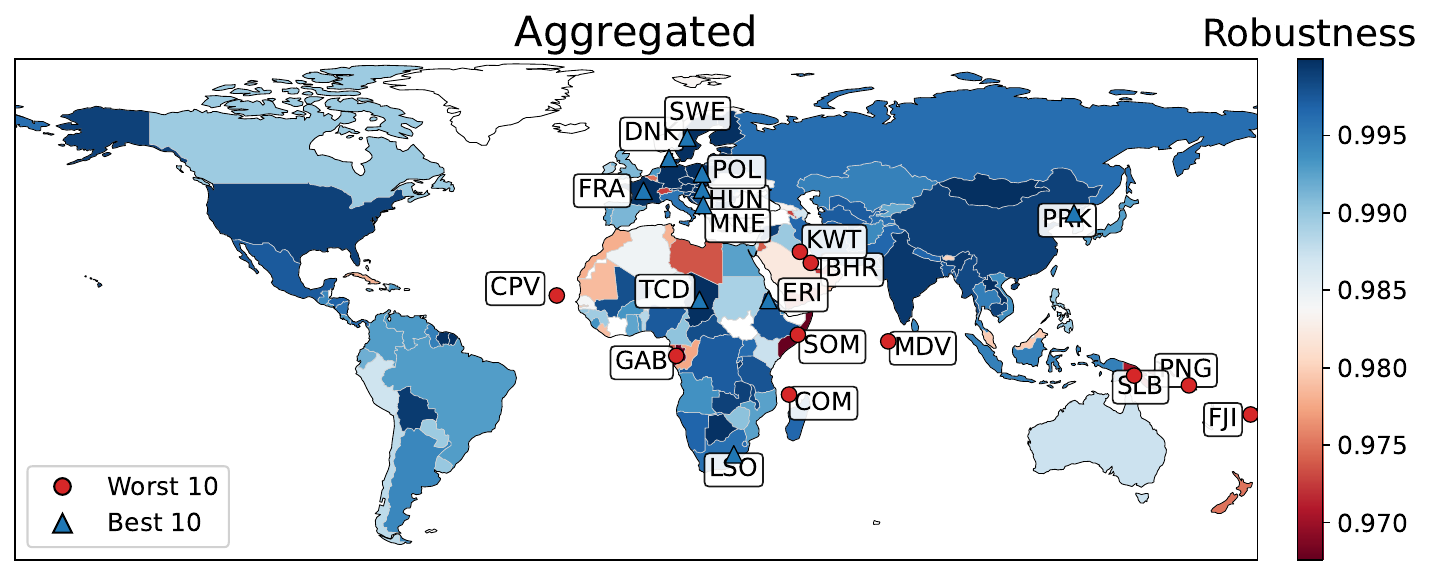}\\
    \includegraphics[width=0.48\linewidth]{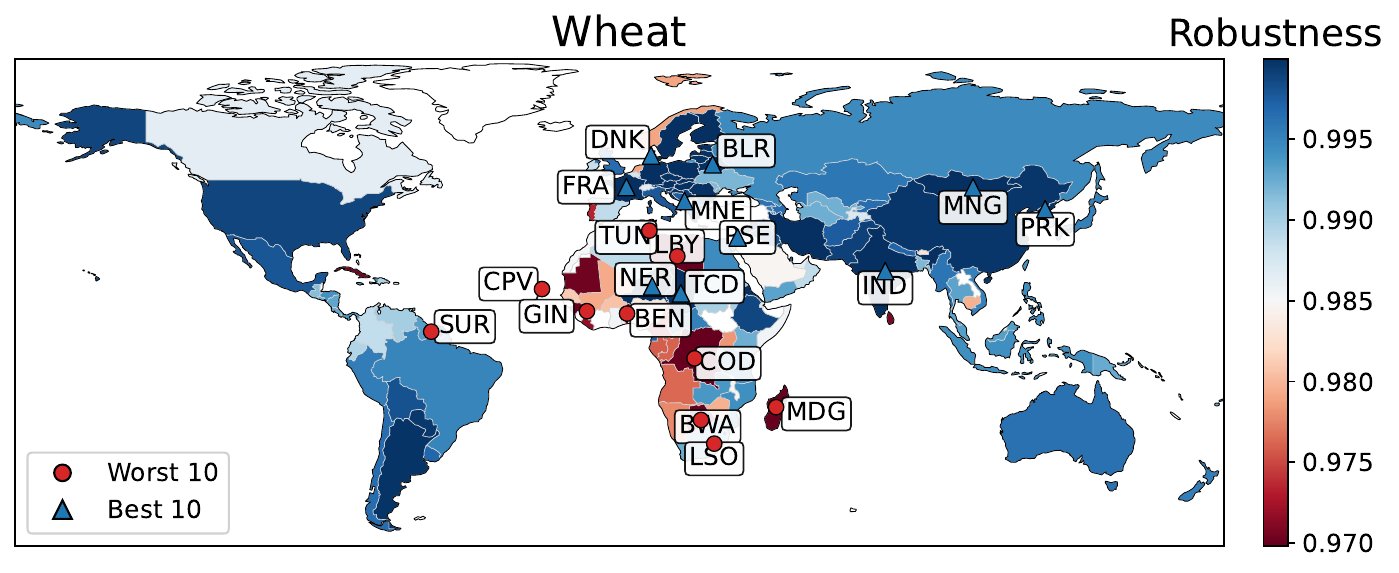}
    \includegraphics[width=0.48\linewidth]{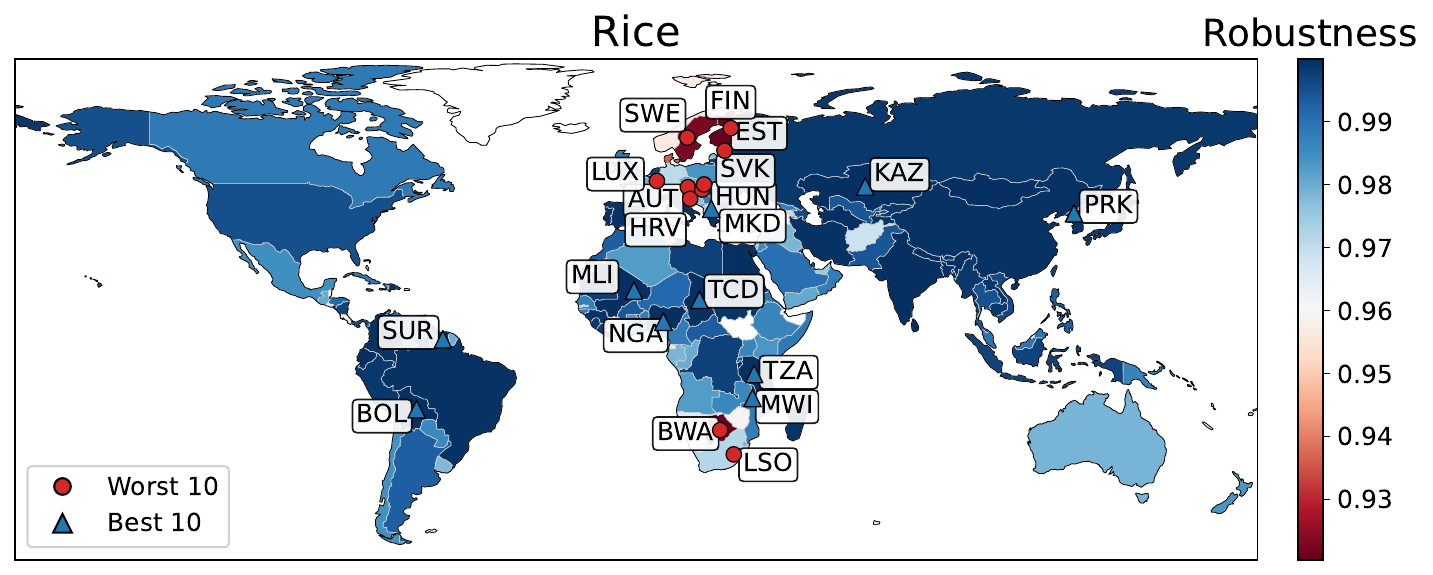}\\
    \includegraphics[width=0.48\linewidth]{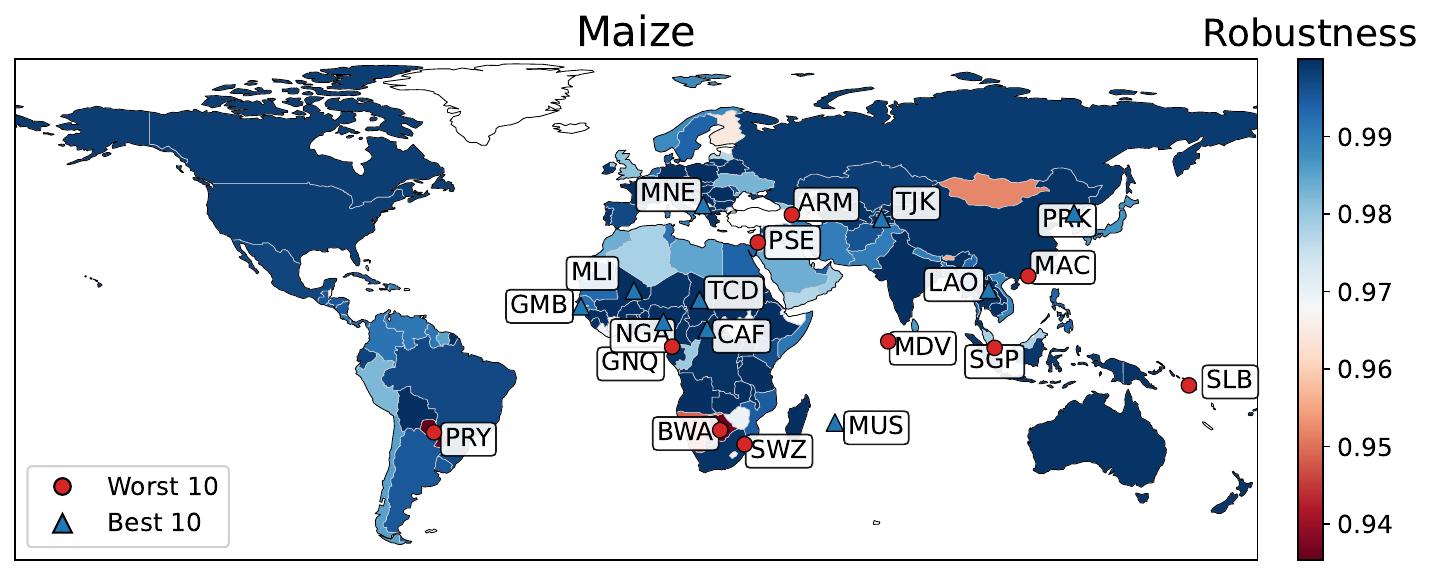}
    \includegraphics[width=0.48\linewidth]{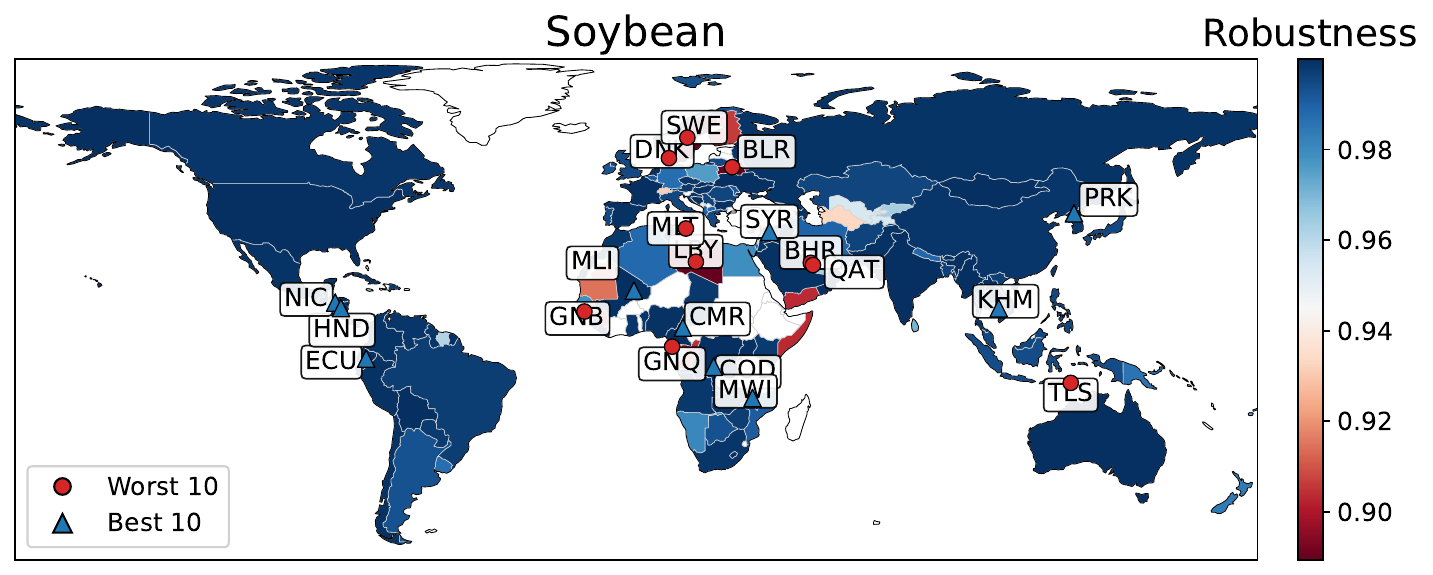}
    \caption{Robustness of economies under production shocks to major staple crops. In each panel, darker blue denotes higher robustness and darker red denotes lower robustness. Red circles identify the 10 least robust economies, whereas blue triangles indicate the 10 most robust economies.}
    \label{Fig:Robustness_Map}
\end{figure}

Fig.~\ref{Fig:Robustness_Map} illustrates the robustness of various economies under production shocks in major producing countries, with the top and bottom ten performers explicitly labeled. To enhance visual clarity, we applied a 5\% left-tail trimming to the color scale. The maps reveal significant spatial variations in robustness across different staple crops. For the aggregated crop set, highly robust economies include North Korea, several European economies, and a few African economies, whereas low-robustness economies are concentrated among small island economies as well as some African and Middle Eastern economies.

Among the four crops, wheat exhibits the highest overall robustness, with values concentrated in a relatively narrow and high range, whereas soybean shows the lowest overall robustness, with a wider spread toward lower values.
In spatial terms, highly robust economies under wheat shocks are distributed across East Asia, parts of Europe, and several African economies, while low-robustness economies display a relatively clear pattern of regional clustering, mainly in Africa the Middle East. For rice, highly robust economies are mainly found in Asia, Africa, and parts of South America, whereas several high income European economies fall into the low-robustness group. In maize, highly robust economies are concentrated mainly in Africa and Asia, while low robustness appears more common among small open and island economies. Soybean shows a less regionally concentrated pattern: highly robust economies are found mainly in Africa, Central America, and parts of Asia, whereas low-robustness economies are more spatially dispersed and associated with deeper low values. 
Together, these results point to substantial cross-crop differences in robustness.

\subsection{Economic robustness across income and regional groups}
\label{Subsection:Group_analysis}

\begin{figure}[h!]
    \centering
    \includegraphics[width=0.48\linewidth]{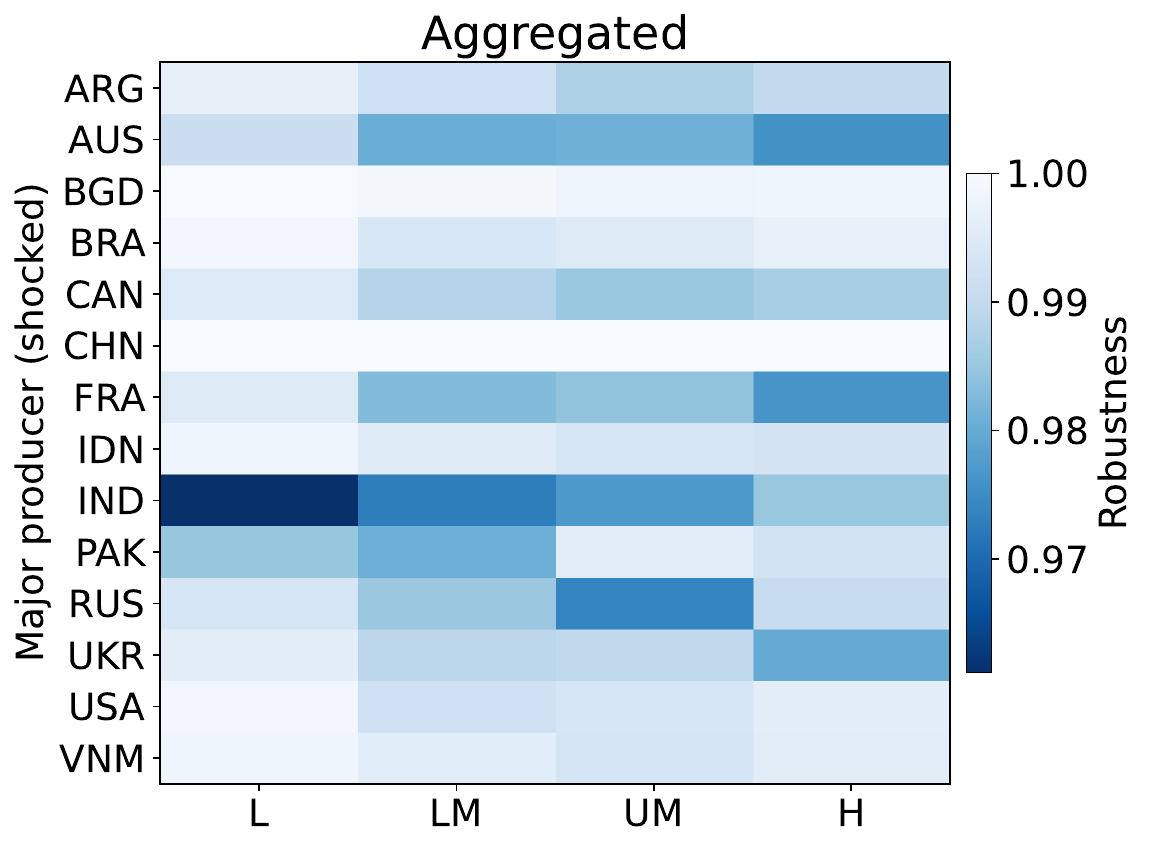}\\
    \includegraphics[width=0.48\linewidth]{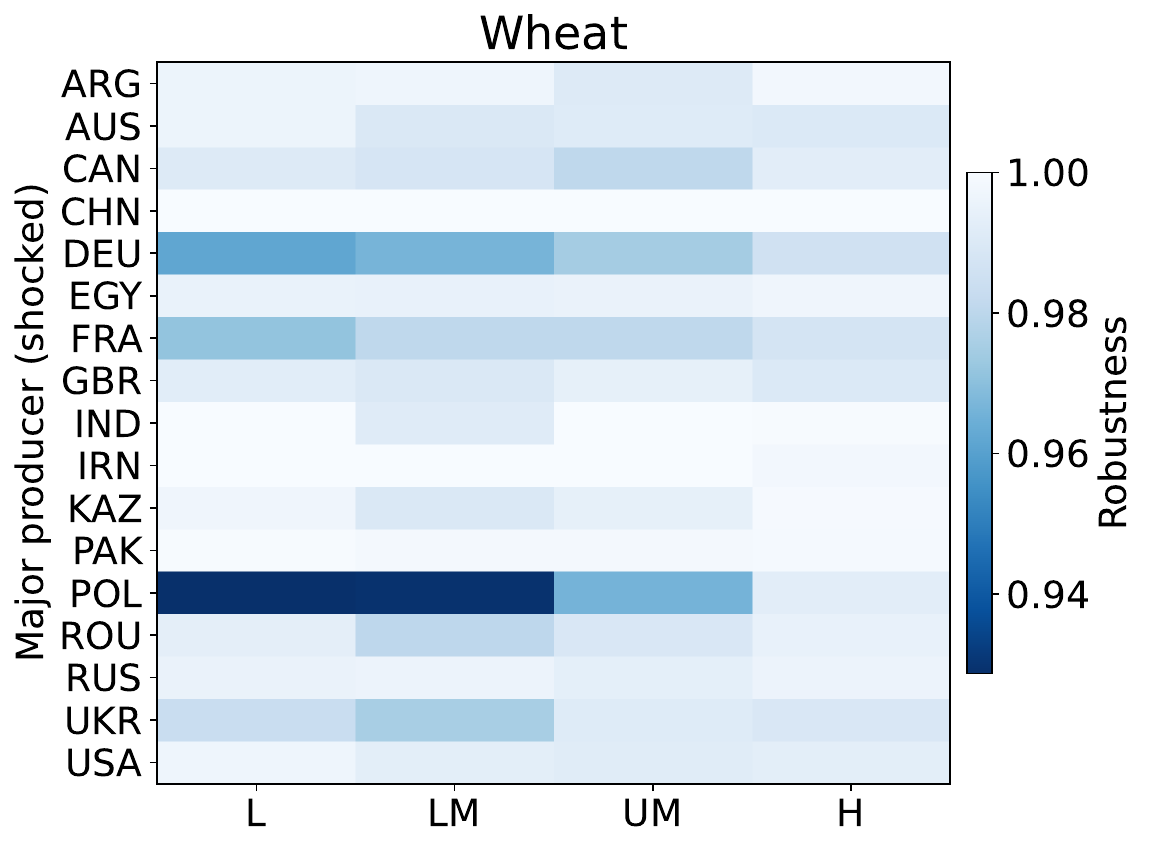}
    \includegraphics[width=0.48\linewidth]{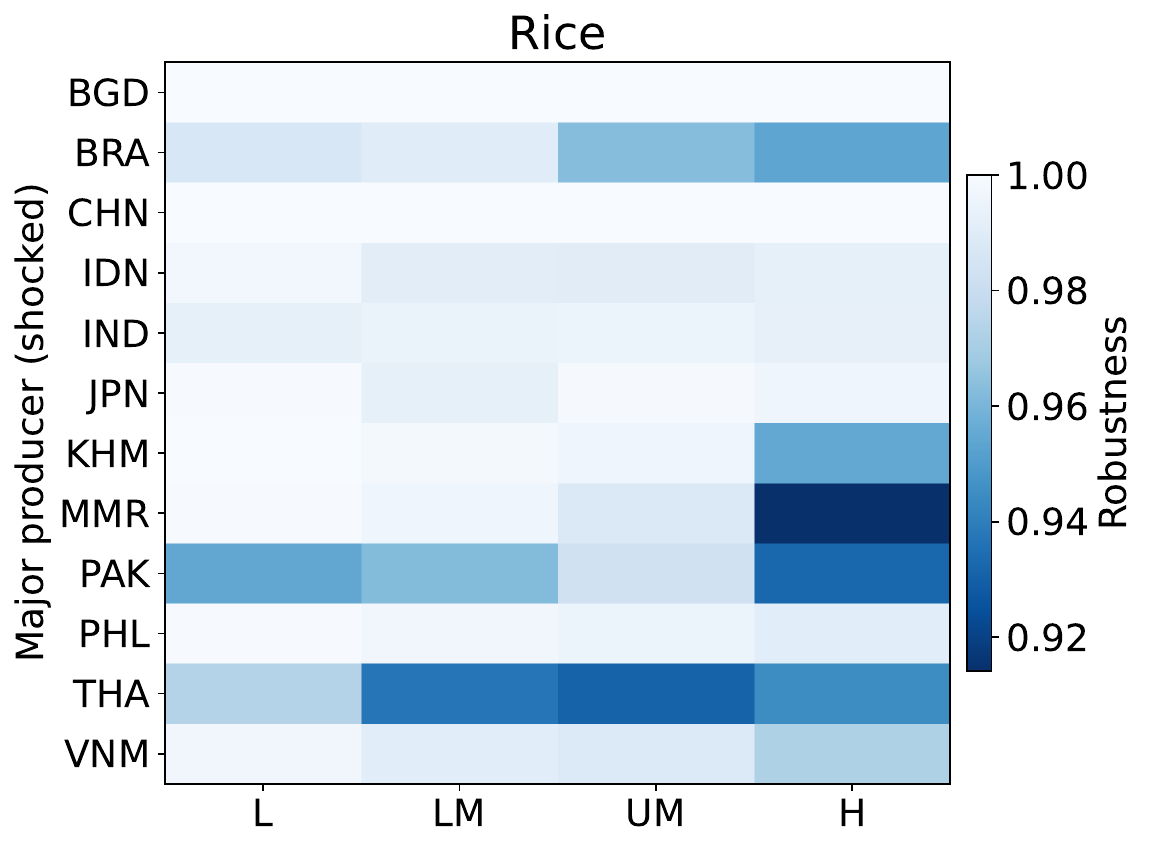}\\
    \includegraphics[width=0.48\linewidth]{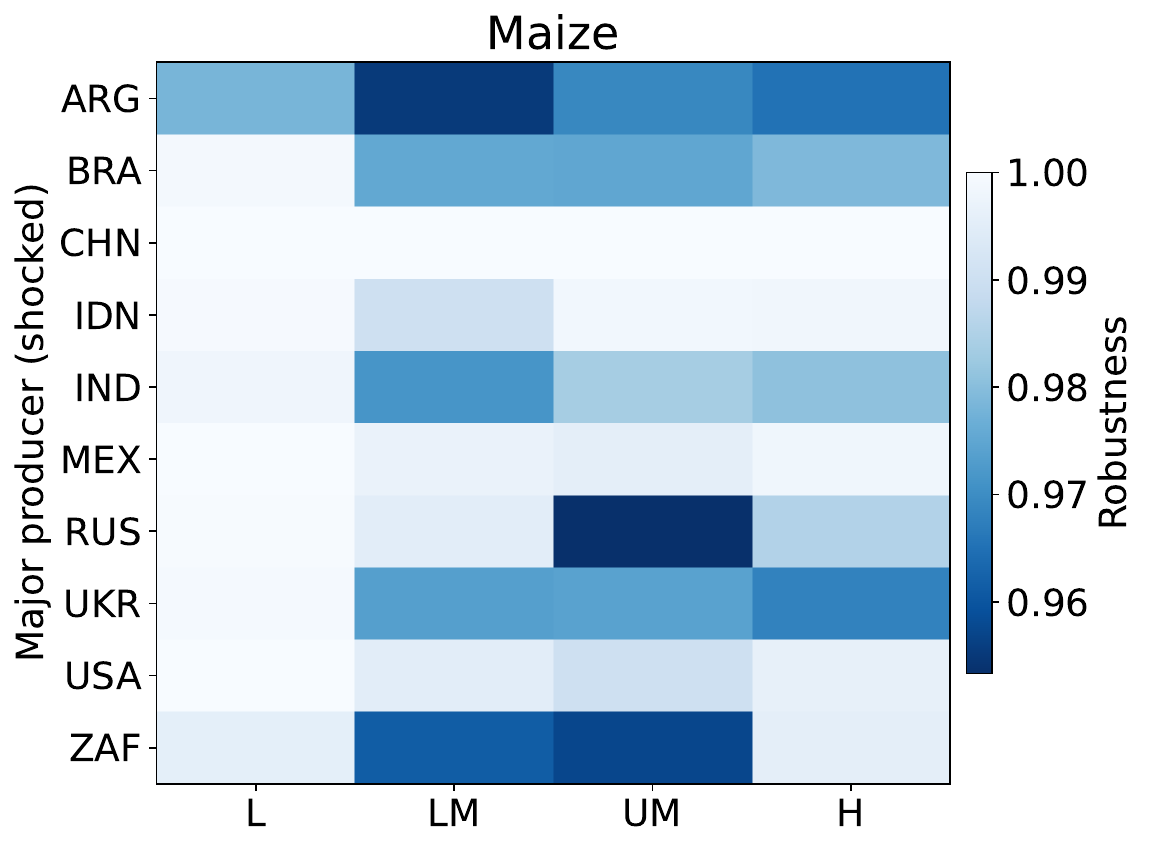}
    \includegraphics[width=0.48\linewidth]{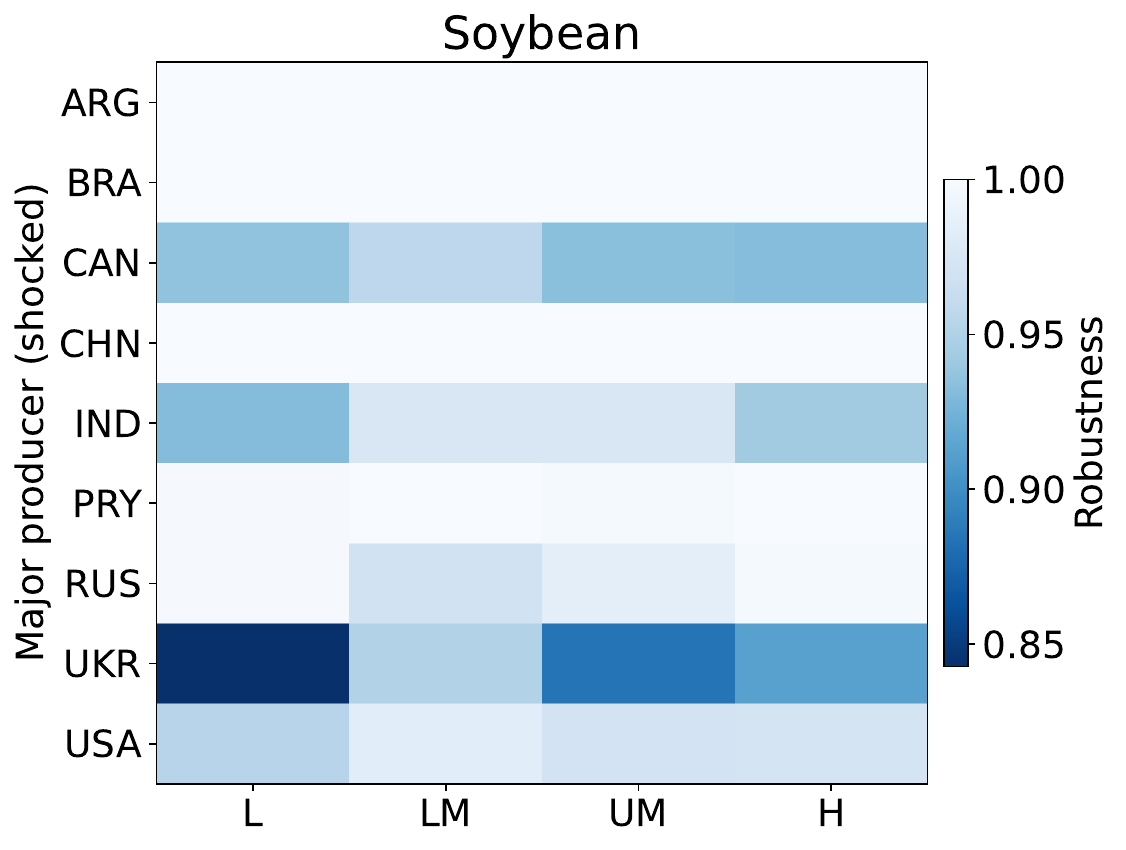}
    \caption{Economic robustness across different income groups. Rows show the shocked major producer economies, while columns show the income groups of affected economies. L, LM, UM, and H denote low-income, lower-middle-income, upper-middle-income, and high-income groups, respectively. Colors indicate the robustness of each income group under shocks originating from different major producers.}
    \label{Fig:Robustness_Income_Group}
\end{figure}

\begin{figure}[h!]
    \centering
    \includegraphics[width=0.48\linewidth]{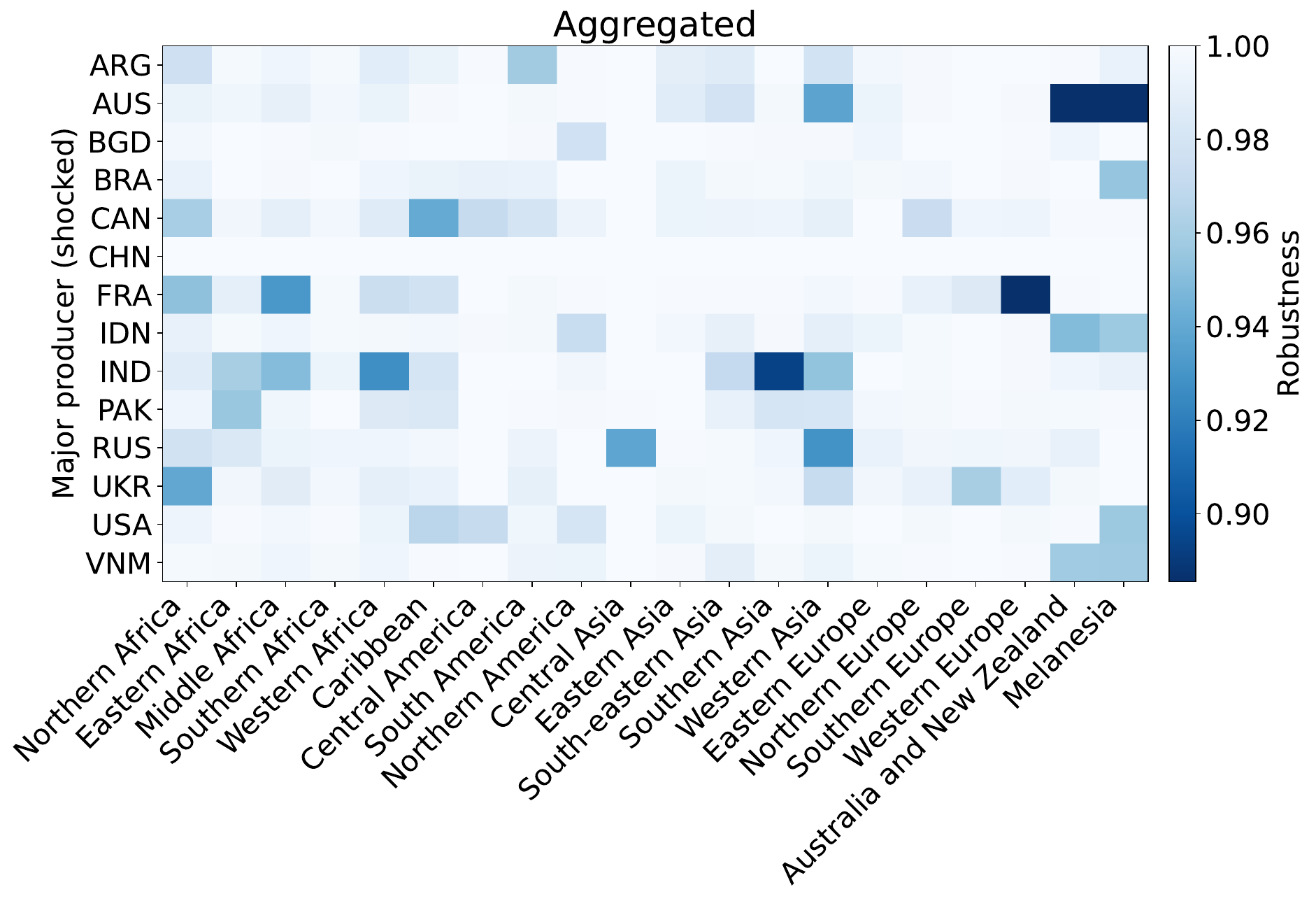}\\
    \includegraphics[width=0.48\linewidth]{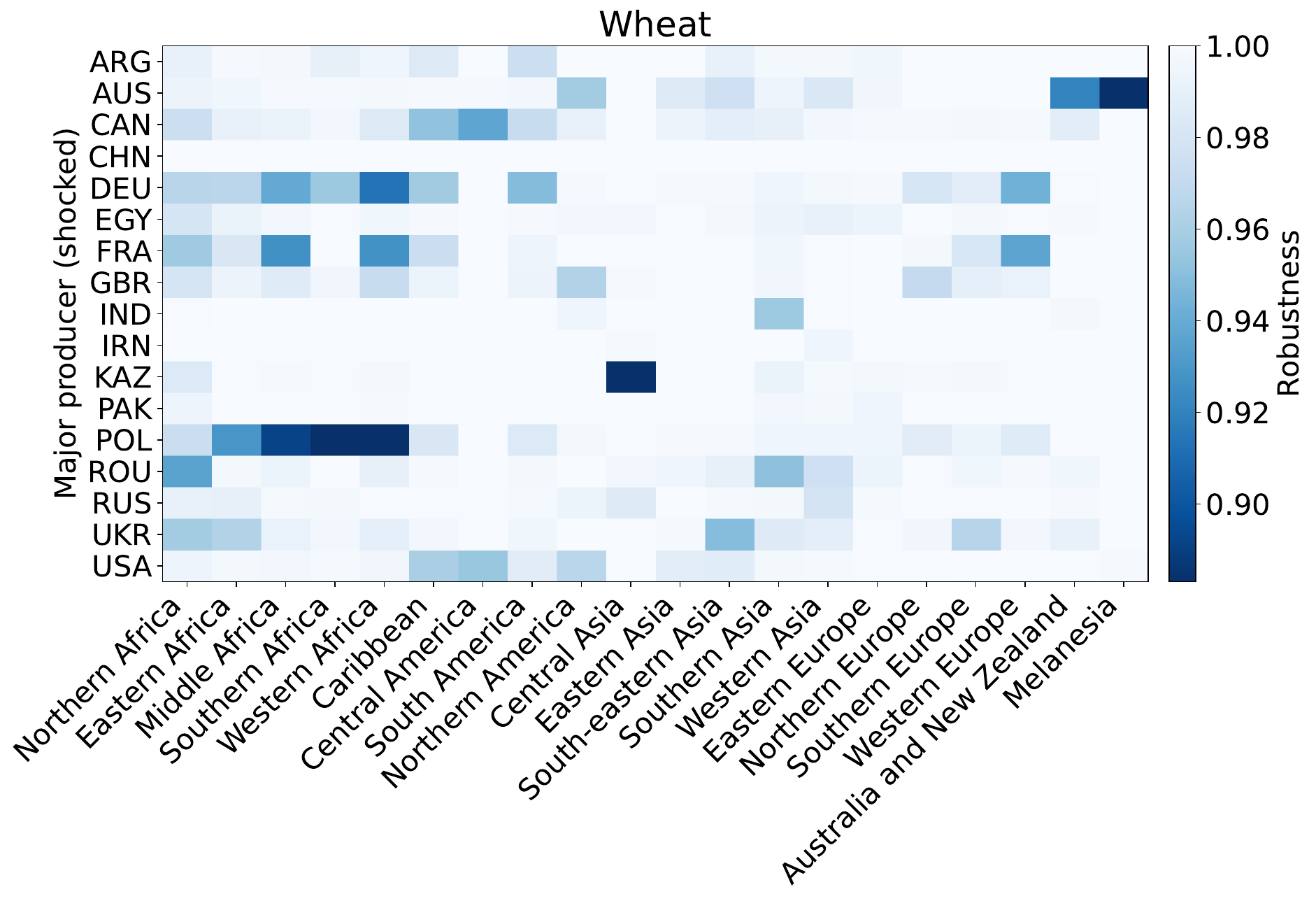}
    \includegraphics[width=0.48\linewidth]{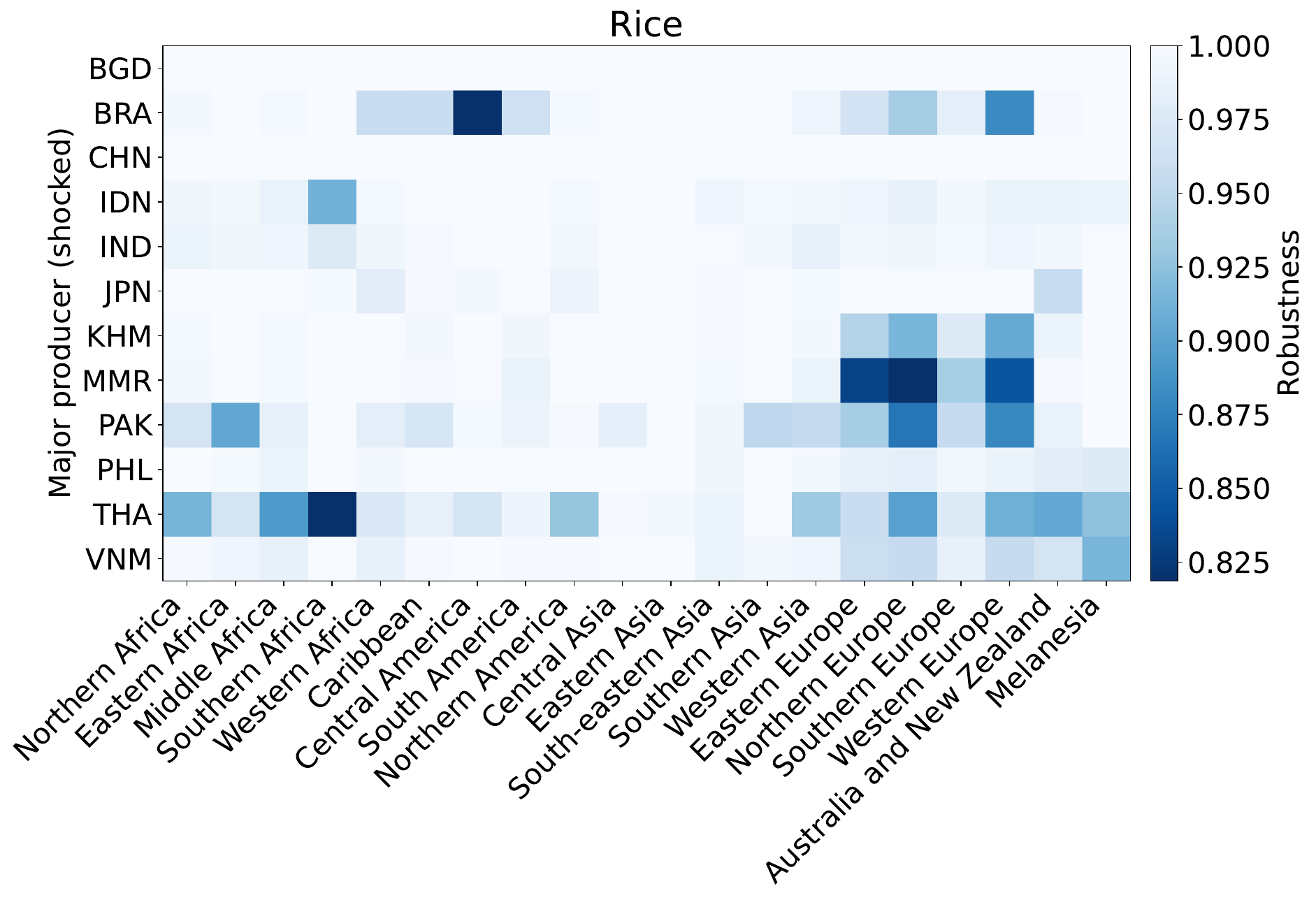}\\
    \includegraphics[width=0.48\linewidth]{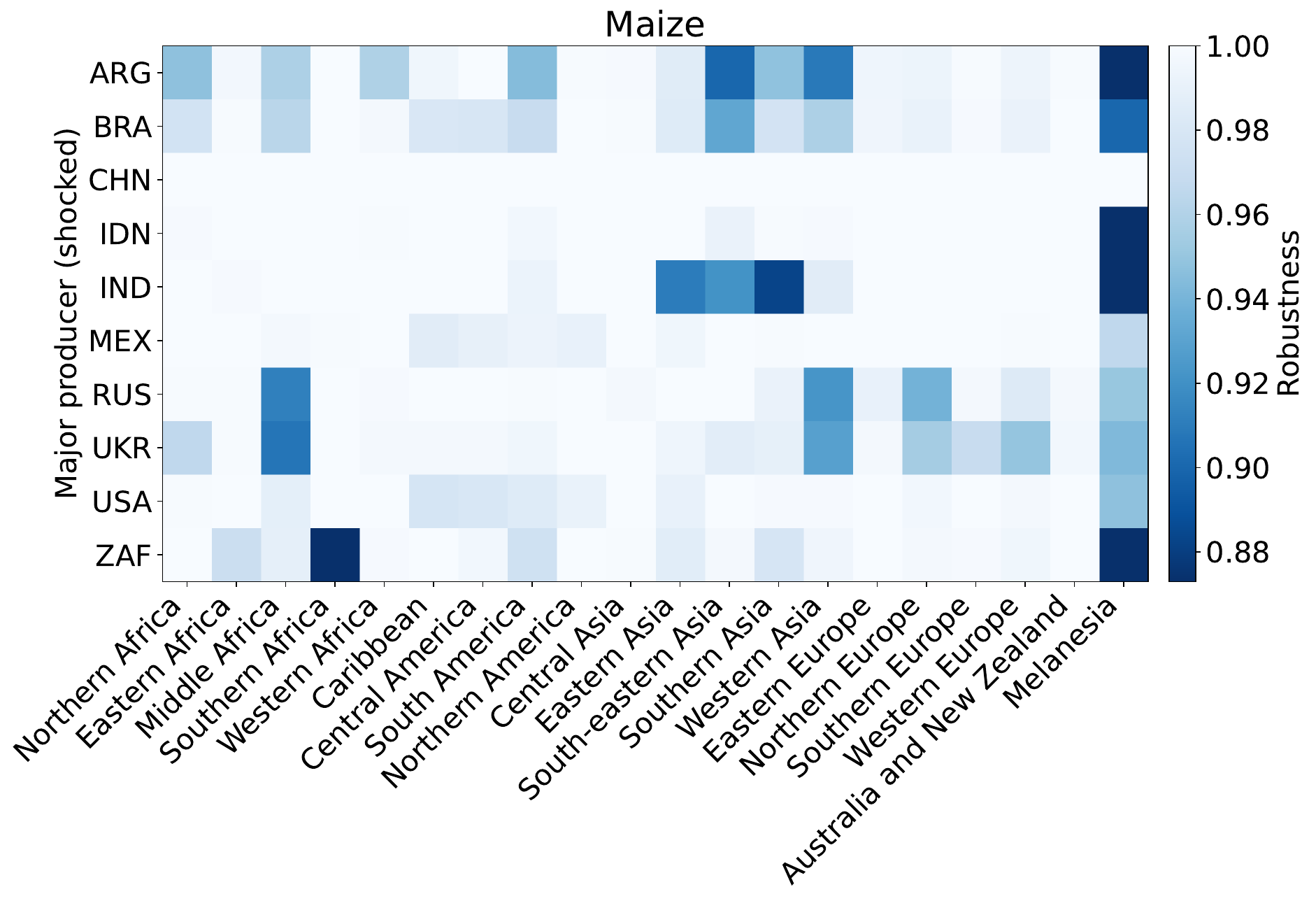}
    \includegraphics[width=0.48\linewidth]{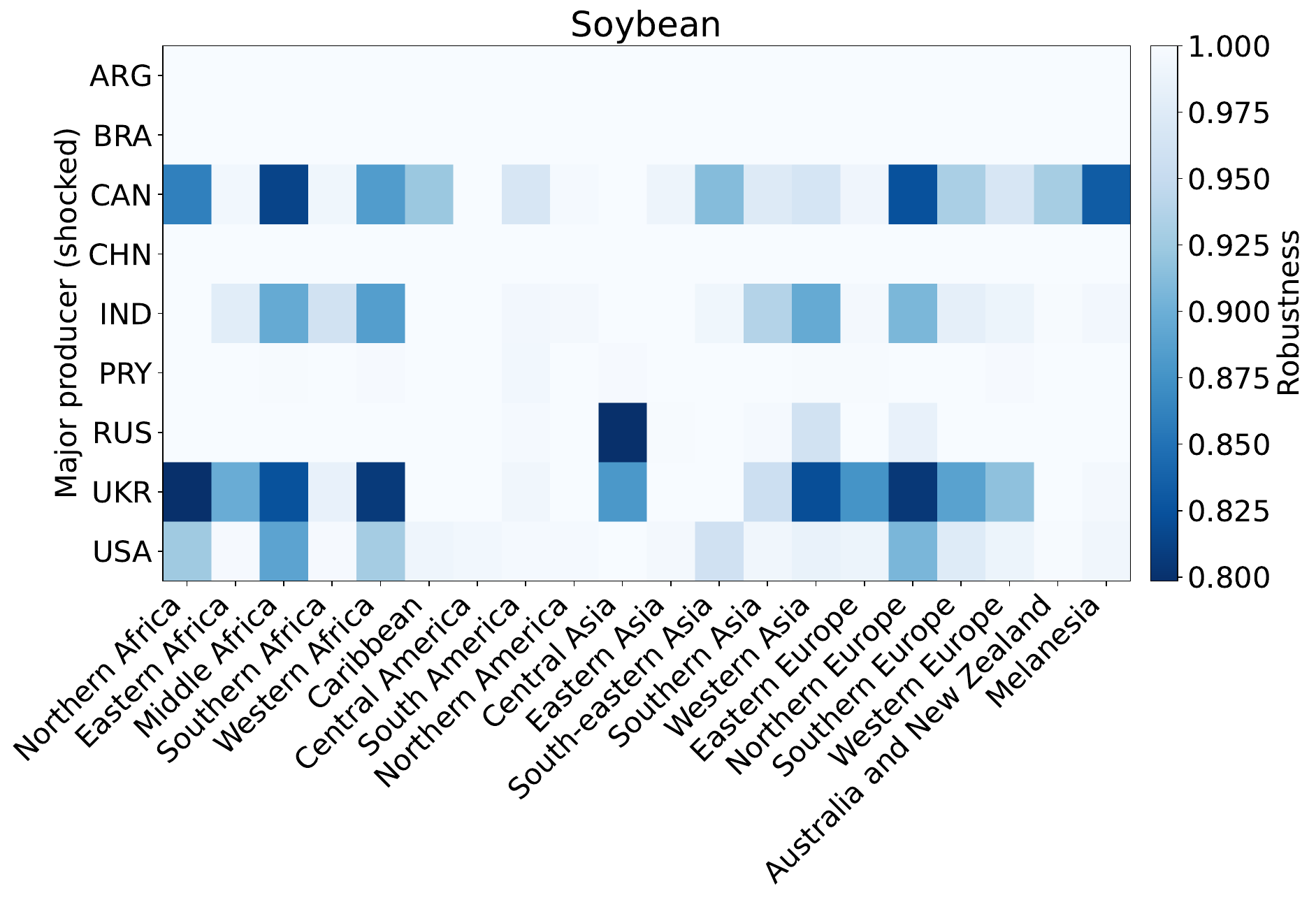}
    \caption{Economic robustness across different region groups. Rows show the shocked major producer economies, while columns show the region groups of affected economies. Colors indicate the robustness of each region group under shocks originating from different major producers.}
    \label{Fig:Robustness_Region_Group}
\end{figure}

To examine how production shocks in major producers affect different categories of economies, we classified countries and regions based on the World Bank’s income levels and the geographic groupings of the United Nations Statistics Division.
Fig.~\ref{Fig:Robustness_Income_Group} and Fig.~\ref{Fig:Robustness_Region_Group} illustrate the supply robustness of these groups. To minimize the distortion caused by extreme outliers, a 1\% left-tail truncation was applied to the visualization.

Economies are categorized into four distinct groups based on their income levels: low income group (L), lower-middle income group (LM), upper-middle income group (UM), and high income group (H). Fig.~\ref{Fig:Robustness_Income_Group} reveal that vertical patterns are relatively weak, suggesting a lack of uniform response within specific income groups. Instead, horizontal patterns are more pronounced, indicating that specific major producers can trigger substantial supply shortfalls across one or even multiple income categories.

For the aggregated crop set, shocks originating from India exert the most significant impact, particularly on the low and lower-middle income groups. Indeed, production declines in most major economies tend to compromise the supply stability of these nations.
For wheat, shocks from Poland are the most disruptive, while the high income group remains largely insulated from production fluctuations in these primary producers. For rice, Thailand’s production shock has the most far-reaching consequences. In the case of maize, the impact of Argentina is evident across all four income levels; meanwhile, Brazil, India, and Ukraine also exert substantial influence, suggesting that the impact of major maize producers is more evenly distributed. For soybeans, the most influential producers are Canada, India, Ukraine, and the United States, with the low income group bearing the most significant impact.

For geographical classification, we adopted a detailed sub-regional classification, while grouping economies from the same broader regions together to highlight regional patterns.
Consequently, although the patterns in Fig.~\ref{Fig:Robustness_Region_Group} appear more dispersed at first glance, several distinct dark-colored matrices emerge. These clusters indicate that certain major producers exert a concentrated impact on specific geographic regions.
For wheat, shocks from Germany, France, Poland, and Ukraine primarily disrupt supply in Africa, while Germany and France also affect European economies; meanwhile, Canada and the United States mainly influence the Americas, highlighting strong regional connectivity. For rice, major producers in Asia predominantly affect Europe, whereas Thailand exerts a significant impact on Africa, and Brazil influences the Americas.
Regarding maize, production declines in major economies generally affect Pacific island nations; specifically, shocks from Argentina, Brazil, and India have substantial impacts on Asian economies, while Russia and Ukraine primarily affect Central Africa, Western Asia, and Europe. In contrast, soybeans do not exhibit a pronounced spatial clustering effect. Among the major soybean producers, Canada, India, Ukraine, and the United States exert relatively strong effects, with Ukraine notably affecting the supply stability of African and European economies.

\subsection{Determinants of economic robustness to production shocks}
\label{Subsection:Determinants}

The group based analysis indicates that economies exhibit heterogeneity in their robustness when facing production shocks originating from major producing economies. To explain the mechanisms underlying these differences, our study further examines the structural factors that determine economic robustness.
We examine the factors determining an economy's supply robustness across three primary dimensions: production, stocks, and trade. Additionally, to account for intrinsic national disparities, we include per capita GDP as a controlled variable.

\begin{figure}[h!]
    \centering
    \includegraphics[width=0.7\linewidth]{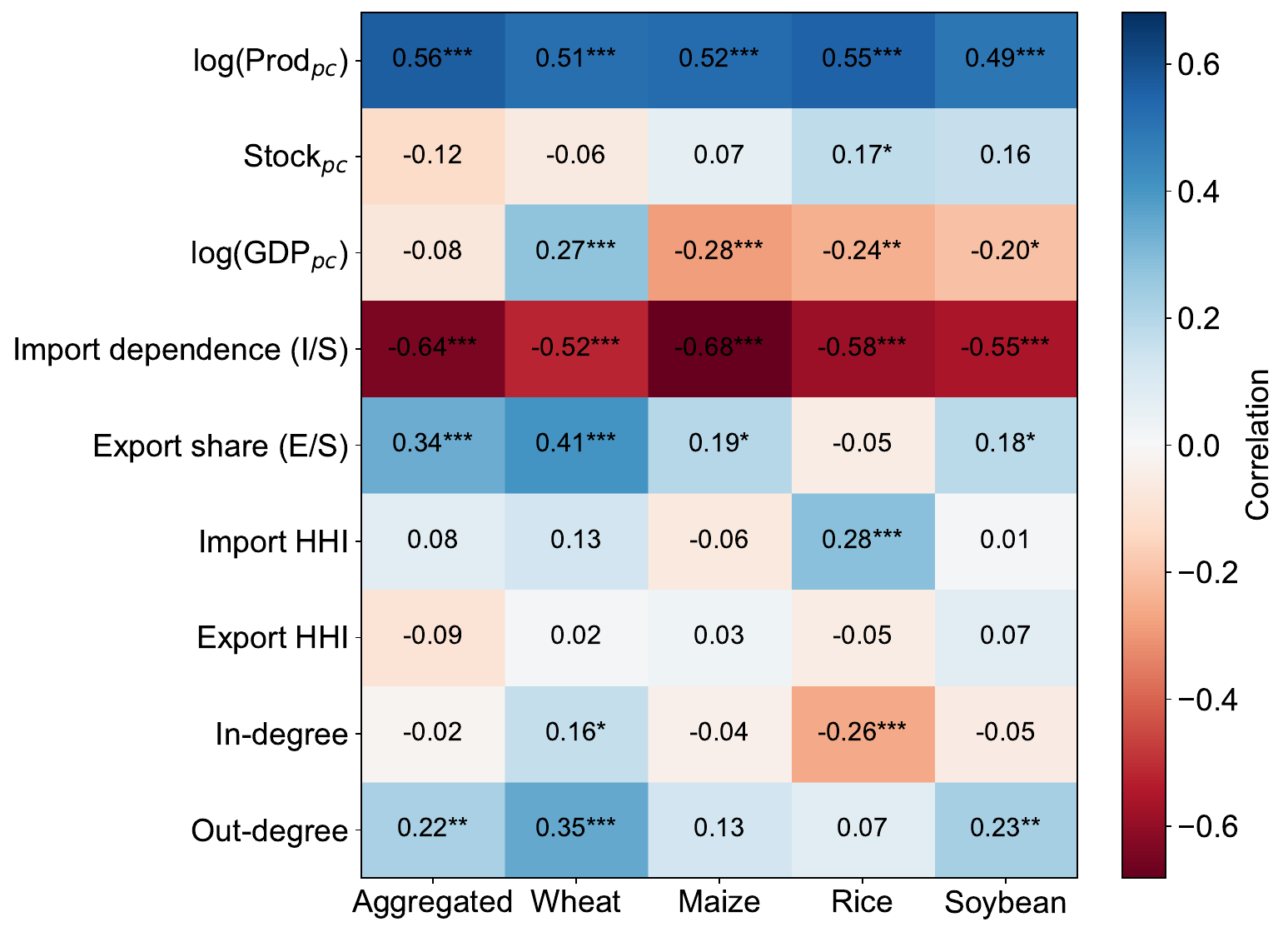}
    \caption{Correlation matrix of staple crop robustness and determinant variables across economies.}
    \label{Fig:Robustness_X_Corr}
\end{figure}

Fig.~\ref{Fig:Robustness_X_Corr} illustrates the correlations between supply robustness and these various economic characteristics across different crop types.
A consistent pattern emerges across all crops: supply robustness is positively correlated with domestic production levels and negatively correlated with import dependency. This relationship is intuitive because higher domestic production reduces an economy's exposure to external risks, making it less susceptible to supply fluctuations in other nations. Conversely, a high reliance on imports increases vulnerability, as a contraction in global supply directly threatens the availability of food within the domestic market.
The influence and significance of other variables vary considerably across different crops. For instance, food stocks exhibit a significant positive correlation only with the supply robustness of rice. While GDP is a significant factor for all crop categories, it only shows a positive correlation with wheat; this likely reflects the fact that high-GDP economies predominantly rely on wheat as their primary staple grain.
Furthermore, the export-to-production ratio is significantly and positively correlated with robustness for the aggregated crop set, wheat, and soybeans. This potentially suggests that economies with higher export capacities may possess stronger domestic production bases, thereby enhancing their supply stability. A similar logic may apply to export intensity. Notably, import concentration for rice is significantly and positively proportional to robustness. This seemingly counterintuitive finding might result from the fact that certain economies, despite being import dependent, maintain highly stable and long-term partnerships with their primary rice suppliers.

To identify the key variables determining the supply robustness of economies, we applied Random Forest to analyze feature importance. Given that the robustness indicators are heavily clustered near 1 with a few extreme outliers at the lower end, we performed a logit transformation on the robustness data to improve its distributional properties:
\begin{equation}
    \operatorname{Logit}(R_i)=\log\left(\frac{R_i}{1-R_i}\right)
\label{Eq:Logit}
\end{equation}
Subsequently, a 5\% winsorization was applied to the left tail to mitigate the influence of these extreme values.

Table~\ref{tab:RF_Importance} reports the random forest feature importance of the variables associated with the robustness of economies. It shows that import dependence is the most important determinant of robustness across all crop systems, followed consistently by per capita production. This indicates that robustness is shaped primarily by the balance between external supply stability and domestic supply capacity. Beyond these two dominant variables, the importance of the remaining determinants varies across crops, suggesting substantial commodity-specific heterogeneity. Stock share is relatively important in the aggregated and wheat models but much less so in rice, maize, and soybean, whereas GDP per capita and in-degree play a more prominent role in maize and soybean.
\begin{table}[htbp]
\centering
\caption{Random forest feature importance for the determinants of economies' robustness.}
\label{tab:RF_Importance}
\begin{tabular*}{\textwidth}{@{\extracolsep{\fill}}lccccc@{}}
\toprule
Determinant variables & Aggregated & Wheat & Rice & Maize & Soybean \\
\midrule
Import dependence & 0.2852 & 0.3260 & 0.3866 & 0.3731 & 0.3726 \\
log(Prodpc) & 0.2065 & 0.1865 & 0.2089 & 0.1371 & 0.1790 \\
log(GDPpc) & 0.0606 & 0.0488 & 0.0537 & 0.1309 & 0.1224 \\
Out-Degree & 0.0417 & 0.0560 & 0.0752 & 0.0343 & 0.0593 \\
Export proportion & 0.0965 & 0.1137 & 0.0720 & 0.0858 & 0.0557 \\
Import HHI & 0.0660 & 0.0769 & 0.0648 & 0.0631 & 0.0505 \\
Stock share & 0.1426 & 0.0917 & 0.0383 & 0.0426 & 0.0269 \\
In-Degree & 0.0667 & 0.0807 & 0.0598 & 0.1054 & 0.0837 \\
Export HHI & 0.0343 & 0.0197 & 0.0406 & 0.0278 & 0.0500 \\
\midrule
$R^2$  & 0.4860 & 0.3736 & 0.4814 & 0.4229 & 0.3432 \\
$\text{RMSE}$ & 1.3457 & 1.7729 & 1.5077 & 2.3300 & 2.3187 \\
\bottomrule
\end{tabular*}
\end{table}

\begin{figure}[h!]
\centering
    \includegraphics[width=0.48\linewidth]{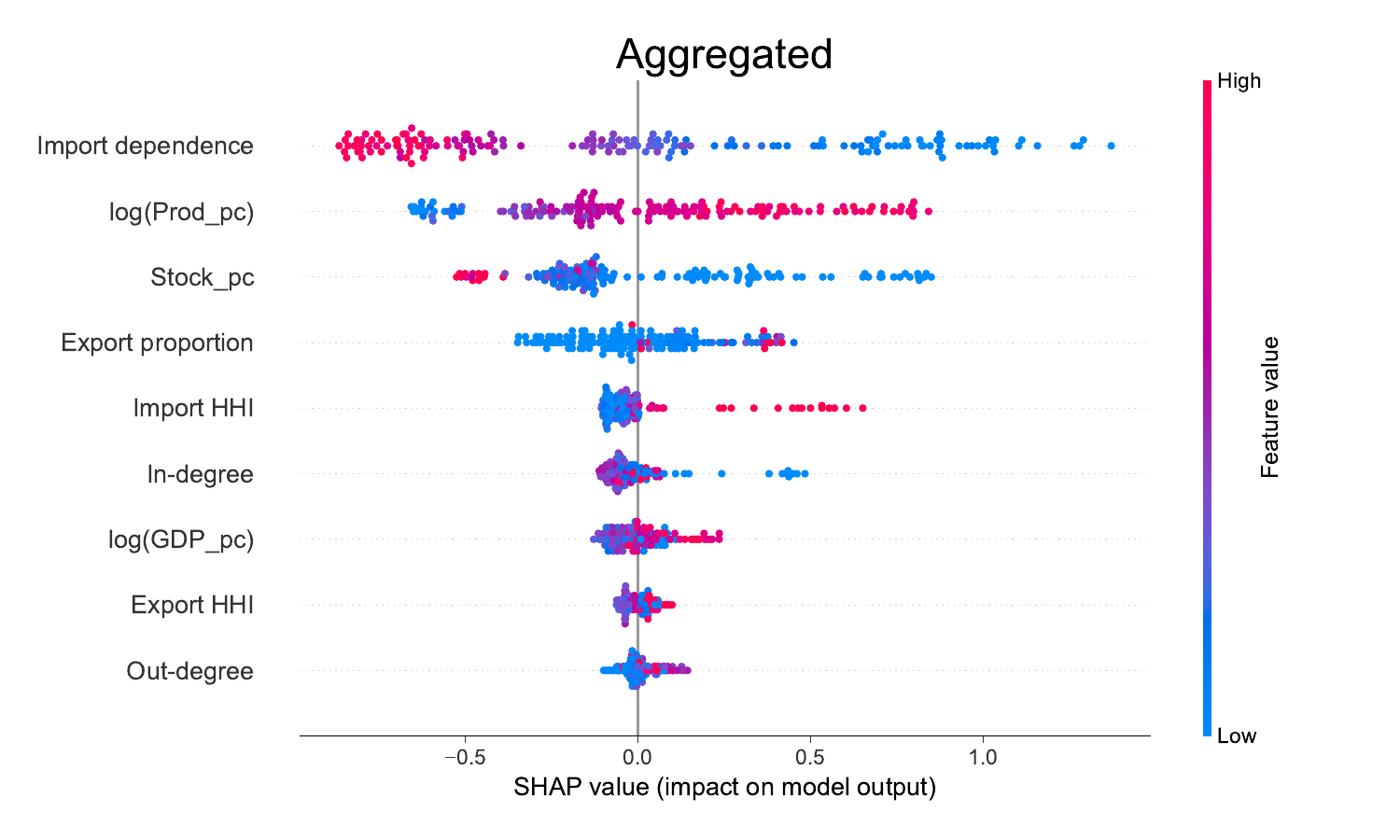}\\
    \includegraphics[width=0.48\linewidth]{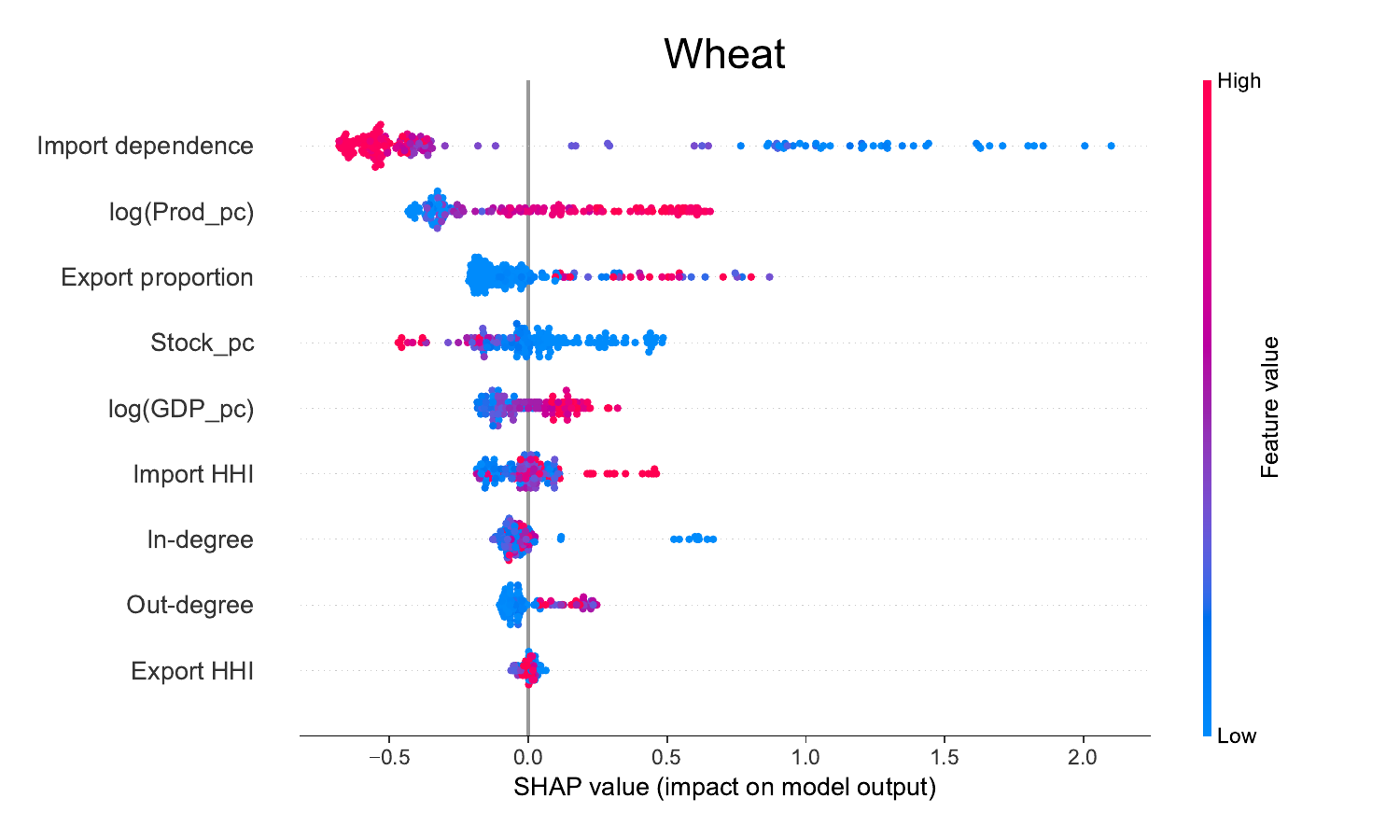}
    \includegraphics[width=0.48\linewidth]{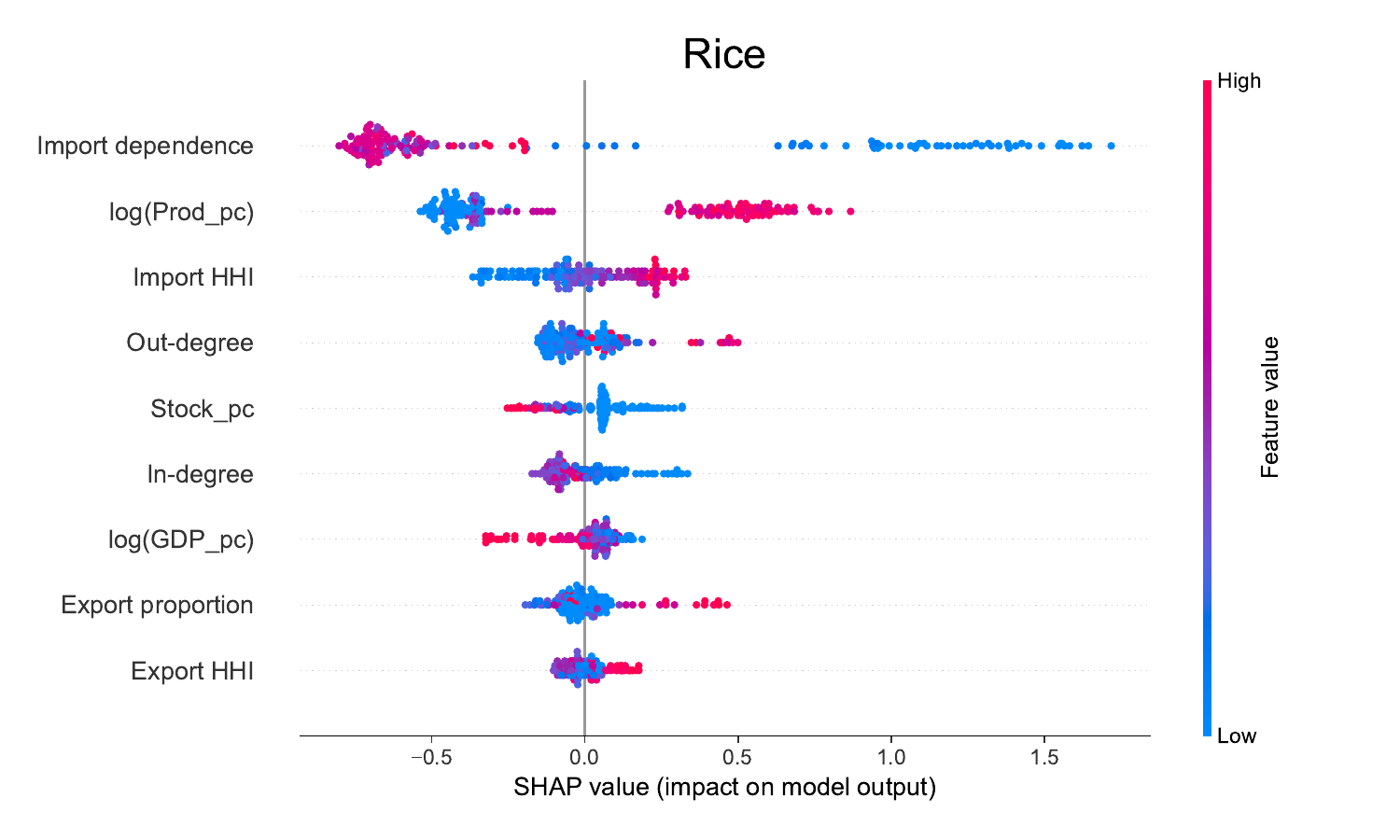}\\
    \includegraphics[width=0.48\linewidth]{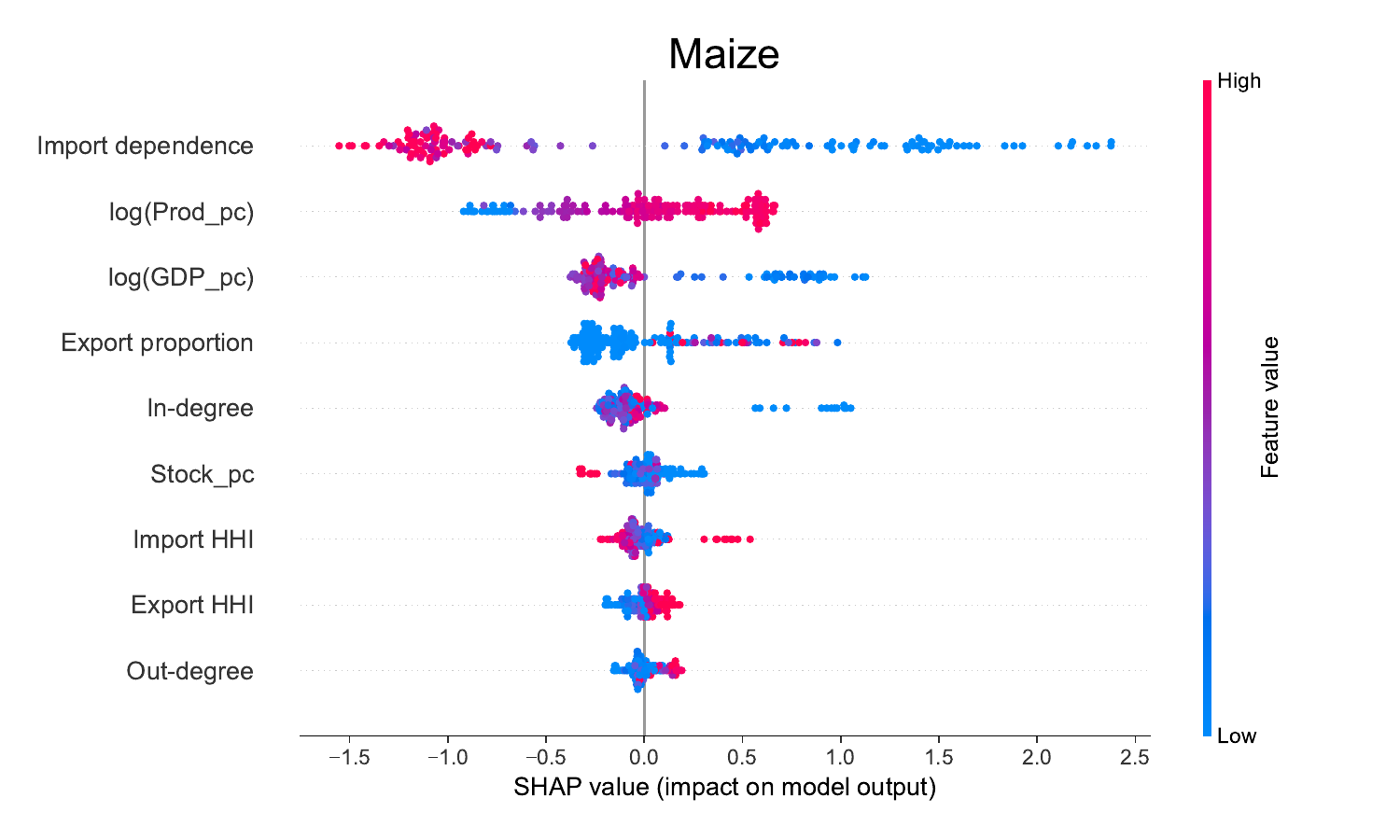}
    \includegraphics[width=0.48\linewidth]{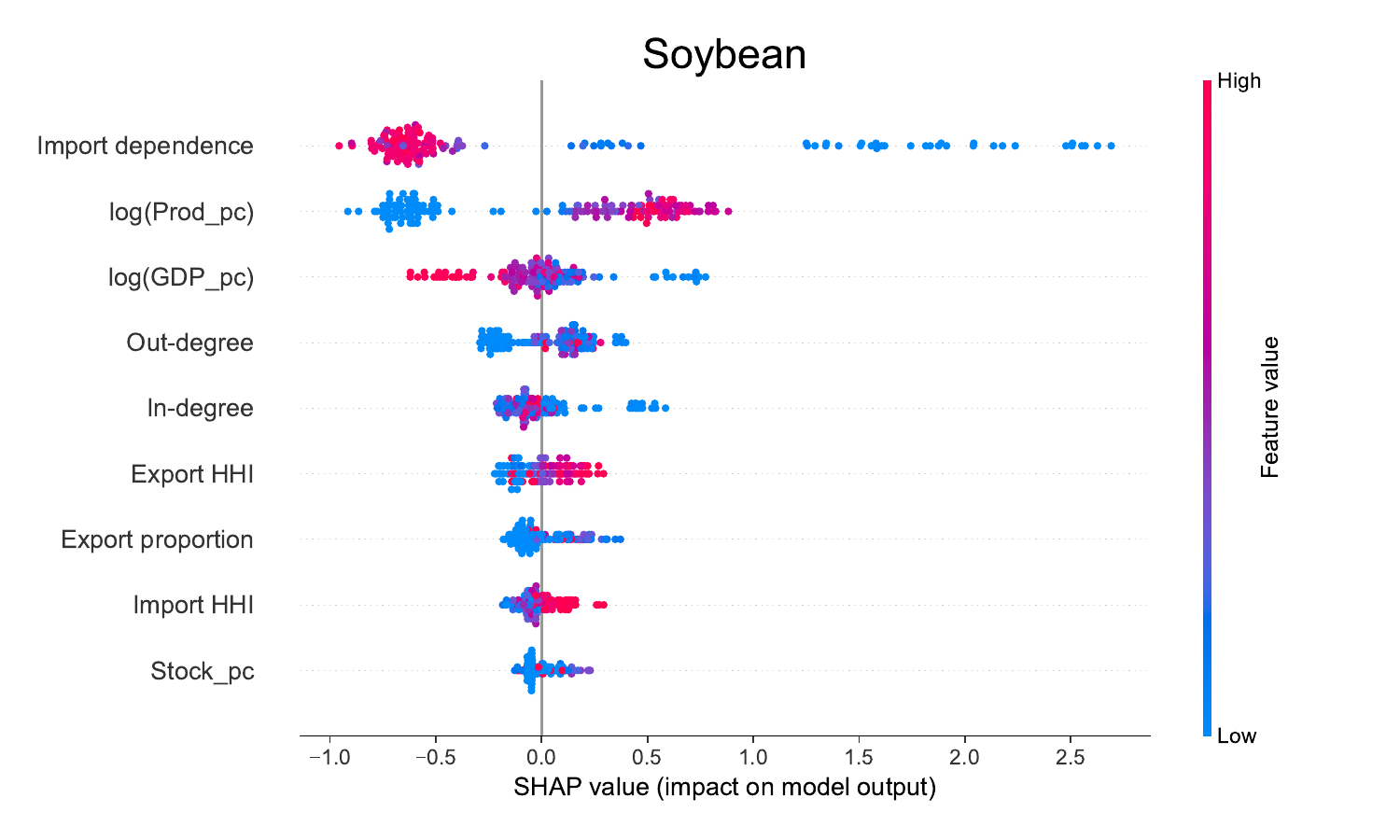}
    \caption{SHAP-based feature analysis of the robustness of economies}
    \label{Fig:Robustness_SHAP}
\end{figure}

To clarify the directional impact of these features on supply robustness, we generated SHAP summary plots as illustrated in Fig.~\ref{Fig:Robustness_SHAP}. In these plots, positive and negative SHAP values represent positive and negative impacts, respectively. Each point represents the contribution of a specific sample for a given feature, where the color transition from blue to red indicates an increasing feature value. Specifically, red points positioned to the right signify that a higher feature value enhances an economy's robustness, while those to the left indicate the opposite.

The SHAP results further clarify the direction of these effects. For all crop groups, high values of import dependence are associated with negative SHAP values, while low values tend to contribute positively to robustness. By contrast, higher domestic production per capita generally shifts SHAP values in the positive direction. 
For maize and soybeans, which are primarily cash crops, the importance of GDP ranks immediately after import dependency and production. Interestingly, higher GDP is associated with lower supply robustness for these two crops. This phenomenon may be attributed to the more intensive use of maize and soybeans as intermediate inputs within high income economies.

\section{Counterfactual Policy Evaluation}
\label{S5:EmpAnal2}

Based on the findings in Section~\ref{Subsection:Determinants}, import dependency and per capita food production emerge as the primary factors exerting a clear and significant influence on an economy's supply robustness. However, production levels are largely dictated by climate, arable land, and production efficiency, making optimization constrained. 
Consequently, we focus our policy optimization on economies characterized by high import dependency. We propose strategic interventions specifically targeting the reserves and trade structures of these economies, with detailed policy configurations presented in Table~\ref{tab:policy_description}.
\begin{table}[htbp]
\centering
\caption{Description of policy scenarios.}
\label{tab:policy_description}
\begin{tabular*}{\textwidth}{@{\extracolsep{\fill}}p{0.10\textwidth}p{0.850\textwidth}@{}}
\toprule
Policy & Descriptive summary \\
\midrule
Policy 1 &
For economies in the top 20\% of import dependence, food stocks are increased by 20\% for those with existing stocks, while economies without reserves are assigned a reserve level equal to the 5th percentile among stock-holding economies. \\
\addlinespace
Policy 2 &
For economies in the top 20\% of import dependence, import allocations are reoptimized toward exporters with higher supply capacity, measured as a weighted index of per capita production and export share using random forest feature importances. \\
\bottomrule
\end{tabular*}
\end{table}

Policy 1 establishes enhanced stock buffers for economies with high import dependency. Policy 2 focuses on restructuring the trade frameworks of these economies. 
We first assessed the supply capacity of each economy, primarily considering their production levels and export shares. Based on the feature importance of production and export ratios derived from a Random Forest model, we then calculated the respective weights:
\begin{equation}
w_{\mathrm{prod}} = \frac{\mathrm{Imp}_{\mathrm{prod}}}{\mathrm{Imp}_{\mathrm{prod}} + \mathrm{Imp}_{\mathrm{exp}}}, 
\qquad
w_{\mathrm{exp}} = \frac{\mathrm{Imp}_{\mathrm{exp}}}{\mathrm{Imp}_{\mathrm{prod}} + \mathrm{Imp}_{\mathrm{exp}}}, 
\label{Eq:Weight}
\end{equation}
Upon obtaining the weights, we calculated the supply capacity ($SC_i$) for each economy:
\begin{equation}
SC_i
= w_{\mathrm{prod}} \cdot \mathrm{norm}\!\left(\log P_i\right)
+ w_{\mathrm{exp}} \cdot \mathrm{norm}\!\left(\mathrm{ExportProportion}_i\right), 
\label{Eq:SC}
\end{equation}
Given the disparate units of production volume and export share, a normalization process was applied to ensure comparability.
After determining the supply capacity for each economy, we redistributed the total imports of each economy to obtain $p_{ij}$:
\begin{equation}
p_{ij}
= \frac{SC_i}{\sum\limits_{k \in \mathcal{N}_j} SC_k}
\label{Eq:p_ij}
\end{equation}
A policy intensity parameter $\delta$ is introduced to control the degree of convergence toward the target structure, yielding the final trade share $s'_{ij}$:
\begin{equation}
s'_{ij} = (1 - \delta)\, s_{ij} + \delta\, p_{ij},
\label{Eq:S_redistribution}
\end{equation}
We set the policy intensity $\delta = 0.1$. 
Once the adjusted trade shares were obtained, we employed the Iterative Proportional Fitting (IPF) algorithm to ensure that the total imports and exports of all economies remained unchanged, ultimately deriving the final trade matrix $F'$.

\begin{figure}[h!]
    \centering
    \includegraphics[width=0.48\linewidth]{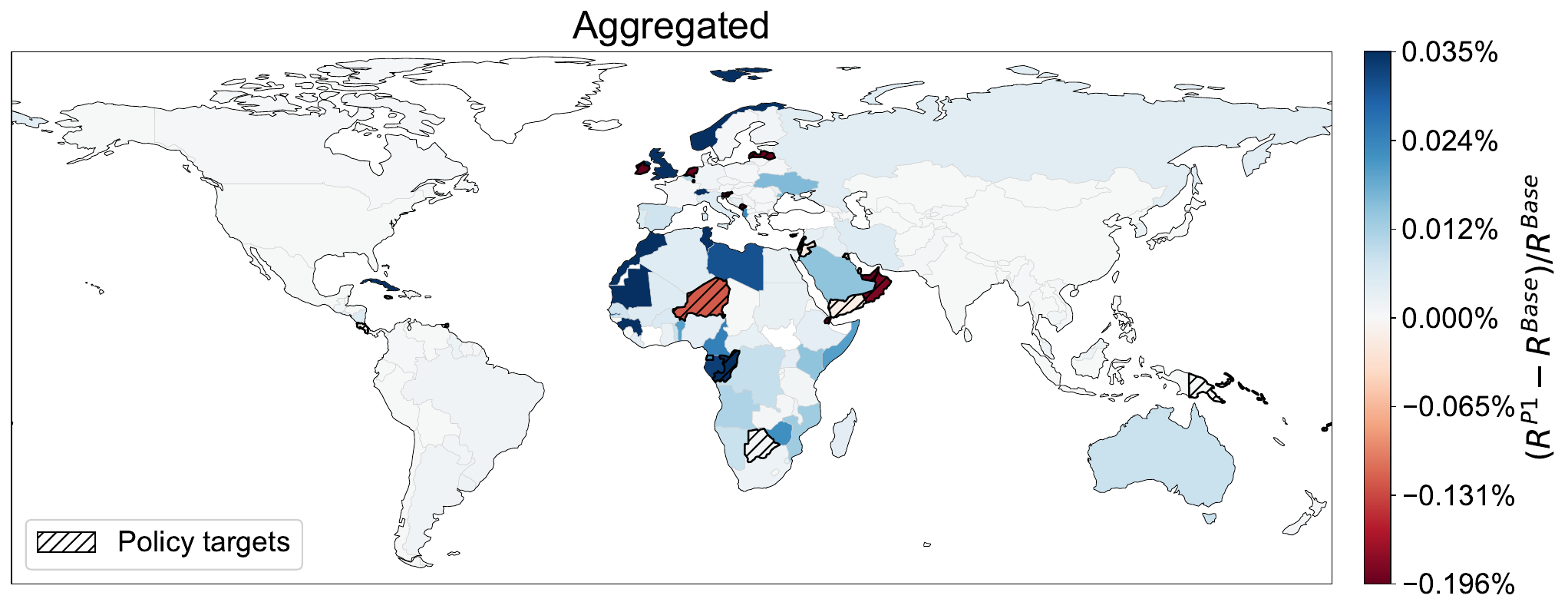}\\
    \includegraphics[width=0.48\linewidth]{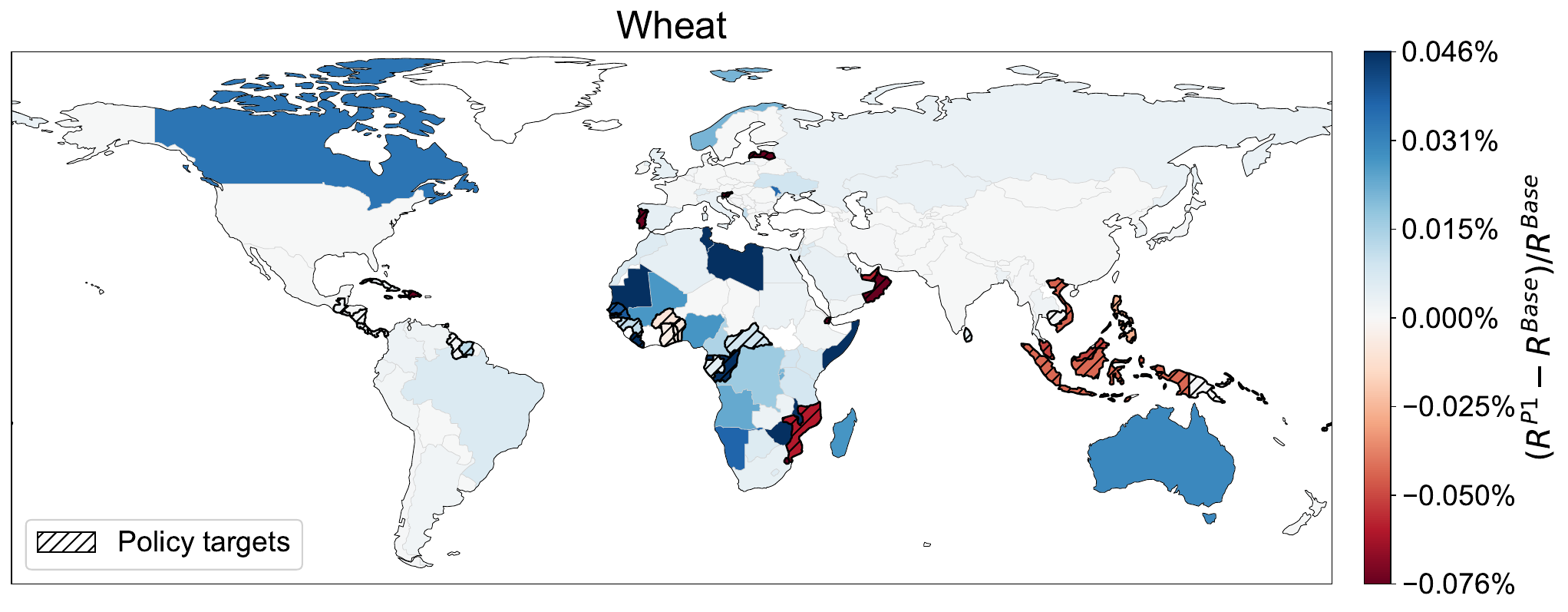}
    \includegraphics[width=0.48\linewidth]{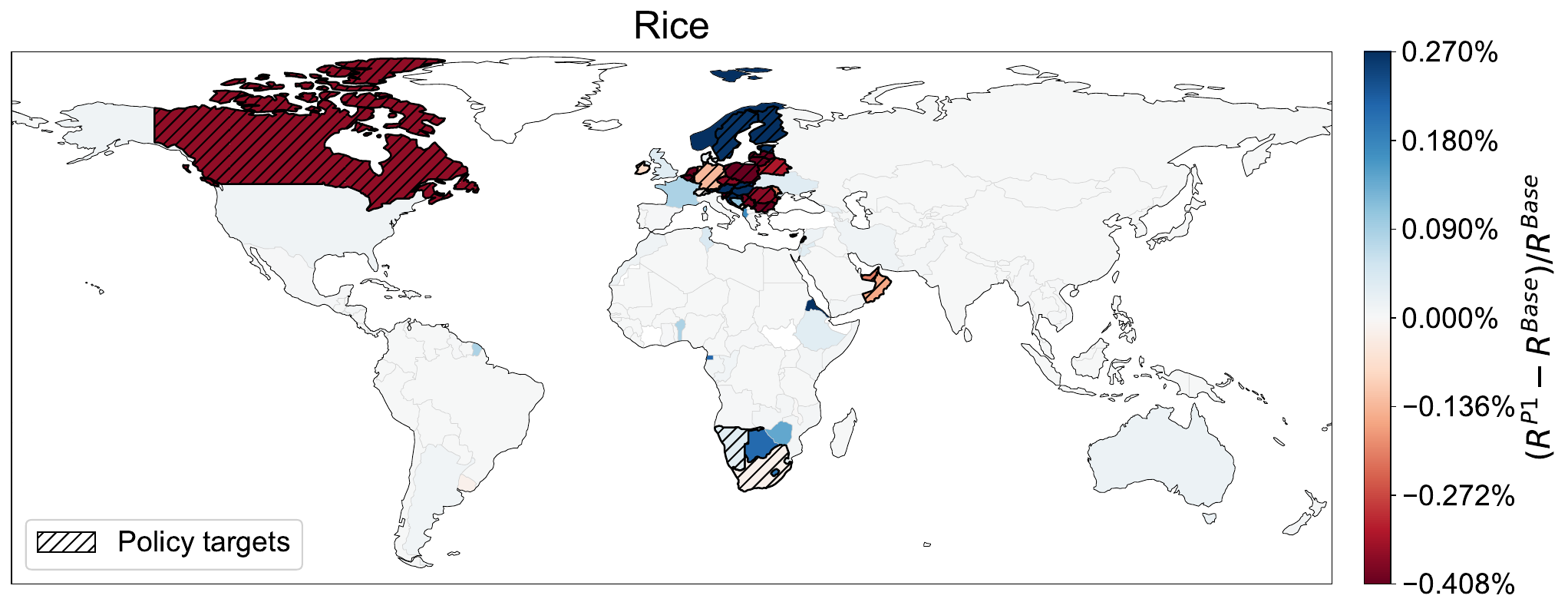}\\
    \includegraphics[width=0.48\linewidth]{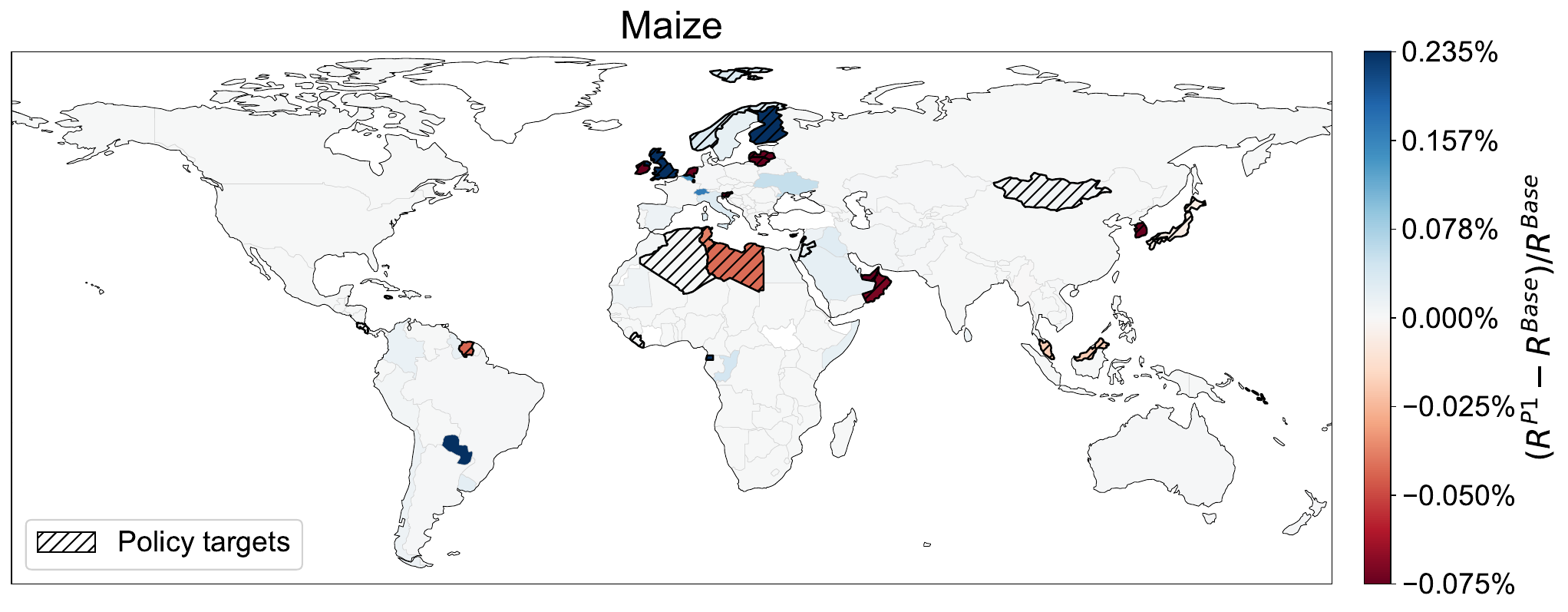}
    \includegraphics[width=0.48\linewidth]{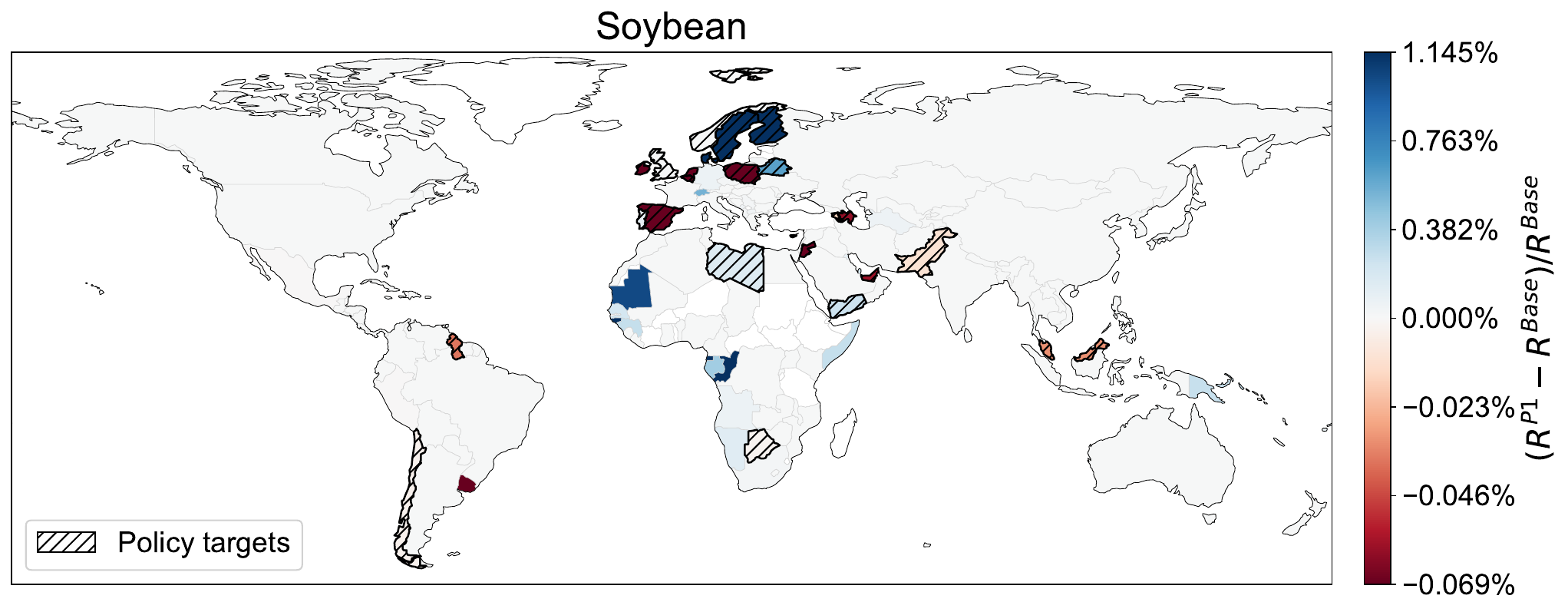}
    \caption{Effects of Policy 1 on robustness. Policy 1 increases food stocks for the top 20\% most import-dependent economies, and colors indicate the proportional change in robustness relative to the baseline.}
    \label{Fig:Policy1_Results}
\end{figure}

\begin{figure}[h!]
    \centering
    \includegraphics[width=0.48\linewidth]{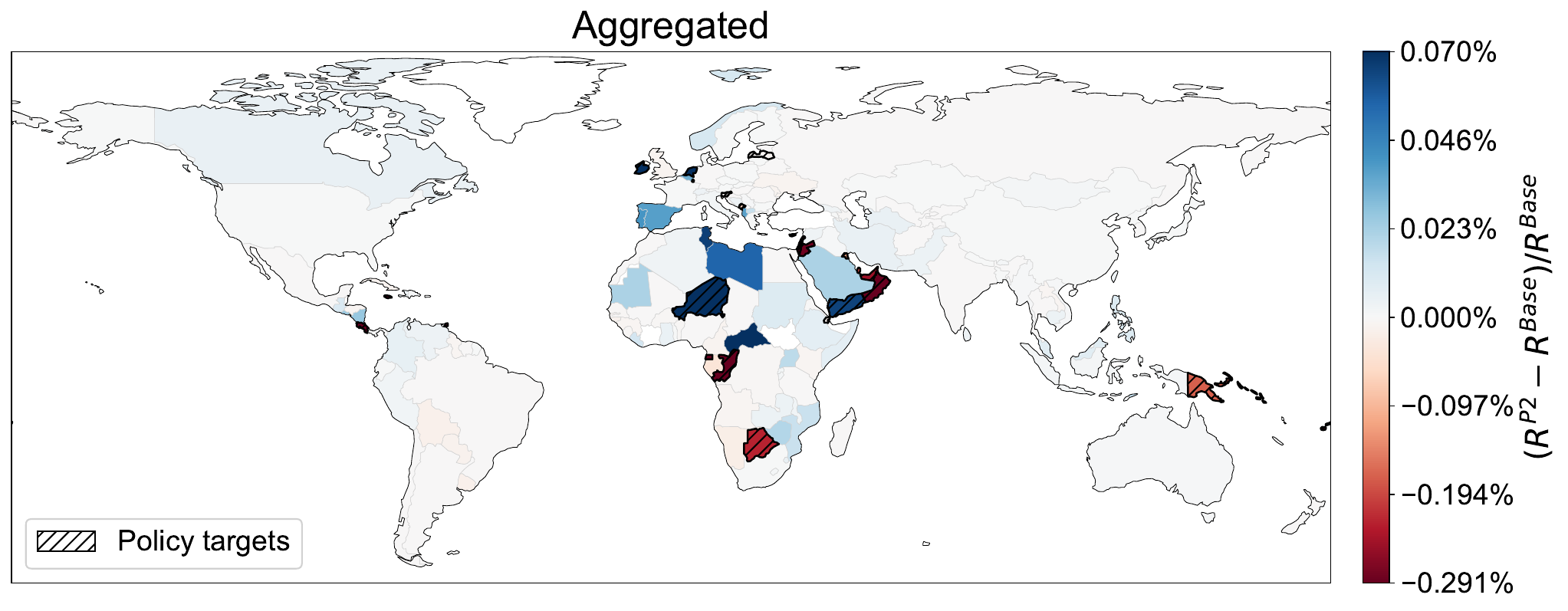}\\
    \includegraphics[width=0.48\linewidth]{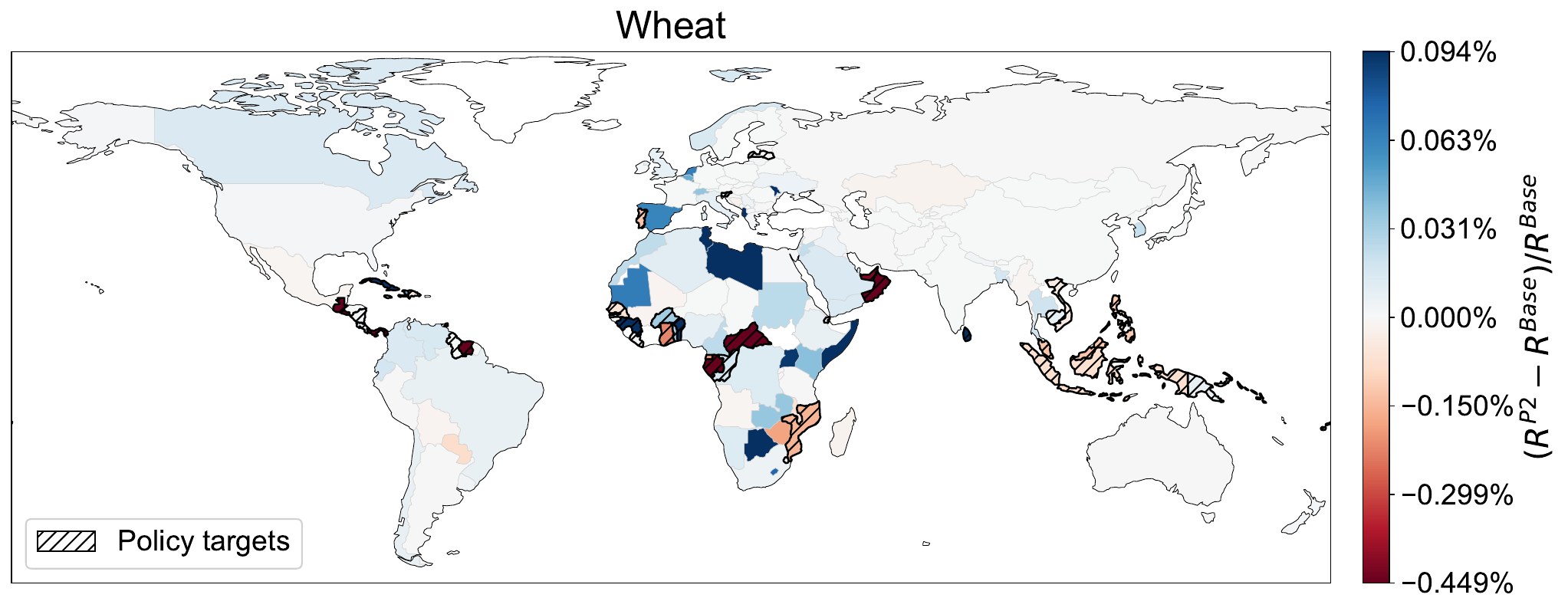}
    \includegraphics[width=0.48\linewidth]{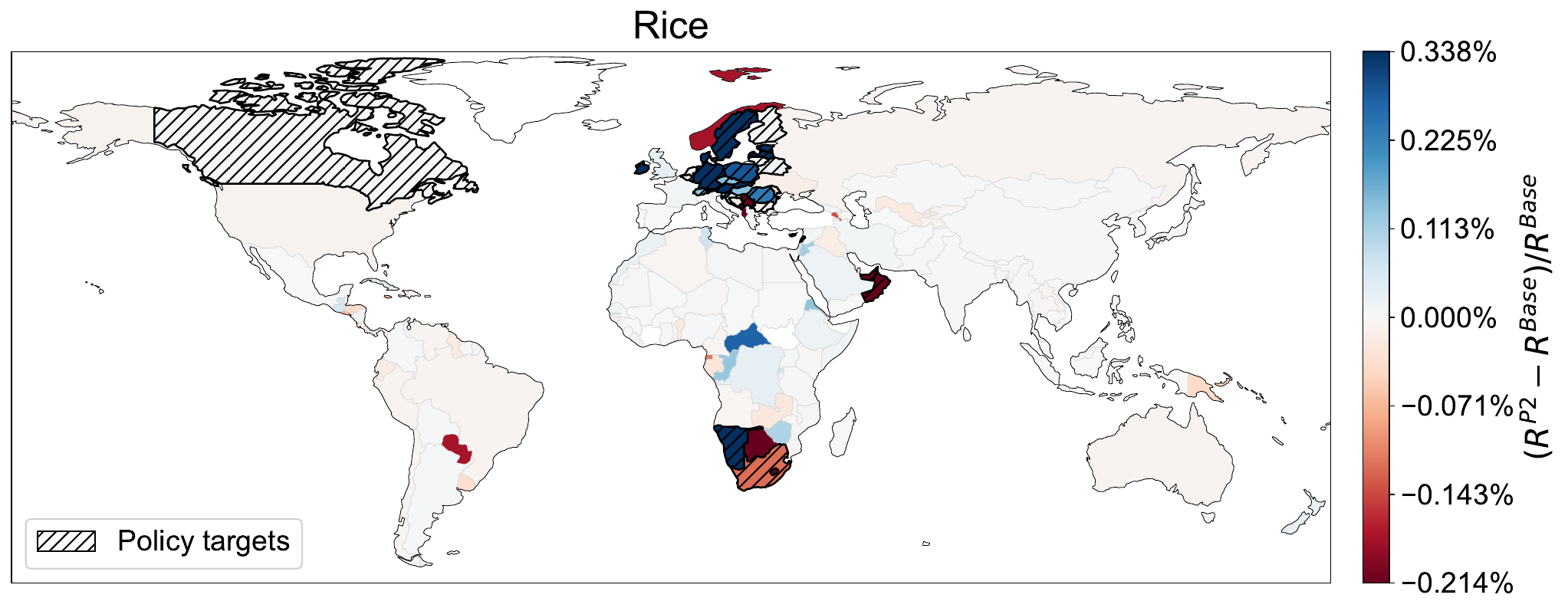}\\
    \includegraphics[width=0.48\linewidth]{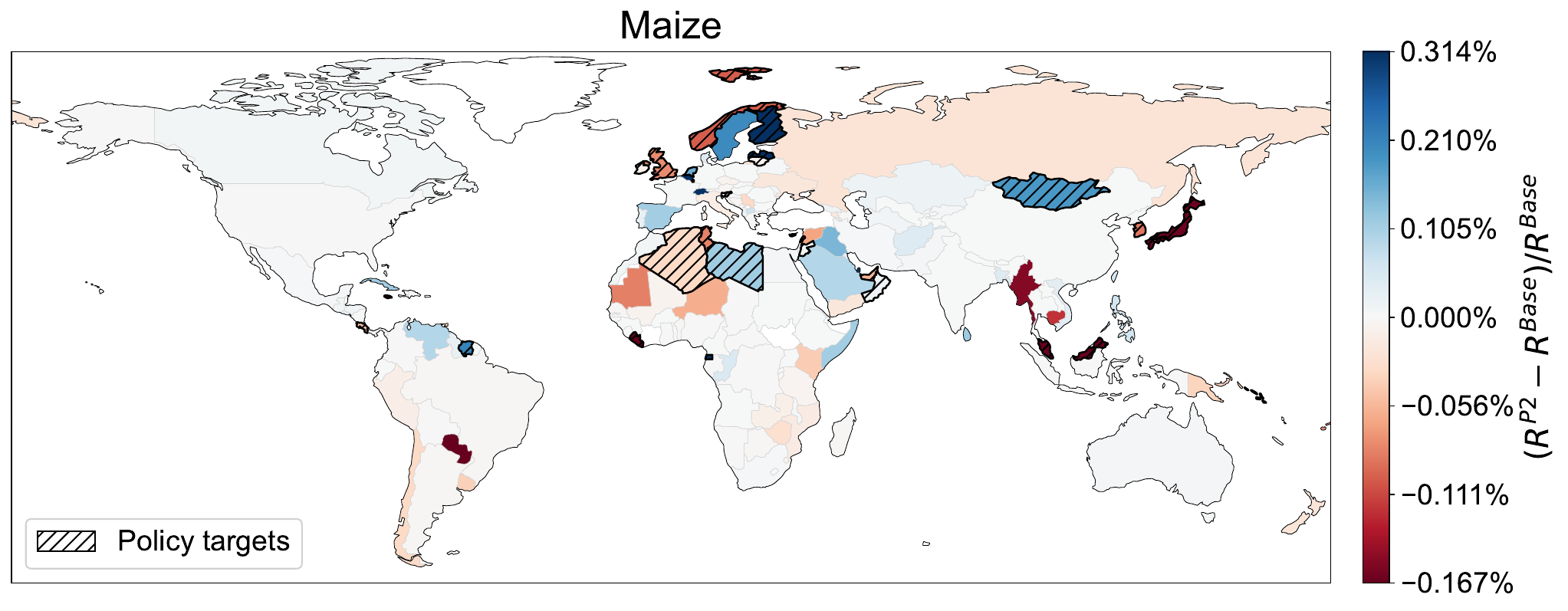}
    \includegraphics[width=0.48\linewidth]{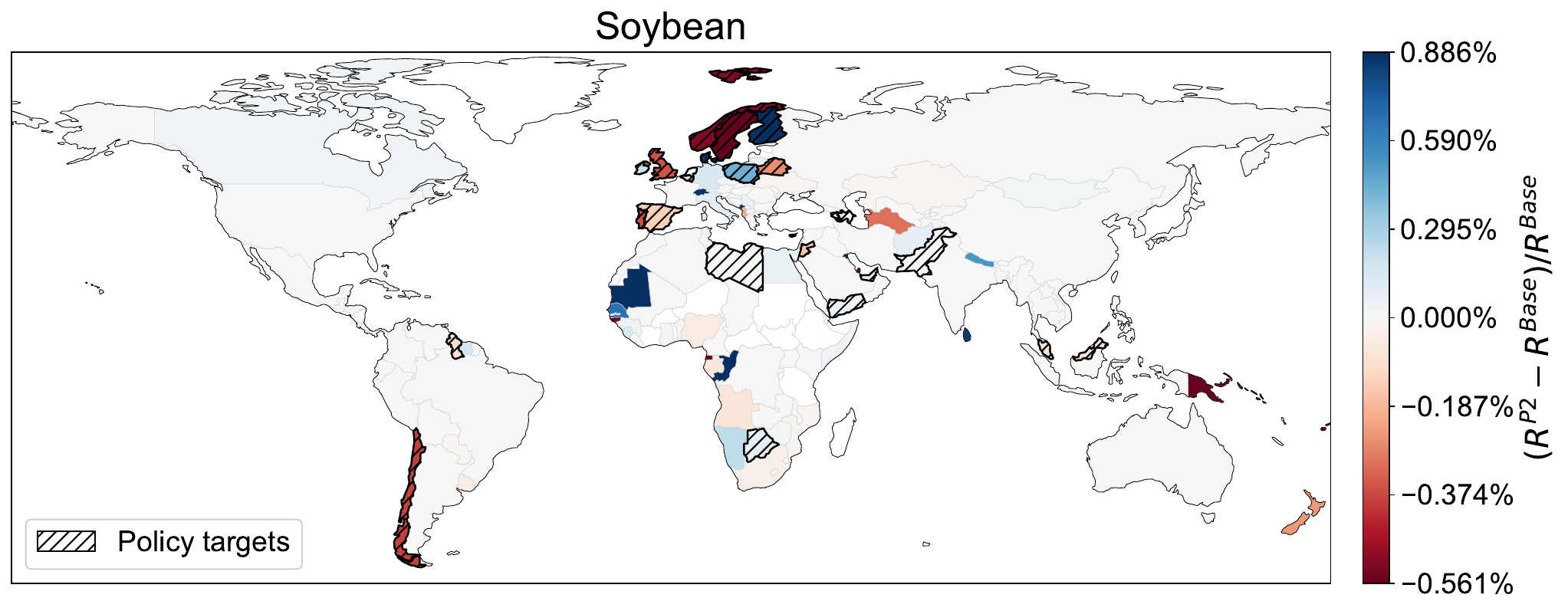}
    \caption{Effects of Policy 2 on robustness. Policy 2 reoptimizes import allocations for the top 20\% most import-dependent economies toward exporters with higher supply capacity, and colors indicate the proportional change in robustness relative to the baseline.}
    \label{Fig:Policy2_Results}
\end{figure}

Fig.~\ref{Fig:Policy1_Results} and Fig.~\ref{Fig:Policy2_Results} illustrate the changes in robustness across various economies following the implementation of Policy 1 and Policy 2, respectively. The shaded economies represent the targets of the policy implementation, which are those in the top 20\% of import dependency. We then evaluate the effectiveness of these two policies.
Overall, the improvements in robustness brought by the two policies are relatively modest and exhibit significant heterogeneity across different crops and economies.

Under Policy 1, increasing stocks does not consistently enhance the robustness of all high import-dependent economies. This outcome is related to our model specification where stock buffers are only applied to offset supply reductions rather than being integrated into the baseline supply. However, for total food aggregates and wheat, increasing the stocks of highly dependent economies generates substantial positive robustness spillovers for other nations. Under the aggregated system, these gains are concentrated mainly in parts of Africa, the Middle East, and Europe, while effects elsewhere remain limited. In the wheat system, the positive responses are more widespread, extending across much of Africa and several traditional import-dependent regions, although some economies in Southeast Asia experience declines. This phenomenon may arise because other countries can achieve more favorable trade structures when high-dependency economies use their own stocks for self-buffering.
Conversely, for rice, increasing these stocks significantly worsens the situation for some economies. Under the rice system, while several European economies and parts of southern Africa show improvements, the large negative response in Canada suggests that the policy substantially reallocates robustness across the network rather than improving it uniformly, even among targeted economies. This may reflect the fact that, for some economies, the original trade-adjustment mechanism provided more favorable supply stabilization.
Compared with wheat and the aggregated food system, maize and soybean exhibit more spatially localized responses to Policy 1. In the maize system, most economies remain close to zero change, while the observed effects are scattered across a small number of locations. This suggests that Policy 1 does not substantially reshape robustness across the broader maize network, although it generates clear improvements in selected economies. In the soybean system, the response is also limited in spatial extent but more concentrated in magnitude, with gains mainly observed in Northern Europe and selected African economies, alongside localized losses in Western Europe, South America, and parts of Asia. Relative to maize, soybean displays stronger robustness improvements in specific economies, indicating a more targeted but uneven policy effect.

Policy 2 focuses on trade-structure adjustment. As shown in Fig.~\ref{Fig:Policy2_Results}, under this policy, most economies still exhibit relatively limited changes, but the spatial effects are generally more widely distributed than under Policy 1, indicating that trade reconfiguration operates through broader network transmission rather than only localized buffering. As with Policy 1, the responses of non-target economies are often as pronounced as, or even more pronounced than, those of the target economies, suggesting that the effects of the intervention are shaped primarily by indirect reallocation within the trade network.
For aggregate crops, Policy 2 yields relatively favorable improvements for economies in Africa, the Middle East, and parts of Europe, while effects elsewhere remain limited. Wheat shows a broadly similar pattern, with positive responses across several African economies, although negative spillovers are still present in some regions.
Rice exhibits the most favorable response under Policy 2. Several European economies and parts of southern Africa show noticeable improvements, suggesting that, compared with Policy 1, rice supply in these economies is better managed through trade adjustments. 
In contrast, trade reconfiguration fails to generate comparably favorable outcomes for maize and soybean. In maize, the observed effects are mixed and spatially dispersed, indicating that the intervention does not substantially improve robustness across the broader system. In soybean, the number of affected economies is relatively limited, but the pattern remains uneven. This may be because Policy 2 encourages economies to concentrate imports from a smaller set of high-capacity suppliers, which can exacerbate systemic vulnerability in crop systems such as maize and soybean that already rely on a limited number of dominant nodes.

\section{Conclusions}
\label{S6:Conclude}

This study constructs a global calorie-based food supply network for wheat, rice, maize, and soybean using 2023 production, reserve, and bilateral trade data for economies worldwide. Building on an existing dynamic shock propagation framework \citep{Marchand-Carr-Dell'Angelo-Fader-Gephart-Kummu-Magliocca-Porkka-Puma-Ratajczak-Rulli-Seekell-Suweis-Tavoni-D'Odorico-2016-EnvironResLett}, we introduce a behavioral rule that economies prioritize domestic supply, and use the model to systematically evaluate the robustness of food supply across economies under production shocks to major producers. We then investigate the structural factors associated with cross-economy variation in robustness and use counterfactual experiments to assess the extent to which reserve enhancement and trade reconfiguration can improve robustness.

The results on economies’ robustness under production shocks to major producers show pronounced heterogeneity across crops and across economies, indicating that the vulnerability structure of the global food supply system is strongly crop-specific. Among the four crops, wheat displays the highest overall robustness, and its low-robustness economies are relatively clustered, mainly in Africa. Soybean, in contrast, exhibits the lowest overall robustness, with low-robustness economies distributed more diffusely across space. For rice, low-robustness economies are concentrated primarily in Europe.
Group-based results further reveal clear heterogeneity across income and geographic categories. Across income groups, production shocks to different major producers lead to differentiated effects on economies at different income levels. Across geographic groups, the effects are more spatially concentrated, suggesting that shocks from major producers tend to have more regionally clustered and targeted consequences.

Taken together, these patterns suggest that cross-economy differences in robustness are closely related to structural factors, particularly trade-based exposure and domestic supply capacity.
The random forest results provide further support for this interpretation. Across crops, import dependence and per capita production emerge as the two most important factors associated with food supply robustness, whereas the roles of other variables are heterogeneous. Import dependence is consistently linked to lower robustness, while per capita production is consistently associated with higher robustness. For maize and soybean in particular, GDP is the next most important factor after import dependence and per capita production, possibly because higher-income economies tend to use these two crops more intensively as intermediate inputs.

Our study further conducts counterfactual experiments by implementing two policy interventions for the top 20\% of economies with the highest import dependency, namely increasing stocks and adjusting trade structures. The results show that both interventions lead to only modest improvements in robustness overall, with effects varying substantially across crops and economies. For the aggregated food system and wheat, both policies produce certain gains, and the positive spillover effects are more evident in Africa. For rice, trade adjustment delivers larger robustness improvements than stock increases. By contrast, for maize and soybean, stock increases generate relatively stronger but spatially concentrated robustness gains, mainly benefiting selected initially vulnerable economies rather than producing broad improvements. Trade adjustment provides little benefit for these two crops and may even generate negative spillovers for some vulnerable economies. Collectively, these findings indicate that neither reserve expansion nor trade reallocation alone is sufficient to fundamentally strengthen the robustness of the global food supply system, and that achieving simultaneous improvements in both national and system-wide robustness remains a challenge for future research.

\appendix
\section{Economy-level relative food supply curves under production shocks to major producers}
\label{Appendix}

Figures~\ref{Fig:Wheat_Qrel_global1pct_1} illustrates the changes in relative supply for various economies when major wheat producers experience production shocks. The horizontal axis represents the shock intensity, quantified as the percentage of global wheat production lost by a major producer. The vertical axis shows the relative supply level. Each line corresponds to a shock scenario originating from a specific major producer, while the dashed black line represents the ensemble average across all scenarios.

\setcounter{figure}{0}
\begin{figure}[h!]
    \centering
    \includegraphics[width=0.24\linewidth]{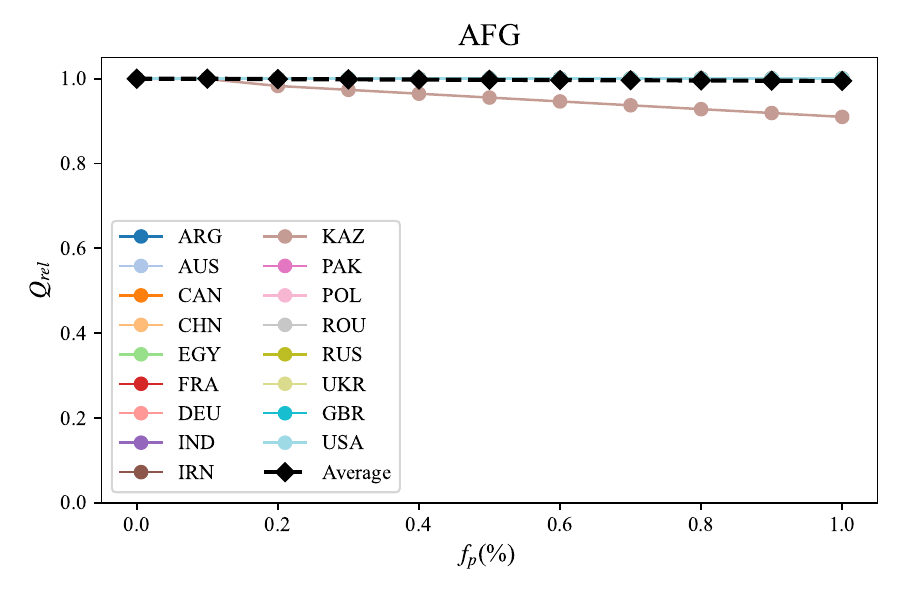}
    \includegraphics[width=0.24\linewidth]{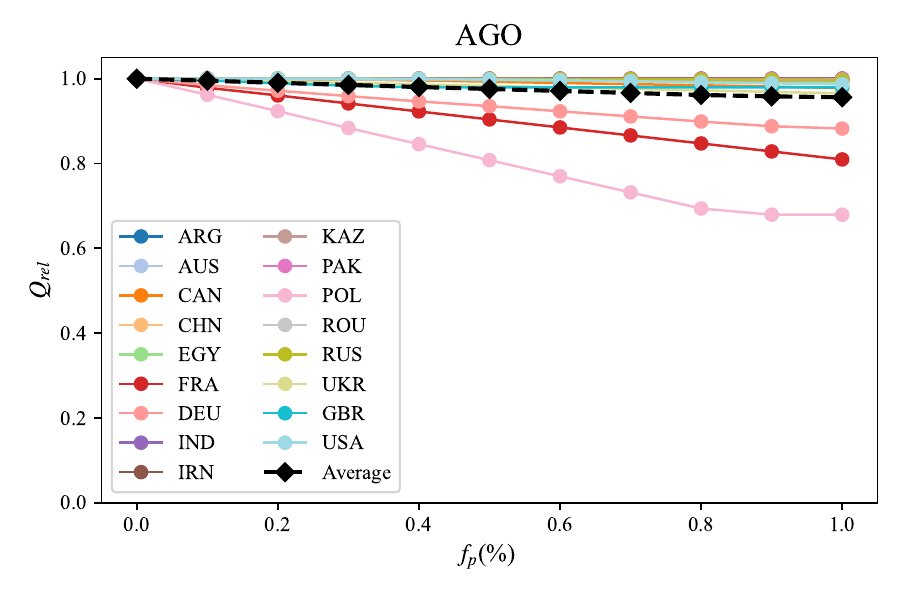}
    \includegraphics[width=0.24\linewidth]{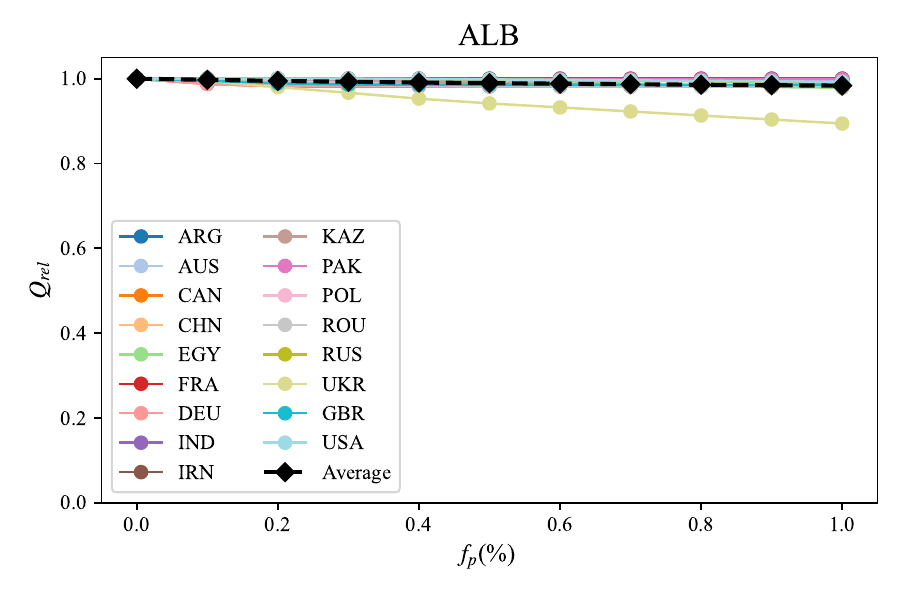}
    \includegraphics[width=0.24\linewidth]{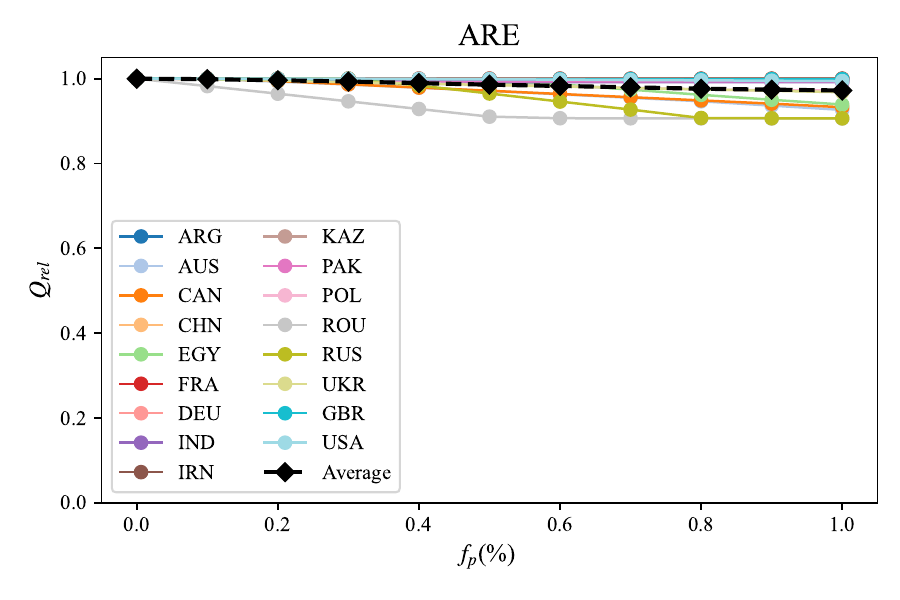}\\
    \includegraphics[width=0.24\linewidth]{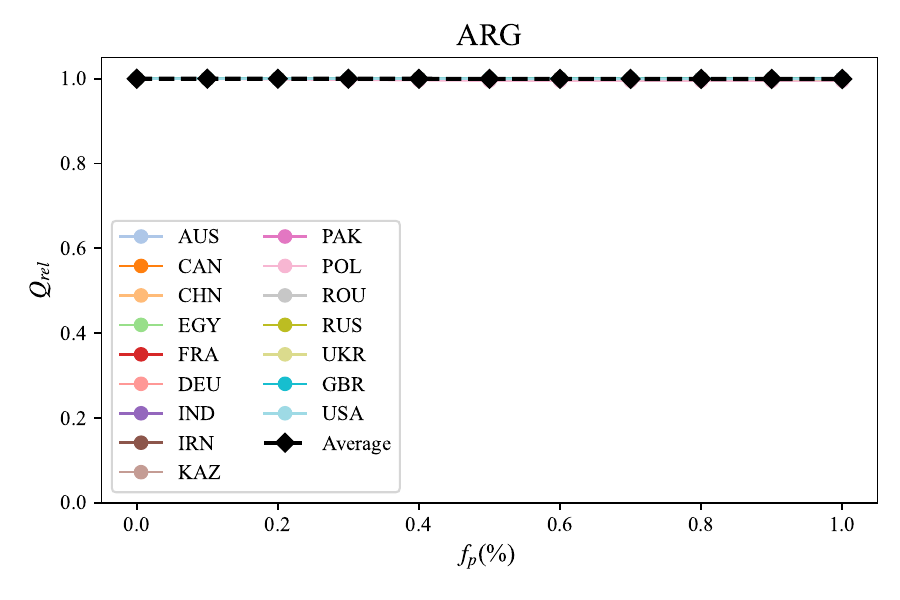}
    \includegraphics[width=0.24\linewidth]{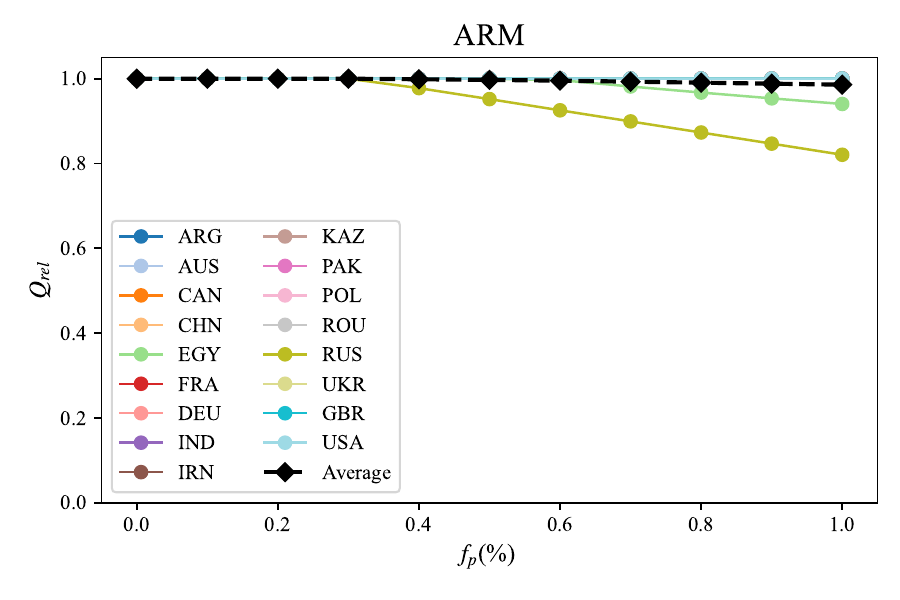}
    \includegraphics[width=0.24\linewidth]{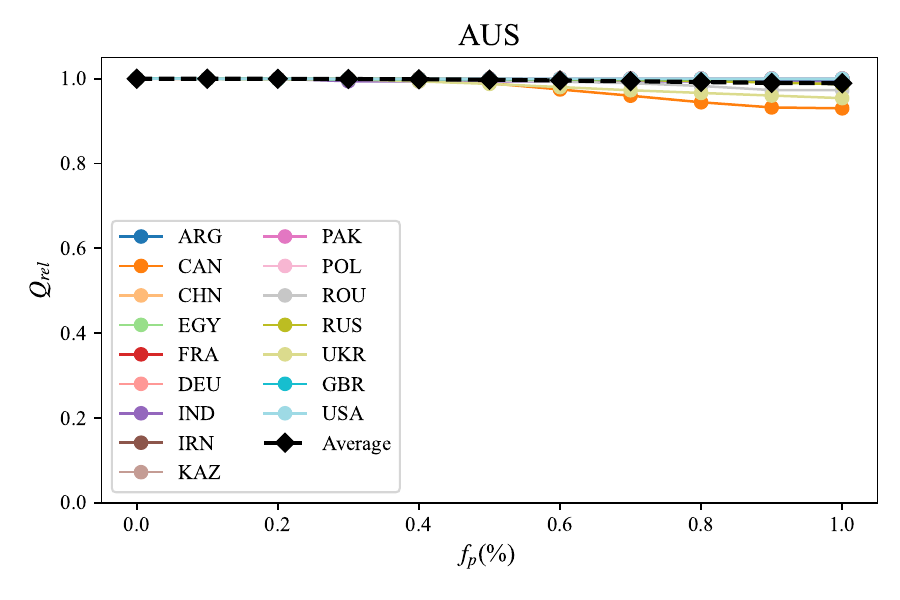}
    \includegraphics[width=0.24\linewidth]{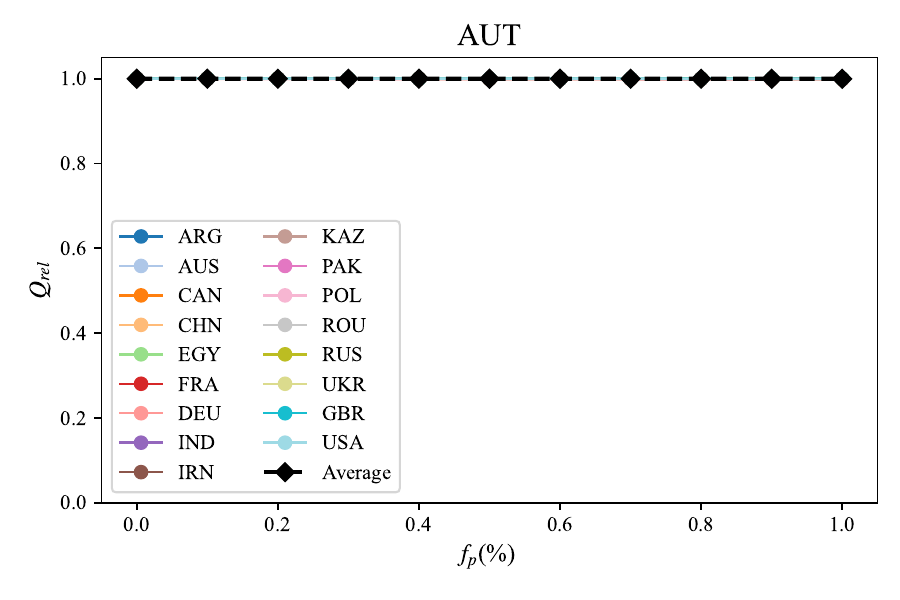}\\
    \includegraphics[width=0.24\linewidth]{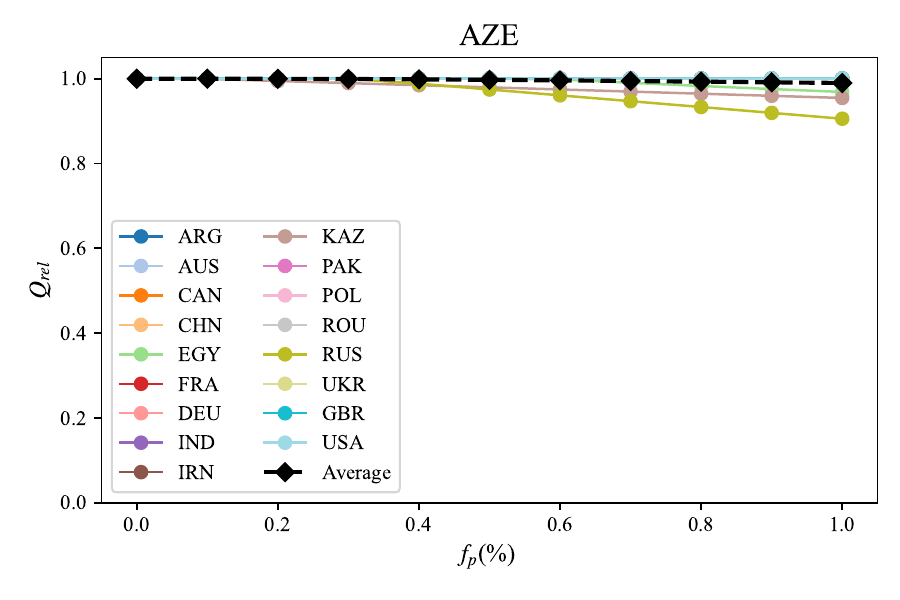}
    \includegraphics[width=0.24\linewidth]{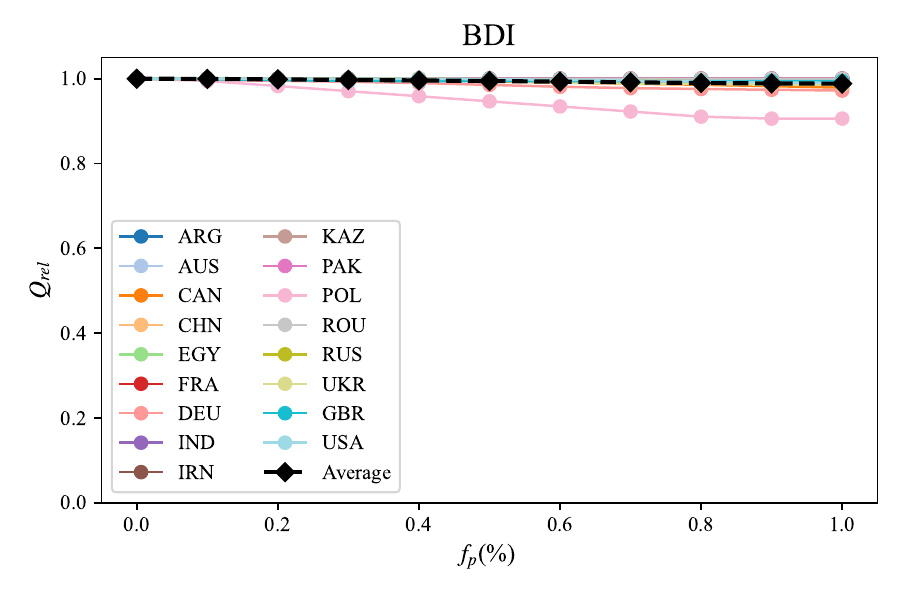}
    \includegraphics[width=0.24\linewidth]{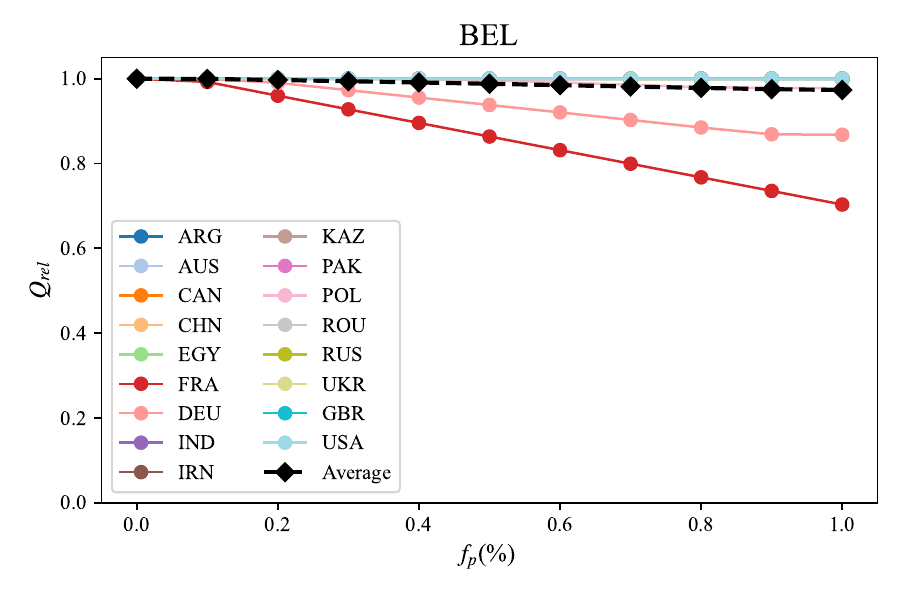}
    \includegraphics[width=0.24\linewidth]{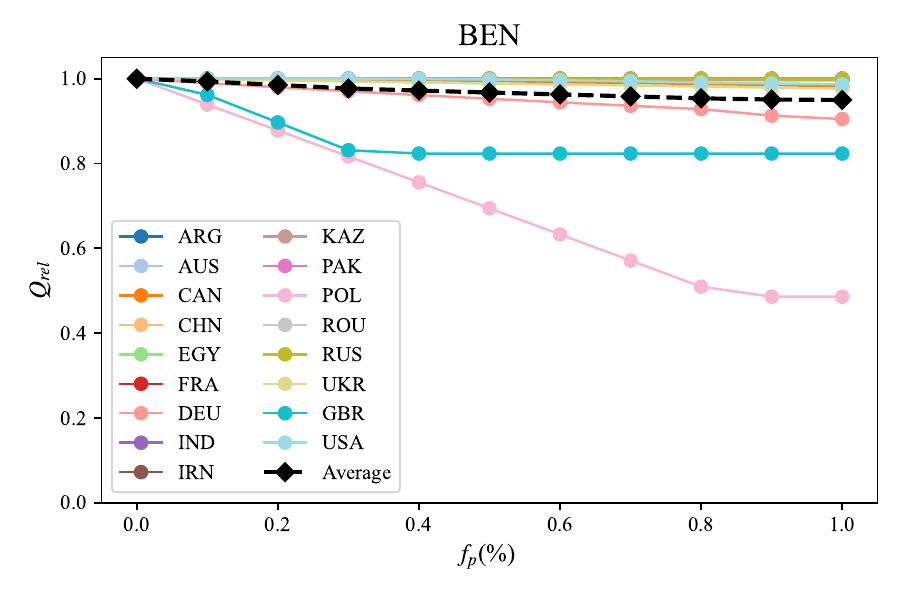}\\
    \includegraphics[width=0.24\linewidth]{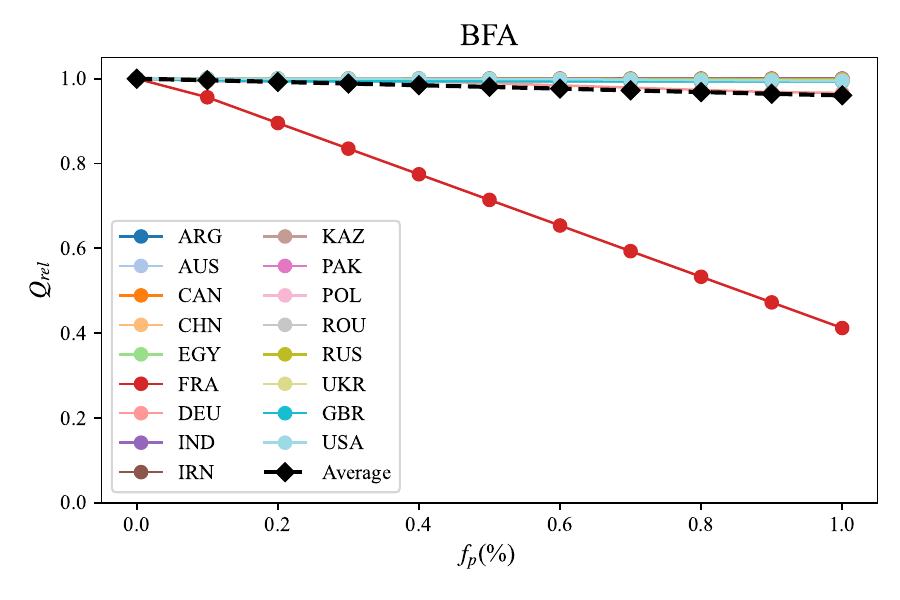}
    \includegraphics[width=0.24\linewidth]{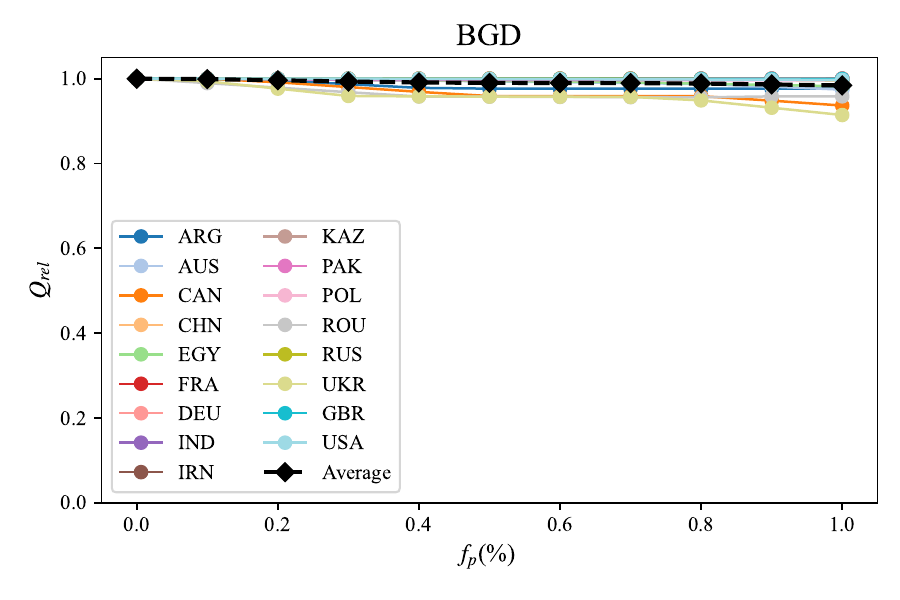}
    \includegraphics[width=0.24\linewidth]{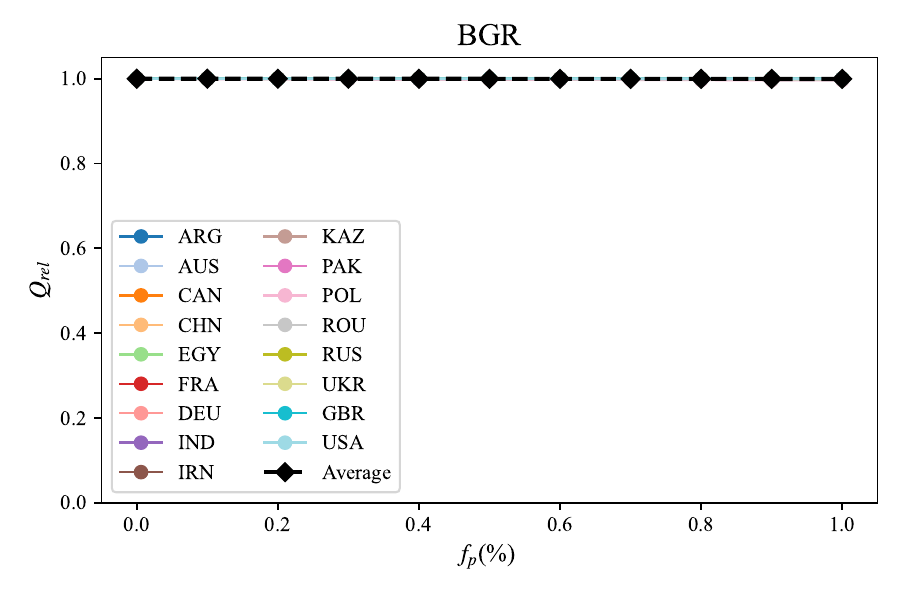}
    \includegraphics[width=0.24\linewidth]{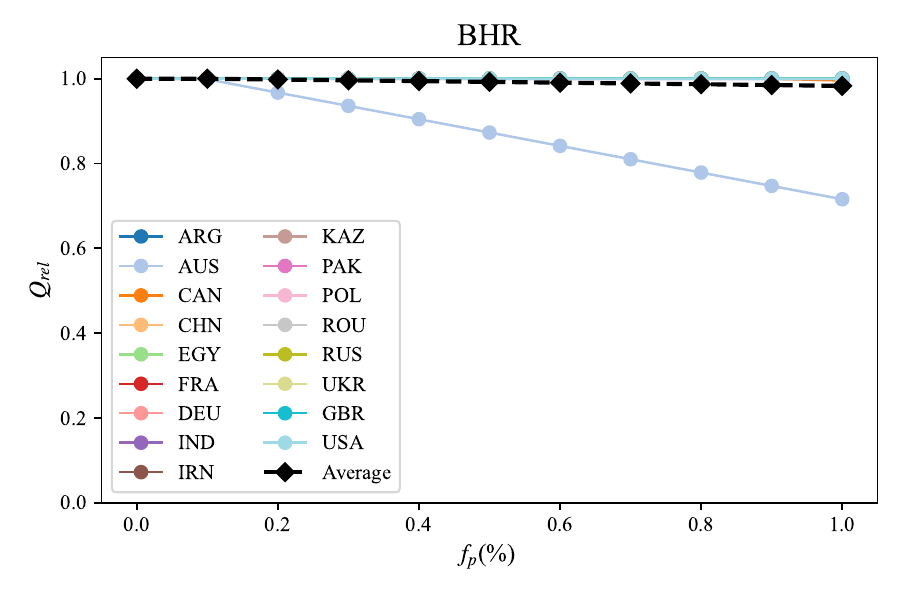}\\
    \includegraphics[width=0.24\linewidth]{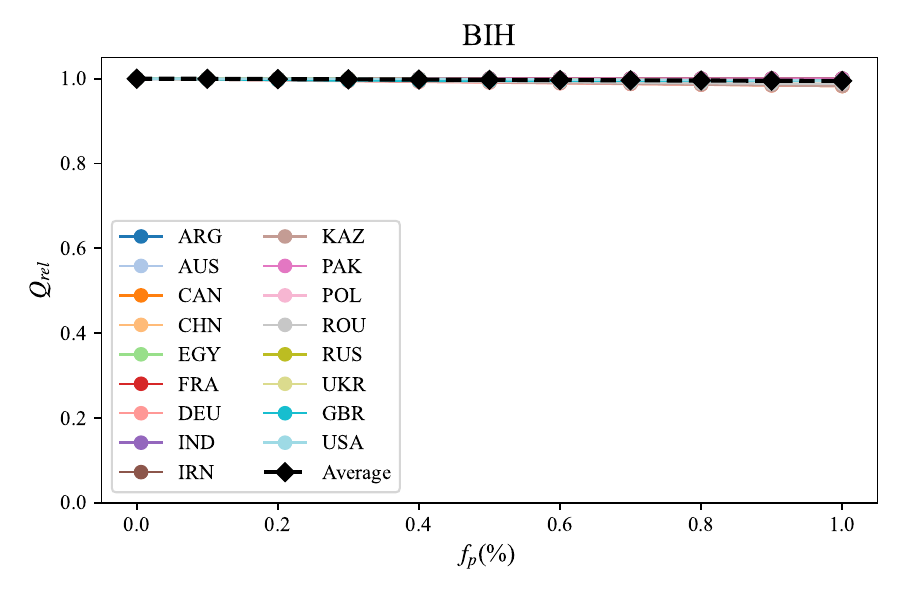}
    \includegraphics[width=0.24\linewidth]{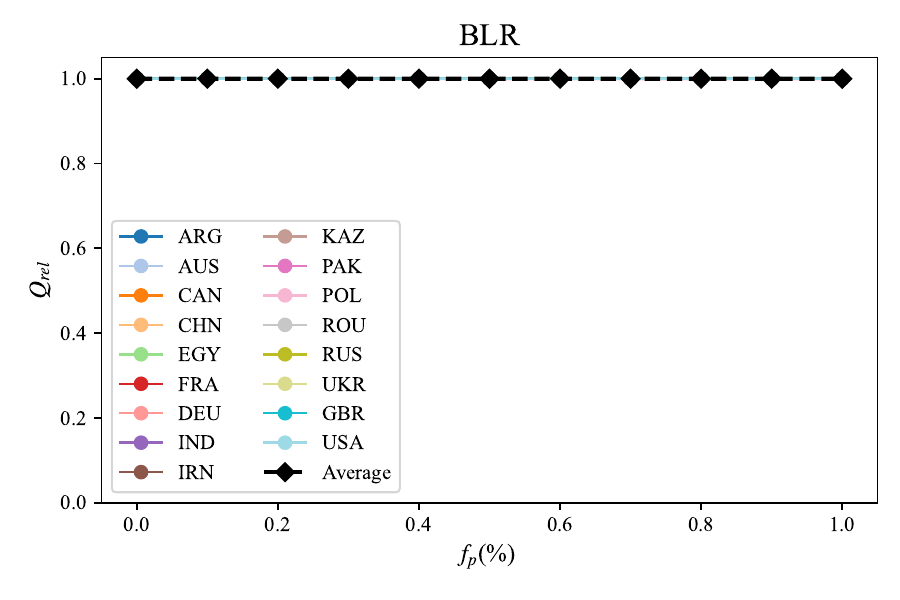}
    \includegraphics[width=0.24\linewidth]{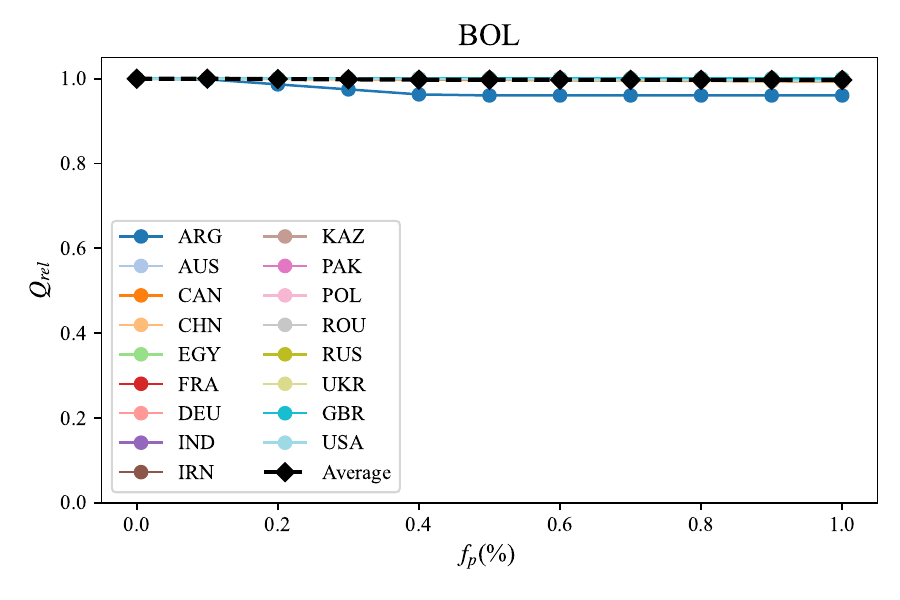}
    \includegraphics[width=0.24\linewidth]{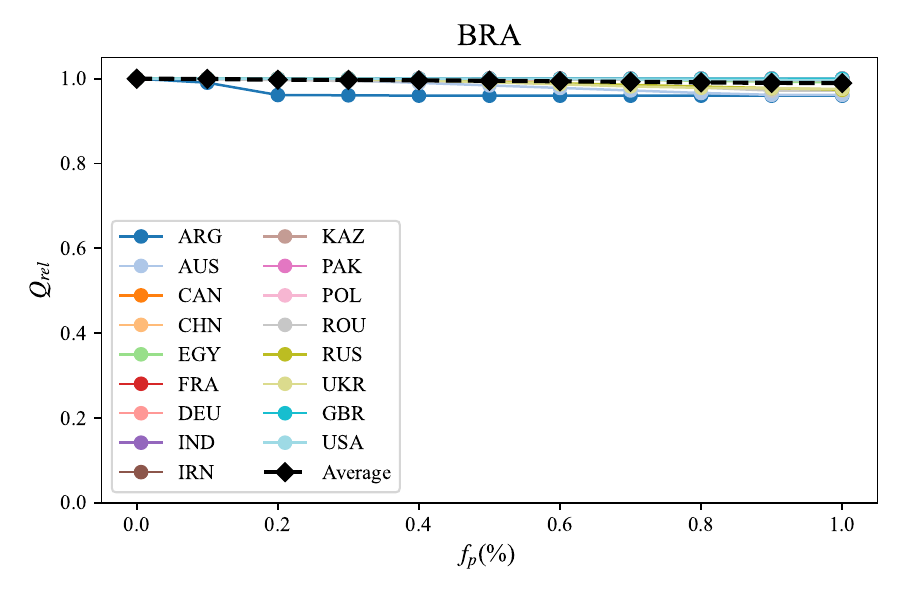}\\
    \includegraphics[width=0.24\linewidth]{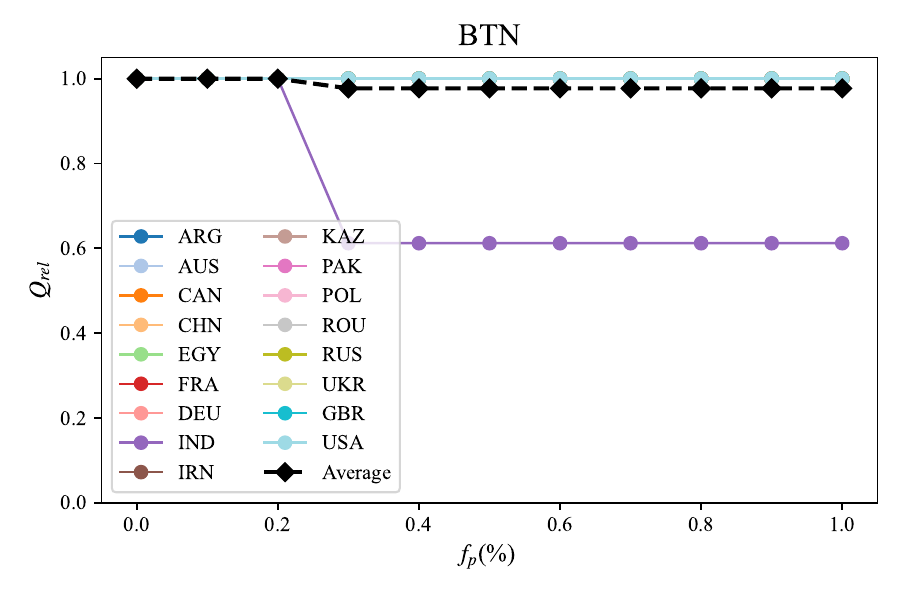}
    \includegraphics[width=0.24\linewidth]{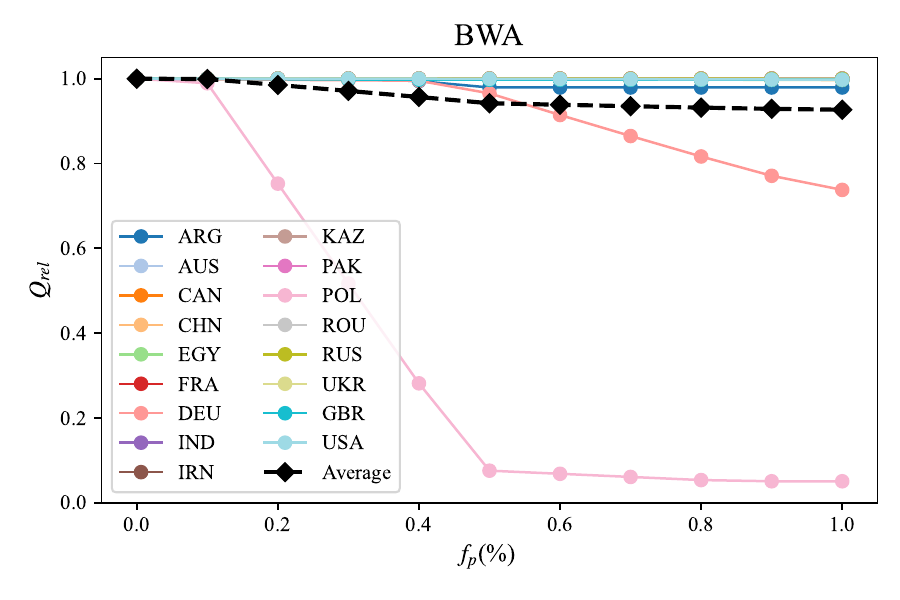}
    \includegraphics[width=0.24\linewidth]{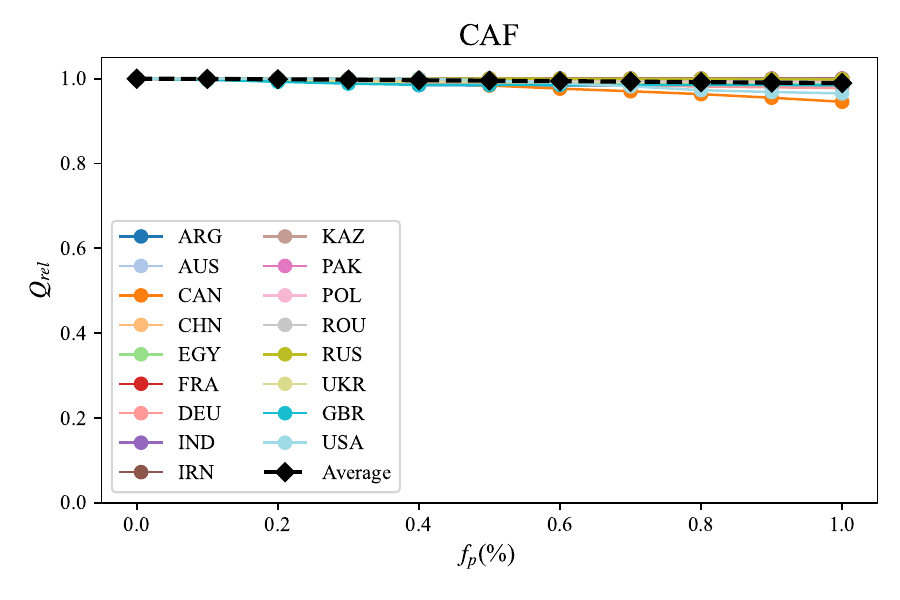}
    \includegraphics[width=0.24\linewidth]{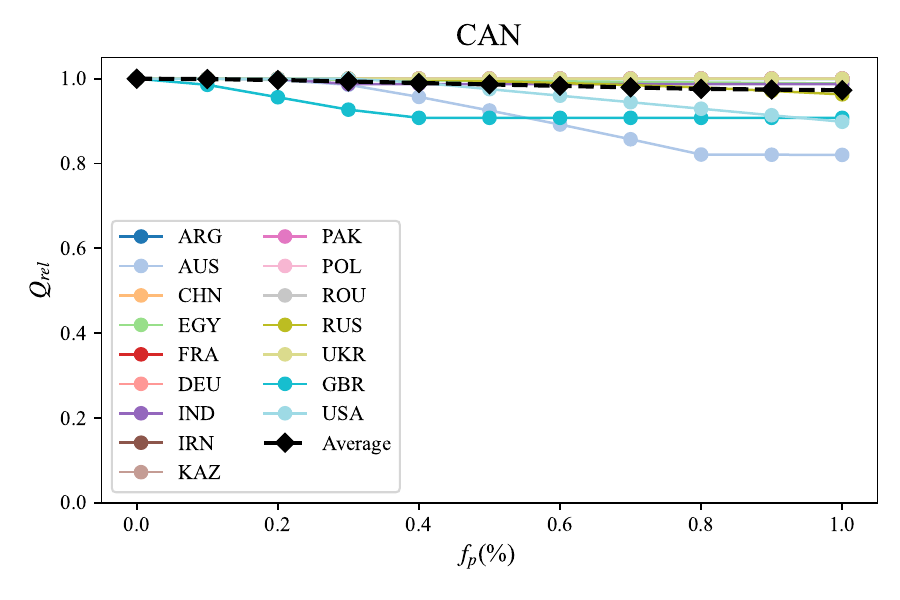}\\
    \includegraphics[width=0.24\linewidth]{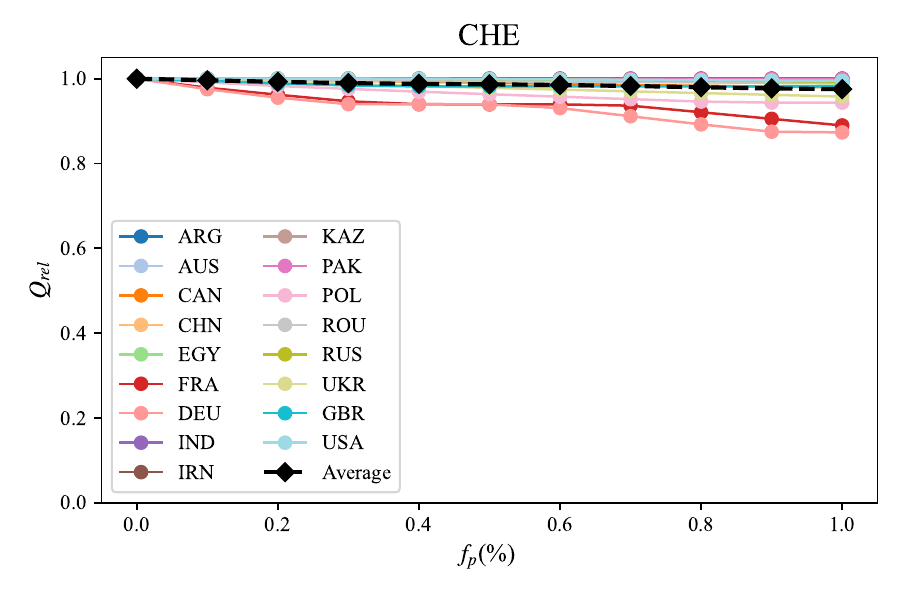}
    \includegraphics[width=0.24\linewidth]{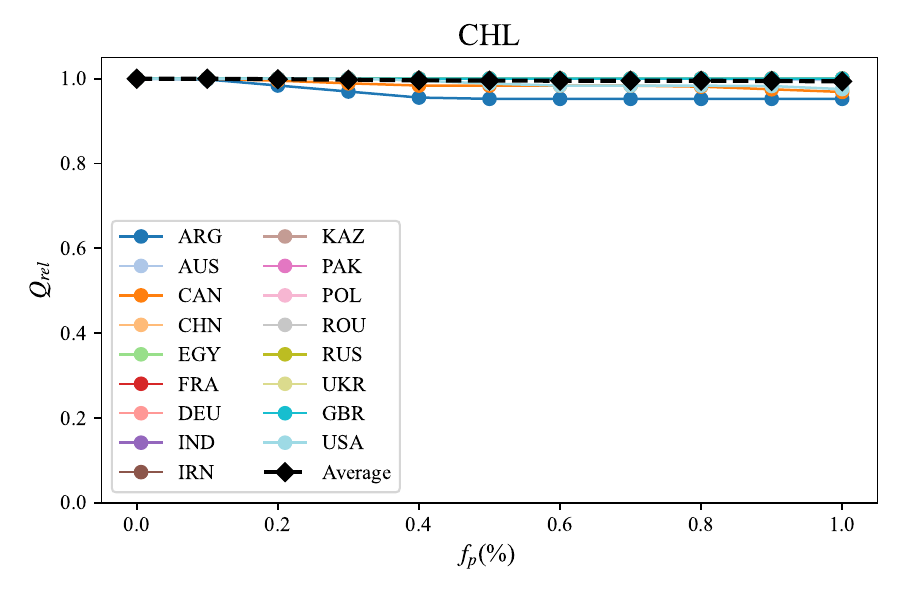}
    \includegraphics[width=0.24\linewidth]{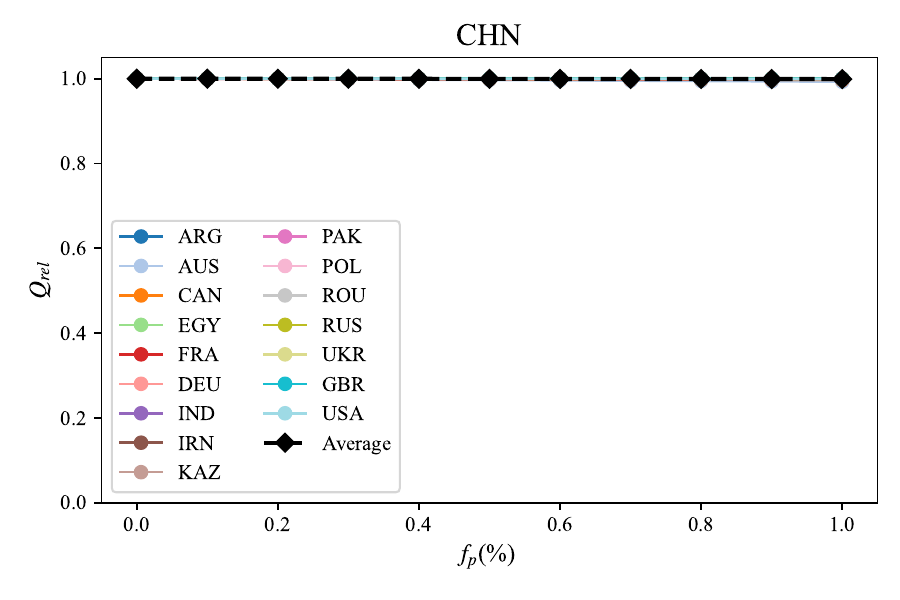}
    \includegraphics[width=0.24\linewidth]{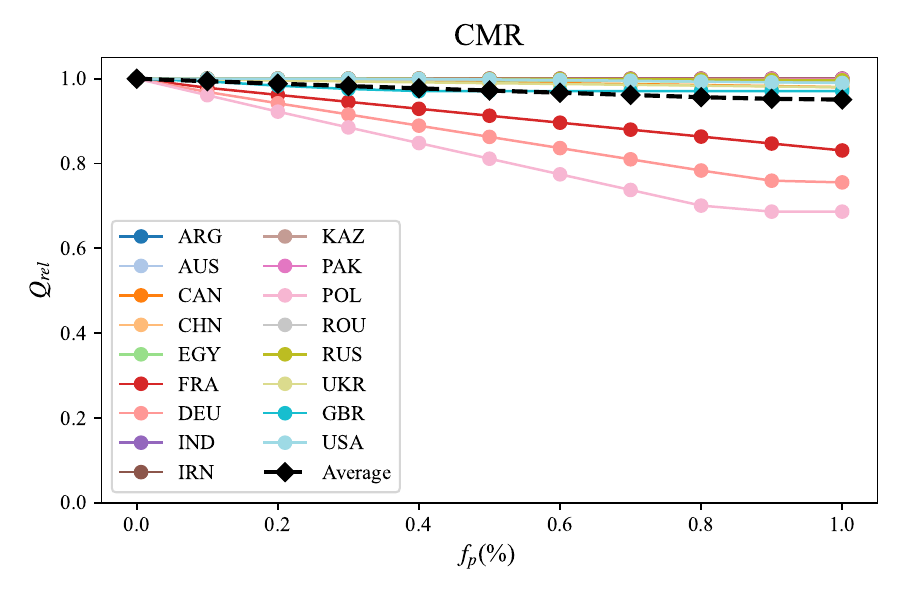}\\
    \includegraphics[width=0.24\linewidth]{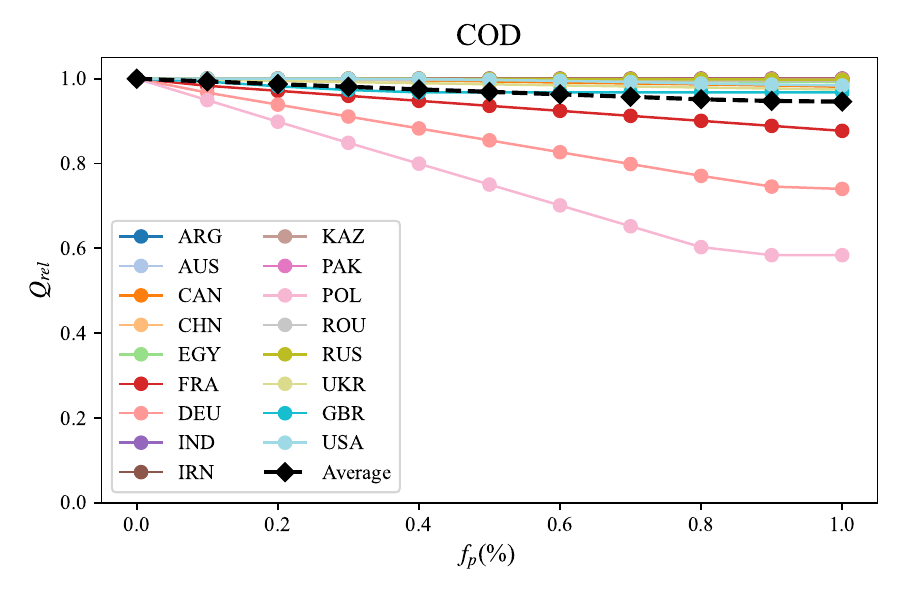}
    \includegraphics[width=0.24\linewidth]{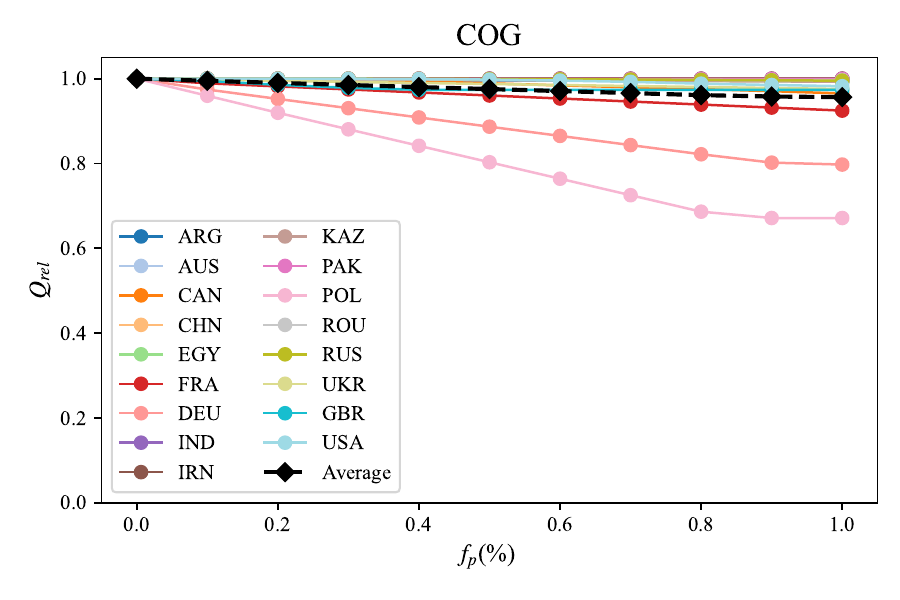}
    \includegraphics[width=0.24\linewidth]{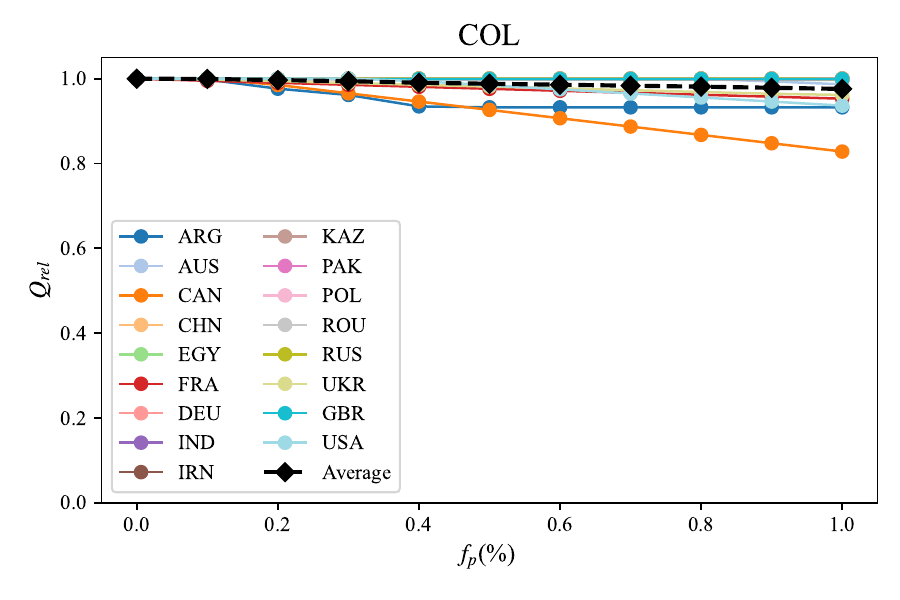}
    \includegraphics[width=0.24\linewidth]{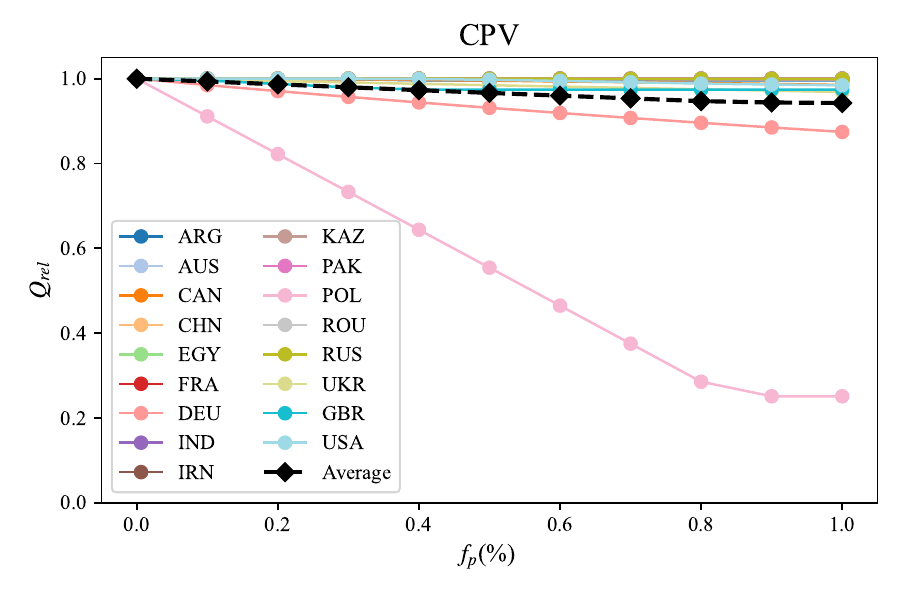}
    \caption{Wheat: $Q_{\mathrm{rel}}$ curves under global 1\% production shock (page 1).}
    \label{Fig:Wheat_Qrel_global1pct_1}
\end{figure}

\begin{figure}[p]\ContinuedFloat
    \centering
    \includegraphics[width=0.24\linewidth]{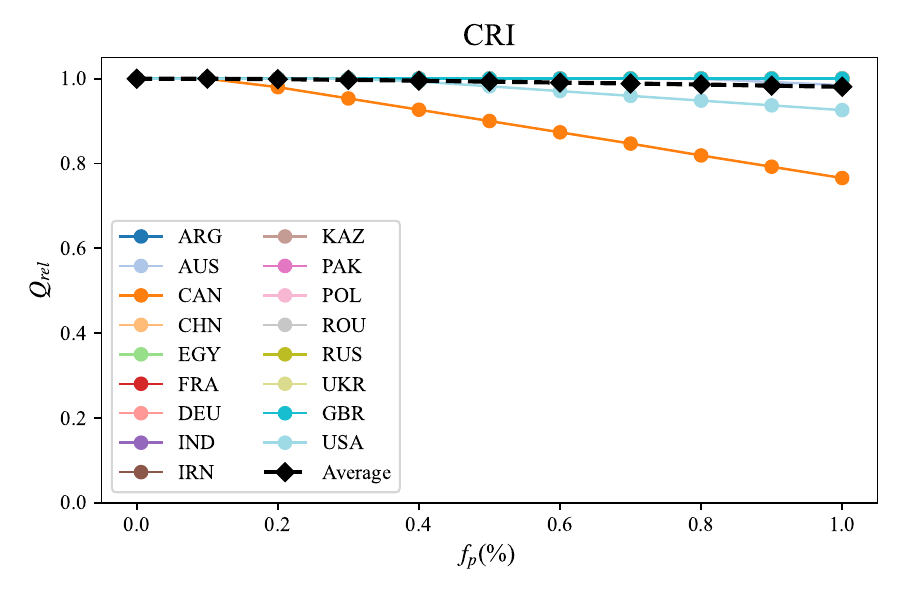}
    \includegraphics[width=0.24\linewidth]{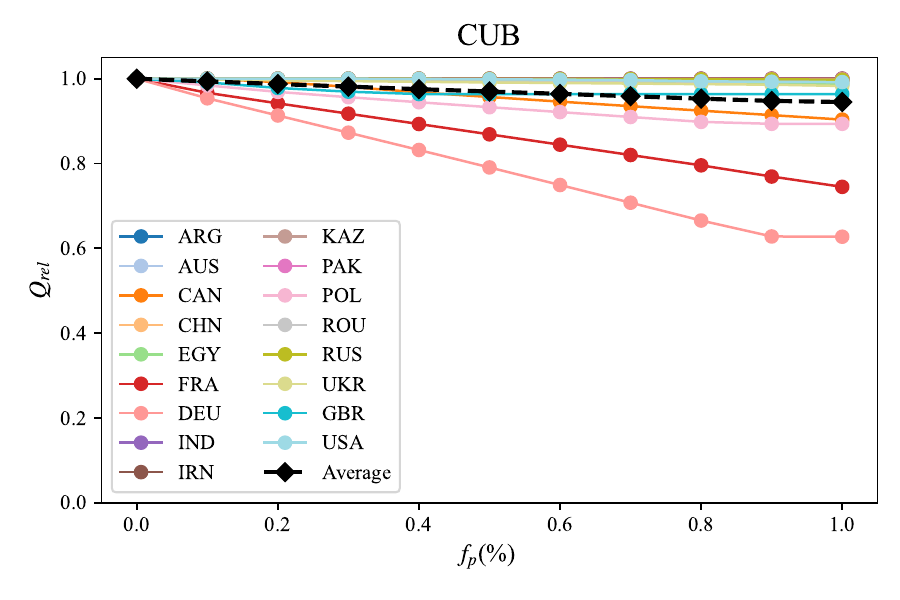}
    \includegraphics[width=0.24\linewidth]{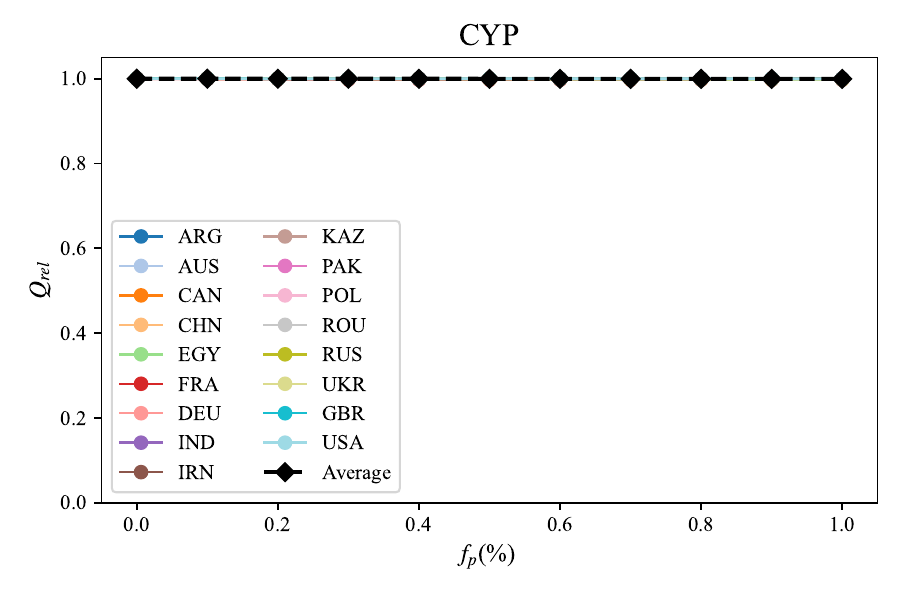}
    \includegraphics[width=0.24\linewidth]{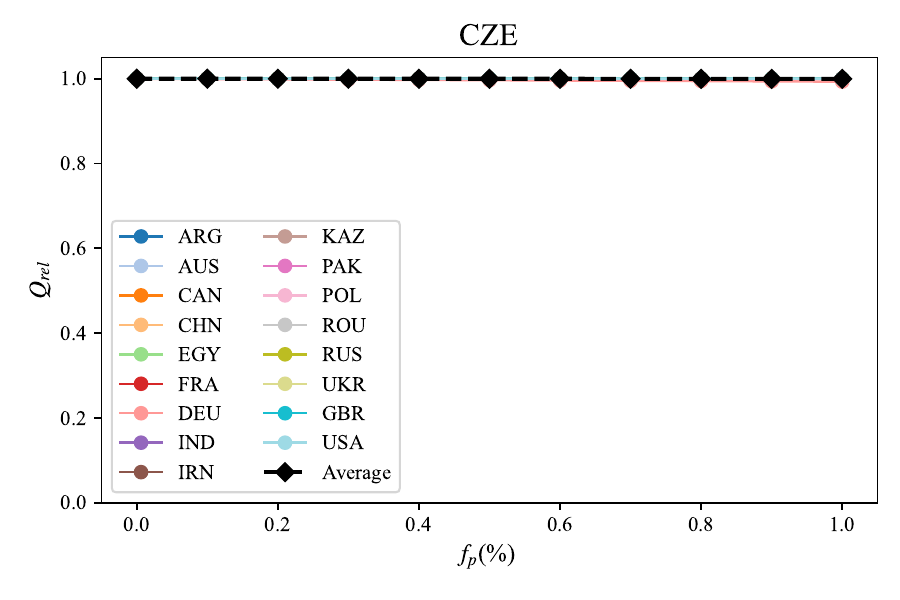}\\
    \includegraphics[width=0.24\linewidth]{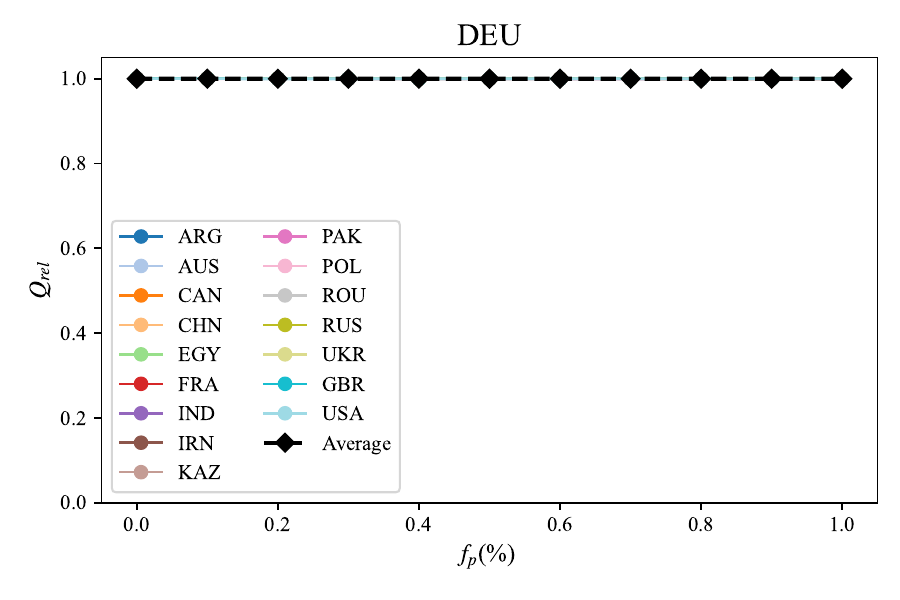}
    \includegraphics[width=0.24\linewidth]{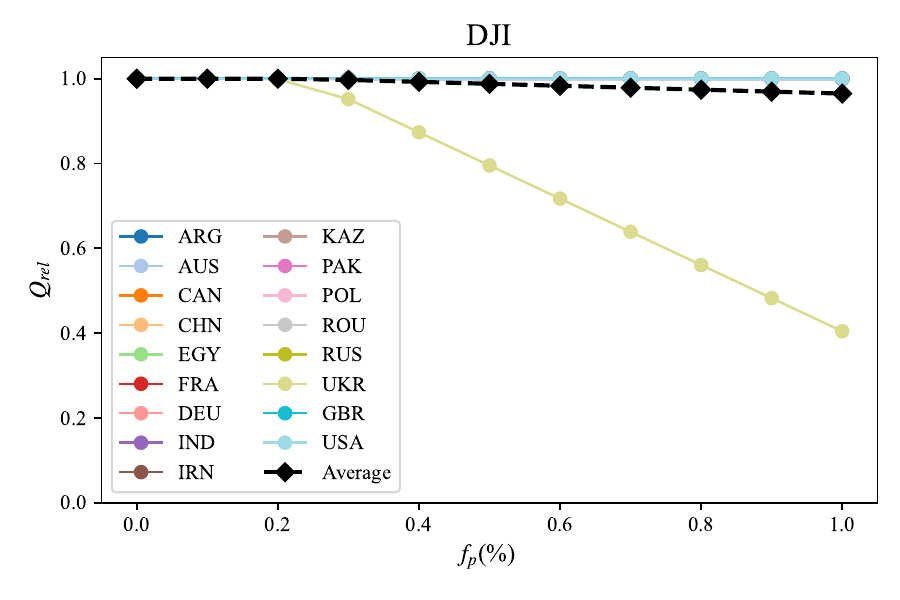}
    \includegraphics[width=0.24\linewidth]{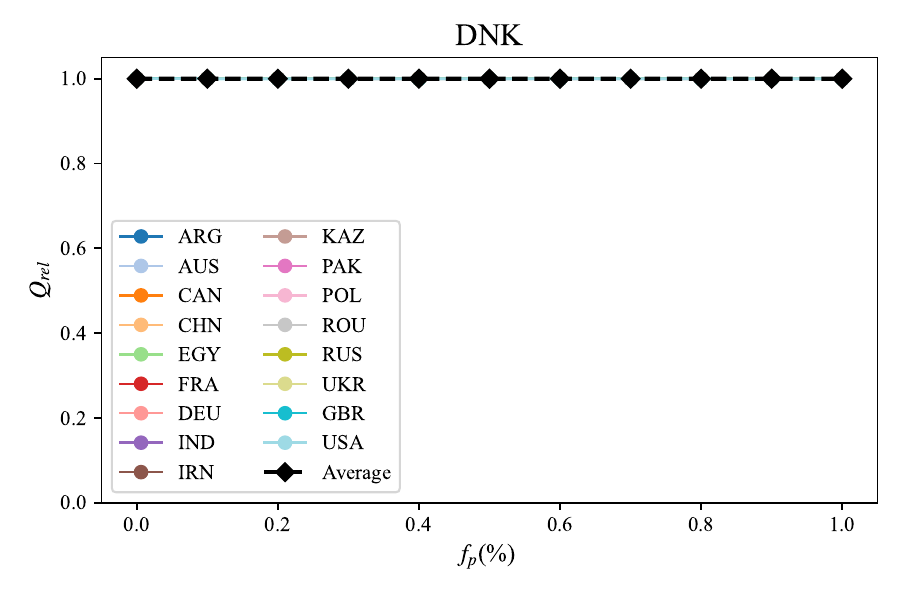}
    \includegraphics[width=0.24\linewidth]{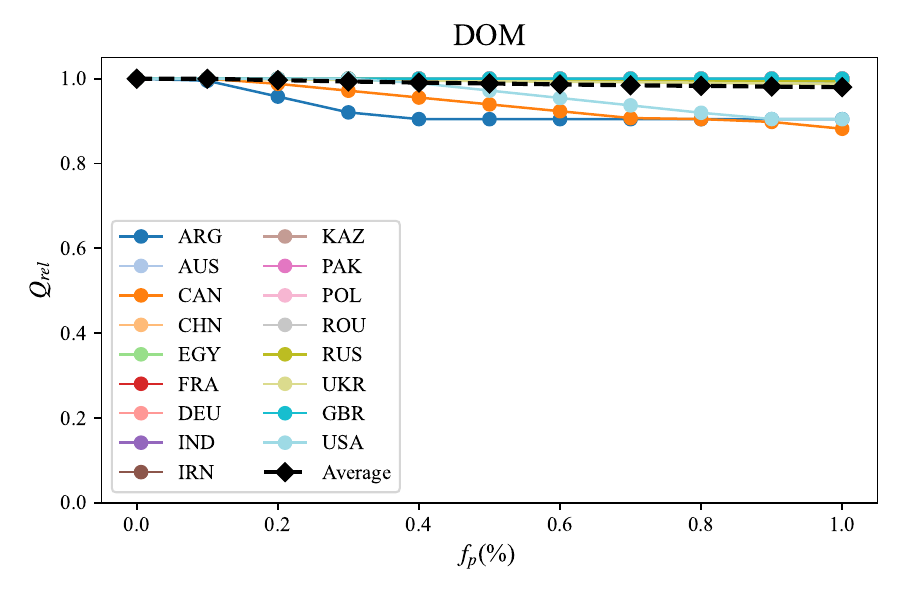}\\
    \includegraphics[width=0.24\linewidth]{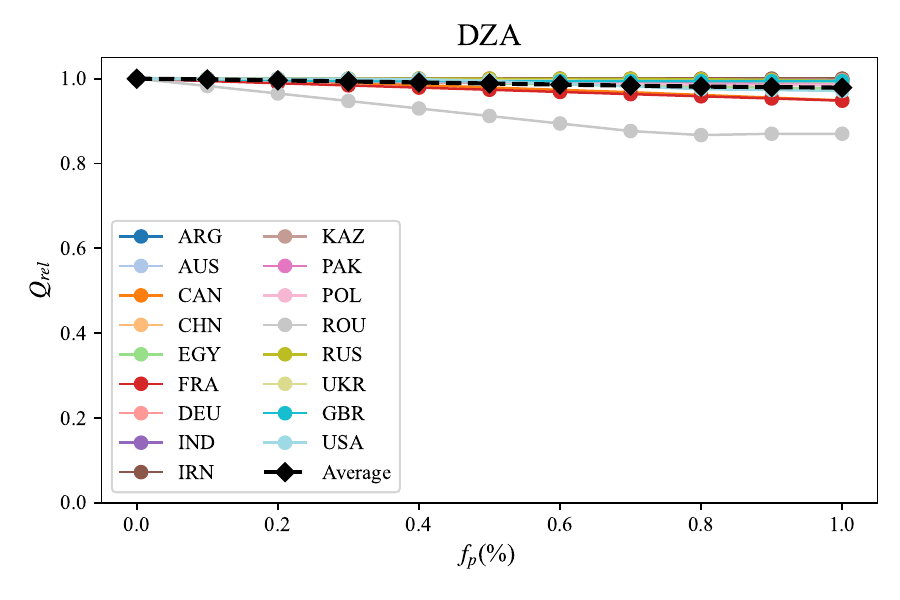}
    \includegraphics[width=0.24\linewidth]{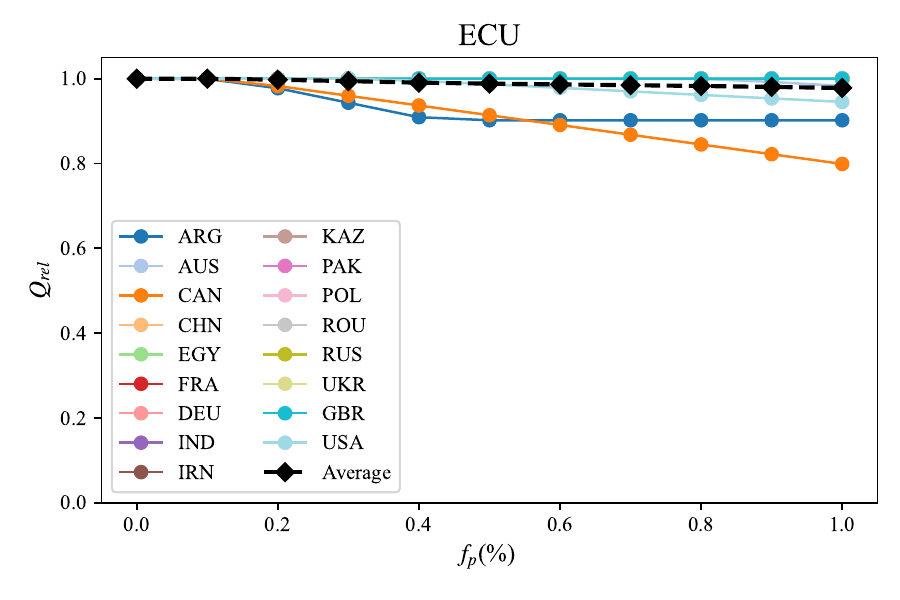}
    \includegraphics[width=0.24\linewidth]{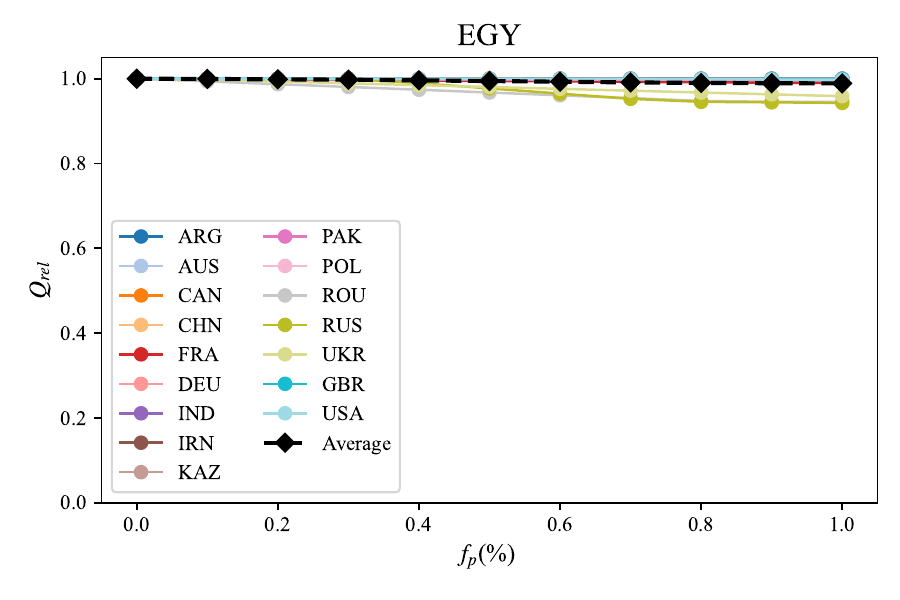}
    \includegraphics[width=0.24\linewidth]{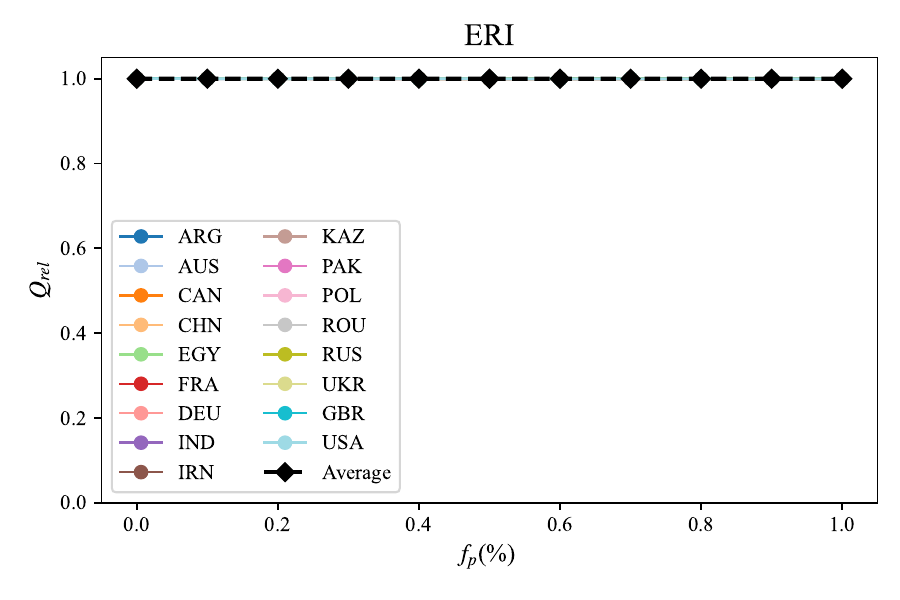}\\
    \includegraphics[width=0.24\linewidth]{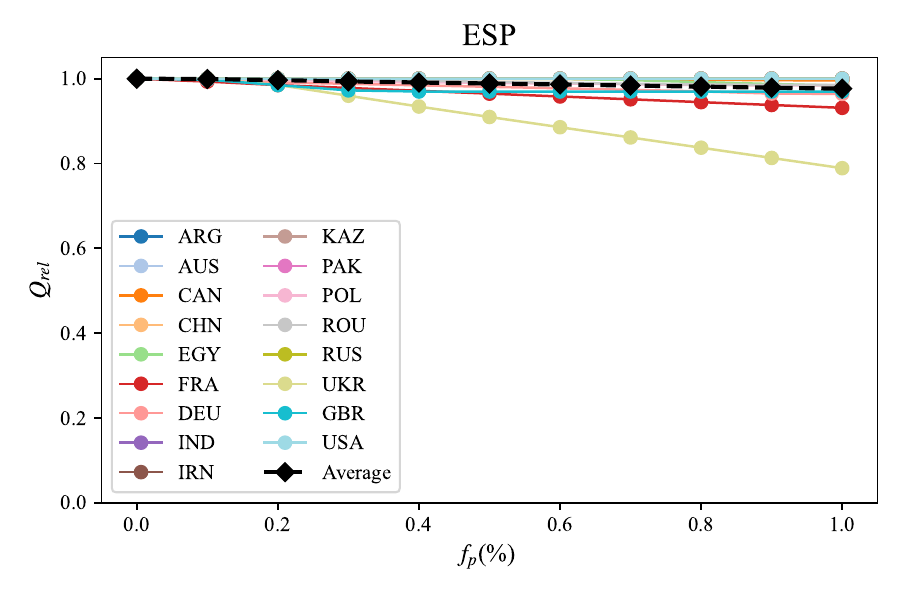}
    \includegraphics[width=0.24\linewidth]{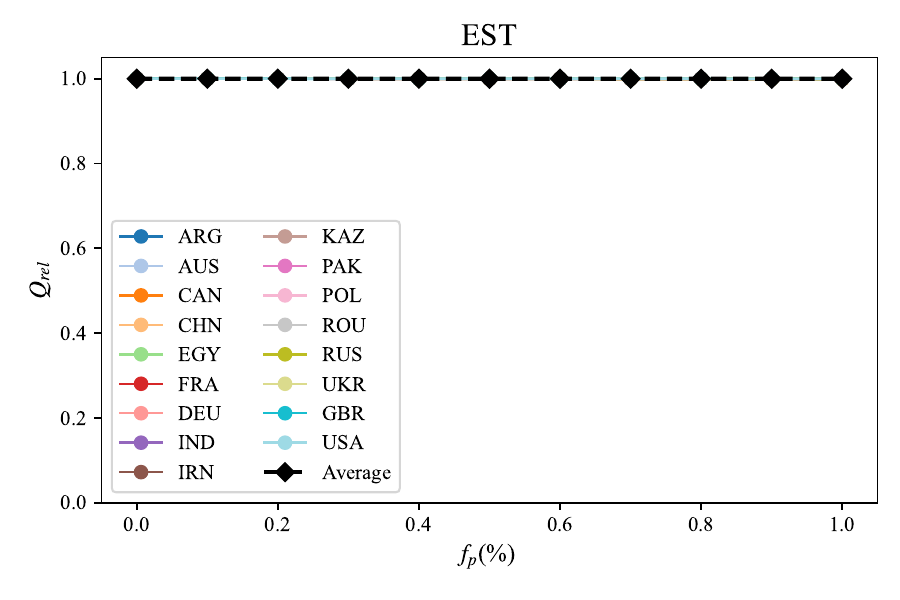}
    \includegraphics[width=0.24\linewidth]{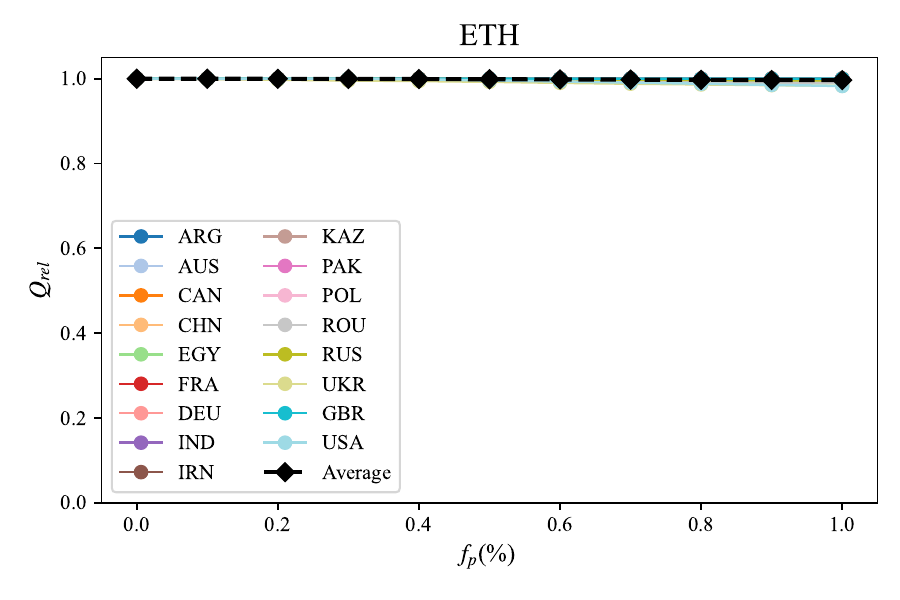}
    \includegraphics[width=0.24\linewidth]{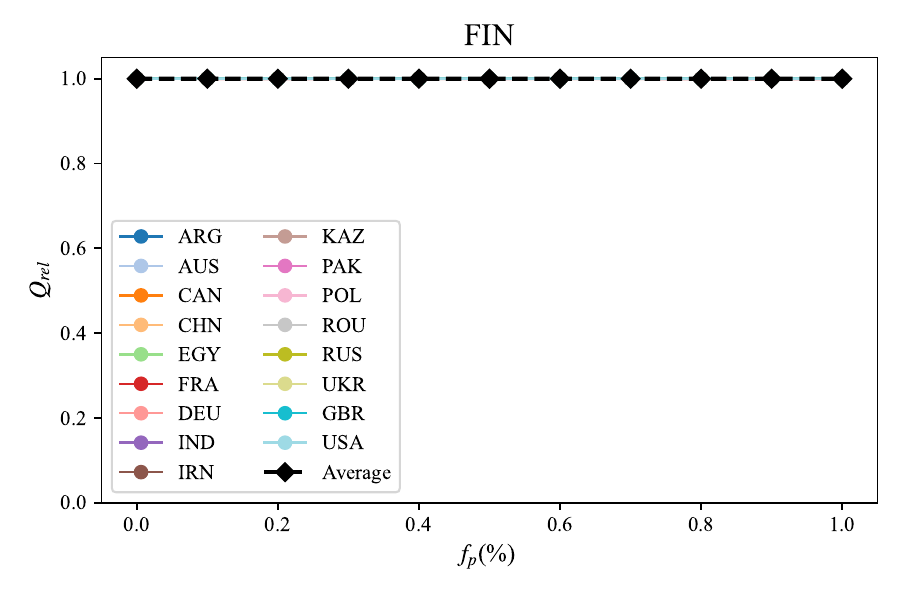}\\
    \includegraphics[width=0.24\linewidth]{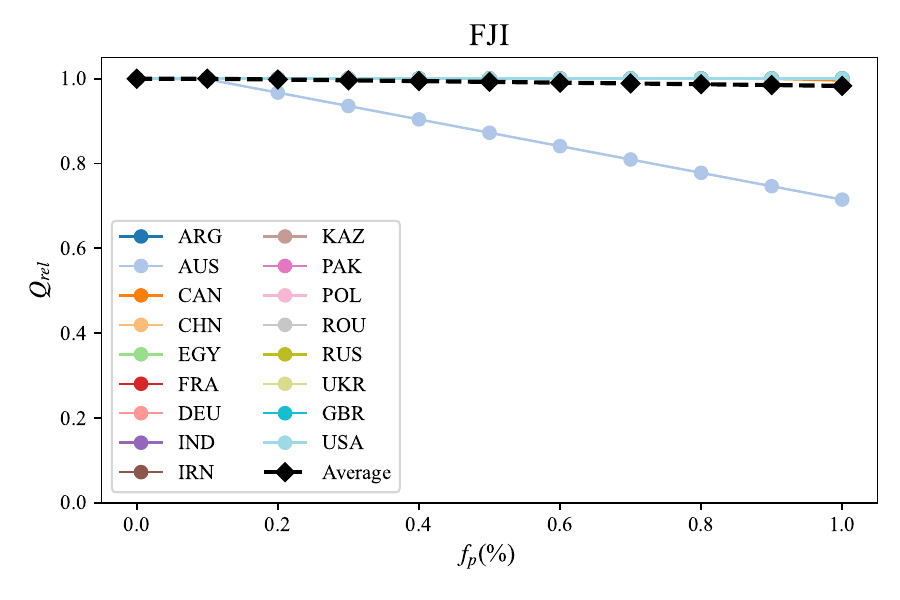}
    \includegraphics[width=0.24\linewidth]{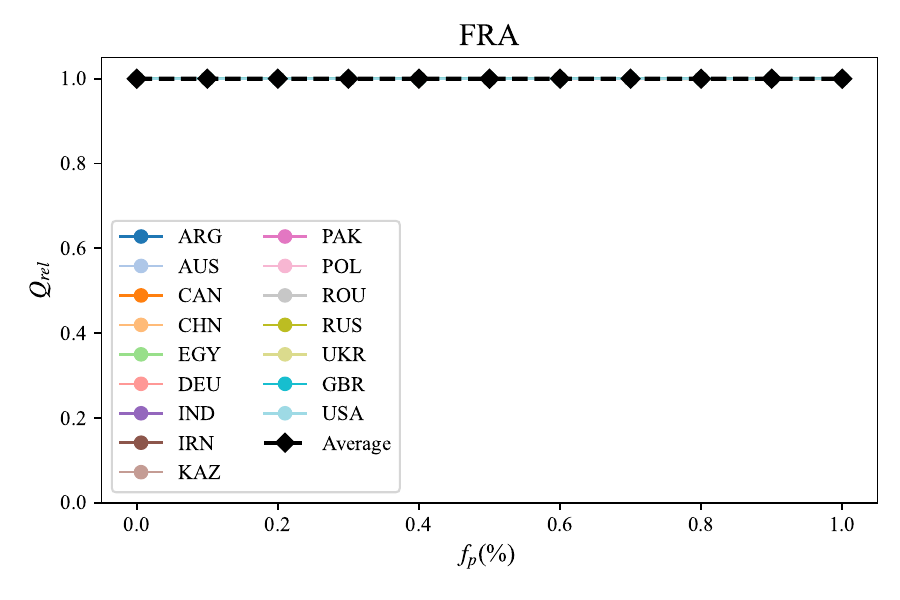}
    \includegraphics[width=0.24\linewidth]{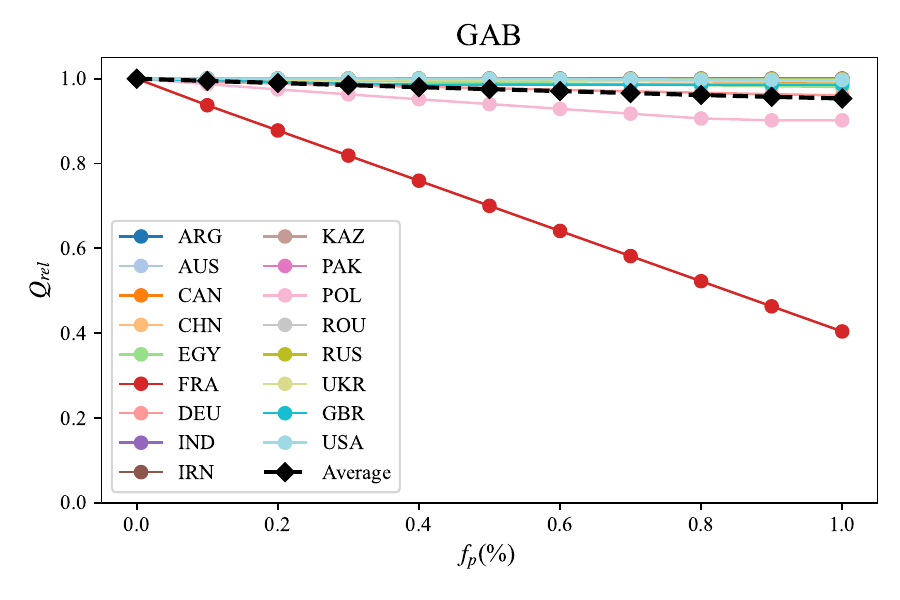}
    \includegraphics[width=0.24\linewidth]{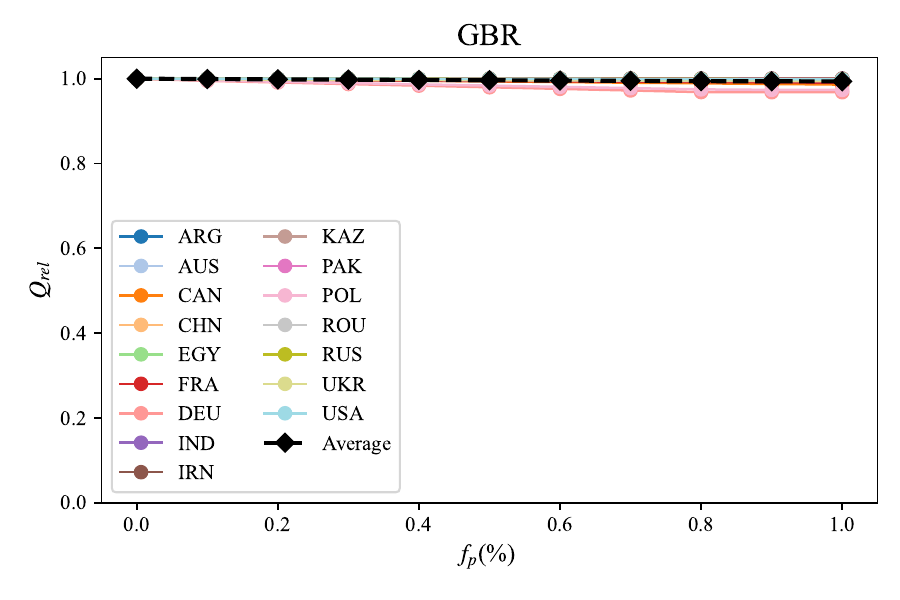}\\
    \includegraphics[width=0.24\linewidth]{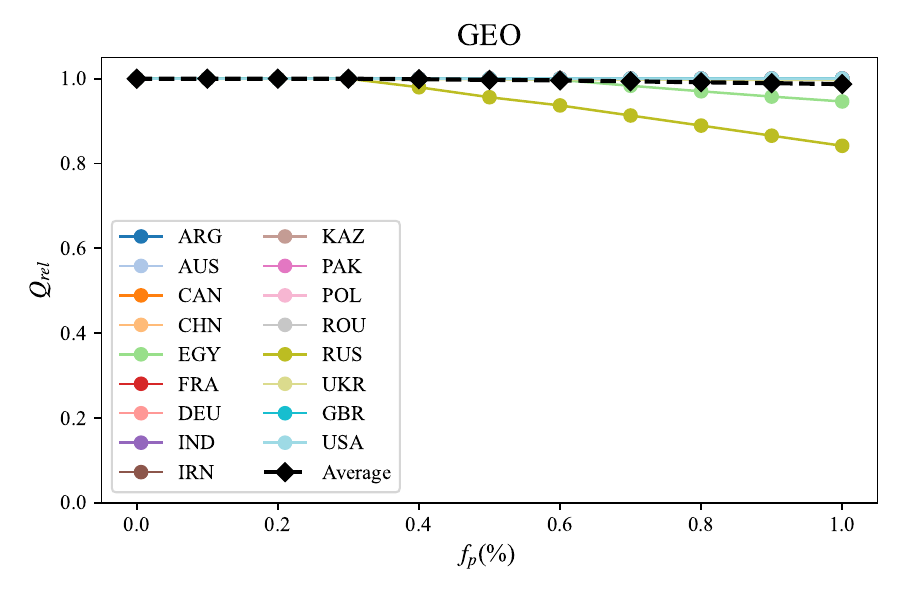}
    \includegraphics[width=0.24\linewidth]{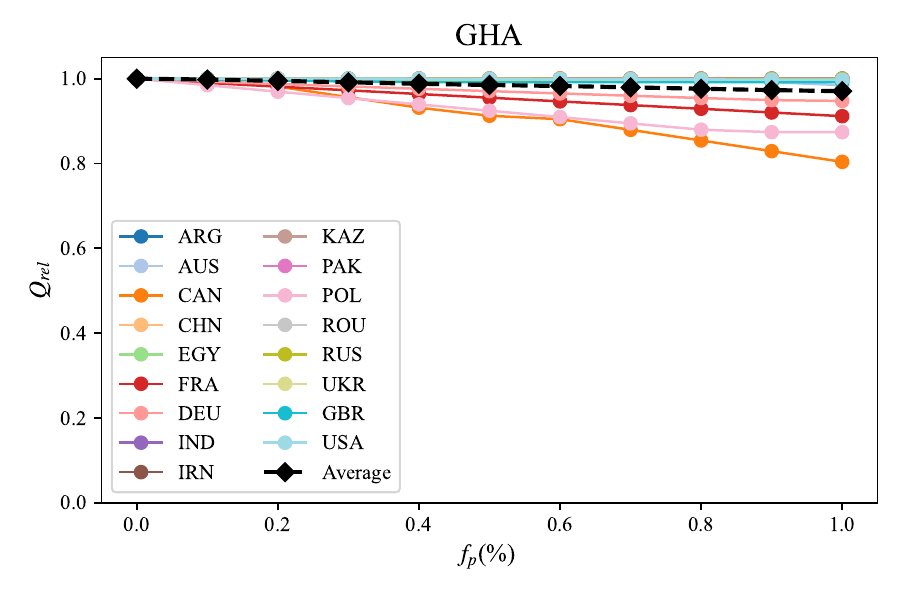}
    \includegraphics[width=0.24\linewidth]{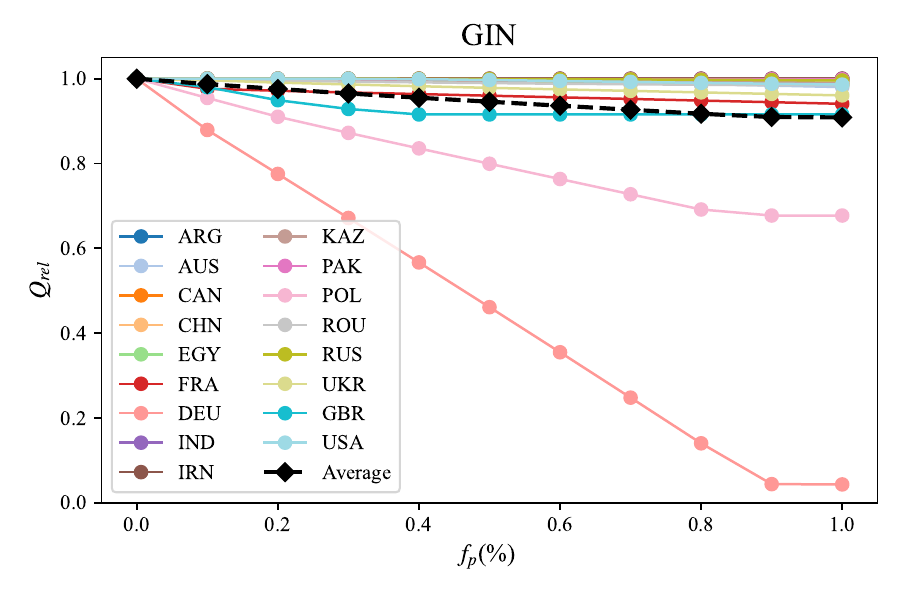}
    \includegraphics[width=0.24\linewidth]{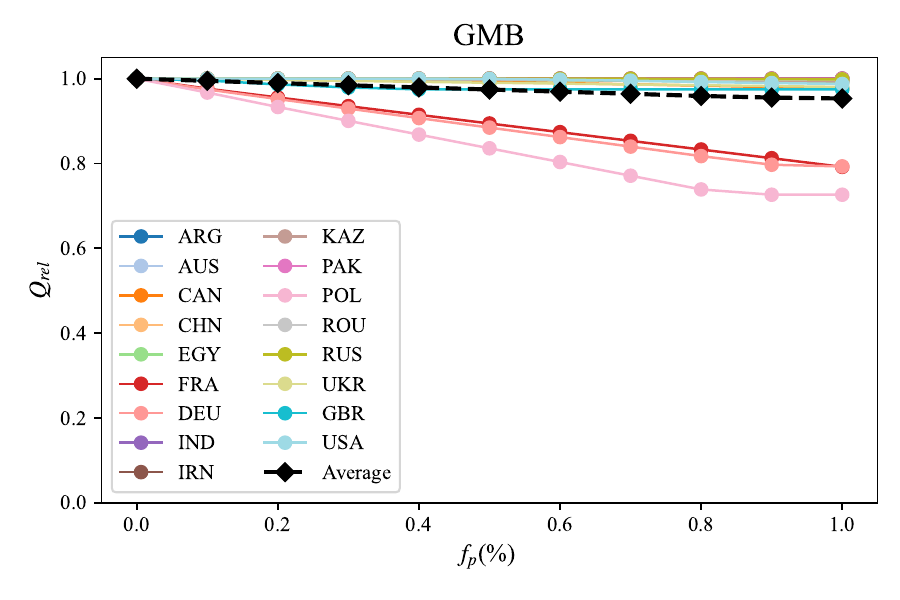}\\
    \includegraphics[width=0.24\linewidth]{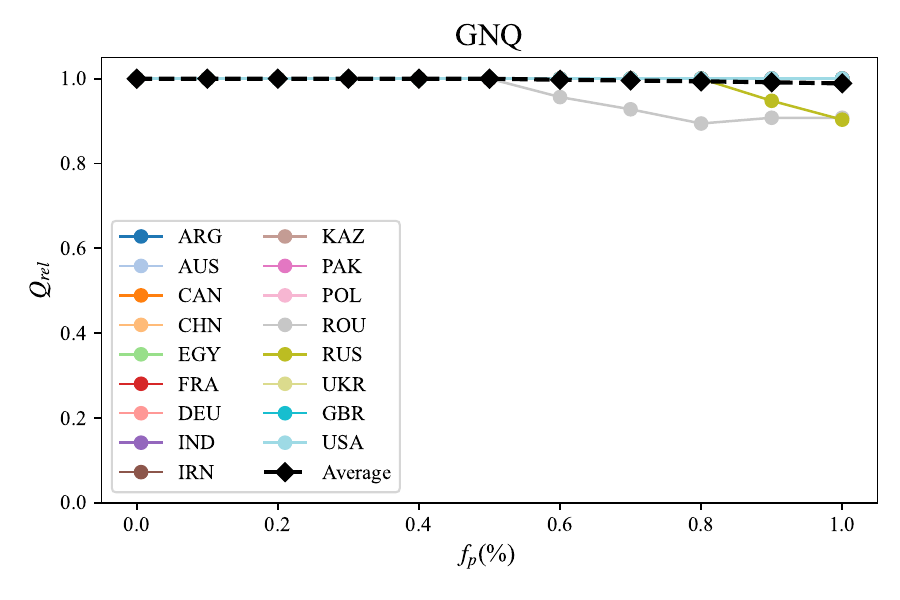}
    \includegraphics[width=0.24\linewidth]{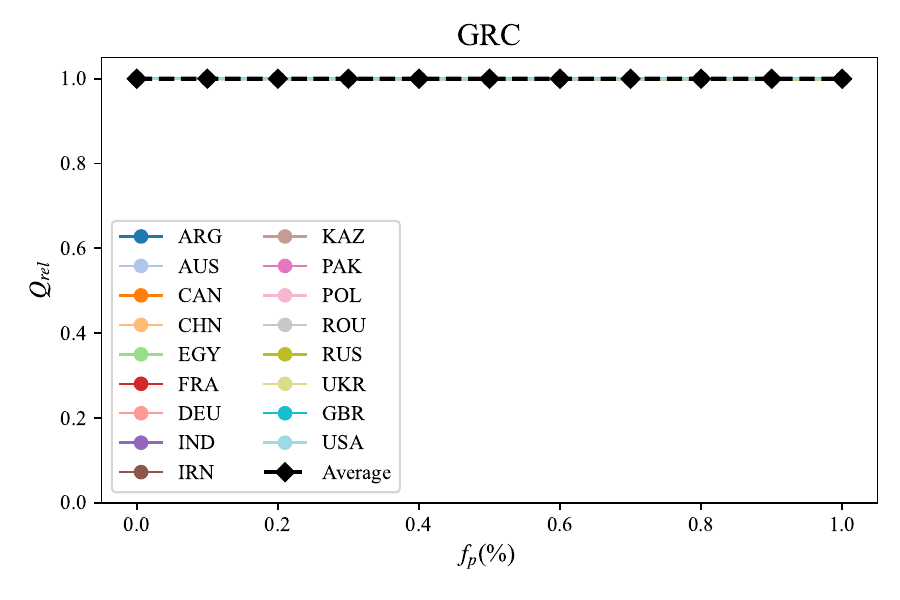}
    \includegraphics[width=0.24\linewidth]{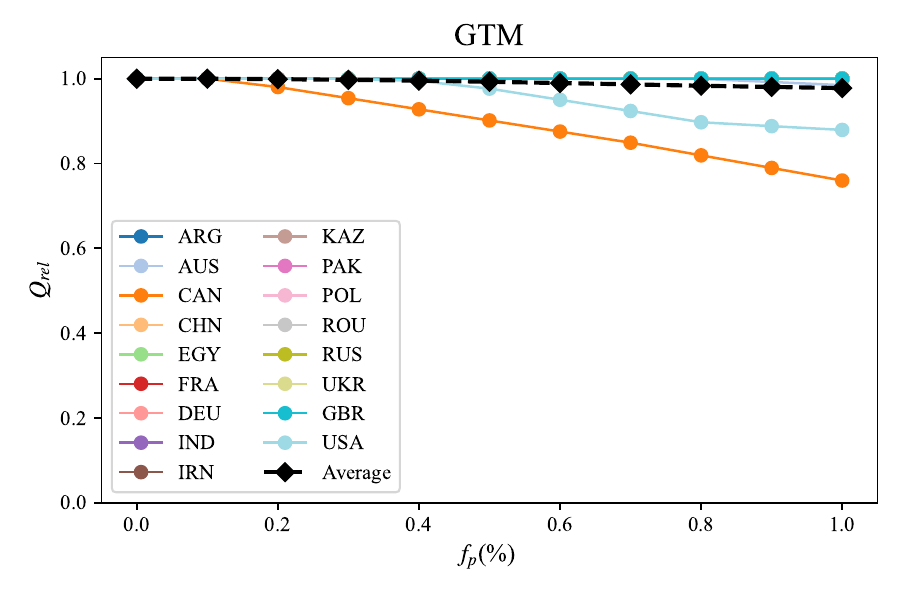}
    \includegraphics[width=0.24\linewidth]{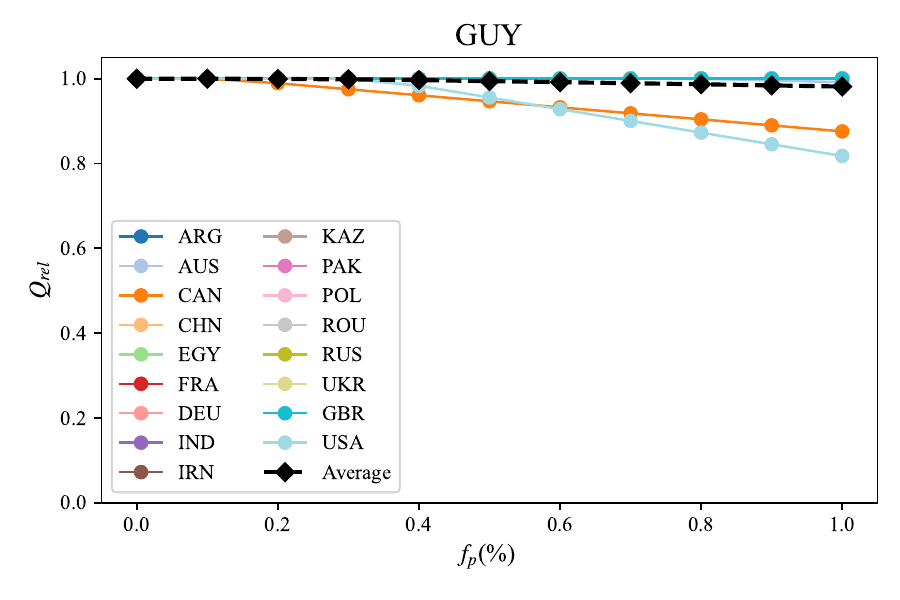}\\
    \includegraphics[width=0.24\linewidth]{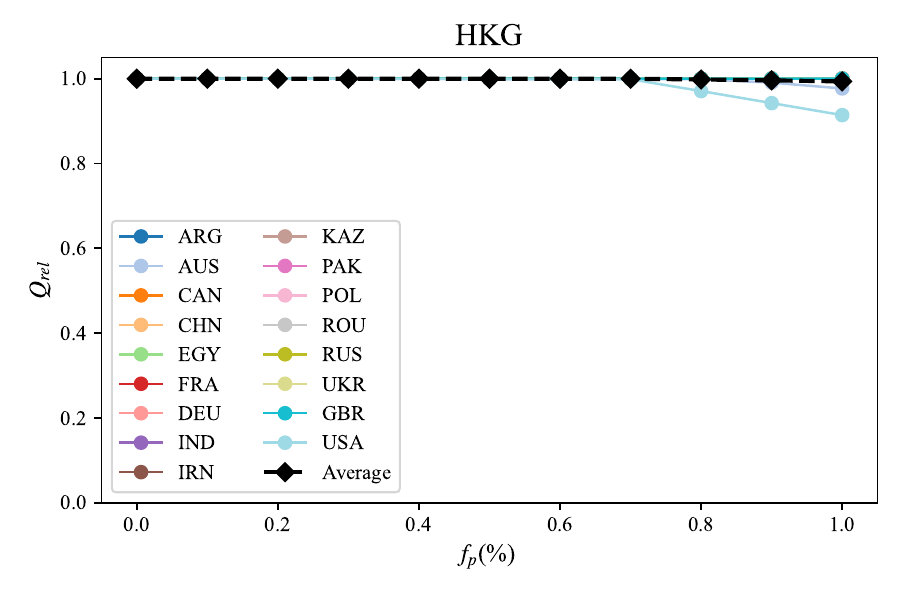}
    \includegraphics[width=0.24\linewidth]{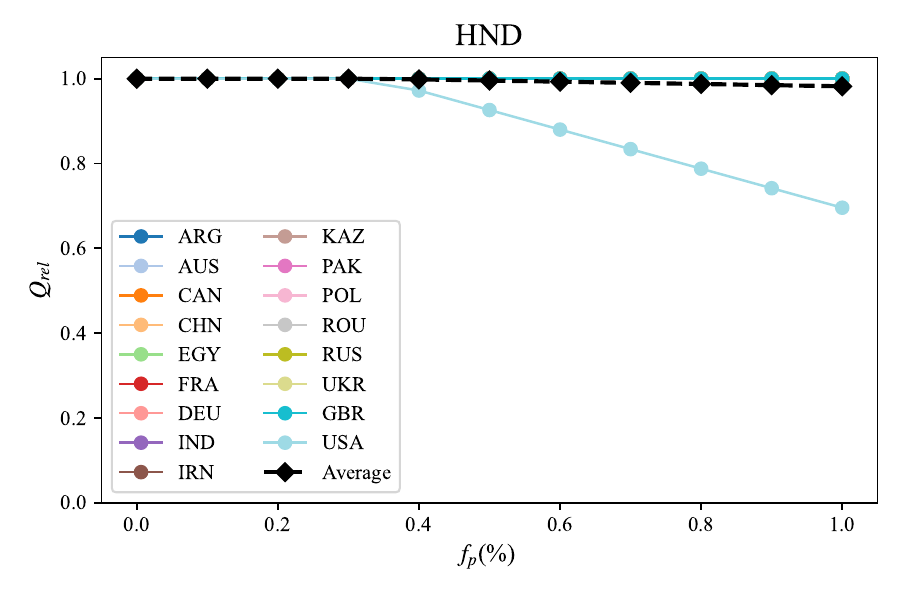}
    \includegraphics[width=0.24\linewidth]{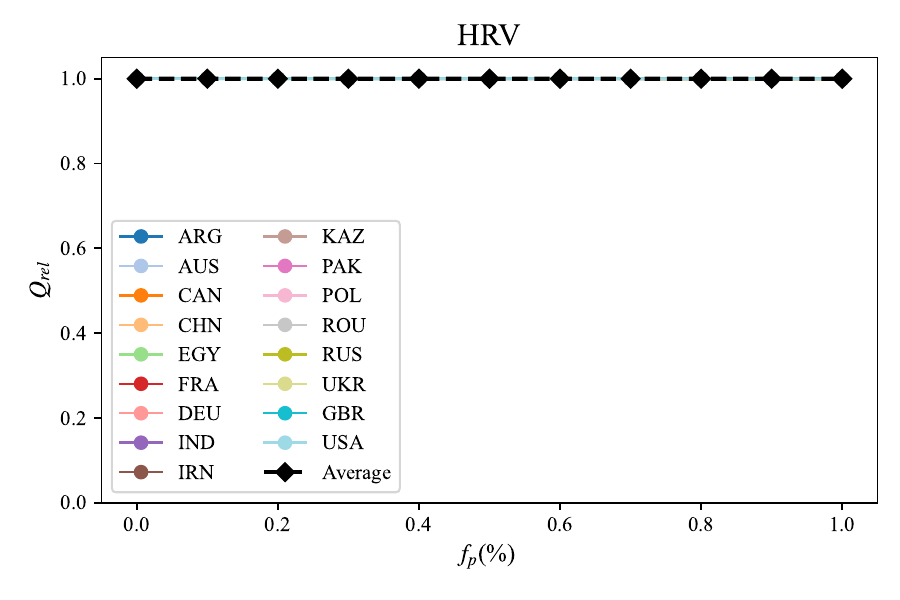}
    \includegraphics[width=0.24\linewidth]{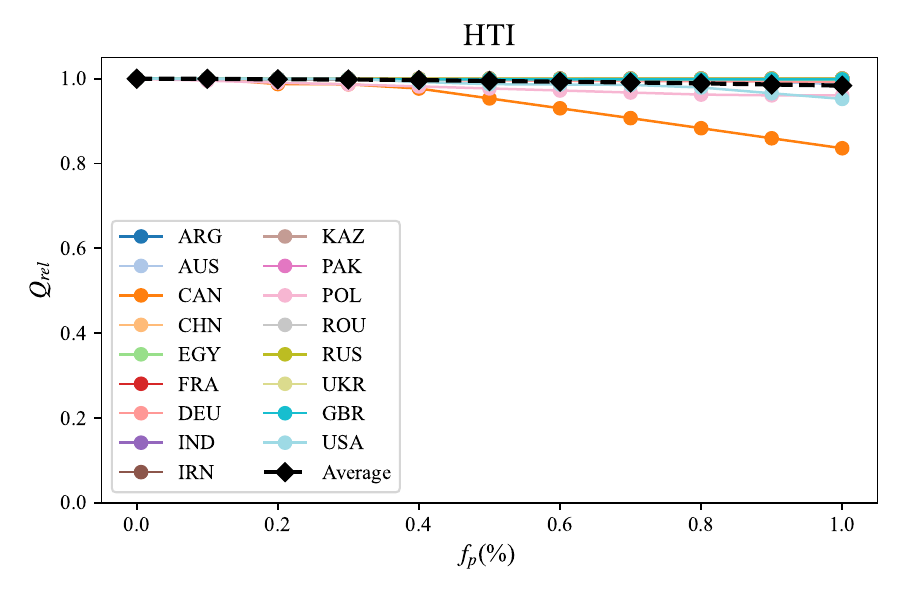}
    \caption{Wheat: $Q_{\mathrm{rel}}$ curves under global 1\% production shock (page 2).}
    \label{Fig:Wheat_Qrel_global1pct_2}
\end{figure}

\begin{figure}[p]\ContinuedFloat
    \centering
    \includegraphics[width=0.24\linewidth]{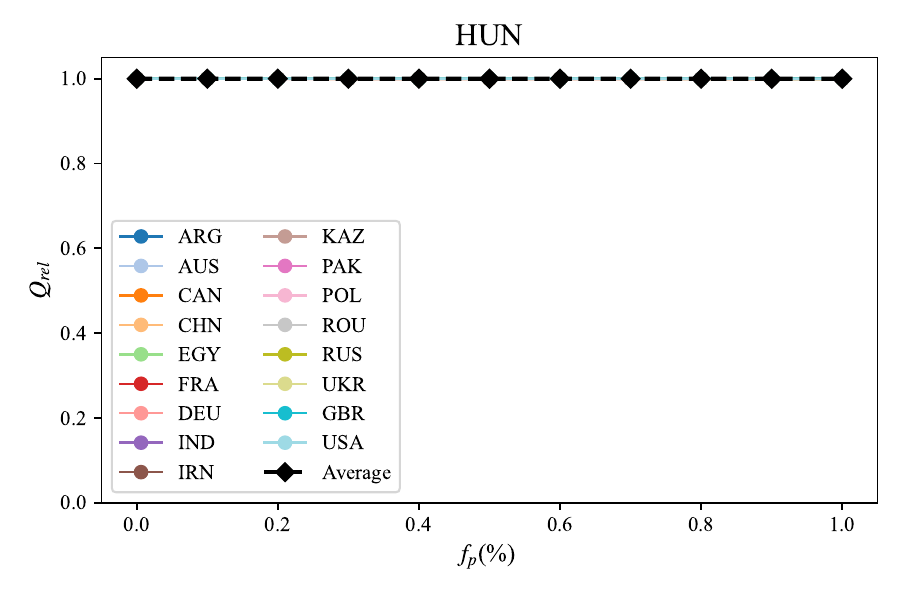}
    \includegraphics[width=0.24\linewidth]{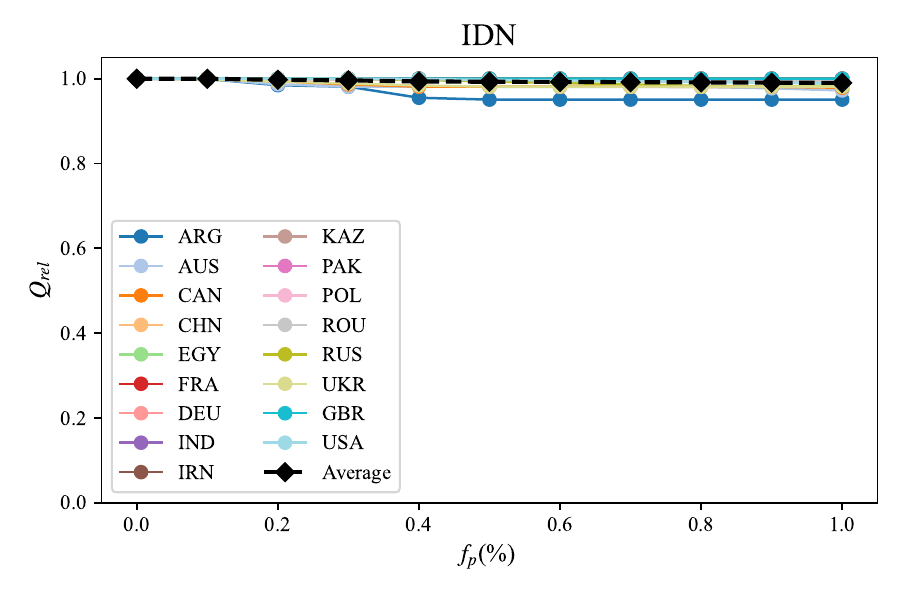}
    \includegraphics[width=0.24\linewidth]{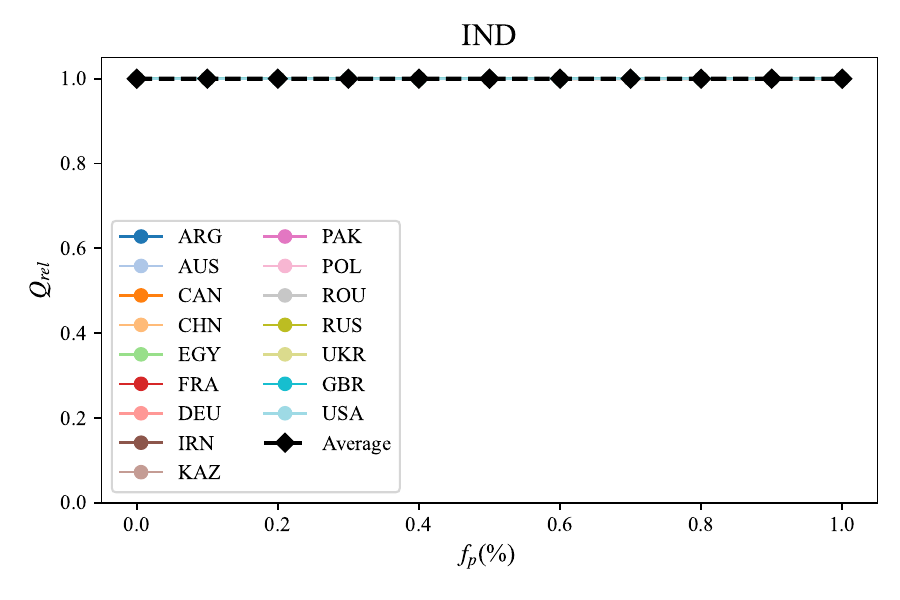}
    \includegraphics[width=0.24\linewidth]{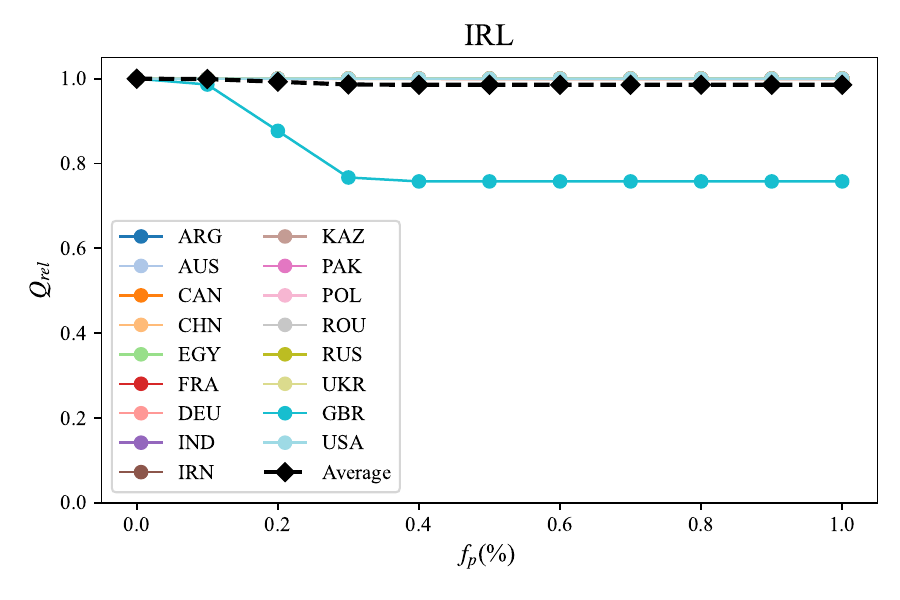}\\
    \includegraphics[width=0.24\linewidth]{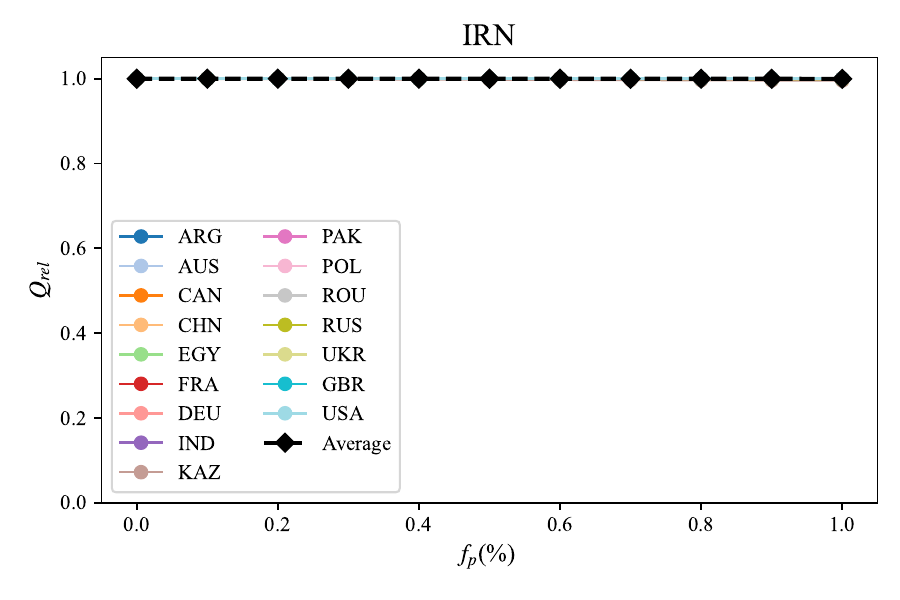}
    \includegraphics[width=0.24\linewidth]{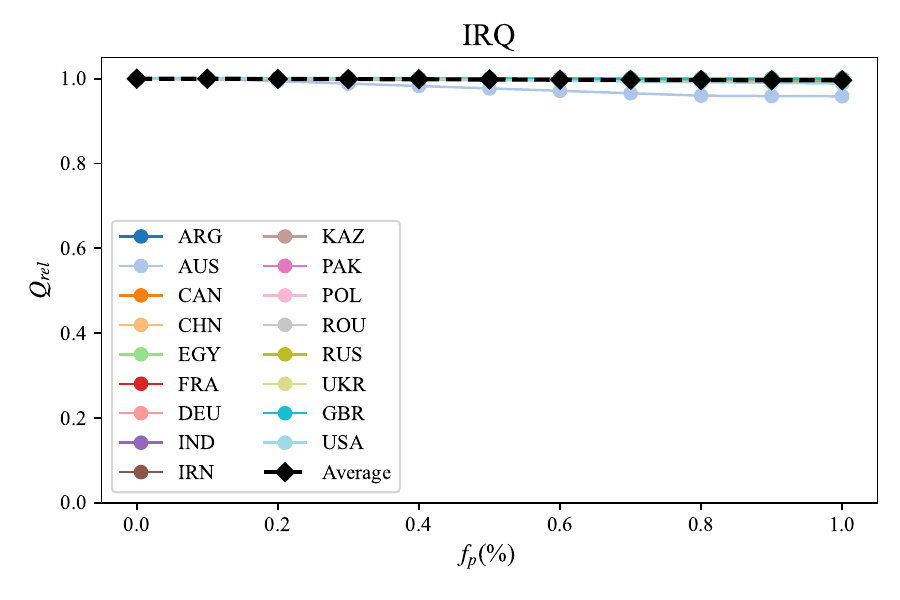}
    \includegraphics[width=0.24\linewidth]{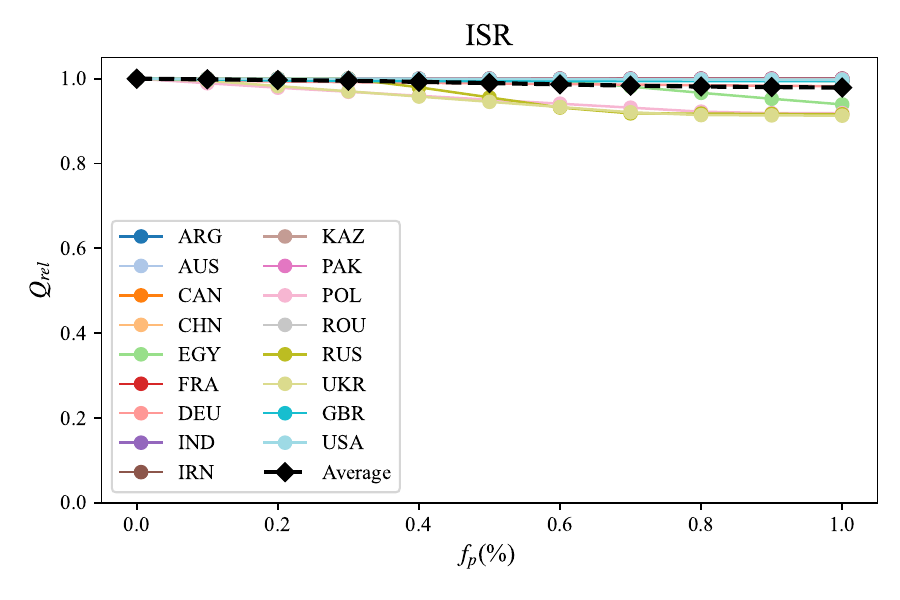}
    \includegraphics[width=0.24\linewidth]{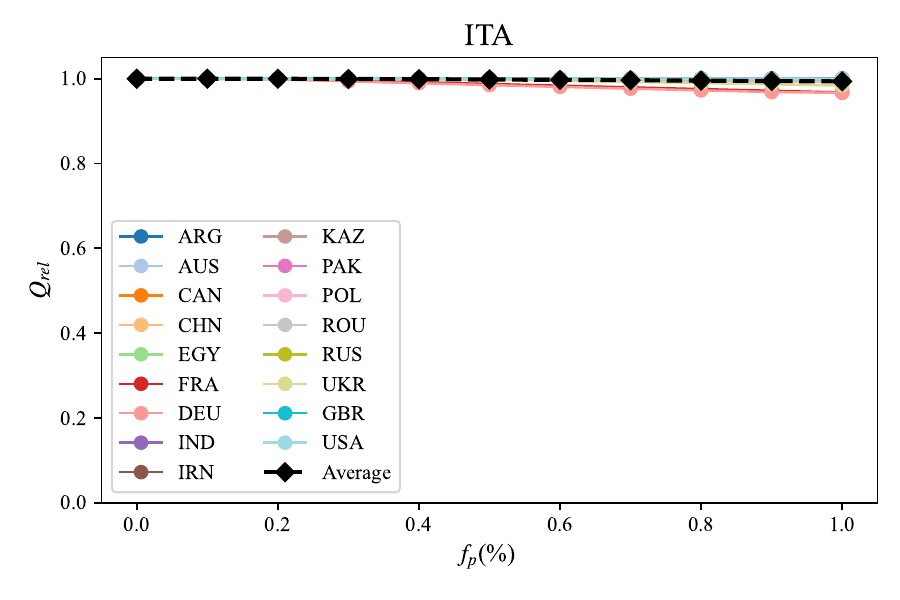}\\
    \includegraphics[width=0.24\linewidth]{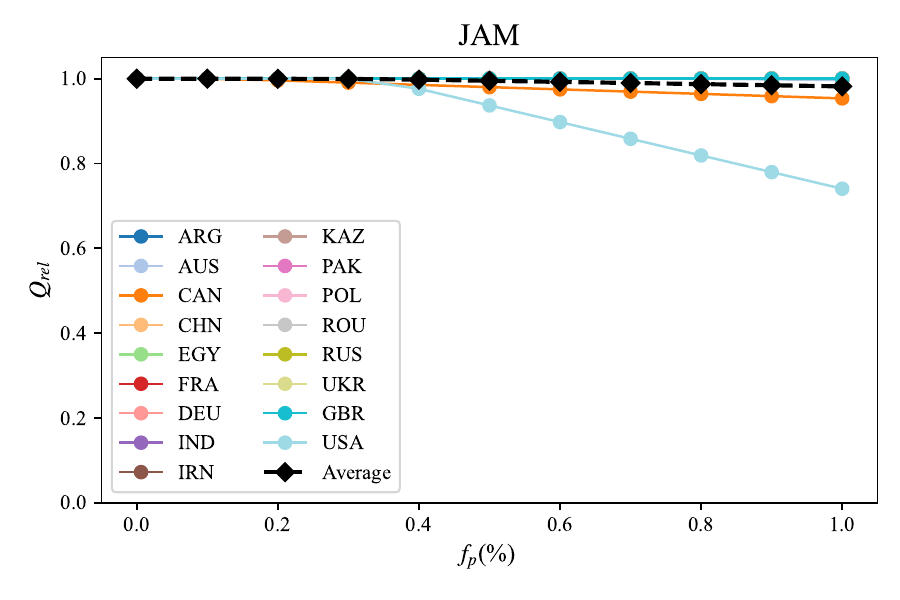}
    \includegraphics[width=0.24\linewidth]{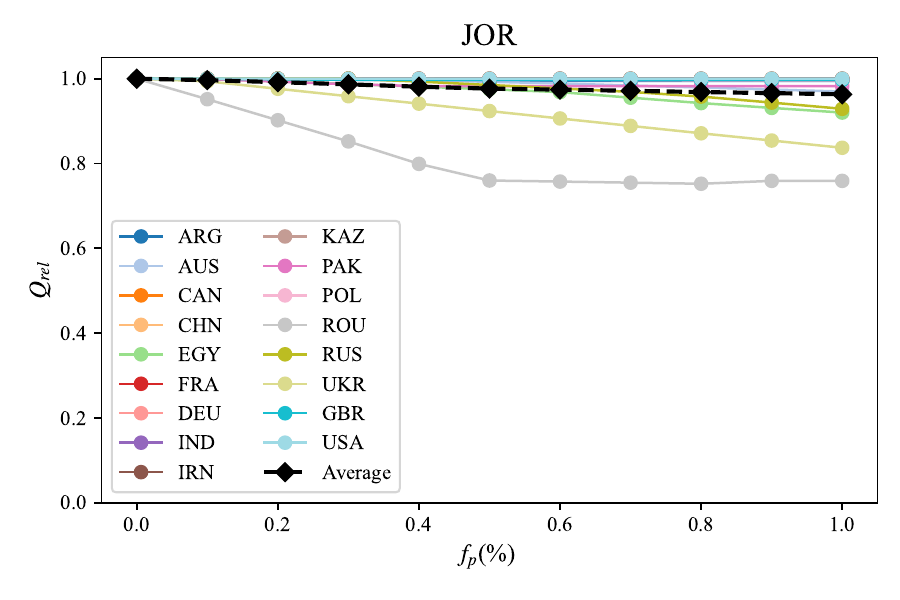}
    \includegraphics[width=0.24\linewidth]{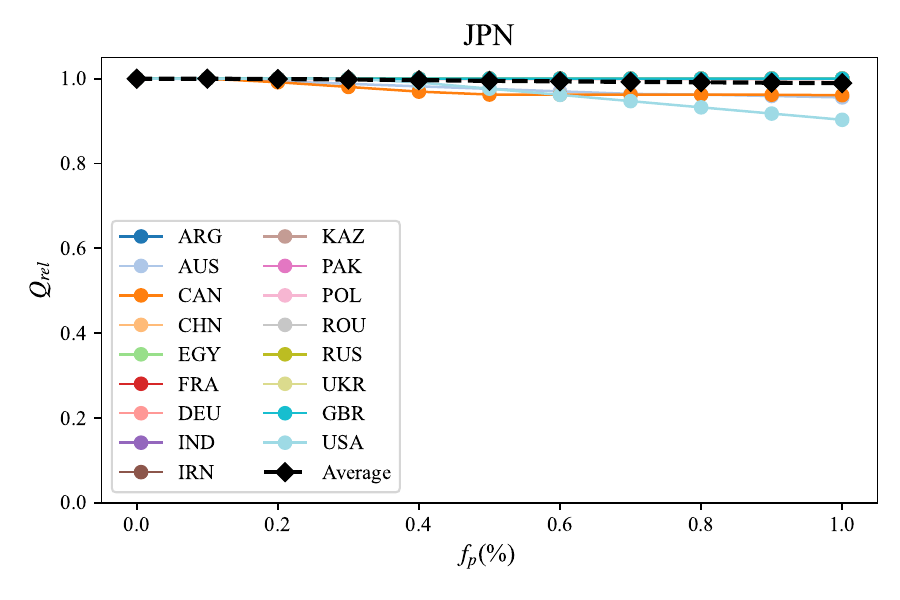}
    \includegraphics[width=0.24\linewidth]{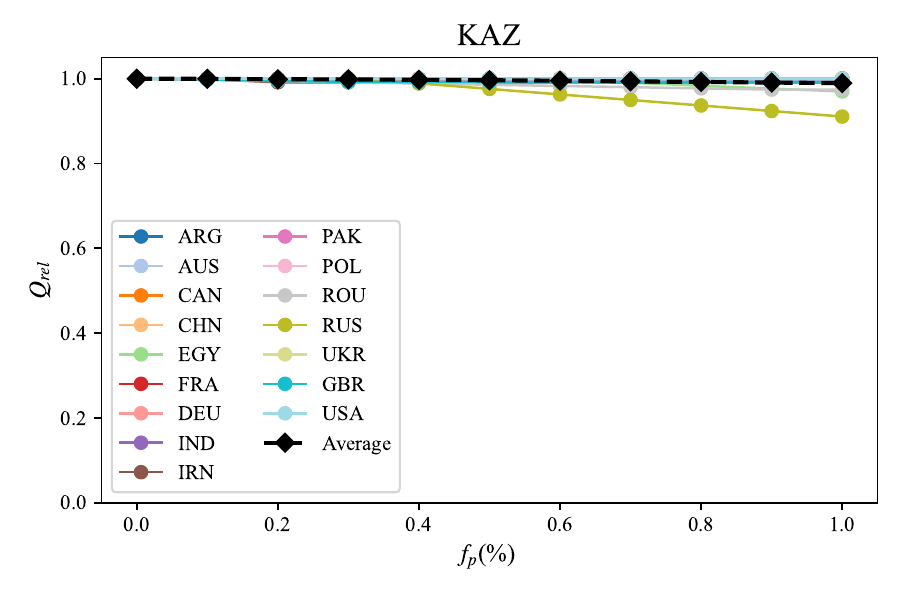}\\
    \includegraphics[width=0.24\linewidth]{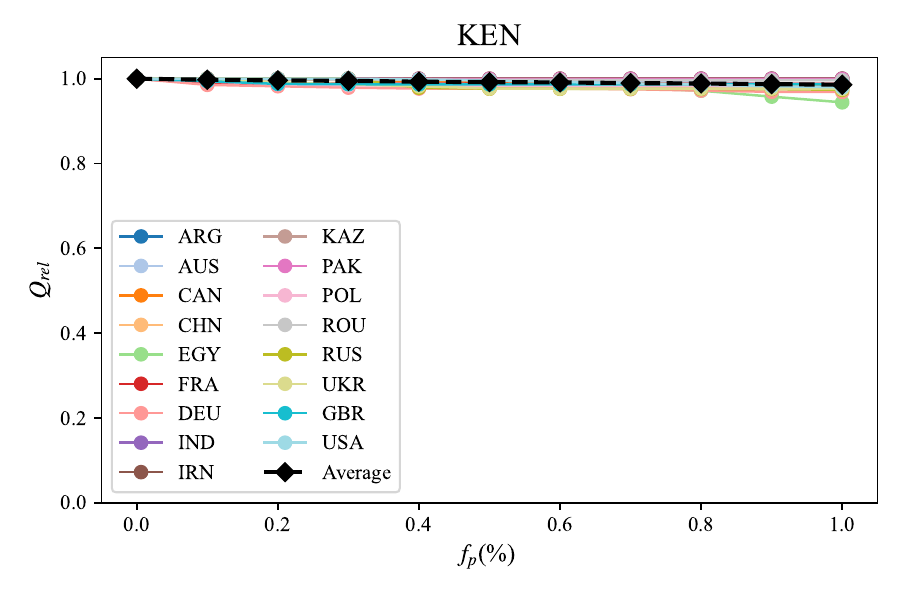}
    \includegraphics[width=0.24\linewidth]{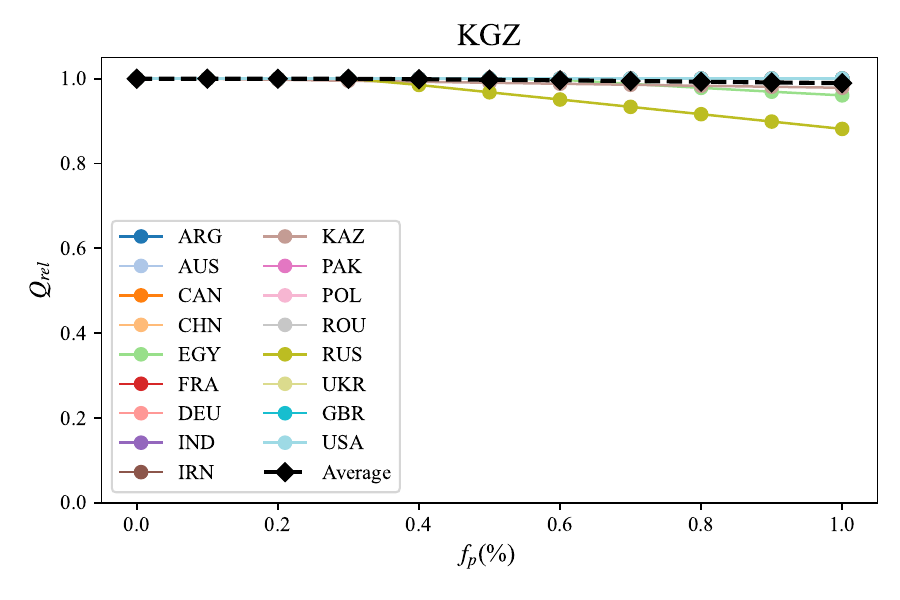}
    \includegraphics[width=0.24\linewidth]{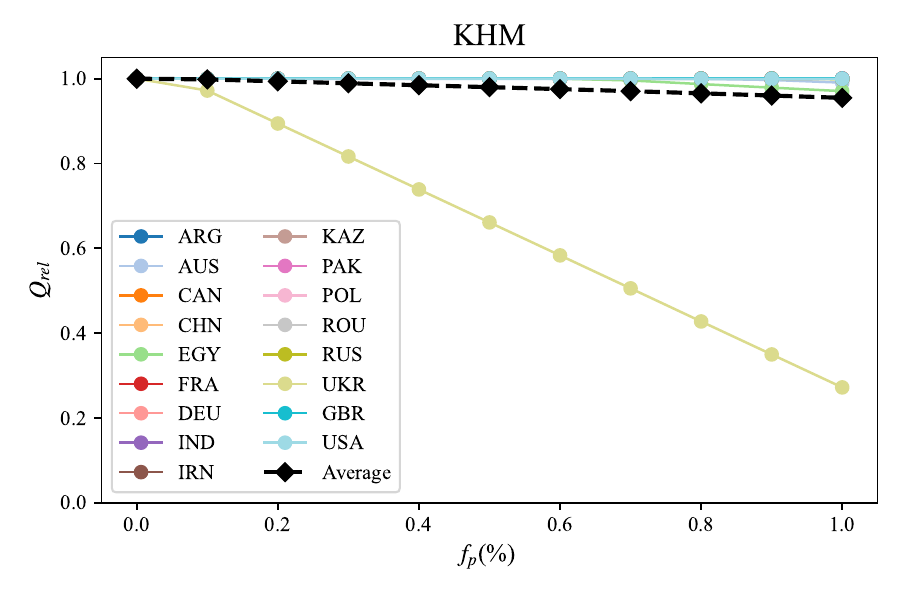}
    \includegraphics[width=0.24\linewidth]{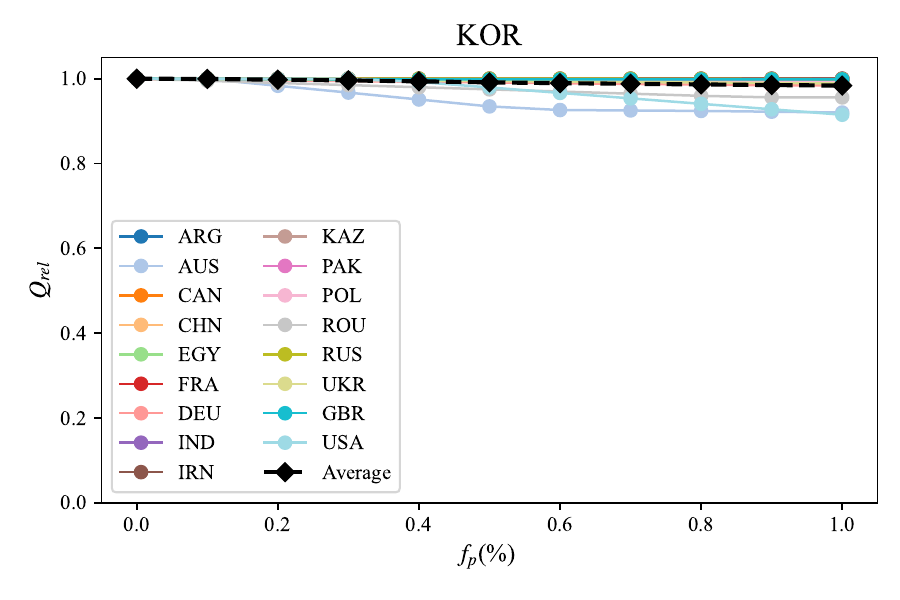}\\
    \includegraphics[width=0.24\linewidth]{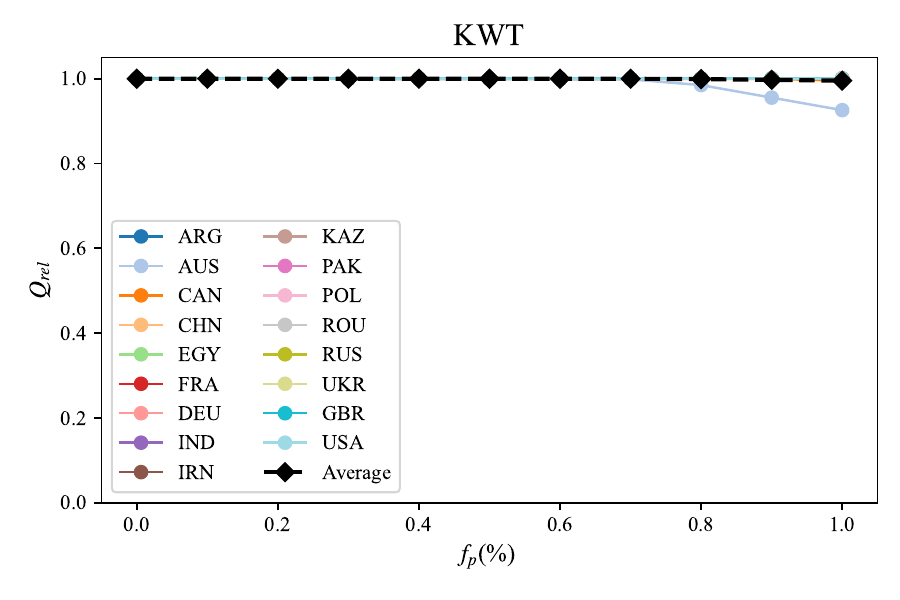}
    \includegraphics[width=0.24\linewidth]{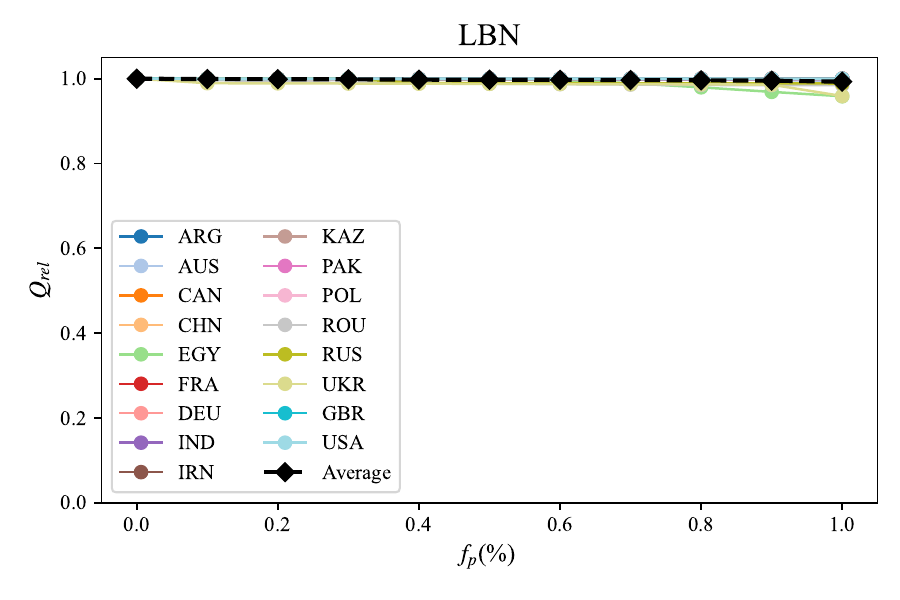}
    \includegraphics[width=0.24\linewidth]{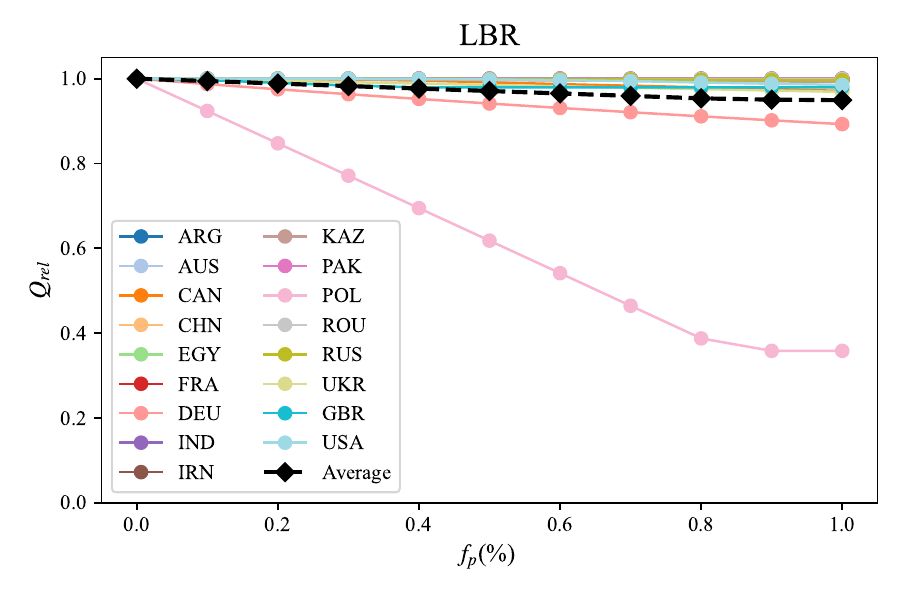}
    \includegraphics[width=0.24\linewidth]{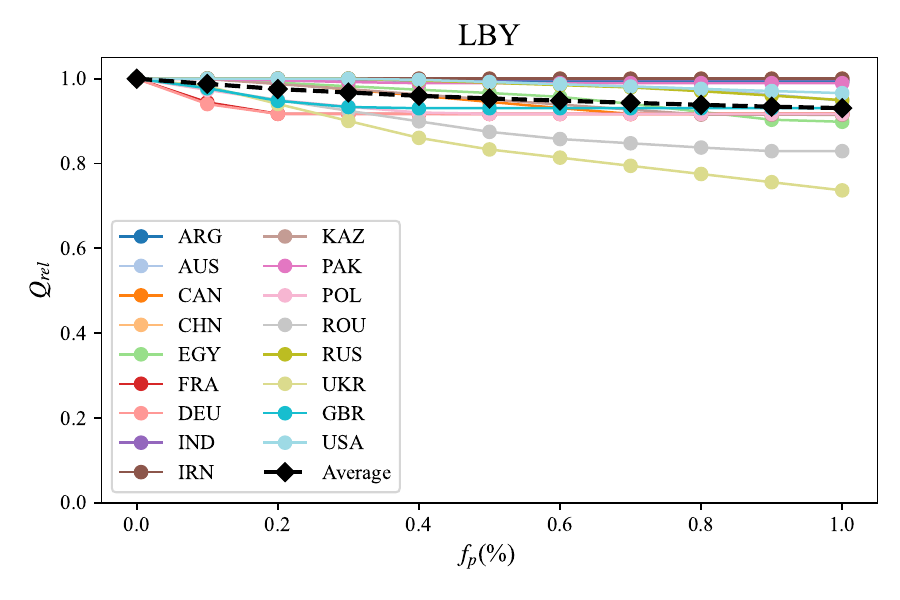}\\
    \includegraphics[width=0.24\linewidth]{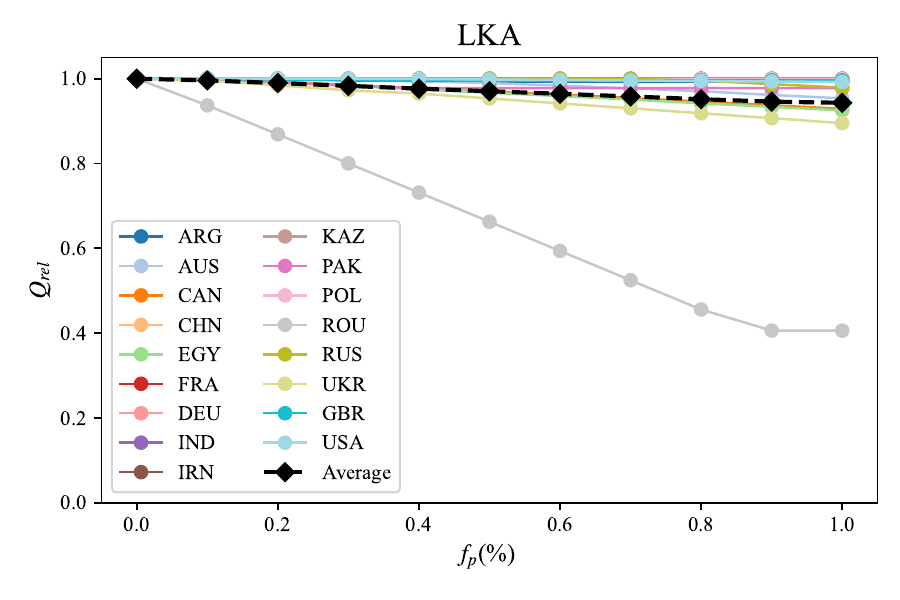}
    \includegraphics[width=0.24\linewidth]{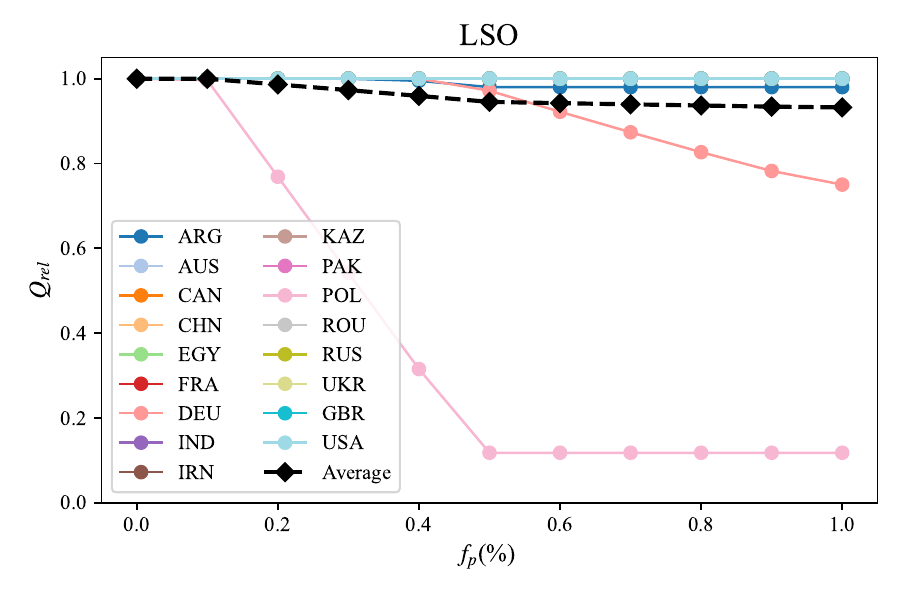}
    \includegraphics[width=0.24\linewidth]{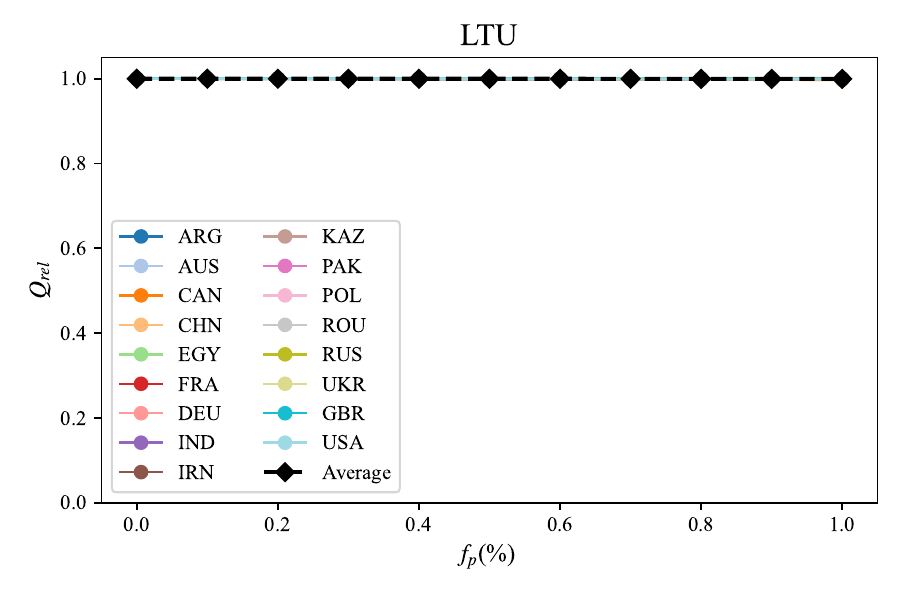}
    \includegraphics[width=0.24\linewidth]{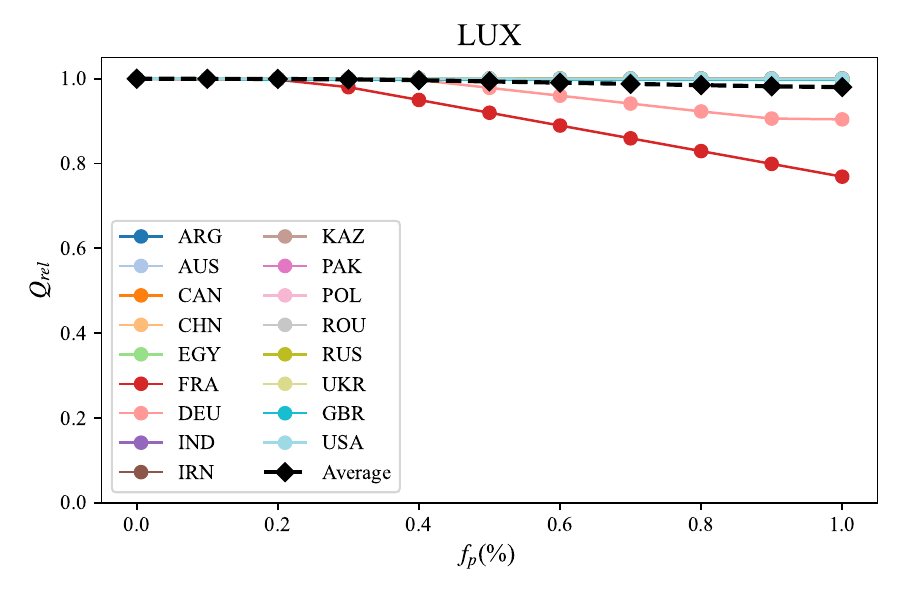}\\
    \includegraphics[width=0.24\linewidth]{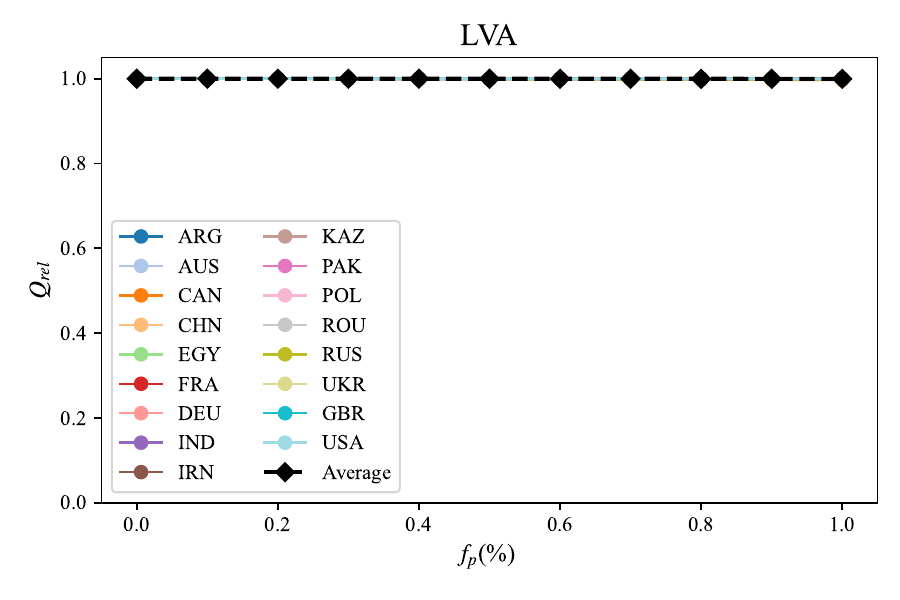}
    \includegraphics[width=0.24\linewidth]{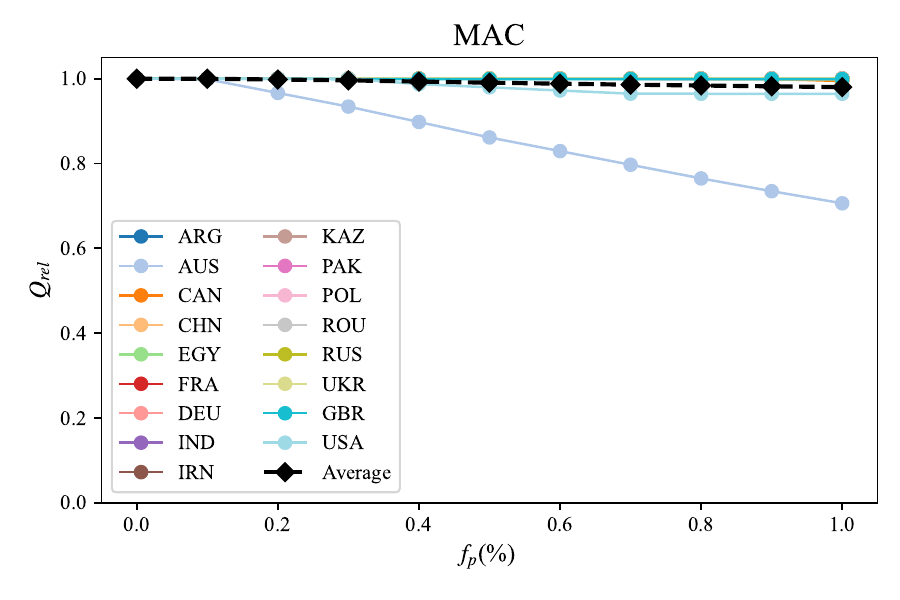}
    \includegraphics[width=0.24\linewidth]{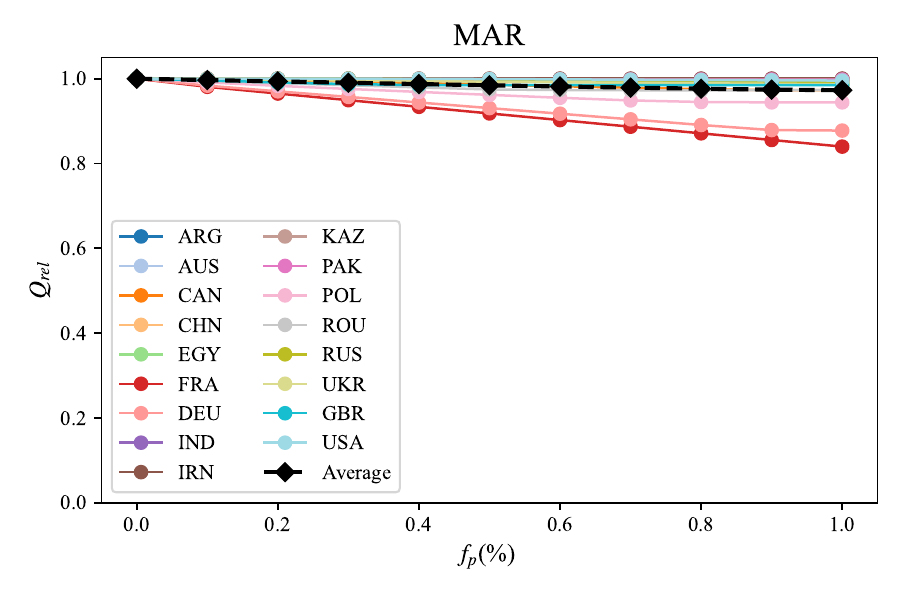}
    \includegraphics[width=0.24\linewidth]{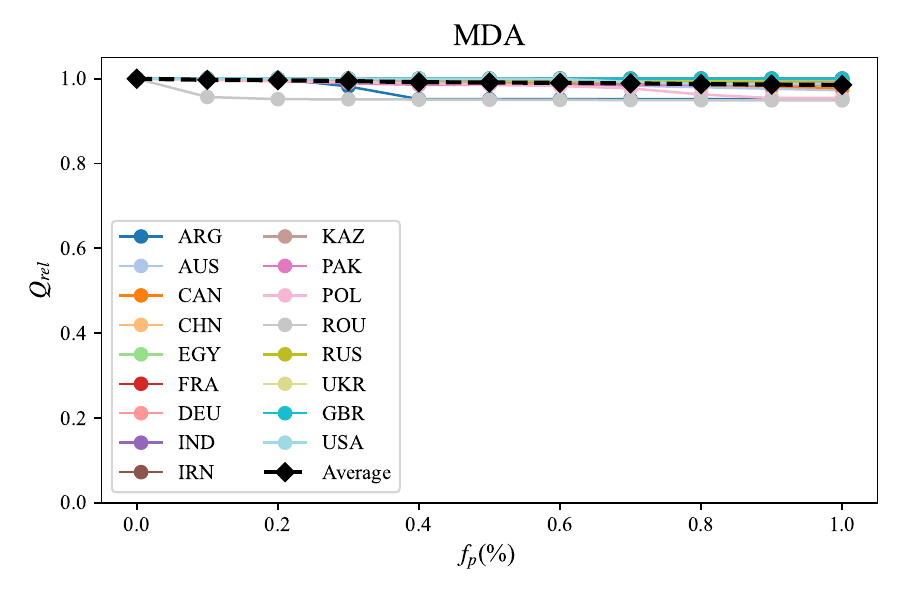}\\
    \includegraphics[width=0.24\linewidth]{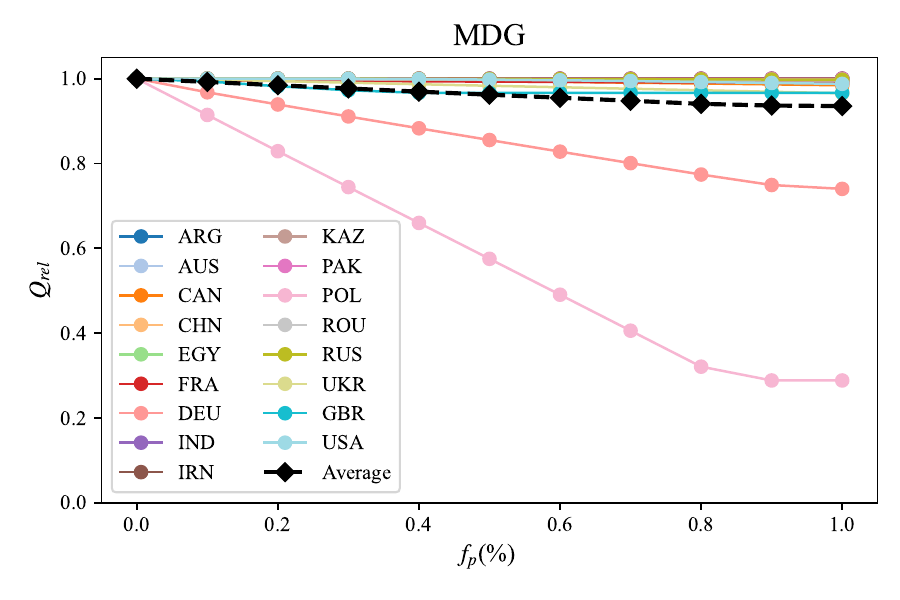}
    \includegraphics[width=0.24\linewidth]{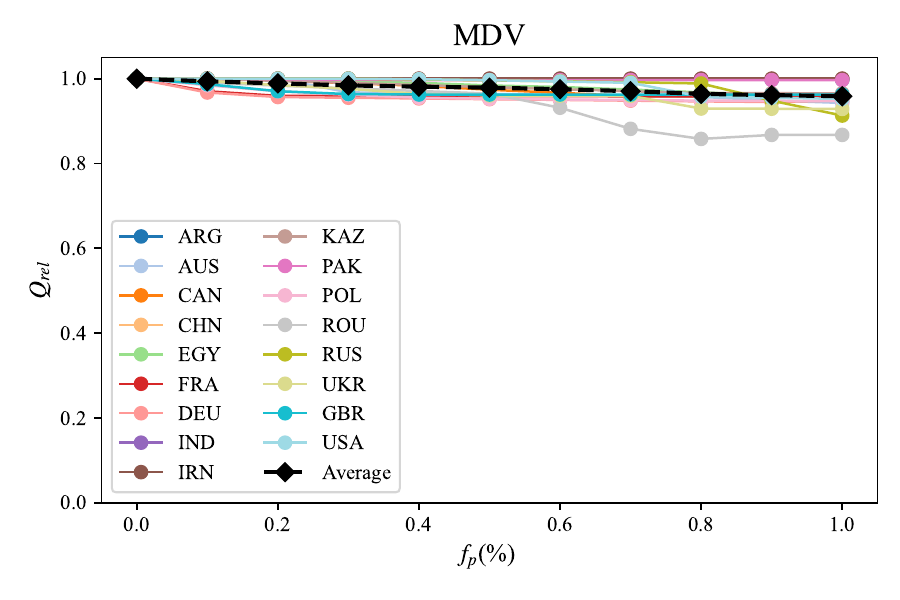}
    \includegraphics[width=0.24\linewidth]{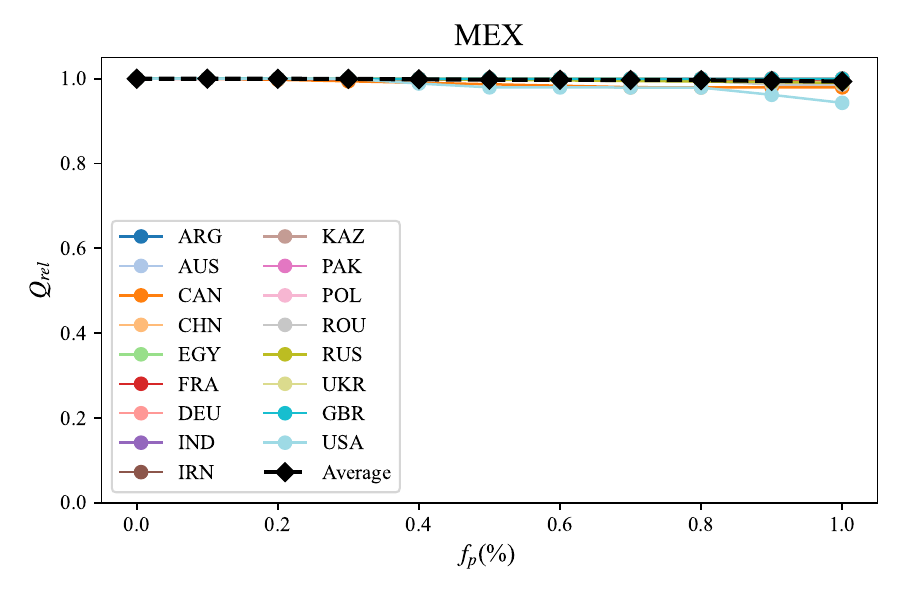}
    \includegraphics[width=0.24\linewidth]{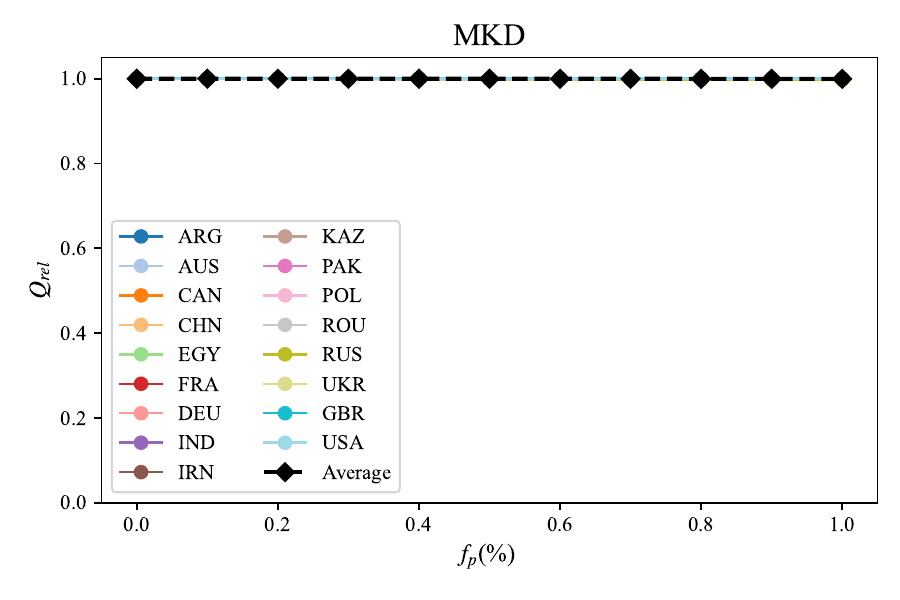}\\
    \caption{Wheat: $Q_{\mathrm{rel}}$ curves under global 1\% production shock (page 3).}
    \label{Fig:Wheat_Qrel_global1pct_3}
\end{figure}

\begin{figure}[p]\ContinuedFloat
    \centering
    \includegraphics[width=0.24\linewidth]{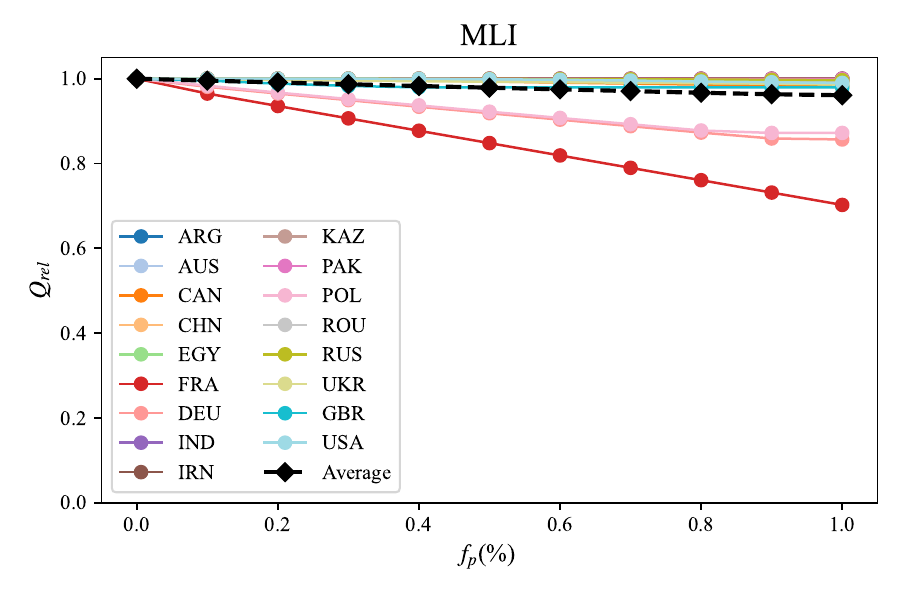}
    \includegraphics[width=0.24\linewidth]{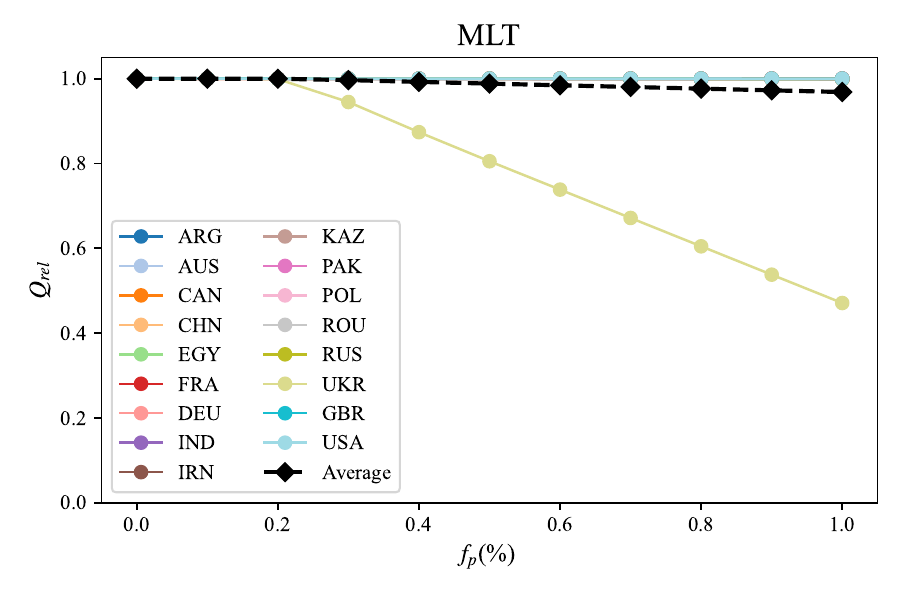}
    \includegraphics[width=0.24\linewidth]{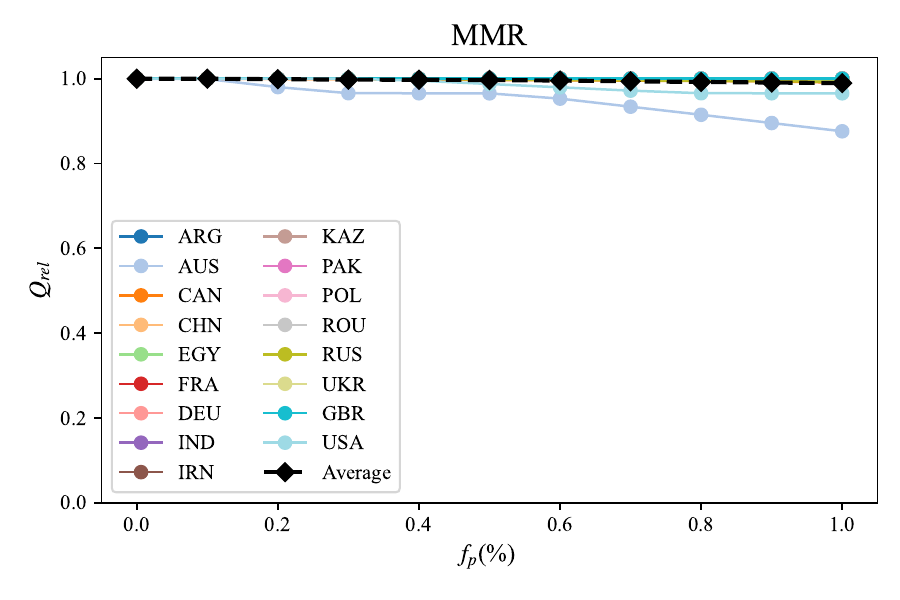}
    \includegraphics[width=0.24\linewidth]{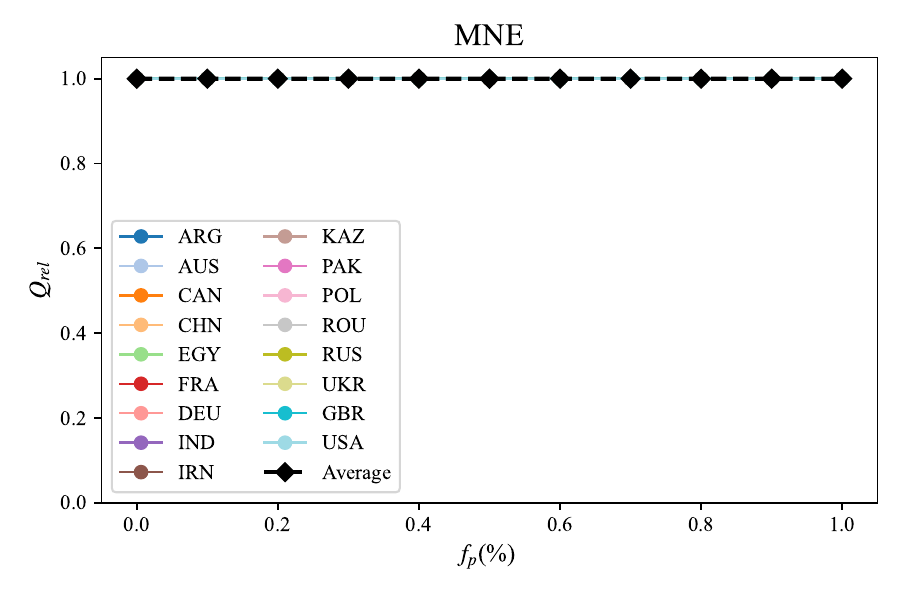}\\
    \includegraphics[width=0.24\linewidth]{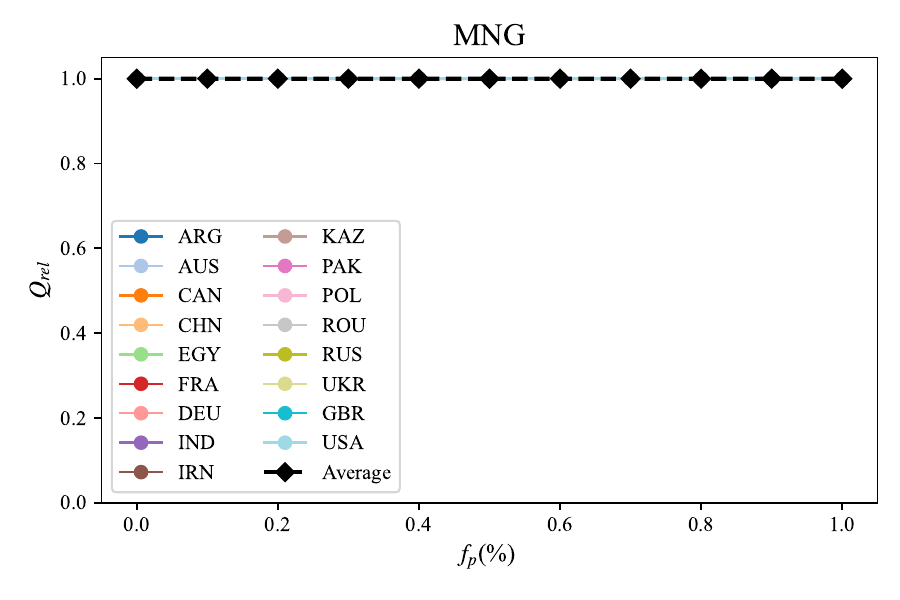}
    \includegraphics[width=0.24\linewidth]{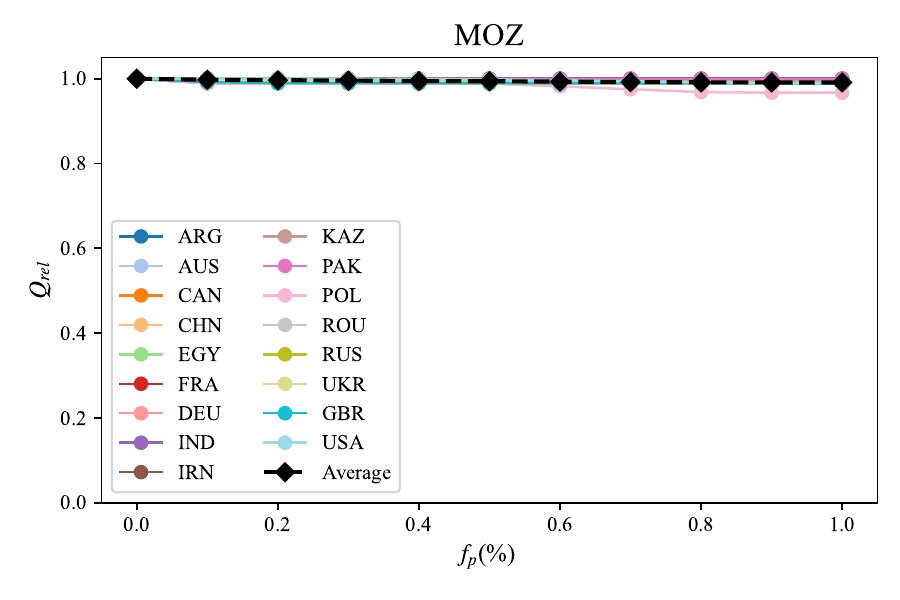}
    \includegraphics[width=0.24\linewidth]{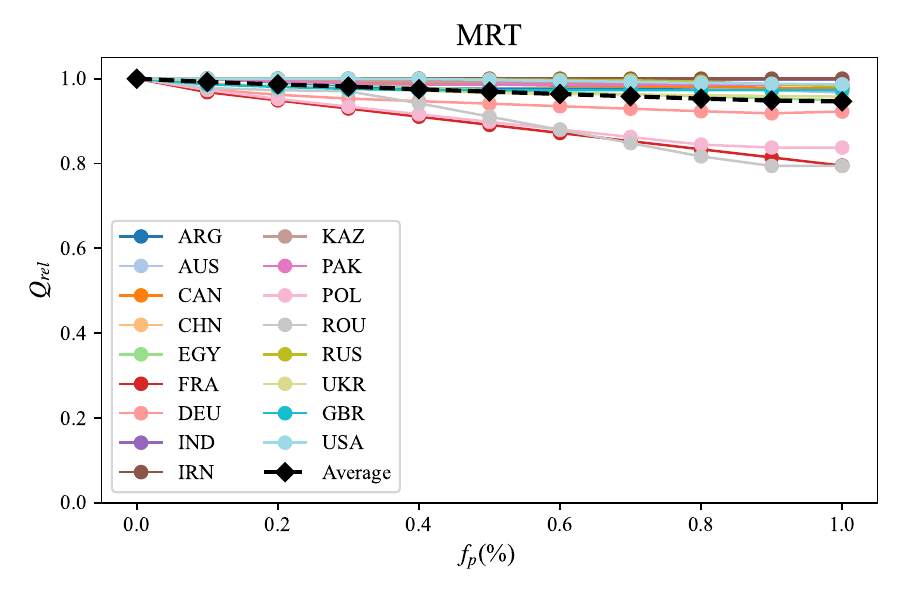}
    \includegraphics[width=0.24\linewidth]{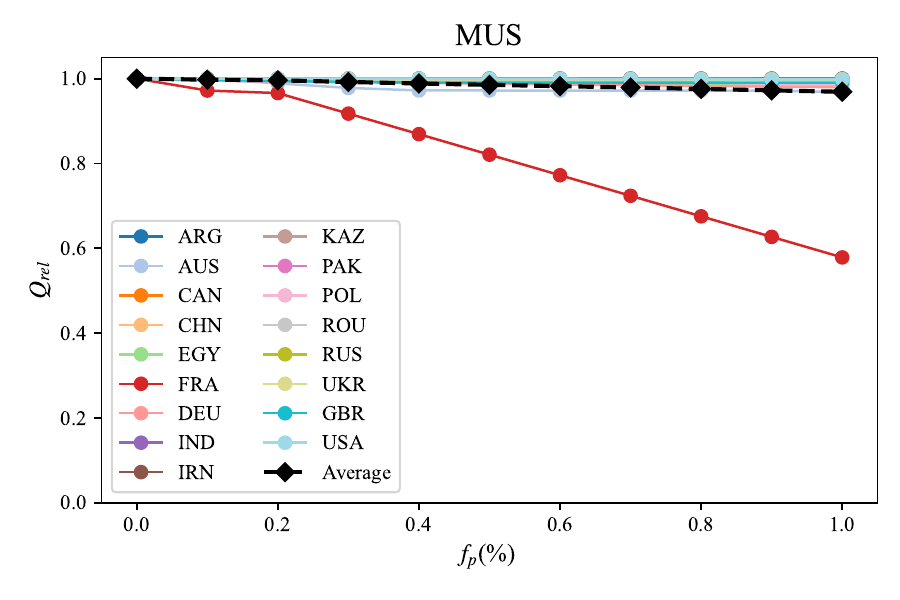}\\
    \includegraphics[width=0.24\linewidth]{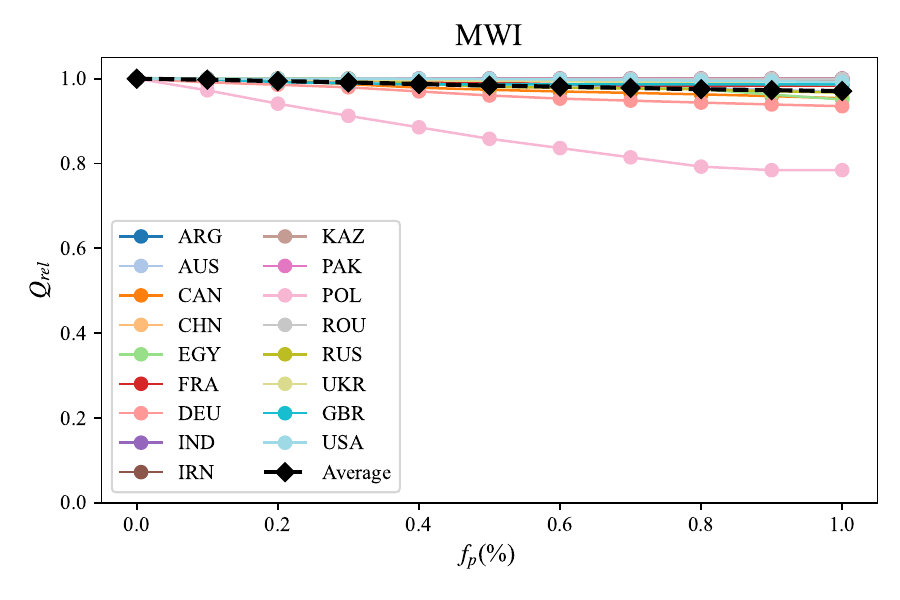}
    \includegraphics[width=0.24\linewidth]{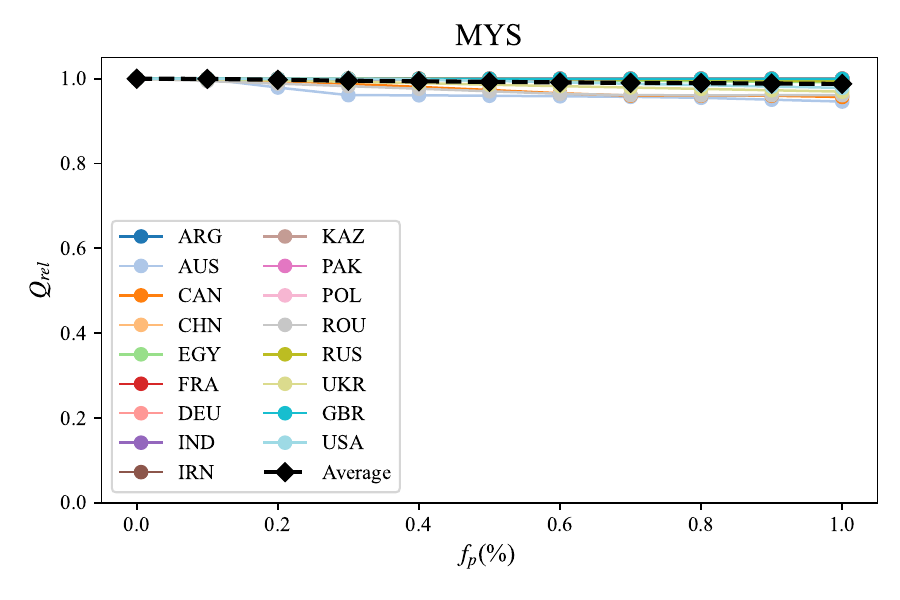}
    \includegraphics[width=0.24\linewidth]{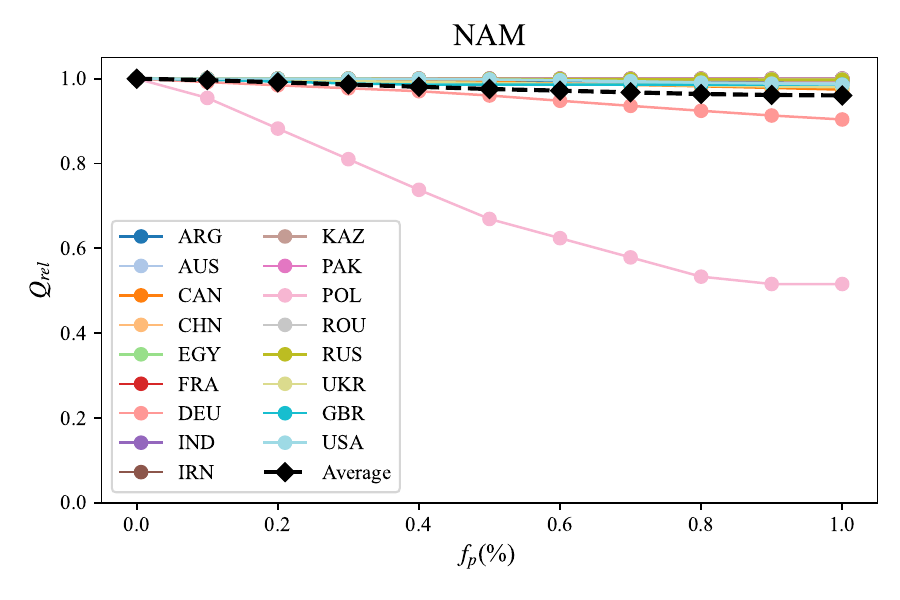}
    \includegraphics[width=0.24\linewidth]{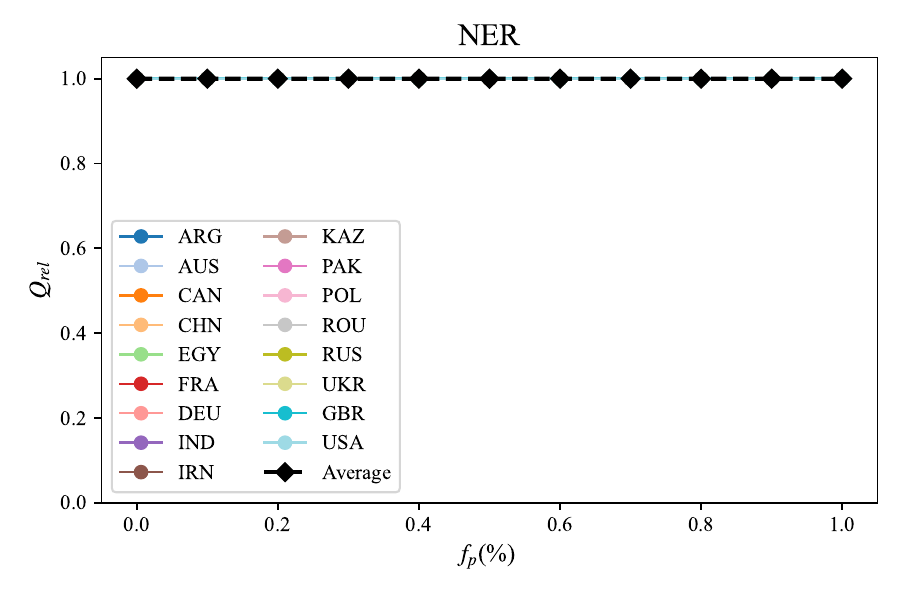}\\
    \includegraphics[width=0.24\linewidth]{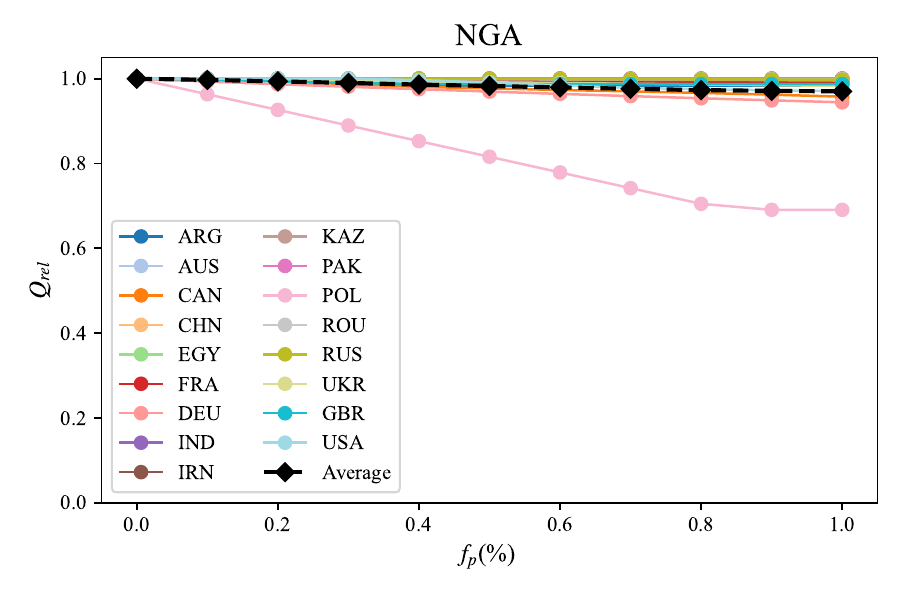}
    \includegraphics[width=0.24\linewidth]{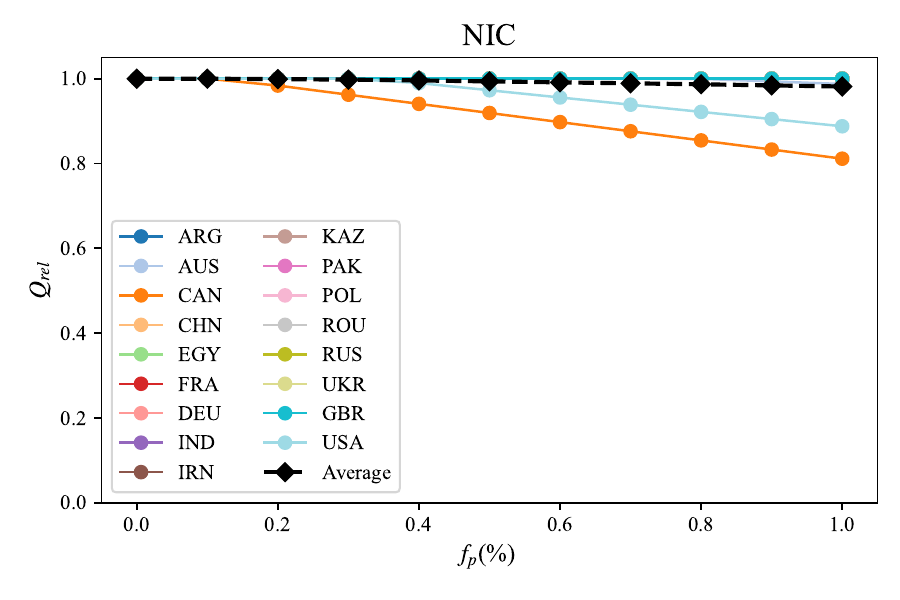}
    \includegraphics[width=0.24\linewidth]{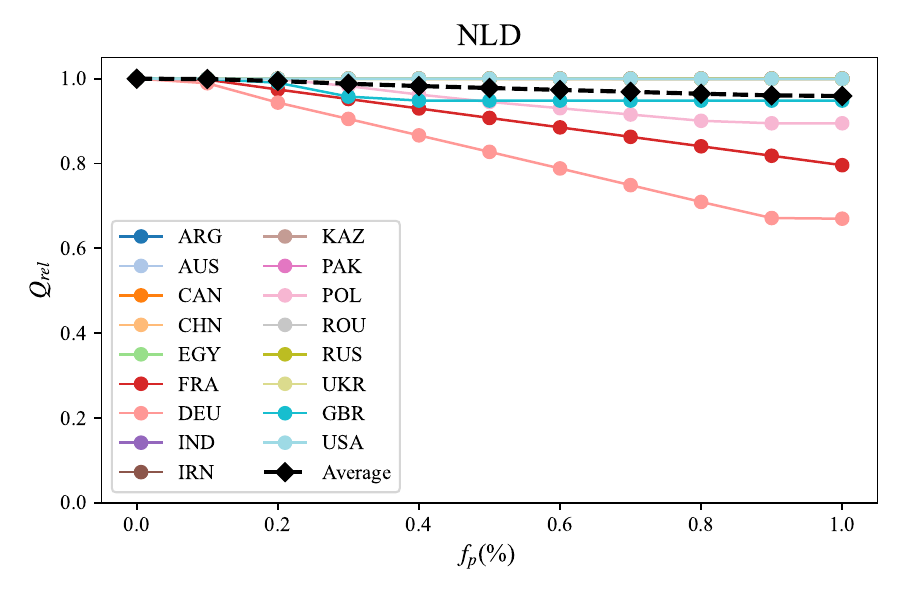}
    \includegraphics[width=0.24\linewidth]{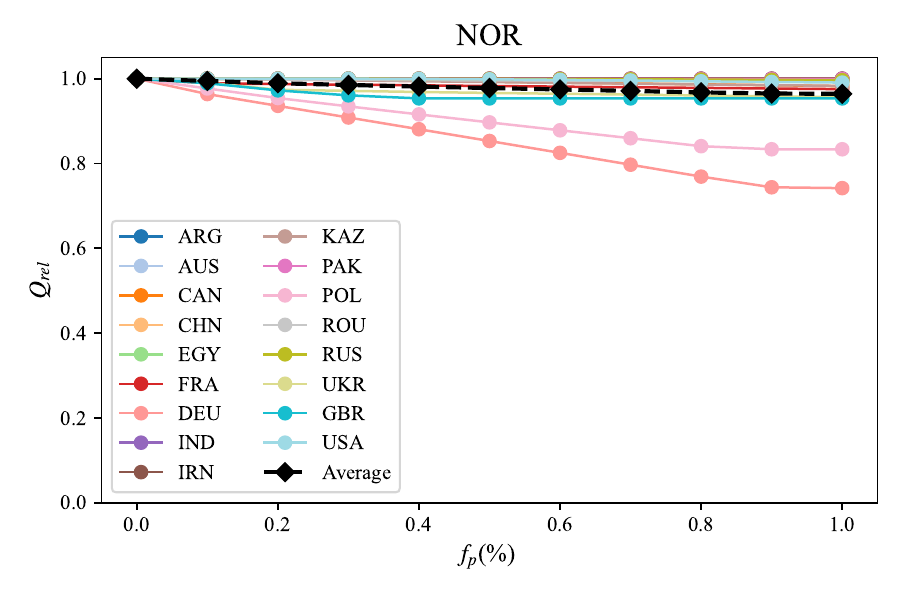}\\
    \includegraphics[width=0.24\linewidth]{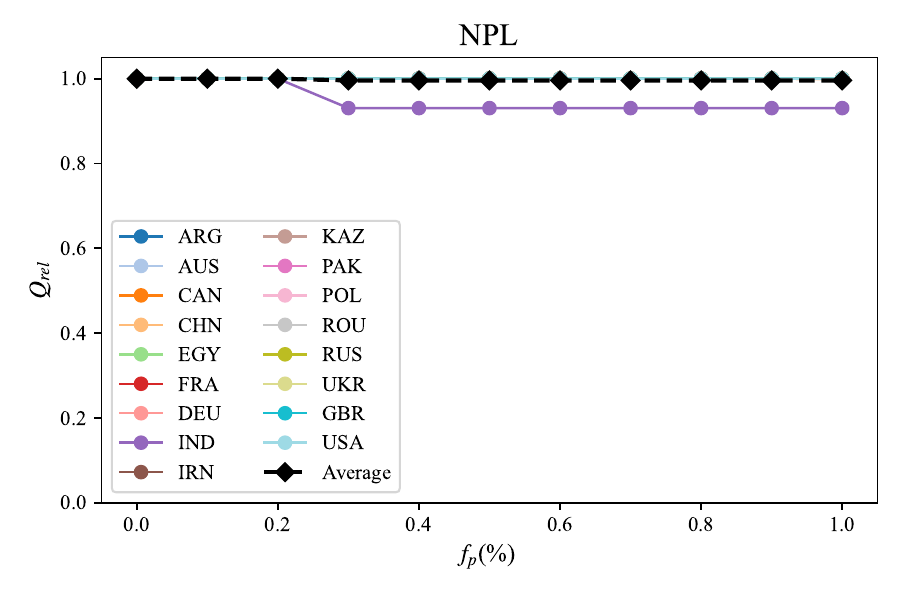}
    \includegraphics[width=0.24\linewidth]{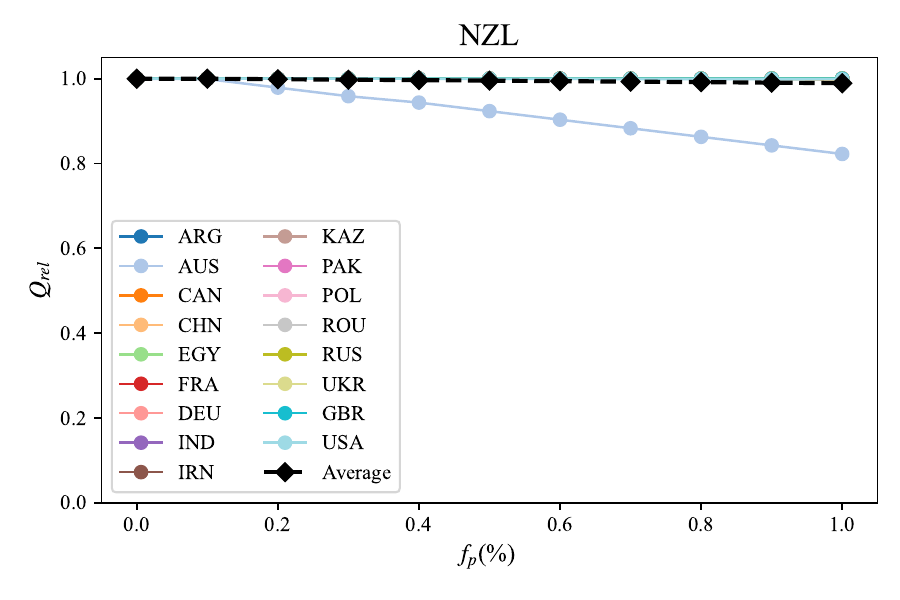}
    \includegraphics[width=0.24\linewidth]{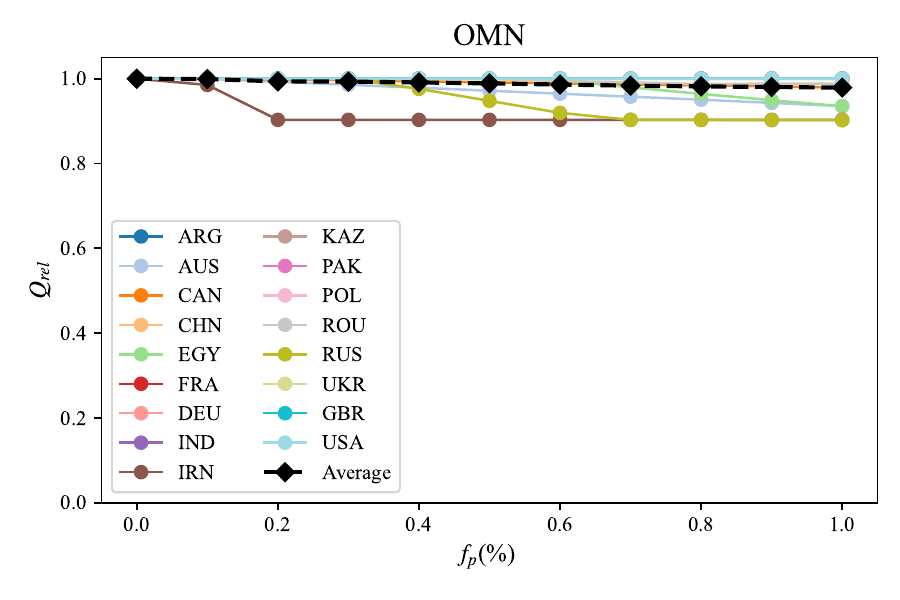}
    \includegraphics[width=0.24\linewidth]{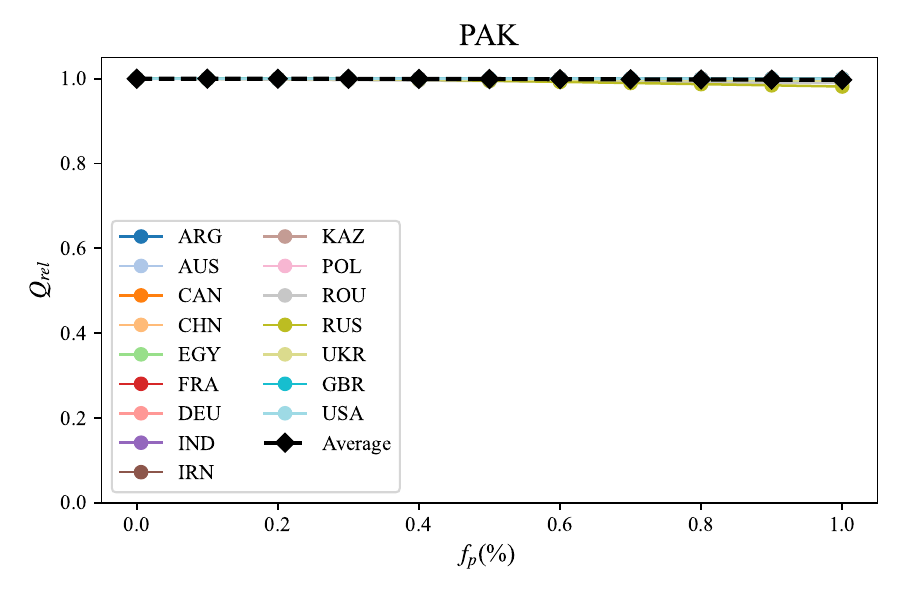}\\
    \includegraphics[width=0.24\linewidth]{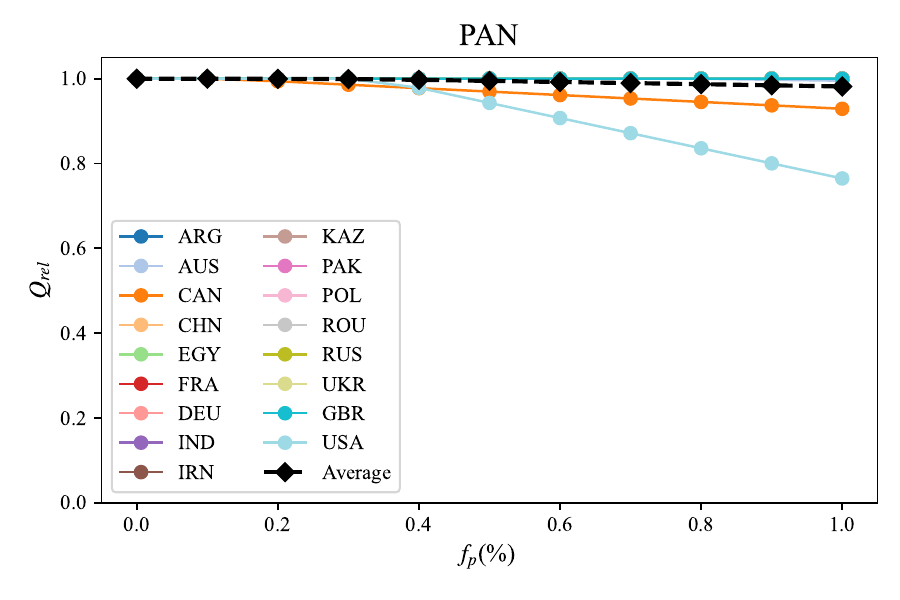}
    \includegraphics[width=0.24\linewidth]{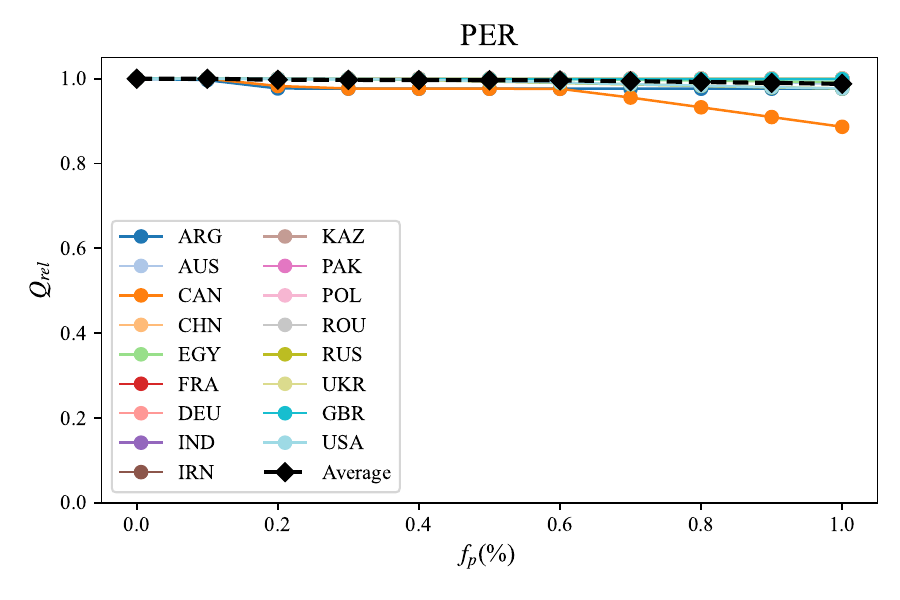}
    \includegraphics[width=0.24\linewidth]{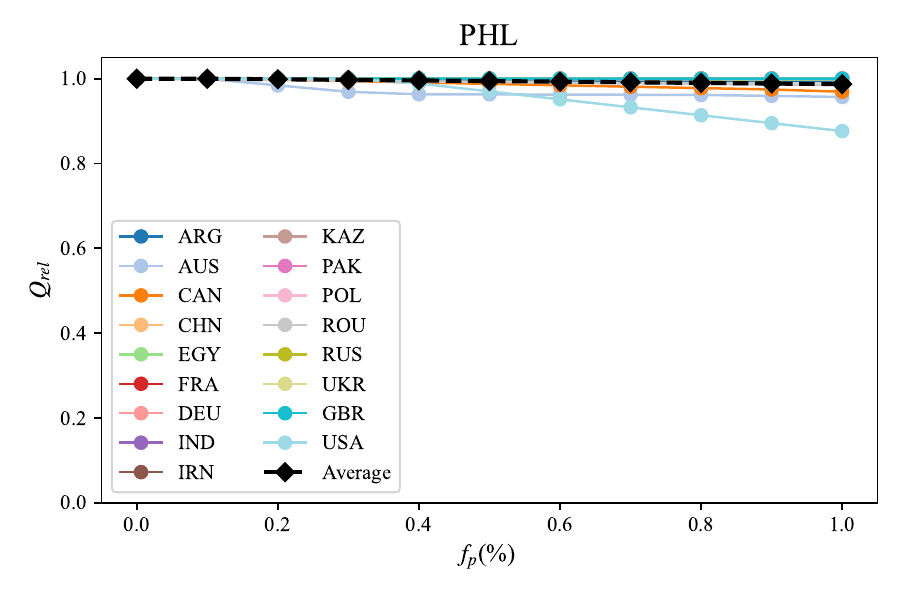}
    \includegraphics[width=0.24\linewidth]{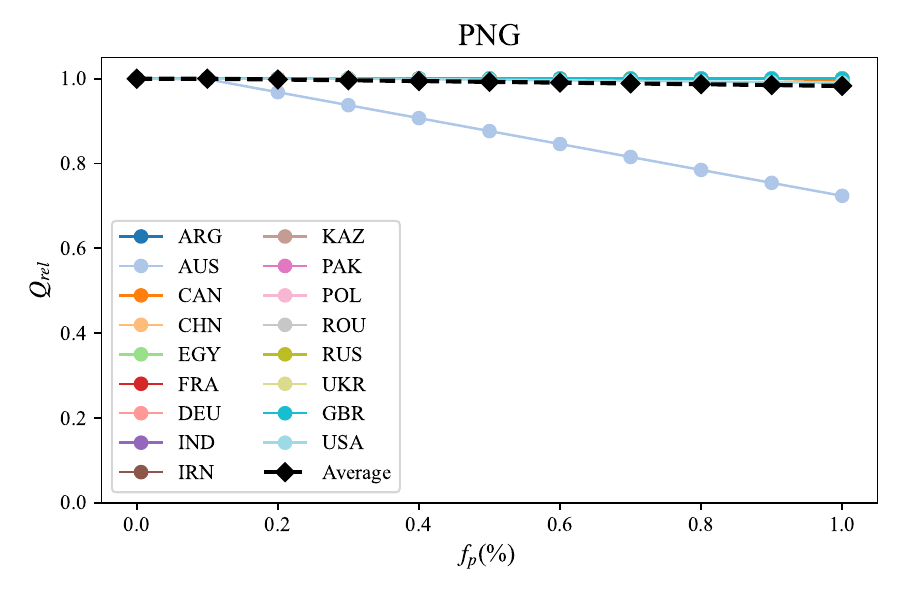}\\
    \includegraphics[width=0.24\linewidth]{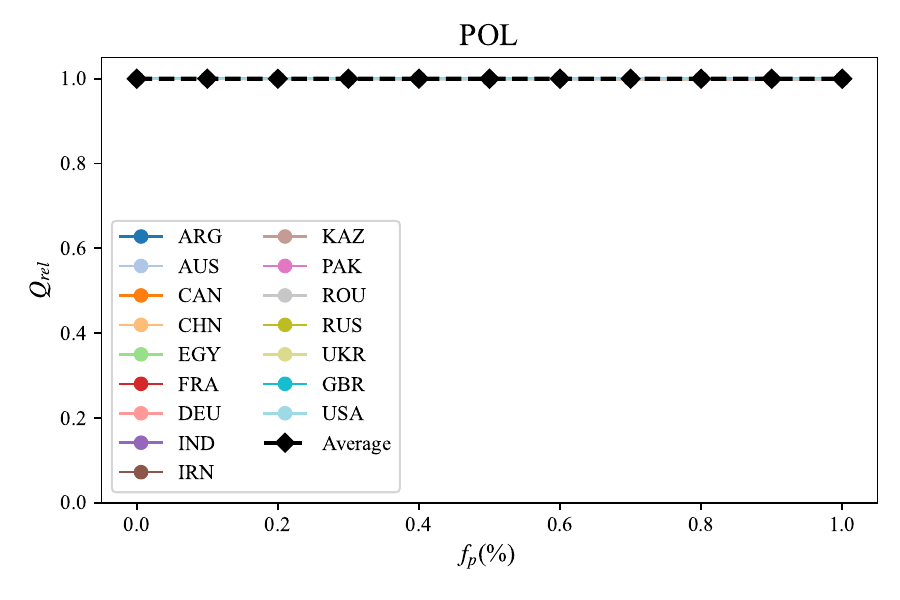}
    \includegraphics[width=0.24\linewidth]{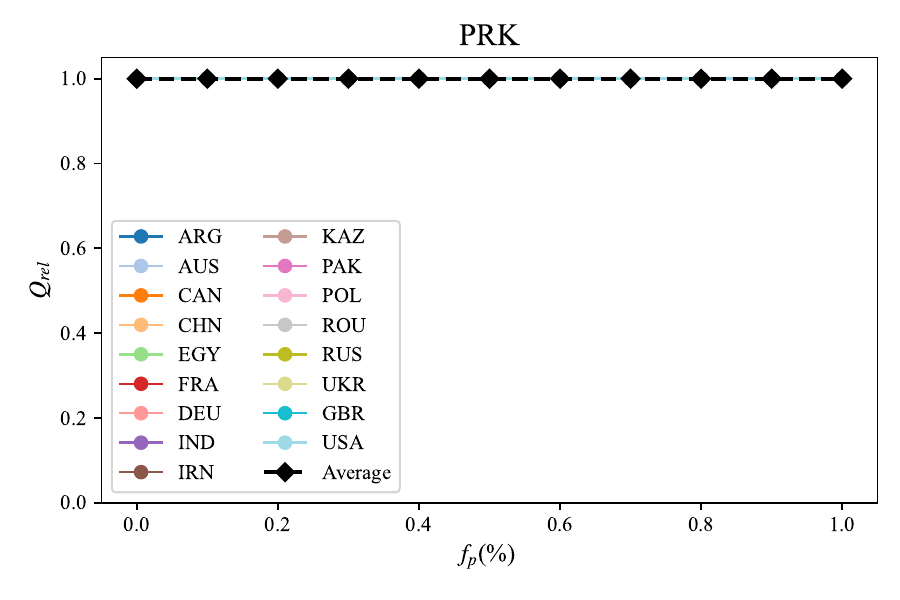}
    \includegraphics[width=0.24\linewidth]{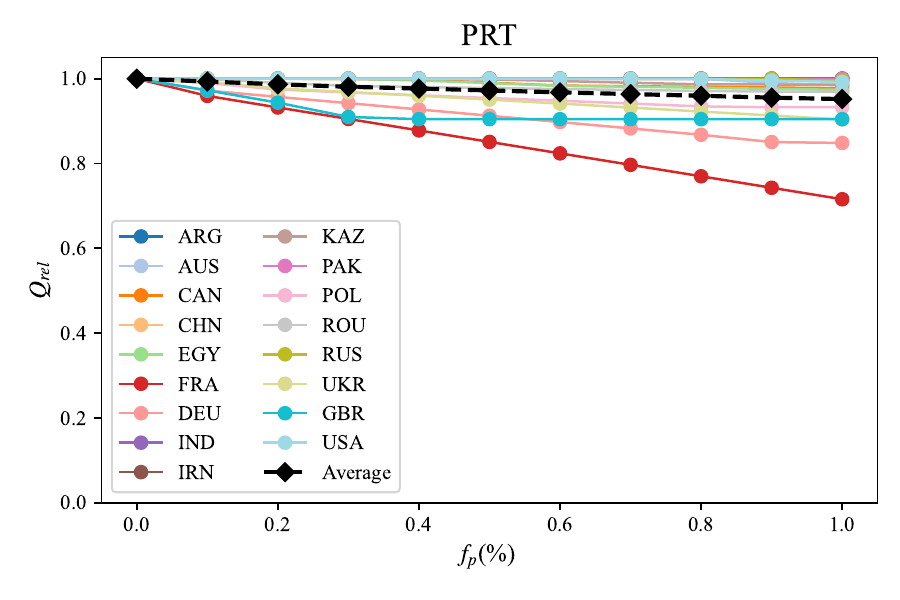}
    \includegraphics[width=0.24\linewidth]{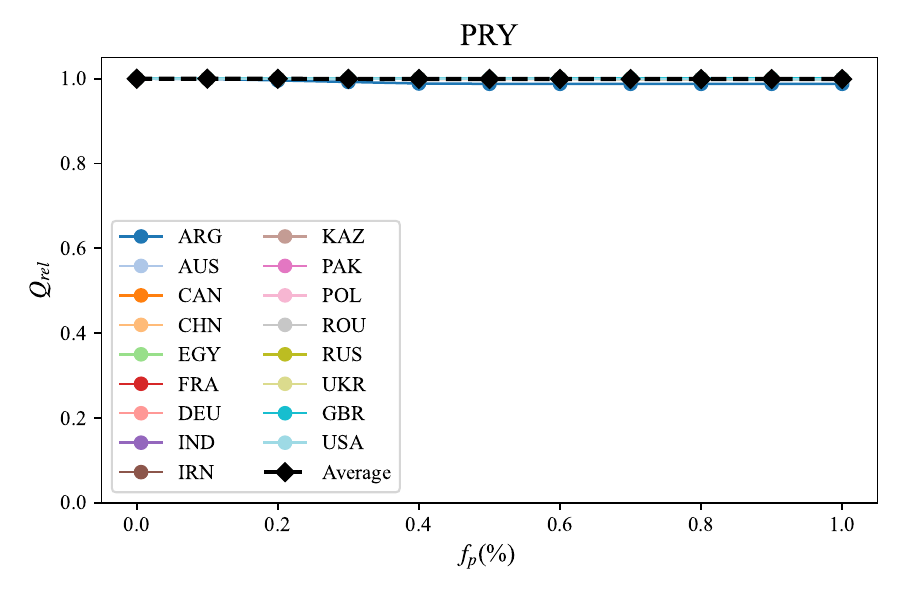}\\
    \includegraphics[width=0.24\linewidth]{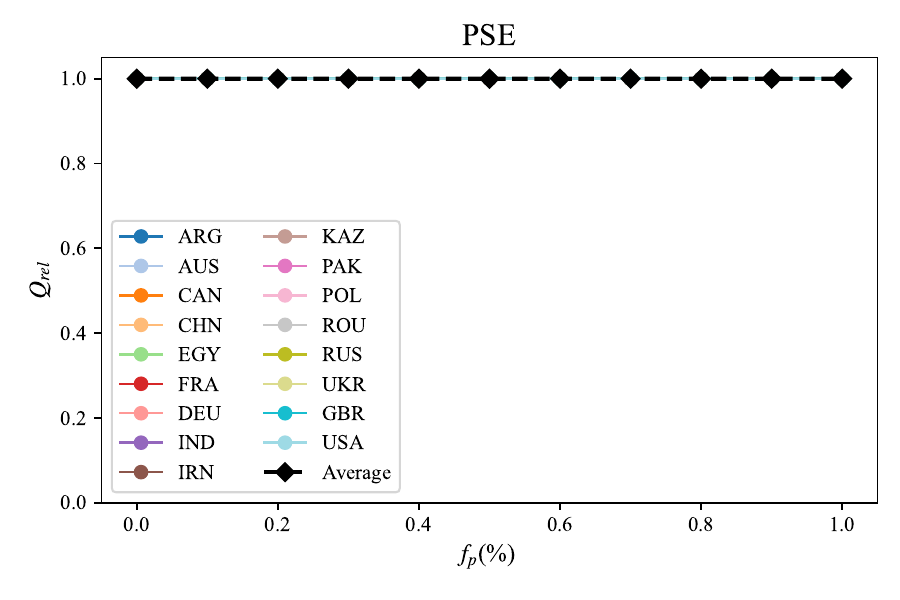}
    \includegraphics[width=0.24\linewidth]{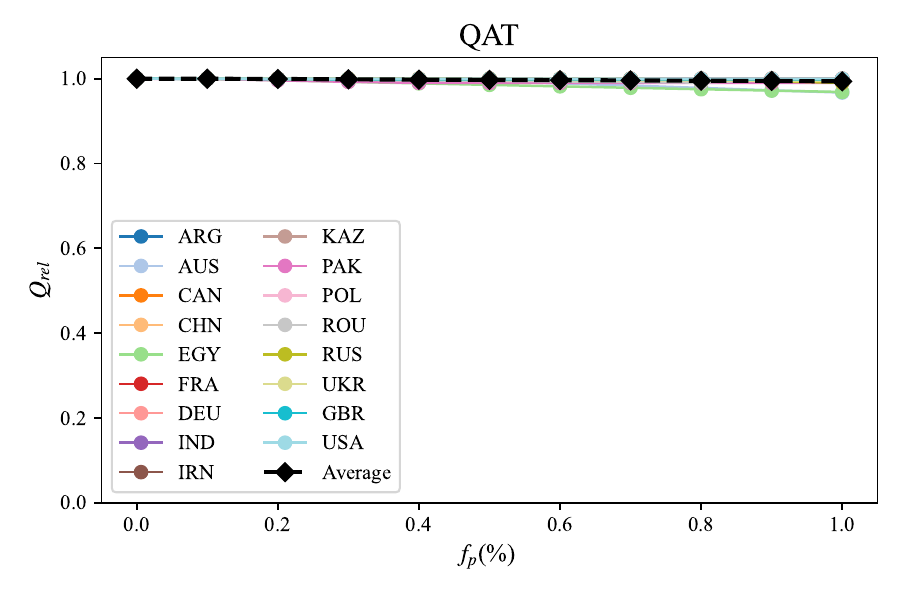}
    \includegraphics[width=0.24\linewidth]{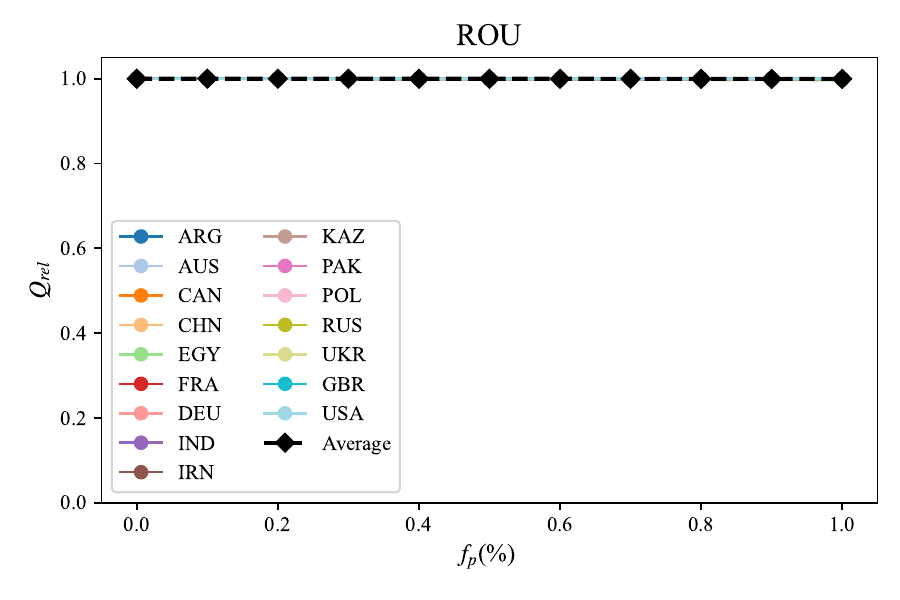}
    \includegraphics[width=0.24\linewidth]{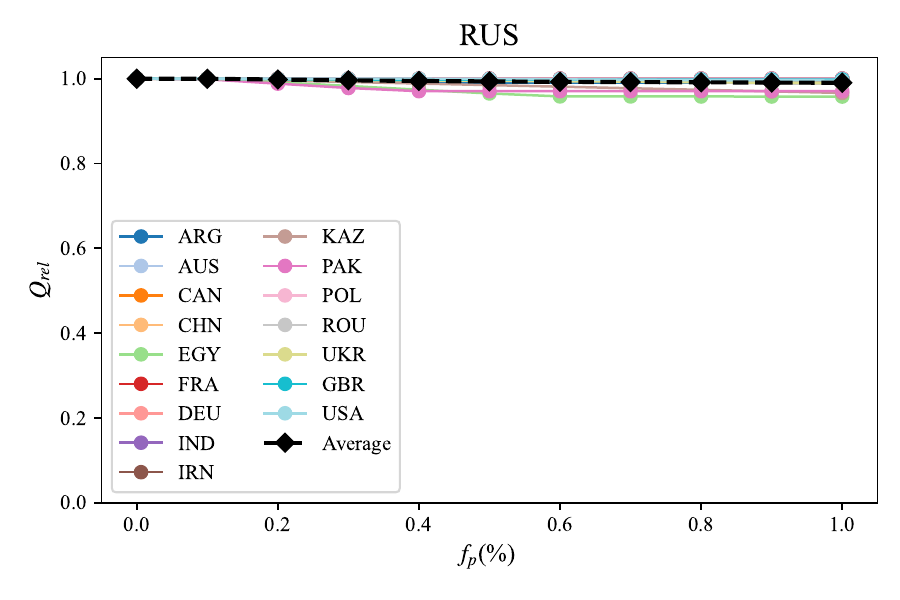}
    \caption{Wheat: $Q_{\mathrm{rel}}$ curves under global 1\% production shock (page 4).}
    \label{Fig:Wheat_Qrel_global1pct_4}
\end{figure}

\begin{figure}[p]\ContinuedFloat
    % \centering
    \includegraphics[width=0.24\linewidth]{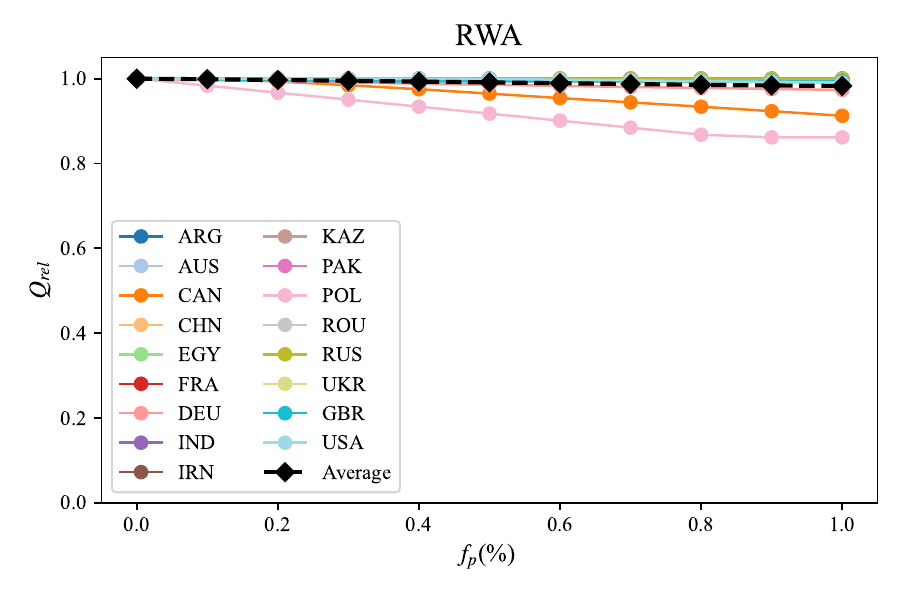}
    \includegraphics[width=0.24\linewidth]{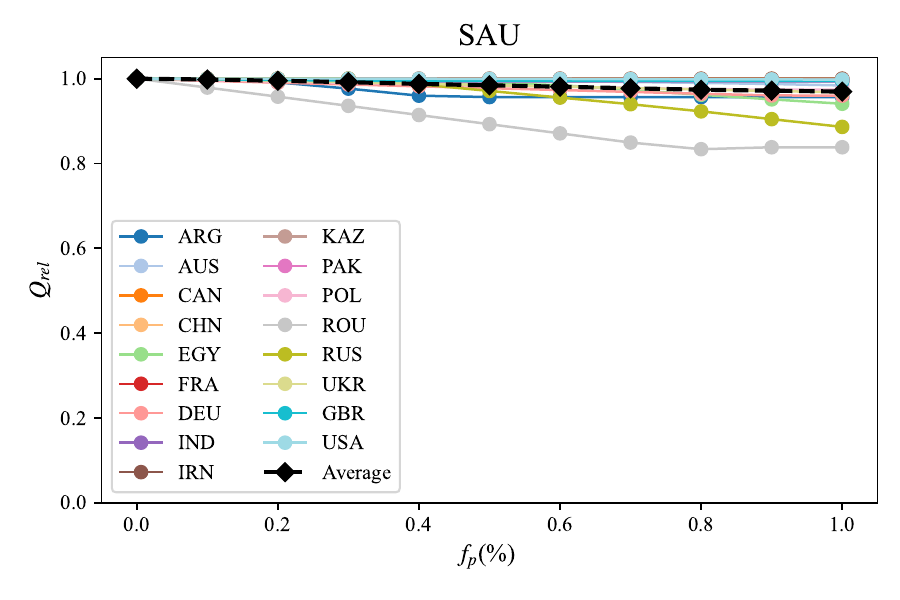}
    \includegraphics[width=0.24\linewidth]{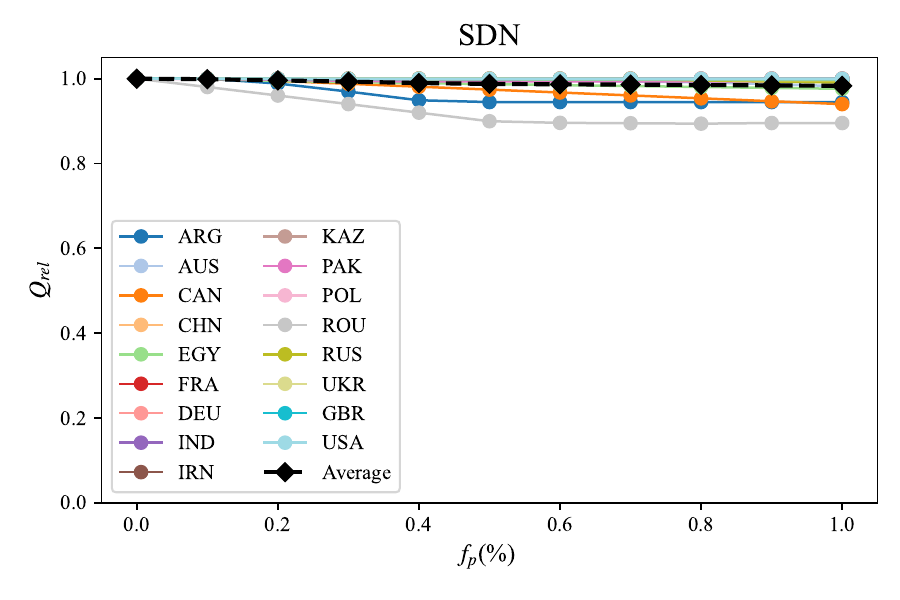}
    \includegraphics[width=0.24\linewidth]{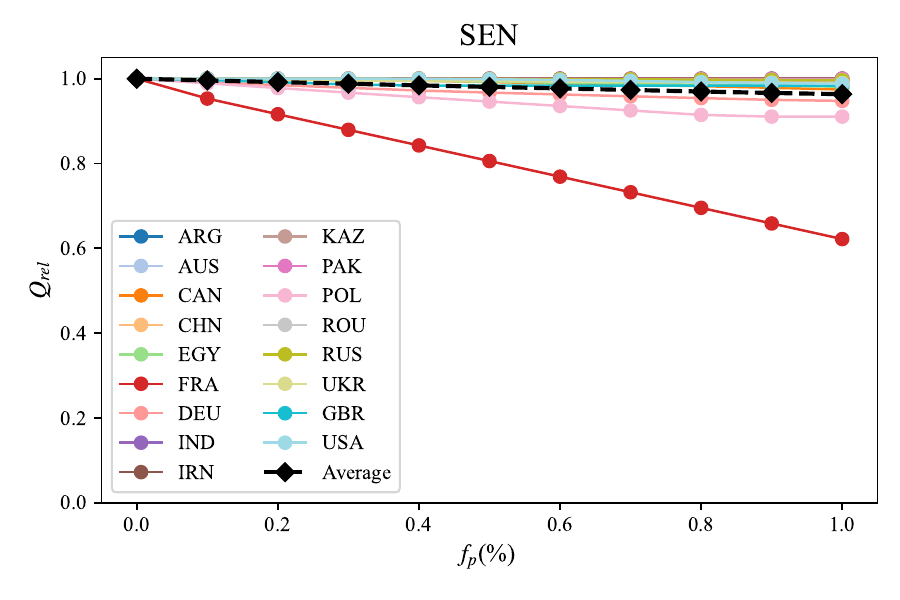}\\
    \includegraphics[width=0.24\linewidth]{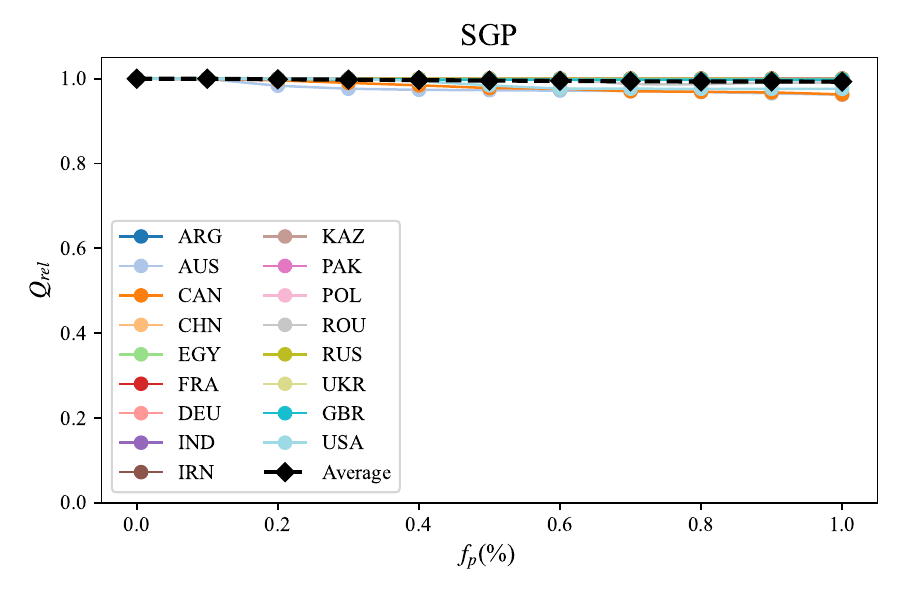}
    \includegraphics[width=0.24\linewidth]{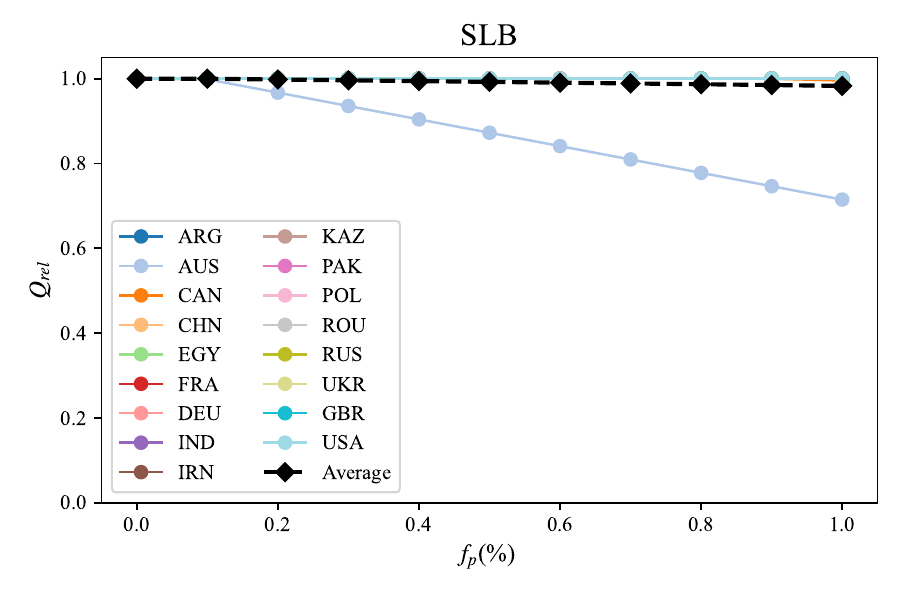}
    \includegraphics[width=0.24\linewidth]{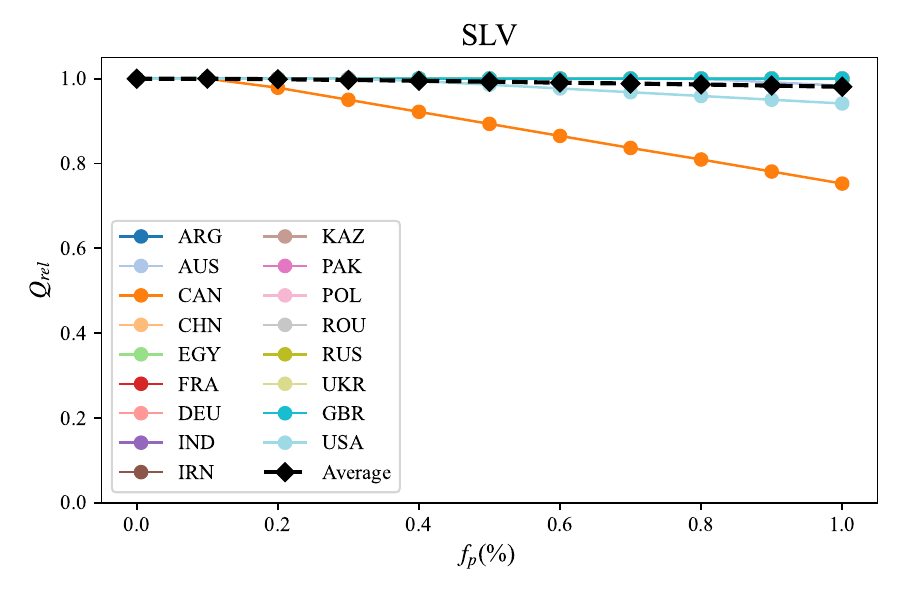}
    \includegraphics[width=0.24\linewidth]{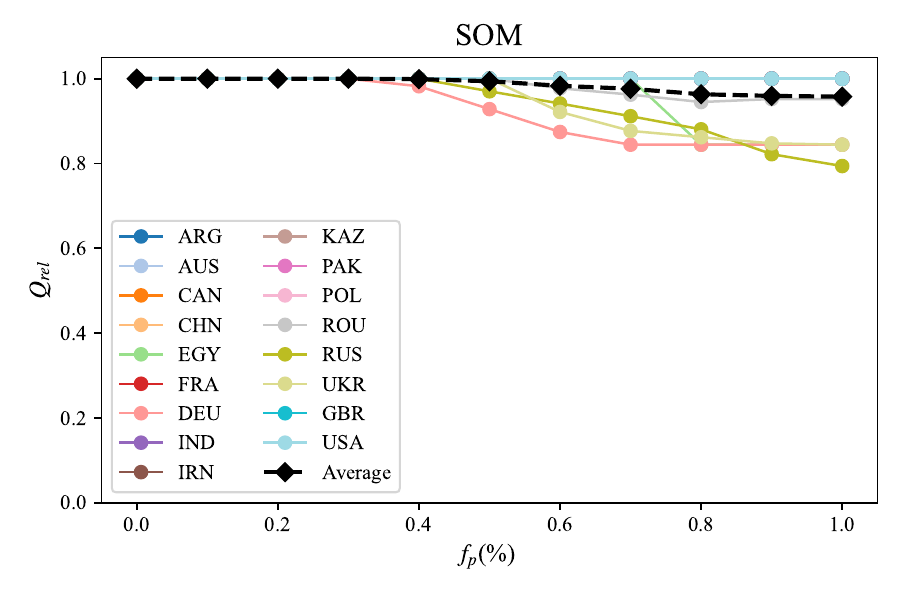}\\
    \includegraphics[width=0.24\linewidth]{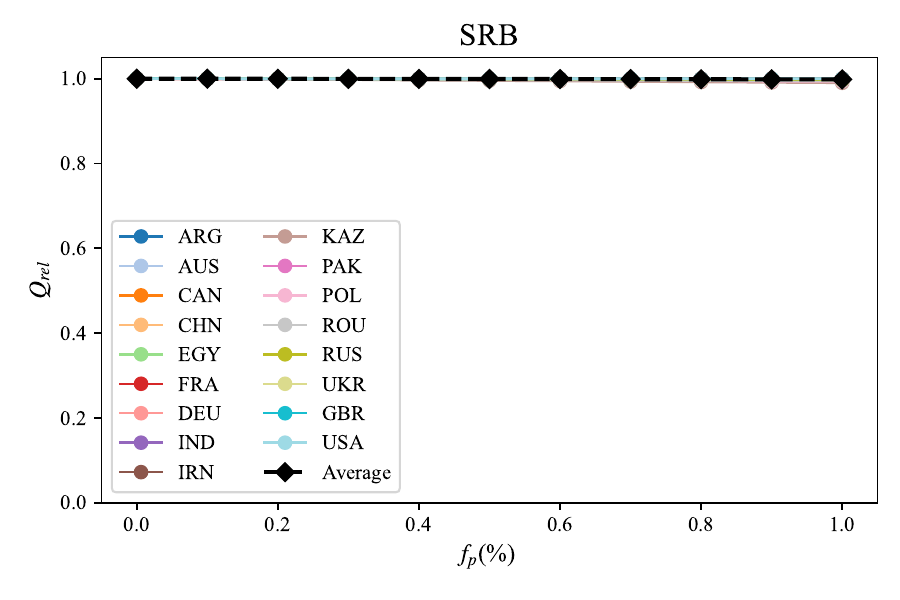}
    \includegraphics[width=0.24\linewidth]{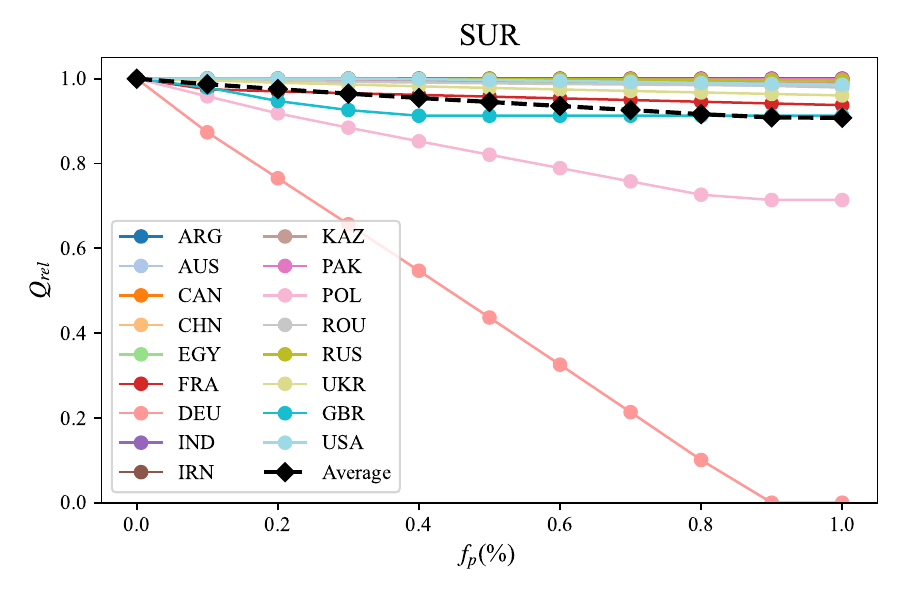}
    \includegraphics[width=0.24\linewidth]{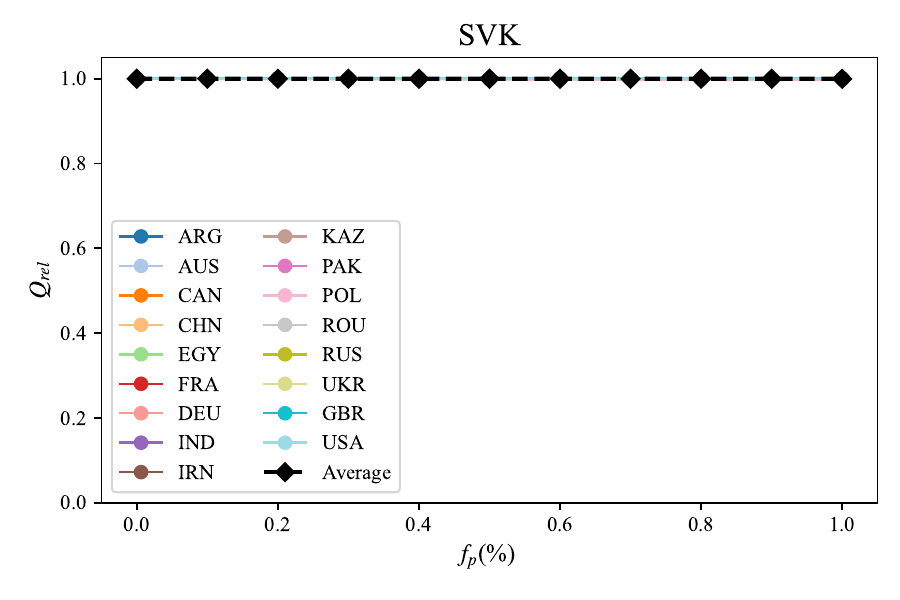}
    \includegraphics[width=0.24\linewidth]{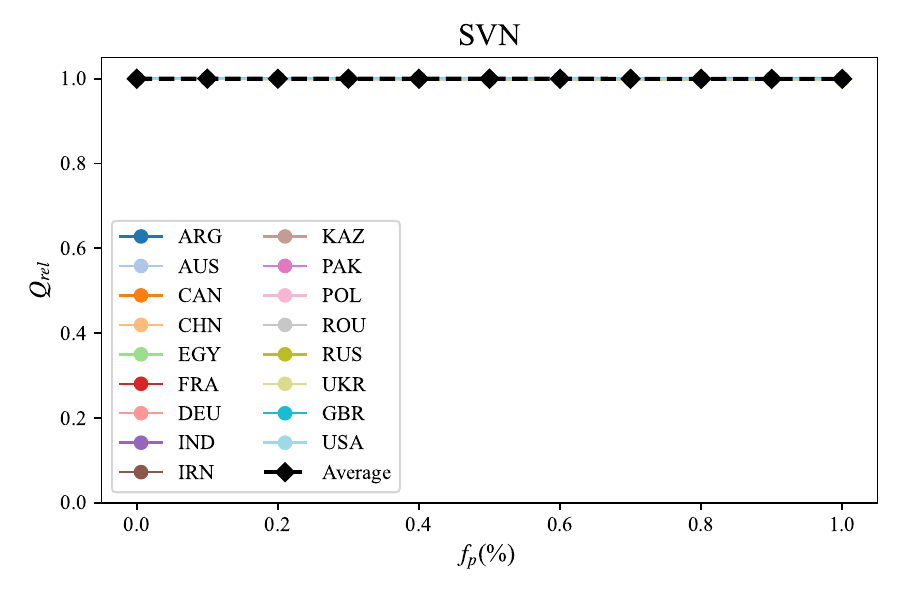}\\
    \includegraphics[width=0.24\linewidth]{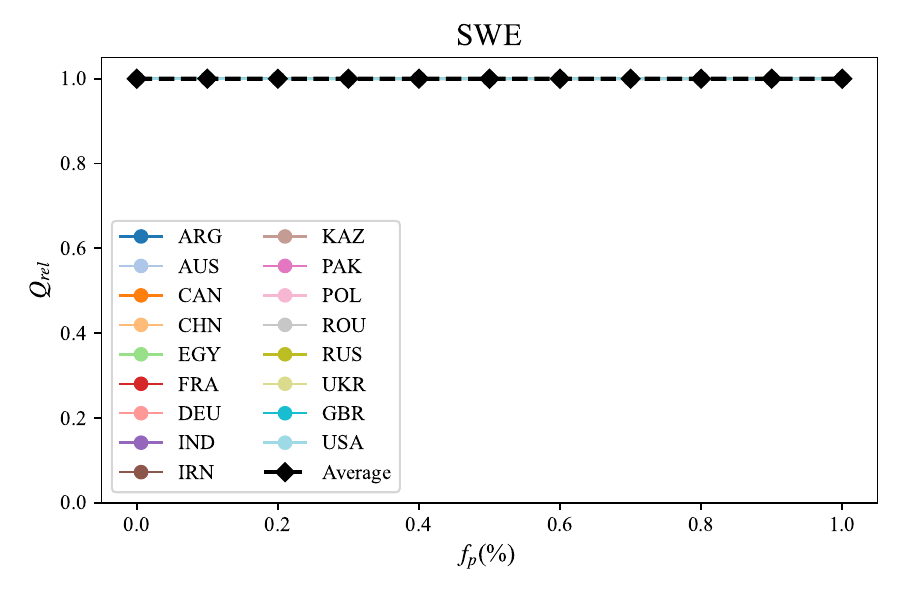}
    \includegraphics[width=0.24\linewidth]{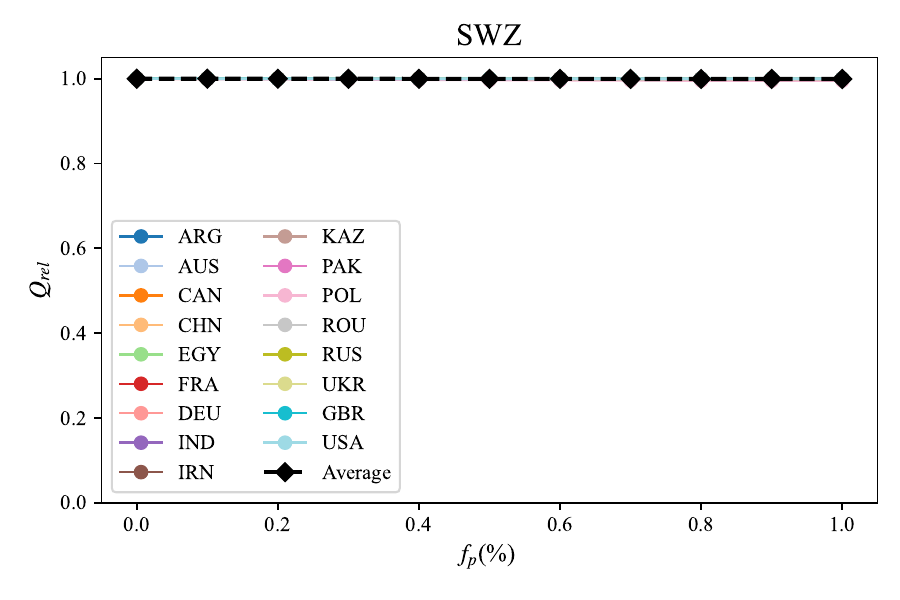}
    \includegraphics[width=0.24\linewidth]{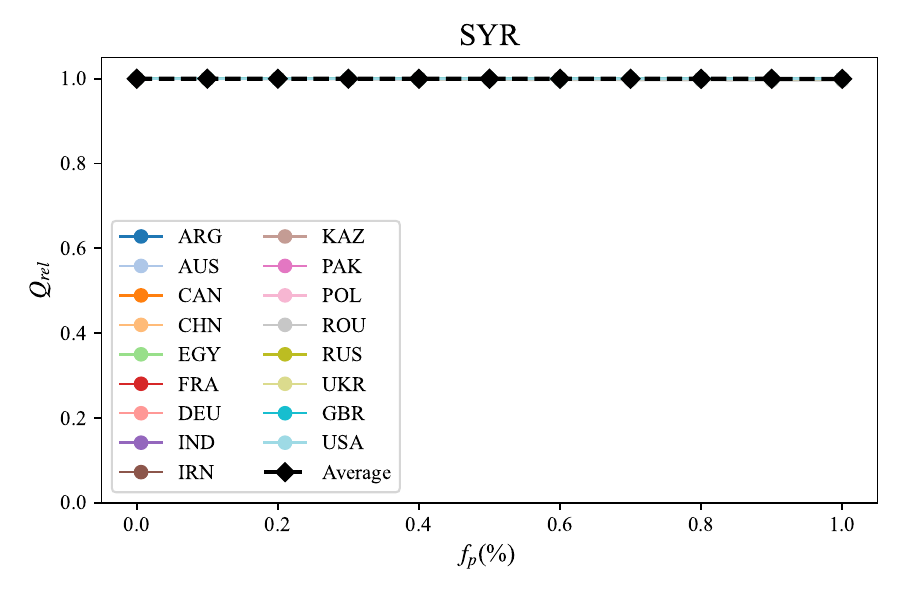}
    \includegraphics[width=0.24\linewidth]{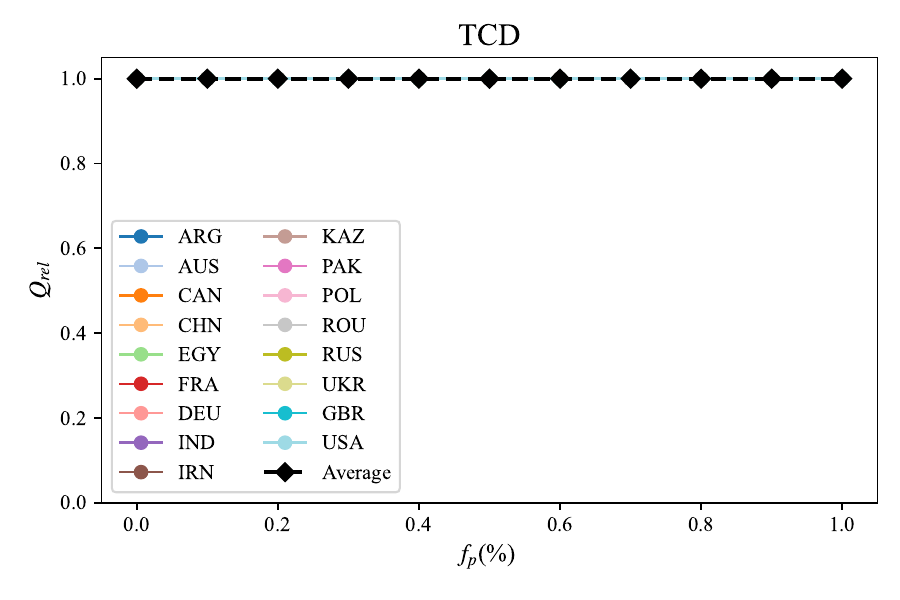}\\
    \includegraphics[width=0.24\linewidth]{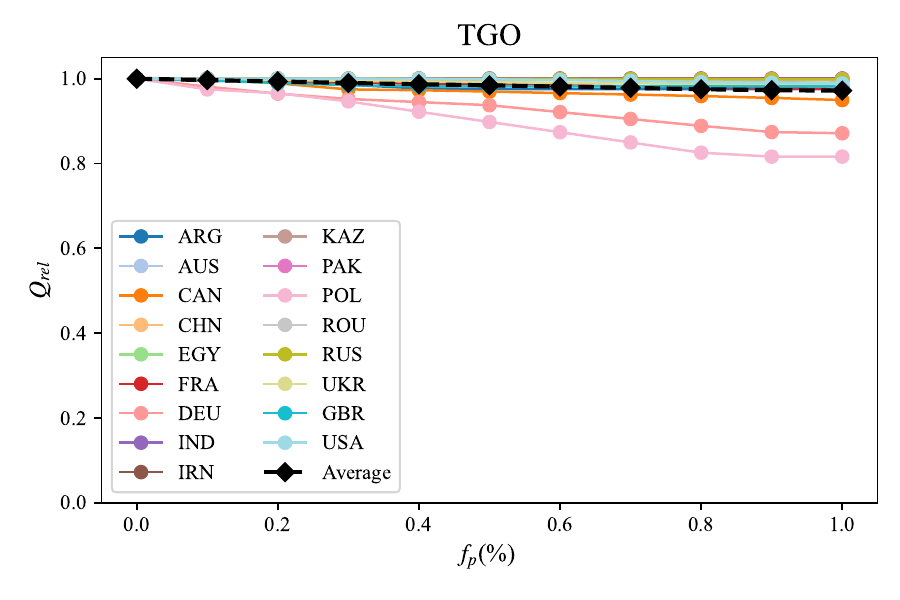}
    \includegraphics[width=0.24\linewidth]{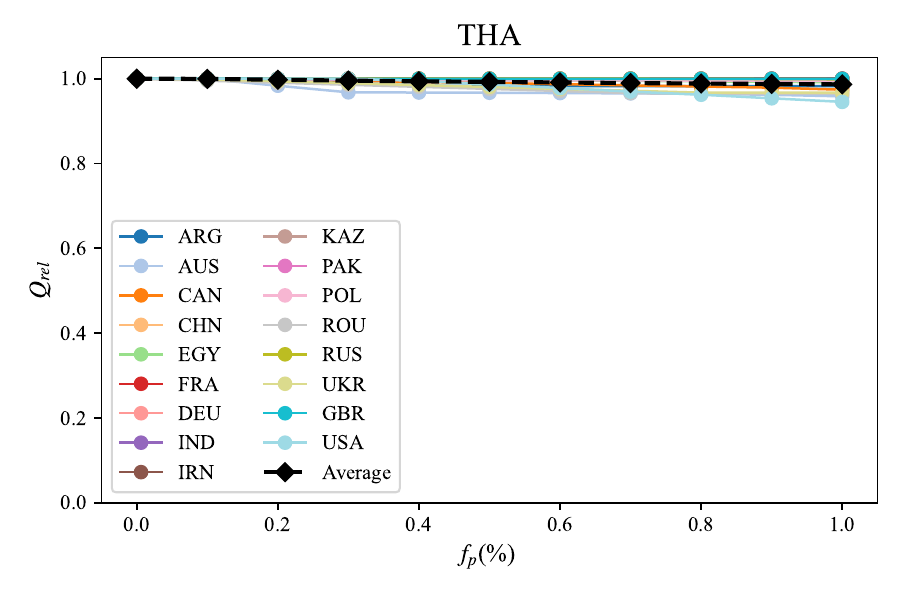}
    \includegraphics[width=0.24\linewidth]{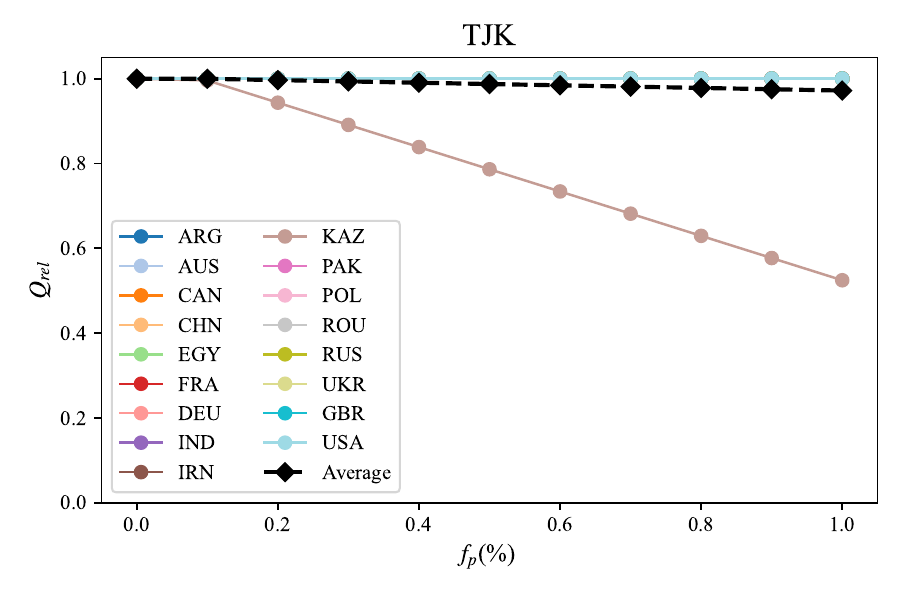}
    \includegraphics[width=0.24\linewidth]{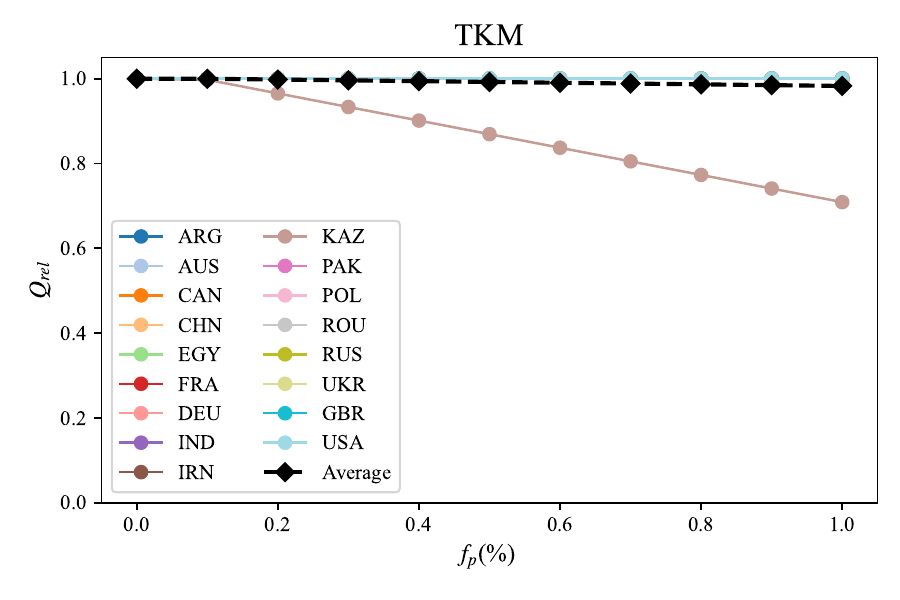}\\
    \includegraphics[width=0.24\linewidth]{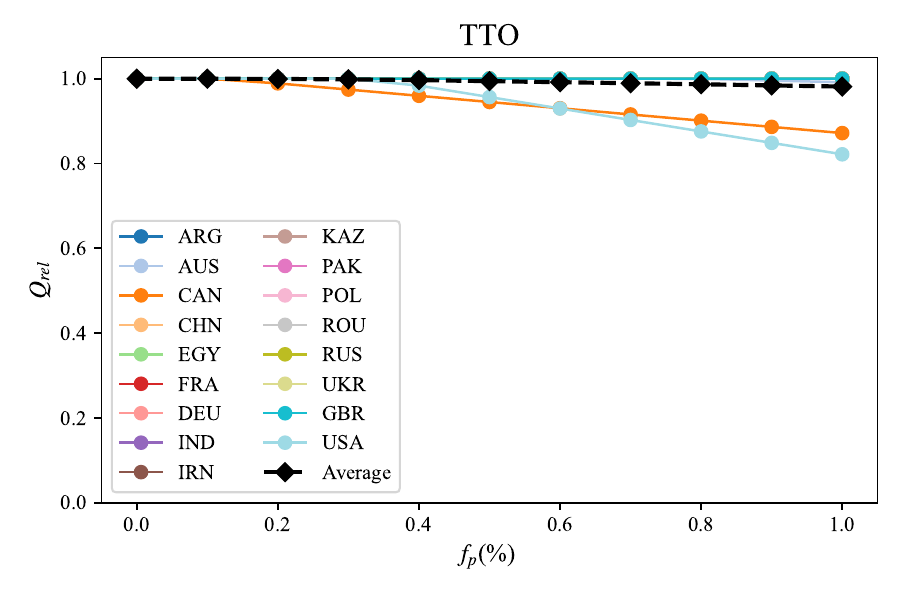}
    \includegraphics[width=0.24\linewidth]{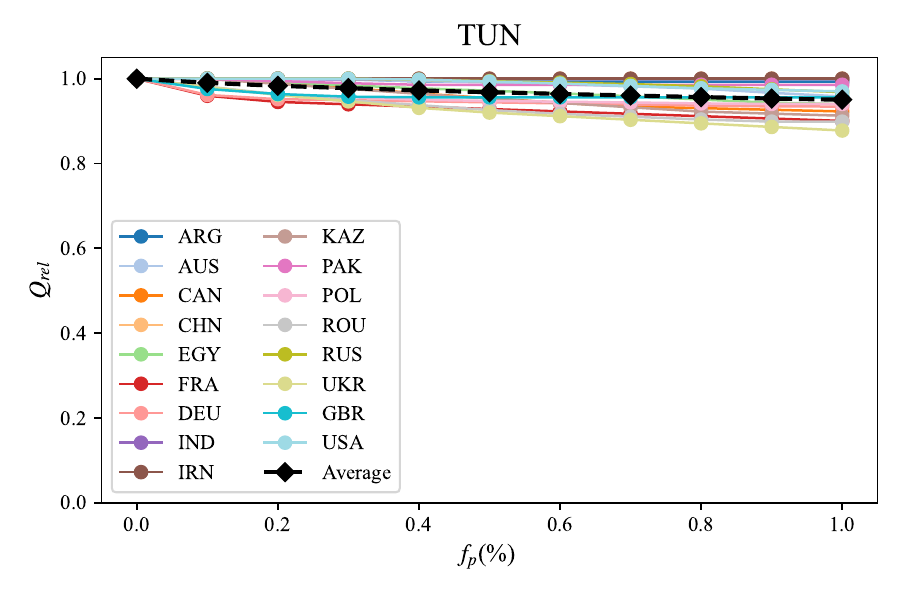}
    \includegraphics[width=0.24\linewidth]{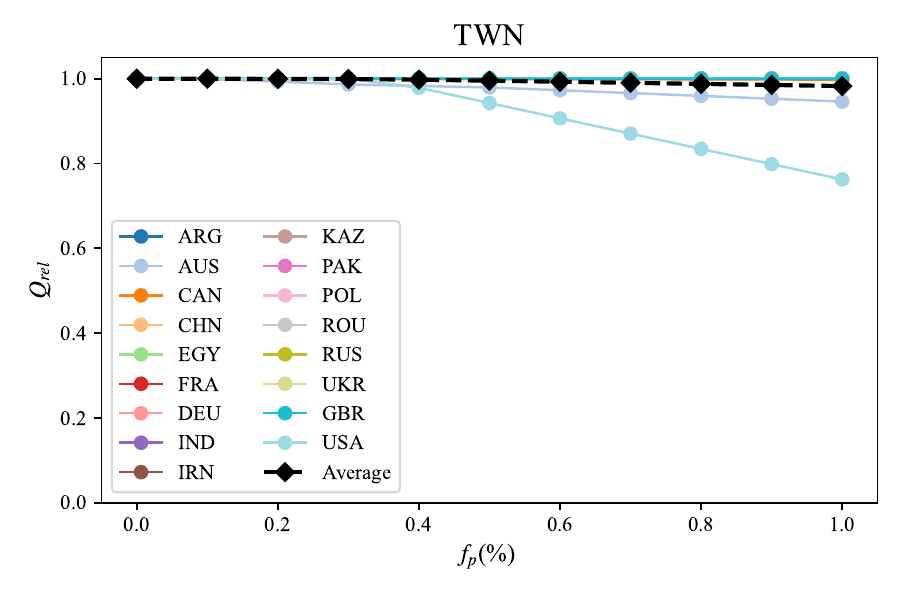}
    \includegraphics[width=0.24\linewidth]{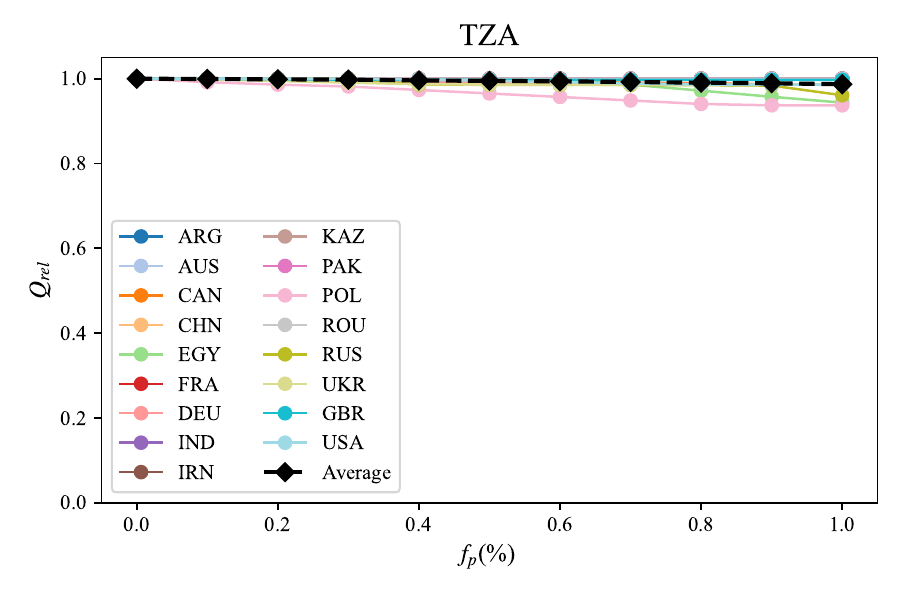}\\
    \includegraphics[width=0.24\linewidth]{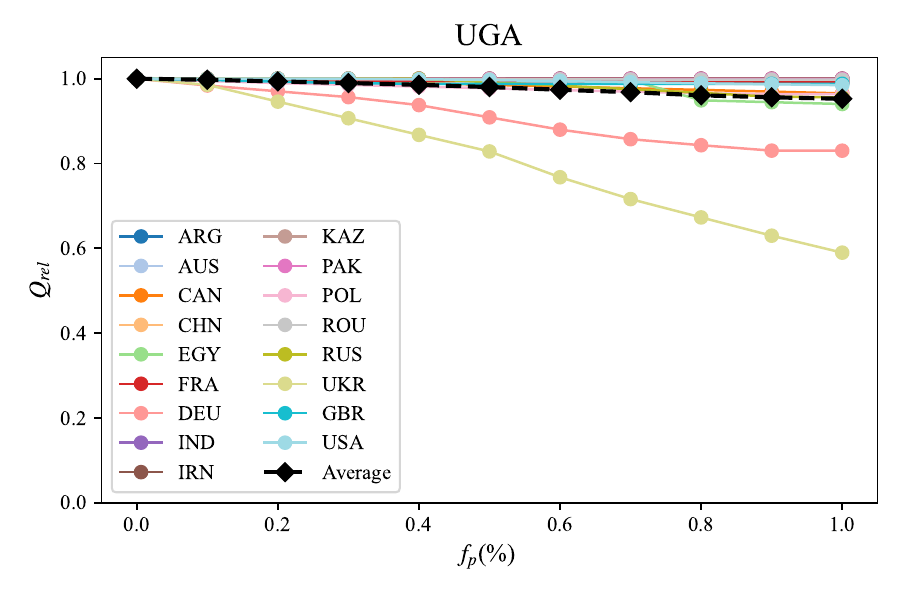}
    \includegraphics[width=0.24\linewidth]{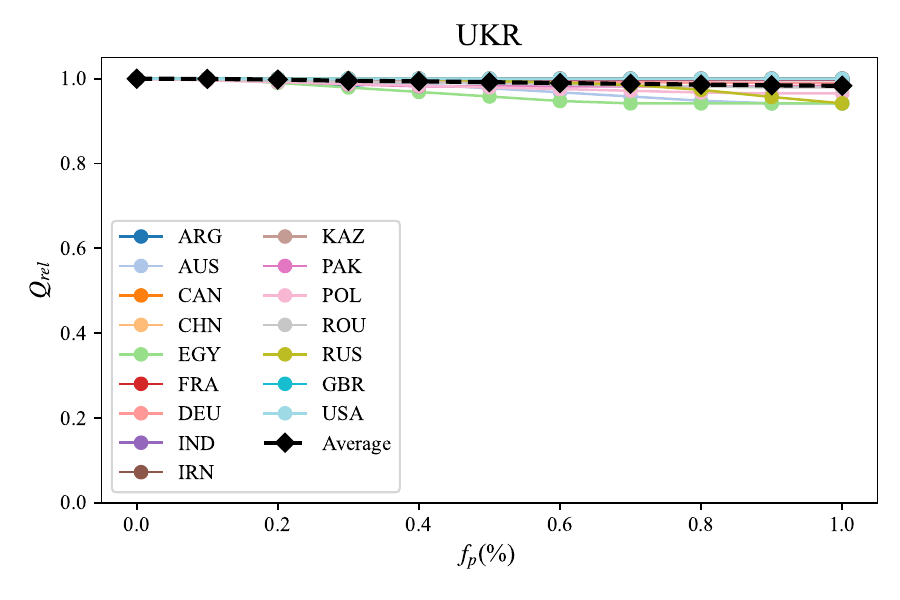}
    \includegraphics[width=0.24\linewidth]{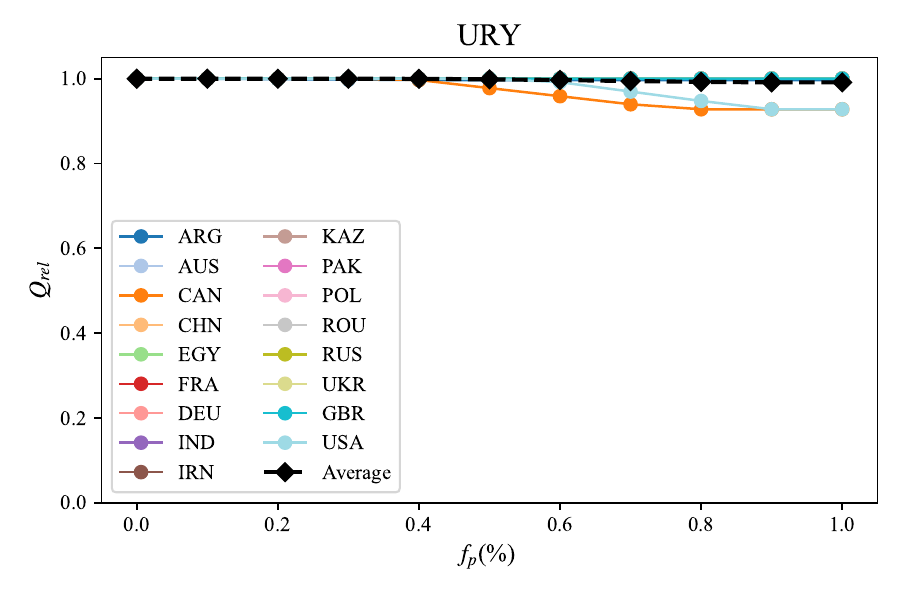}
    \includegraphics[width=0.24\linewidth]{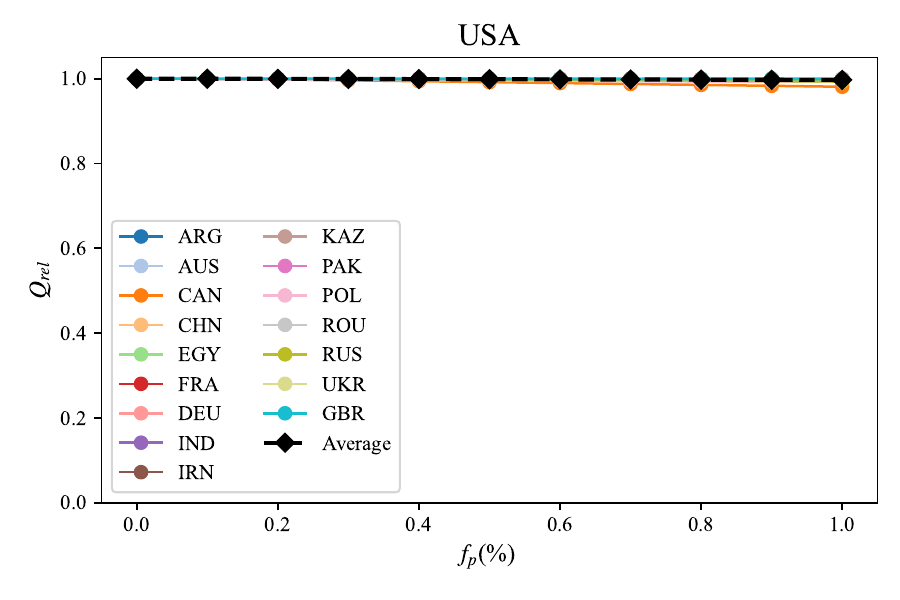}\\
    \includegraphics[width=0.24\linewidth]{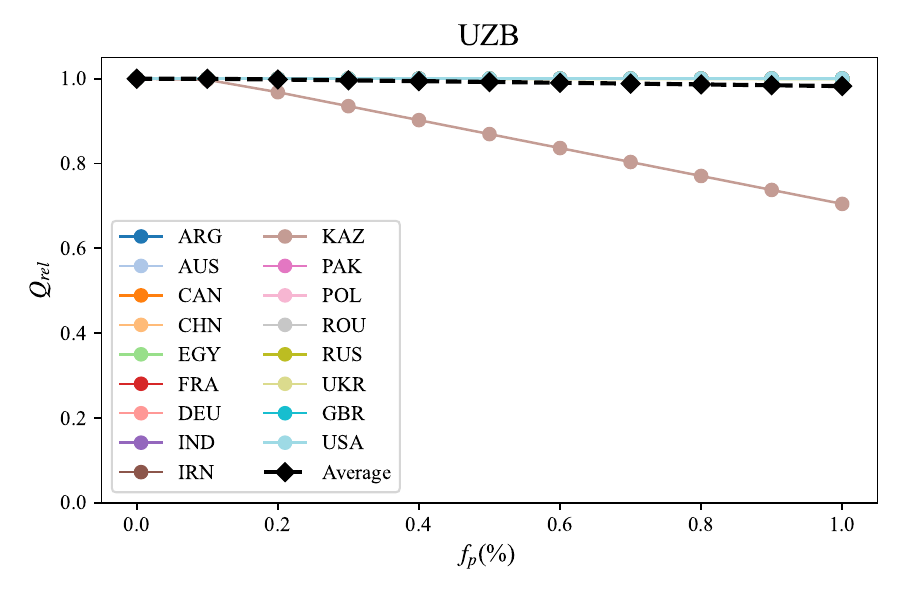}
    \includegraphics[width=0.24\linewidth]{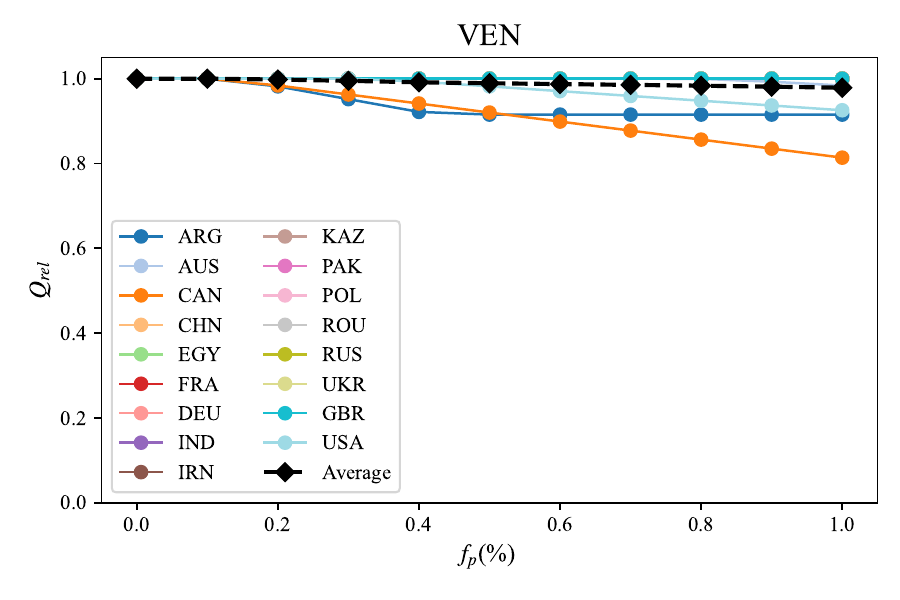}
    \includegraphics[width=0.24\linewidth]{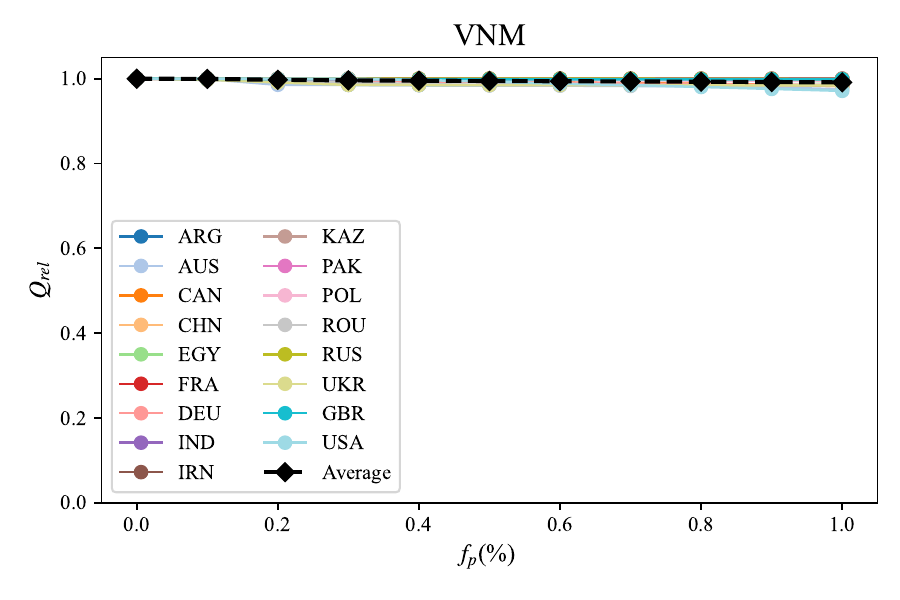}
    \includegraphics[width=0.24\linewidth]{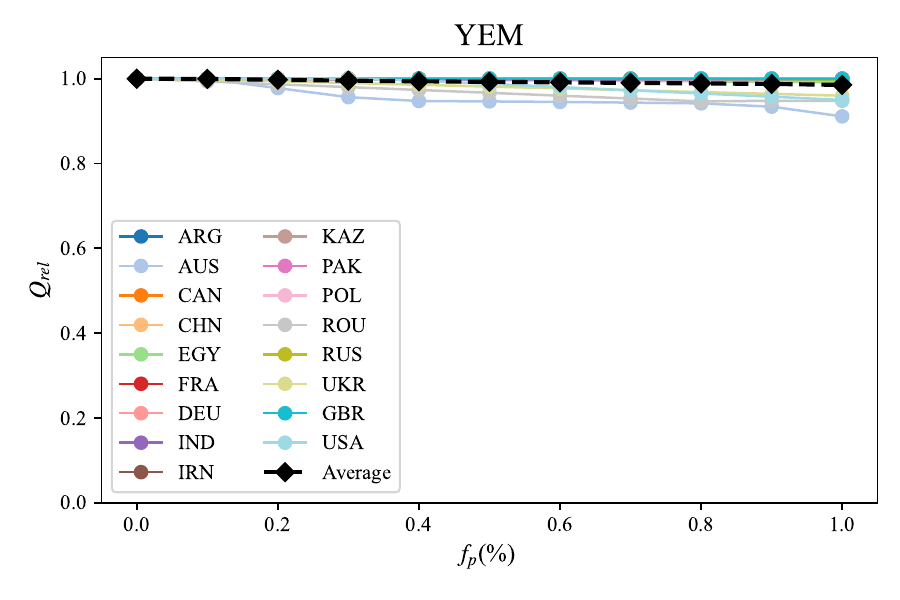}
    \caption{Wheat: $Q_{\mathrm{rel}}$ curves under global 1\% production shock (page 5).}
    \label{Fig:Wheat_Qrel_global1pct_5}
\end{figure}

\begin{figure}[h]\ContinuedFloat
    \centering
    \includegraphics[width=0.24\linewidth]{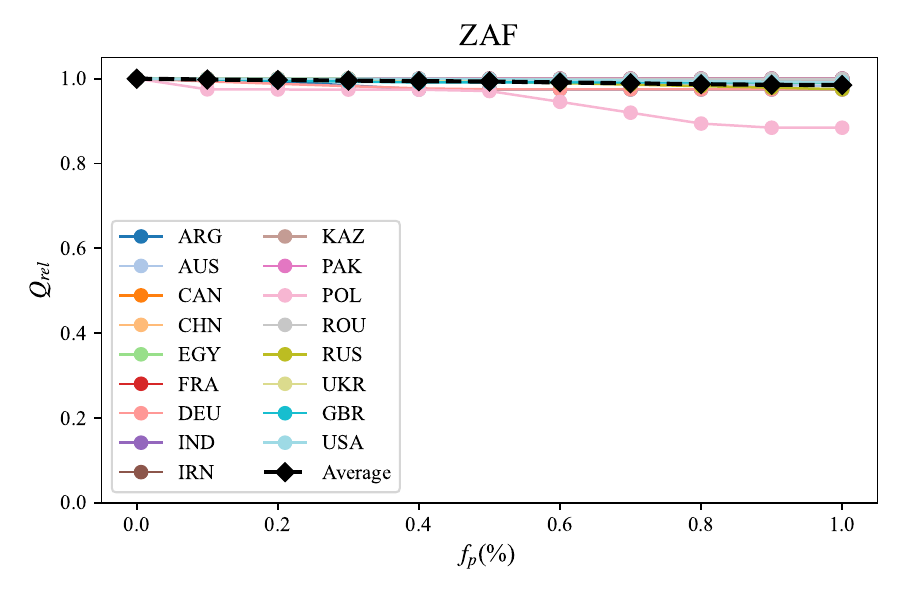}
    \includegraphics[width=0.24\linewidth]{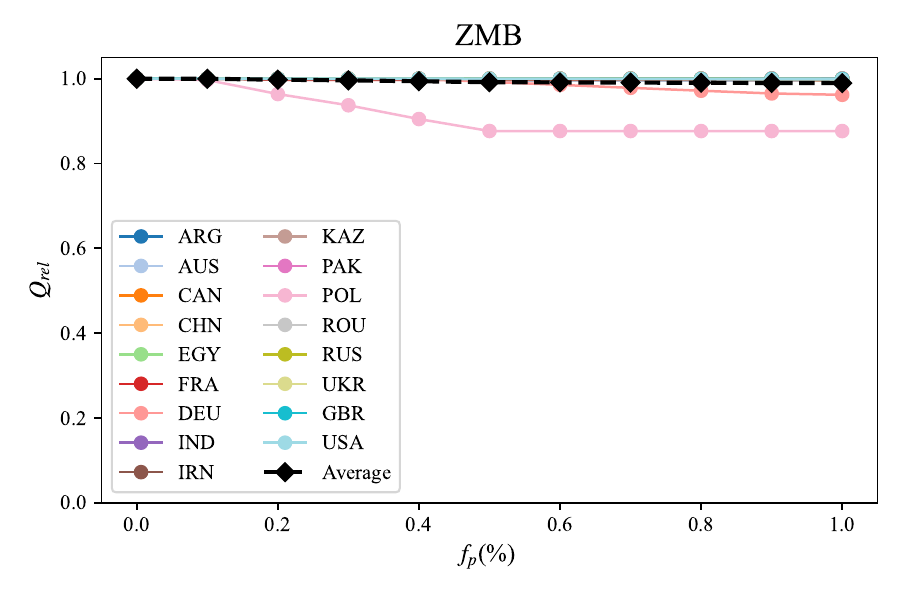}
    \includegraphics[width=0.24\linewidth]{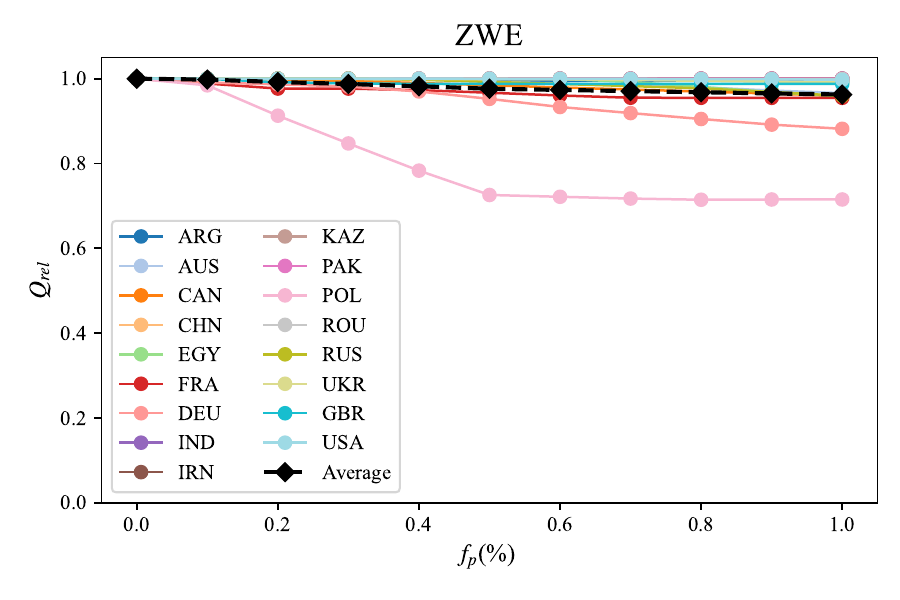}
    \caption{Wheat: $Q_{\mathrm{rel}}$ curves under global 1\% production shock (page 6).}
    \label{Fig:Wheat_Qrel_global1pct_6}
\end{figure}

% \newpage
% \bibliography{BibAll, BibNew, BibRCE, Reference}

\begin{thebibliography}{28}
\expandafter\ifx\csname natexlab\endcsname\relax\def\natexlab#1{#1}\fi
\providecommand{\url}[1]{\texttt{#1}}
\providecommand{\href}[2]{#2}
\providecommand{\path}[1]{#1}
\providecommand{\DOIprefix}{doi:}
\providecommand{\ArXivprefix}{arXiv:}
\providecommand{\URLprefix}{URL: }
\providecommand{\Pubmedprefix}{pmid:}
\providecommand{\doi}[1]{\href{http://dx.doi.org/#1}{\path{#1}}}
\providecommand{\Pubmed}[1]{\href{pmid:#1}{\path{#1}}}
\providecommand{\bibinfo}[2]{#2}
\ifx\xfnm\relax \def\xfnm[#1]{\unskip,\space#1}\fi
%Type = Article
\bibitem[{Abay et~al.(2023)Abay, Breisinger, Glauber, Kurdi, Laborde and
  Siddig}]{Abay-Breisinger-Glauber-Kurdi-Laborde-Siddig-2023-GlobFoodSecur-AgricPolicy}
\bibinfo{author}{Abay, K.A.}, \bibinfo{author}{Breisinger, C.},
  \bibinfo{author}{Glauber, J.}, \bibinfo{author}{Kurdi, S.},
  \bibinfo{author}{Laborde, D.}, \bibinfo{author}{Siddig, K.},
  \bibinfo{year}{2023}.
\newblock \bibinfo{title}{The {R}ussia-{U}kraine war: implications for global
  and regional food security and potential policy responses}.
\newblock \bibinfo{journal}{Glob. Food Secur.} \bibinfo{volume}{36},
  \bibinfo{pages}{100675}.
\newblock \DOIprefix\doi{10.1016/j.gfs.2023.100675}.
%Type = Article
\bibitem[{An et~al.(2016)An, Qiu and Zheng}]{An-Qiu-Zheng-2016-FoodPolicy}
\bibinfo{author}{An, H.}, \bibinfo{author}{Qiu, F.}, \bibinfo{author}{Zheng,
  Y.}, \bibinfo{year}{2016}.
\newblock \bibinfo{title}{How do export controls affect price transmission and
  volatility spillovers in the ukrainian wheat and flour markets?}
\newblock \bibinfo{journal}{Food Policy} \bibinfo{volume}{62},
  \bibinfo{pages}{142--150}.
\newblock \DOIprefix\doi{10.1016/j.foodpol.2016.06.002}.
%Type = Article
\bibitem[{Bouet and Debucquet(2012)}]{Bouet-Debucquet-2012-RevWorldEcon}
\bibinfo{author}{Bouet, A.}, \bibinfo{author}{Debucquet, D.L.},
  \bibinfo{year}{2012}.
\newblock \bibinfo{title}{Food crisis and export taxation: the cost of
  non-cooperative trade policies}.
\newblock \bibinfo{journal}{Rev. World Econ.} \bibinfo{volume}{148},
  \bibinfo{pages}{209--233}.
\newblock \DOIprefix\doi{10.1007/s10290-011-0108-8}.
%Type = Article
\bibitem[{Burkholz and
  Schweitzer(2019)}]{Burkholz-Schweitzer-2019-EnvironResLett}
\bibinfo{author}{Burkholz, R.}, \bibinfo{author}{Schweitzer, F.},
  \bibinfo{year}{2019}.
\newblock \bibinfo{title}{International crop trade networks: the impact of
  shocks and cascades}.
\newblock \bibinfo{journal}{Environ. Res. Lett.} \bibinfo{volume}{14},
  \bibinfo{pages}{114013}.
\newblock \DOIprefix\doi{10.1088/1748-9326/ab4864}.
%Type = Article
\bibitem[{Chen et~al.(2024)Chen, Tu, An, Wu, Lin and
  Gong}]{Chen-Tu-An-Wu-Lin-Gong-2024-CommunEarthEnviron}
\bibinfo{author}{Chen, B.}, \bibinfo{author}{Tu, Y.}, \bibinfo{author}{An, J.},
  \bibinfo{author}{Wu, S.}, \bibinfo{author}{Lin, C.}, \bibinfo{author}{Gong,
  P.}, \bibinfo{year}{2024}.
\newblock \bibinfo{title}{Quantification of losses in agriculture production in
  eastern ukraine due to the russia-ukraine war}.
\newblock \bibinfo{journal}{Commun. Earth Environ.} \bibinfo{volume}{5},
  \bibinfo{pages}{336}.
\newblock \DOIprefix\doi{10.1038/s43247-024-01488-3}.
%Type = Article
\bibitem[{Church et~al.(2017)Church, Haigh, Widhalm, de~Jalon, Babin, Carlton,
  Dunn, Fagan, Knutson and
  Prokopy}]{Church-Haigh-Widhalm-deJalon-Babin-Carlton-Dunn-Fagan-Knutson-Prokopy-2017-CLIMRISKMANAG}
\bibinfo{author}{Church, S.P.}, \bibinfo{author}{Haigh, T.},
  \bibinfo{author}{Widhalm, M.}, \bibinfo{author}{de~Jalon, S.G.},
  \bibinfo{author}{Babin, N.}, \bibinfo{author}{Carlton, J.S.},
  \bibinfo{author}{Dunn, M.}, \bibinfo{author}{Fagan, K.},
  \bibinfo{author}{Knutson, C.L.}, \bibinfo{author}{Prokopy, L.S.},
  \bibinfo{year}{2017}.
\newblock \bibinfo{title}{Agricultural trade publications and the 2012
  midwestern u.s. drought: a missed opportunity for climate risk
  communication}.
\newblock \bibinfo{journal}{CLIM. RISK MANAG.} \bibinfo{volume}{15},
  \bibinfo{pages}{45--60}.
\newblock \DOIprefix\doi{10.1016/j.crm.2016.10.006}.
%Type = Article
\bibitem[{Distefano et~al.(2018)Distefano, Laio, Ridolfi and
  Schiavo}]{Distefano-Laio-Ridolfi-Schiavo-2018-PLoSOne}
\bibinfo{author}{Distefano, T.}, \bibinfo{author}{Laio, F.},
  \bibinfo{author}{Ridolfi, L.}, \bibinfo{author}{Schiavo, S.},
  \bibinfo{year}{2018}.
\newblock \bibinfo{title}{Shock transmission in the international food trade
  network}.
\newblock \bibinfo{journal}{PLoS One} \bibinfo{volume}{13},
  \bibinfo{pages}{e0200639}.
\newblock \DOIprefix\doi{10.1371/journal.pone.0200639}.
%Type = Article
\bibitem[{Falkendal et~al.(2021)Falkendal, Otto, Schewe, Jagermeyr, Konar,
  Kummu, Watkins and
  Puma}]{Falkendal-Otto-Schewe-Jagermeyr-Konar-Kummu-Watkins-Puma-2021-NatFood}
\bibinfo{author}{Falkendal, T.}, \bibinfo{author}{Otto, C.},
  \bibinfo{author}{Schewe, J.}, \bibinfo{author}{Jagermeyr, J.},
  \bibinfo{author}{Konar, M.}, \bibinfo{author}{Kummu, M.},
  \bibinfo{author}{Watkins, B.}, \bibinfo{author}{Puma, M.J.},
  \bibinfo{year}{2021}.
\newblock \bibinfo{title}{Grain export restrictions during {COVID}-19 risk food
  insecurity in many low- and middle-income countries}.
\newblock \bibinfo{journal}{Nat. Food} \bibinfo{volume}{2},
  \bibinfo{pages}{11--14}.
\newblock \DOIprefix\doi{10.1038/s43016-020-00211-7}.
%Type = Article
\bibitem[{Fonteijn et~al.(2024)Fonteijn, van Oort and
  Hengeveld}]{Fonteijn-vanOort-Hengeveld-2024-JASSS}
\bibinfo{author}{Fonteijn, H.M.J.}, \bibinfo{author}{van Oort, P.A.J.},
  \bibinfo{author}{Hengeveld, G.M.}, \bibinfo{year}{2024}.
\newblock \bibinfo{title}{Darts: evolving resilience of the global food system
  to production and trade shocks}.
\newblock \bibinfo{journal}{J. Artif. Soc. Soc. Simul.} \bibinfo{volume}{27},
  \bibinfo{pages}{5307}.
\newblock \DOIprefix\doi{10.18564/jasss.5307}.
%Type = Article
\bibitem[{Giordani et~al.(2016)Giordani, Rocha and
  Ruta}]{Giordani-Rocha-Ruta-2016-JIntEcon}
\bibinfo{author}{Giordani, P.E.}, \bibinfo{author}{Rocha, N.},
  \bibinfo{author}{Ruta, M.}, \bibinfo{year}{2016}.
\newblock \bibinfo{title}{Food prices and the multiplier effect of trade
  policy}.
\newblock \bibinfo{journal}{J. Int. Econ.} \bibinfo{volume}{101},
  \bibinfo{pages}{102--122}.
\newblock \DOIprefix\doi{10.1016/j.jinteco.2016.04.001}.
%Type = Article
\bibitem[{Grassia et~al.(2022)Grassia, Mangioni, Schiavo and
  Traverso}]{Grassia-Mangioni-Schiavo-Traverso-2022-SciRep}
\bibinfo{author}{Grassia, M.}, \bibinfo{author}{Mangioni, G.},
  \bibinfo{author}{Schiavo, S.}, \bibinfo{author}{Traverso, S.},
  \bibinfo{year}{2022}.
\newblock \bibinfo{title}{Insights into countries' exposure and vulnerability
  to food trade shocks from network-based simulations}.
\newblock \bibinfo{journal}{Sci. Rep.} \bibinfo{volume}{12},
  \bibinfo{pages}{4644}.
\newblock \DOIprefix\doi{10.1038/s41598-022-08419-2}.
%Type = Article
\bibitem[{Headey(2011)}]{Headey-2011-FoodPolicy}
\bibinfo{author}{Headey, D.}, \bibinfo{year}{2011}.
\newblock \bibinfo{title}{Rethinking the global food crisis: the role of trade
  shocks}.
\newblock \bibinfo{journal}{Food Policy} \bibinfo{volume}{36},
  \bibinfo{pages}{136--146}.
\newblock \DOIprefix\doi{10.1016/j.foodpol.2010.10.003}.
%Type = Article
\bibitem[{Heslin et~al.(2020)Heslin, Puma, Marchand, Carr, Dell'Angelo,
  D'Odorico, Gephart, Kummu, Porkka, Rulli, Seekell, Suweis and
  Tavoni}]{Heslin-Puma-Marchand-Carr-Dell'Angelo-D'Odorico-Gephart-Kummu-Porkka-Rulli-Seekell-Suweis-Tavoni-2020-FrontSustainFoodSyst}
\bibinfo{author}{Heslin, A.}, \bibinfo{author}{Puma, M.J.},
  \bibinfo{author}{Marchand, P.}, \bibinfo{author}{Carr, J.A.},
  \bibinfo{author}{Dell'Angelo, J.}, \bibinfo{author}{D'Odorico, P.},
  \bibinfo{author}{Gephart, J.A.}, \bibinfo{author}{Kummu, M.},
  \bibinfo{author}{Porkka, M.}, \bibinfo{author}{Rulli, M.C.},
  \bibinfo{author}{Seekell, D.A.}, \bibinfo{author}{Suweis, S.},
  \bibinfo{author}{Tavoni, A.}, \bibinfo{year}{2020}.
\newblock \bibinfo{title}{Simulating the cascading effects of an extreme
  agricultural production shock: global implications of a contemporary us dust
  bowl event}.
\newblock \bibinfo{journal}{Front. Sustain. Food Syst.} \bibinfo{volume}{4},
  \bibinfo{pages}{26}.
\newblock \DOIprefix\doi{10.3389/fsufs.2020.00026}.
%Type = Article
\bibitem[{Huang et~al.(2025)Huang, Forero, Wagner-Medina, Diaz, Tremma,
  Fargetton and
  Lowenberg-DeBoer}]{Huang-Forero-WagnerMedina-Diaz-Tremma-Fargetton-LowenbergDeBoer-2025-GlobFoodSecur-AgricPolicy}
\bibinfo{author}{Huang, I.Y.}, \bibinfo{author}{Forero, O.A.},
  \bibinfo{author}{Wagner-Medina, Erika, V.}, \bibinfo{author}{Diaz, H.F.},
  \bibinfo{author}{Tremma, O.}, \bibinfo{author}{Fargetton, X.},
  \bibinfo{author}{Lowenberg-DeBoer, J.}, \bibinfo{year}{2025}.
\newblock \bibinfo{title}{Resilience of food supply systems to sudden shocks: a
  global review and narrative synthesis}.
\newblock \bibinfo{journal}{Glob. Food Secur.} \bibinfo{volume}{44},
  \bibinfo{pages}{100842}.
\newblock \DOIprefix\doi{10.1016/j.gfs.2025.100842}.
%Type = Article
\bibitem[{Hunt et~al.(2021)Hunt, Femia, Werrell, Christian, Otkin, Basara,
  Anderson, White, Hain, Randall and
  McGaughey}]{Hunt-Femia-Werrell-Christian-Otkin-Basara-Anderson-White-Hain-Randall-McGaughey-2021-WeatherClimExtremes}
\bibinfo{author}{Hunt, E.}, \bibinfo{author}{Femia, F.},
  \bibinfo{author}{Werrell, C.}, \bibinfo{author}{Christian, J.I.},
  \bibinfo{author}{Otkin, J.A.}, \bibinfo{author}{Basara, J.},
  \bibinfo{author}{Anderson, M.}, \bibinfo{author}{White, T.},
  \bibinfo{author}{Hain, C.}, \bibinfo{author}{Randall, R.},
  \bibinfo{author}{McGaughey, K.}, \bibinfo{year}{2021}.
\newblock \bibinfo{title}{Agricultural and food security impacts from the 2010
  russia flash drought}.
\newblock \bibinfo{journal}{Weather Clim. Extremes} \bibinfo{volume}{34},
  \bibinfo{pages}{100383}.
\newblock \DOIprefix\doi{10.1016/j.wace.2021.100383}.
%Type = Article
\bibitem[{Jia et~al.(2024)Jia, Xia, Li, Yu, Wu, Li, Su, Wang, Chen and
  Liu}]{Jia-Xia-Li-Yu-Wu-Li-Su-Wang-Chen-Liu-2024-CommunEarthEnviron}
\bibinfo{author}{Jia, N.}, \bibinfo{author}{Xia, Z.}, \bibinfo{author}{Li, Y.},
  \bibinfo{author}{Yu, X.}, \bibinfo{author}{Wu, X.}, \bibinfo{author}{Li, Y.},
  \bibinfo{author}{Su, R.}, \bibinfo{author}{Wang, M.}, \bibinfo{author}{Chen,
  R.}, \bibinfo{author}{Liu, J.}, \bibinfo{year}{2024}.
\newblock \bibinfo{title}{The russia-ukraine war reduced food production and
  exports with a disparate geographical impact worldwide}.
\newblock \bibinfo{journal}{Commun. Earth Environ.} \bibinfo{volume}{5},
  \bibinfo{pages}{765}.
\newblock \DOIprefix\doi{10.1038/s43247-024-01915-5}.
%Type = Article
\bibitem[{Kafando and Sakurai(2025)}]{Kafando-Sakurai-2025-JAgricEcon}
\bibinfo{author}{Kafando, W.A.}, \bibinfo{author}{Sakurai, T.},
  \bibinfo{year}{2025}.
\newblock \bibinfo{title}{Effects and mechanisms of armed conflict on
  agricultural production: spatial evidence from terrorist violence in burkina
  faso}.
\newblock \bibinfo{journal}{J. Agric. Econ.} \bibinfo{volume}{76},
  \bibinfo{pages}{24--44}.
\newblock \DOIprefix\doi{10.1111/1477-9552.12613}.
%Type = Article
\bibitem[{Kuhla et~al.(2025)Kuhla, Kubiczek and
  Otto}]{Kuhla-Kubiczek-Otto-2025-EcolEcon}
\bibinfo{author}{Kuhla, K.}, \bibinfo{author}{Kubiczek, P.},
  \bibinfo{author}{Otto, C.}, \bibinfo{year}{2025}.
\newblock \bibinfo{title}{Understanding agricultural market dynamics in times
  of crisis: the dynamic agent-based network model agrimate}.
\newblock \bibinfo{journal}{Ecol. Econ.} \bibinfo{volume}{231},
  \bibinfo{pages}{108546}.
\newblock \DOIprefix\doi{10.1016/j.ecolecon.2025.108546}.
%Type = Article
\bibitem[{Kummu et~al.(2020)Kummu, Kinnunen, Lehikoinen, Porkka, Queiroz, Roos,
  Troell and
  Well}]{Kummu-Kinnunen-Lehikoinen-Porkka-Queiroz-Roos-Troell-Well-2020-GlobFoodSecur-AgricPolicy}
\bibinfo{author}{Kummu, M.}, \bibinfo{author}{Kinnunen, P.},
  \bibinfo{author}{Lehikoinen, E.}, \bibinfo{author}{Porkka, M.},
  \bibinfo{author}{Queiroz, C.}, \bibinfo{author}{Roos, E.},
  \bibinfo{author}{Troell, M.}, \bibinfo{author}{Well, C.},
  \bibinfo{year}{2020}.
\newblock \bibinfo{title}{Interplay of trade and food system resilience: gains
  on supply diversity over time at the cost of trade independency}.
\newblock \bibinfo{journal}{Glob. Food Secur.} \bibinfo{volume}{24},
  \bibinfo{pages}{100360}.
\newblock \DOIprefix\doi{10.1016/j.gfs.2020.100360}.
%Type = Article
\bibitem[{Lesk et~al.(2016)Lesk, Rowhani and
  Ramankutty}]{Lesk-Rowhani-Ramankutty-2016-Nature}
\bibinfo{author}{Lesk, C.}, \bibinfo{author}{Rowhani, P.},
  \bibinfo{author}{Ramankutty, N.}, \bibinfo{year}{2016}.
\newblock \bibinfo{title}{Influence of extreme weather disasters on global crop
  production}.
\newblock \bibinfo{journal}{Nature} \bibinfo{volume}{529},
  \bibinfo{pages}{84--+}.
\newblock \DOIprefix\doi{10.1038/nature16467}.
%Type = Article
\bibitem[{Lobell et~al.(2011)Lobell, Schlenker and
  Costa-Roberts}]{Lobell-Schlenker-CostaRoberts-2011-Science}
\bibinfo{author}{Lobell, D.B.}, \bibinfo{author}{Schlenker, W.},
  \bibinfo{author}{Costa-Roberts, J.}, \bibinfo{year}{2011}.
\newblock \bibinfo{title}{Climate trends and global crop production since
  1980}.
\newblock \bibinfo{journal}{Science} \bibinfo{volume}{333},
  \bibinfo{pages}{616--620}.
\newblock \DOIprefix\doi{10.1126/science.1204531}.
%Type = Article
\bibitem[{Marchand et~al.(2016)Marchand, Carr, Dell'Angelo, Fader, Gephart,
  Kummu, Magliocca, Porkka, Puma, Ratajczak, Rulli, Seekell, Suweis, Tavoni and
  D'Odorico}]{Marchand-Carr-Dell'Angelo-Fader-Gephart-Kummu-Magliocca-Porkka-Puma-Ratajczak-Rulli-Seekell-Suweis-Tavoni-D'Odorico-2016-EnvironResLett}
\bibinfo{author}{Marchand, P.}, \bibinfo{author}{Carr, J.A.},
  \bibinfo{author}{Dell'Angelo, J.}, \bibinfo{author}{Fader, M.},
  \bibinfo{author}{Gephart, J.A.}, \bibinfo{author}{Kummu, M.},
  \bibinfo{author}{Magliocca, N.R.}, \bibinfo{author}{Porkka, M.},
  \bibinfo{author}{Puma, M.J.}, \bibinfo{author}{Ratajczak, Z.},
  \bibinfo{author}{Rulli, M.C.}, \bibinfo{author}{Seekell, D.A.},
  \bibinfo{author}{Suweis, S.}, \bibinfo{author}{Tavoni, A.},
  \bibinfo{author}{D'Odorico, P.}, \bibinfo{year}{2016}.
\newblock \bibinfo{title}{Reserves and trade jointly determine exposure to food
  supply shocks}.
\newblock \bibinfo{journal}{Environ. Res. Lett.} \bibinfo{volume}{11},
  \bibinfo{pages}{095009}.
\newblock \DOIprefix\doi{10.1088/1748-9326/11/9/095009}.
%Type = Article
\bibitem[{Rude and An(2015)}]{Rude-An-2015-FoodPolicy}
\bibinfo{author}{Rude, J.}, \bibinfo{author}{An, H.}, \bibinfo{year}{2015}.
\newblock \bibinfo{title}{Explaining grain and oilseed price volatility: the
  role of export restrictions}.
\newblock \bibinfo{journal}{Food Policy} \bibinfo{volume}{57},
  \bibinfo{pages}{83--92}.
\newblock \DOIprefix\doi{10.1016/j.foodpol.2015.09.002}.
%Type = Article
\bibitem[{Schneider et~al.(2011)Schneider, Moreira, Andrade, Havlin and
  Herrmann}]{Schneider-Moreira-Andrade-Havlin-Herrmann-2011-ProcNatlAcadSciUSA}
\bibinfo{author}{Schneider, C.M.}, \bibinfo{author}{Moreira, A.A.},
  \bibinfo{author}{Andrade, J.S.}, \bibinfo{author}{Havlin, S.},
  \bibinfo{author}{Herrmann, H.J.}, \bibinfo{year}{2011}.
\newblock \bibinfo{title}{Mitigation of malicious attacks on networks}.
\newblock \bibinfo{journal}{Proc. Natl. Acad. Sci. U.S.A.}
  \bibinfo{volume}{108}, \bibinfo{pages}{3838--3841}.
\newblock \DOIprefix\doi{10.1073/pnas.1009440108}.
%Type = Article
\bibitem[{Simon(1996)}]{Simon-1996-FoodPolicy}
\bibinfo{author}{Simon, M.}, \bibinfo{year}{1996}.
\newblock \bibinfo{title}{Food security: A post-modern perspective}.
\newblock \bibinfo{journal}{Food Policy} \bibinfo{volume}{21},
  \bibinfo{pages}{155--170}.
\newblock \DOIprefix\doi{https://doi.org/10.1016/0306-9192(95)00074-7}.
%Type = Article
\bibitem[{Tu et~al.(2019)Tu, Suweis and
  D'Odorico}]{Tu-Suweis-D'Odorico-2019-NatSustain}
\bibinfo{author}{Tu, C.}, \bibinfo{author}{Suweis, S.},
  \bibinfo{author}{D'Odorico, P.}, \bibinfo{year}{2019}.
\newblock \bibinfo{title}{Impact of globalization on the resilience and
  sustainability of natural resources}.
\newblock \bibinfo{journal}{Nat. Sustain.} \bibinfo{volume}{2},
  \bibinfo{pages}{283--289}.
\newblock \DOIprefix\doi{10.1038/s41893-019-0260-z}.
%Type = Article
\bibitem[{Wassenius et~al.(2023)Wassenius, Porkka, Nystrom and
  Jorgensen}]{Wassenius-Porkka-Nystrom-Jorgensen-2023-GlobFoodSecur-AgricPolicy}
\bibinfo{author}{Wassenius, E.}, \bibinfo{author}{Porkka, M.},
  \bibinfo{author}{Nystrom, M.}, \bibinfo{author}{Jorgensen, P.S.},
  \bibinfo{year}{2023}.
\newblock \bibinfo{title}{A global analysis of potential self-sufficiency and
  diversity displays diverse supply risks}.
\newblock \bibinfo{journal}{Glob. Food Secur.} \bibinfo{volume}{37},
  \bibinfo{pages}{100673}.
\newblock \DOIprefix\doi{10.1016/j.gfs.2023.100673}.
%Type = Article
\bibitem[{Wright(2011)}]{Wright-2011-ApplEconPerspectPolicy}
\bibinfo{author}{Wright, B.D.}, \bibinfo{year}{2011}.
\newblock \bibinfo{title}{The economics of grain price volatility}.
\newblock \bibinfo{journal}{Appl. Econ. Perspect. Policy} \bibinfo{volume}{33},
  \bibinfo{pages}{32--58}.
\newblock \DOIprefix\doi{10.1093/aepp/ppq033}.

\end{thebibliography}
% \bibliographystyle{plain}

% \end{CJK*}
\end{document}